\documentclass[12pt]{article}
\usepackage{epsf}
\usepackage{cite}
\usepackage{amsmath}
\usepackage{graphics}

\begin{document}

\setlength{\textheight}{21.5cm}
\setlength{\oddsidemargin}{0.cm}
\setlength{\evensidemargin}{0.cm}
\setlength{\topmargin}{0.cm}
\setlength{\footskip}{1cm}
\setlength{\arraycolsep}{2pt}

\renewcommand{\thefootnote}{\#\arabic{footnote}}
\setcounter{footnote}{0}

\newcommand{\gtrsim}{ \mathop{}_{\textstyle \sim}^{\textstyle >} }
\newcommand{\lesssim}{ \mathop{}_{\textstyle \sim}^{\textstyle <} }
\newcommand{\rem}[1]{{\bf #1}}
\renewcommand{\thefootnote}{\fnsymbol{footnote}}
\setcounter{footnote}{0}
\begin{titlepage}
\def\thefootnote{\fnsymbol{footnote}}

\begin{center}
%\hfill hep-ph/yymmnnn\\
%\hfill June 2007\\
\vskip .5in
\bigskip
\bigskip
{\Large \bf Quiver Gauge Theory and Conformality at the 
TeV Scale}

\vskip .45in

{\bf Paul H. Frampton $^{(a)}$ and Thomas W. Kephart $^{(b)}$}

\vskip .45in

{\it $^{(a)}$ Department of Physics and Astronomy, University of North Carolina,\\
Chapel Hill, NC 27599.} 

\vskip .20in

{\it $^{(b)}$ Department of Physics and Astronomy, Vanderbilt University,\\
Nashville, TN 37235.}

\end{center}

\vskip .4in
\begin{abstract}

This review
describes the conformality approach to extending the standard model
of particle phenomenology using an assumption of no conformal
anomaly at high energy. Topics include quiver gauge theory, the
conformality approach to phenomenology, strong-electroweak
unification at 4 TeV, cancellation of quadratic divergences,
cancellation of U(1) anomalies, and a dark matter candidate.
\end{abstract}
\end{titlepage}

\renewcommand{\thepage}{\arabic{page}}
\setcounter{page}{1}
\renewcommand{\thefootnote}{\#\arabic{footnote}}

\newpage

\begin{center}

{\Large \bf Table of Contents}

\end{center}

\large

\vskip 0.45in

{\bf 1. Introduction}   

1.1 Beyond the standard model 

{\bf 2. Quiver gauge theories}

2.1 Orbifolding 

{\bf 3. Conformality phenomenology} 

3.1 Experimental evidence for conformality.

{\bf 4. Tabulation of simplest abelian quivers}

{\bf 5. Chiral fermions}

{\bf 6. Model building} 

6.1 Abelian model building

6.1.1 Abelian non-SUSY models

6.1.2 ${\cal N}=1$ chiral $Z_n$ models. 

6.1.3 ${\cal N}=0, 1$ chiral models for abelian product groups.

6.1.4 Abelian SUSY models.

6.2 Non abelian model building

6.2.1 Non-Abelian non-SUSY models

6.3 Non-abelian groups with order $g \leq 31$

6.3.1 The Scalar Sector

6.3.2 Spontaneous symmetry breaking

{\bf 7. Unification} 

7.1 Grand unification at 4 TeV. 

{\bf 8. Quadratic divergences}

8.1 Is there a global symmetry?

{\bf 9. U(1) Factors}

9.1 Conformality of U(1) couplings.

{\bf 10. Dark matter candidate} 

{\bf 11. Proton decay} 

{\bf 12. Conclusions}

\bigskip
\bigskip

Bibliography

\newpage

\section{Introduction}

\bigskip

The standard model of particle phenomenology is a gauge
field theory based on the gauge group $SU_C(3) \times SU_L(2) \times U_Y(1)$
and with three families of quarks and leptons. The electroweak
symmetry $SU_L(2) \times U_Y(1)$ is spontaneously broken by a Higgs mechanism
to the electromagnetic symmetry $U_{EM}(1)$, leaving one Higgs boson.

This model has successfully explained all experimental data
(with the exception of neutrino masses which can be accommodated
by extension). At higher energy than yet explored, the
proliferation of parameters (19 without neutrino mass, 28 with)
strongly suggests new physics beyond the standard model.

In this review we discuss such an extension based on four dimensional
conformal invariance at high energy and inspired initially
by the duality between gauge theory and superstring theory.

Such conformality model building is a less explored but equally motivated alternative to other
directions of model building such as extra dimensions
or supersymmetry.

\bigskip

\subsection{Beyond the standard model}

\bigskip

Particle phenomenology is in an especially exciting time, mainly
because of the anticipated data in a new energy regime expected
from the Large Hadron Collider (LHC), to be commissioned at the
CERN Laboratory in 2007. This new data is long overdue. The 
Superconducting Supercollider (SSC) could have provided such data
long ago were it not for its political demise in 1993.

Except for the remarkable experimental data concerning neutrino 
masses and mixings which has been obtained since 1998, 
particle physics has been data starved for the last thirty years.
The standard model invented in the '60s and '70s has been confirmed and
reconfirmed. Consequently, theory has ventured into speculative
areas such as string theory, extra dimensions and supersymmetry.
While these ideas are of great interest and theoretically consistent
there is no {\it direct} evidence from experiment for them.
Here we describe a more recent, post 1998, direction known as
conformality. First, to set the stage, we shall discuss why
the conformality approach which is, in our opinion, competitive
with the other three approaches, remained unstudied
for the twenty years up to 1998.

A principal motivation underlying model building, beyond the
standard model, over the last thirty years has been the
{\it hierarchy problem} which is a special case of {\it naturalness}.
This idea stems from Wilson\cite{Wilson1971} in the late '60s.
The definition of naturalness is that a theory should not contain
any unexplained very large (or very small in the inverse)
dimensionless numbers. The adjustment needed to achieve such
naturalness violating numbers is called {\it fine tuning}.
The naturalness situation can be especially acute in gauge
field theories because even after fine tuning at tree level,
{\it i.e.}, the classical lagrangian, the fine tuning may need to 
be repeated an infinite number of times order by order
in the loop expansion during the renormalization process.
While such a theory can be internally consistent it
violates naturalness. Thus naturalness is not only an aesthetic
criterion but one which the vast majority of the community
feel must be imposed on any acceptable extension of the
standard model; ironically, one exception is Wilson himself
\cite{Wilson2004}.

When the standard model of Glashow\cite{Glashow1961} was rendered
renormalizable by appending the Higgs mechanism\cite{Weinberg1967,Salam1968}
it was soon realized that it fell into trouble with naturalness,
specifically through the hierarchy problem. In particular, the
scalar propogator has quadratically divergent radiative corrections
whch imply that a bare Higgs mass $m_H^2$ will be corrected
by an amount $\Lambda^2/m_H^2$ where $\Lambda$ is the cut off
scale corresponding to new physics. Unlike logarithmic
divergences, which can be absorbed in the usual renormalization
process, the quadratic divergences create an unacceptable
fine tuning: for example, if the cut off is at the conventional
grand unification scale $\Lambda \sim 10^{16}$ GeV and $m_H \sim 100$
GeV, we are confronted with a preposterous degree of fine tuning
to one part in $10^{28}$.

As already noted, this hierarchy problem was stated most forcefully
by Wilson who said, in private discussions, that scalar fields
are forbidden in gauge field theories. Between the late '60s and 1974,
it was widely recognized that the scalar fields of the standard model
created this serious hierarchy problem but no one knew what to do
about it.

The next big progress to the hierarchy problem came in 1974 with
the invention\cite{Wess1974} of supersymmetry. This led to the Minimally Supersymmetric
Standard Model (MSSM) which elegantly answered Wilson's objection
since quadratic divergences are cancelled between bosons and
fermions, with only logarithmic divergences surviving. Further it was
proved \cite{Haag1975,Haag19752} that the MSSM and straightforward
generalizations were the unique way to proceed. 
Not surprisingly, the MSSM immediately became
overwhelmingly popular. It has been estimated \cite{Woit2006}
that there are 35,000 papers existing on supersymmetry, more
than an average of one thousand papers per year since its
inception. This approach continued to seem "unique" until 1998. Since
the MSSM has over one hundred free parameters, many
possiblities needed to be investigated and exclusion plots
constructed. During this period, two properties beyond
naturalness rendered the MSSM even more appealing: an
improvement in unification properties and a candidate
for cosmological dark matter.

Before jumping to 1998, it is necessary to mention an unconnected
deveopment in 1983 which was the study of Yang-Mills theory
with extended ${\cal N}=4$ supersymmetry (the MSSM has
${\cal N}=1$ supersymmetry). This remarkable theory, though
phenomenologically quite unrealistic as it allows no
chiral fermions and all matter fields are in adjoint representions,
is finite 
\cite{Mandelstam1983,Mandelstam19832,Mandelstam19833} 
to all orders of perturbation theory
and conformally invariant. Between 1983 and 1997, the relationship
between the ${\cal N}=4$ gauge theory and either string theory,
also believed to be finite, or the standard model remained unclear.

The perspective changed in 1997-98
initially through the insight of Maldacena\cite{Maldacena1997}
who showed a {\it duality} between ${\cal N} = 4$ gauge theory
and the superstring in ten spacetime dimensions. Further
the ${\cal N} = 4$ supersymmetry can be broken by orbifolding
down to ${\cal N} = 0$ models with no supersymmetry at all.
It was conjectured \cite{first}
by one of the authors in 1998
that such nonsupersymmetric orbifolded models can be finite
and conformally invariant, hence the name conformality.

Conformality models have been investigated far less completely
than supersymmetric ones but it is already
clear that supersymmetry is ``not as unique'' as previously
believed. No-go theorems can have not only explicit
assumptions which need to be violated to avoid the theorem
but unconcious implicit assumptions which require further
progress even to appreciate: in 1975 the implicit assumption
was that the gauge group is simple, or if semi-simple
may be regarded as a product of theories each with a simple
gauge group. Naturalness, by cancellation of quadratic
divergences, accurate unification and a dark matter
candidate exist in conformality.

It becomes therefore a concern that the design of the
LHC has been influenced by the requirement of testing
the MSSM. The LHC merits
an investment of theoretical work to check if the
LHC is adequately designed to test conformality which now
seems equally as likely as supersymmetry, although we
fully expect the detectors
ATLAS and CMS to be sufficiently
all purpose to capture any
physics beyond the standad model at the TeV scale.

\newpage

\section{Quiver gauge theories}

Quiver gauge theories possess a gauge group which is generically
a product of $U(N_i)$ factors with matter fields in bifundamental
representations. They have been studied in the physics literature
since the 1980s where they were used in composite model building.
They have attracted much renewed attention because of their
natural appearance in the duality between superstrings and gauge theories.

The best known such duality gives rise to a highly supersymmetric
(${\cal N}=4$) gauge theory with a single $SU(N)$ gauge group with matter
in adjoint representations. In this case one can drop with impunity
the $U(1)$ of $U(N)$ because the matter fields are uncharged under it.
In the quiver theories with less supersymmetry (${\cal N} \le 2$)
it is usually necessary to keep such $U(1)$s. 

Quiver gauge theories are taylor made for particle physics model building. While an $SU(N)$
gauge theory is typically anomalous in  for arbitrary choice of fermions, 
choosing the fermions to lie in a quiver insures anomaly cancelation. 
Furthermore the fermions in a quiver arrange themselves
in bifundamental representations  of the product gauge group. This nicely coinsides with the fact
that all known fundamental fermions are in bifundamental, fundamental, or 
singlet representations of the gauge group. The study of quiver gauge 
theories goes back to the earliest days of gauge theories and the standard 
model. Other notable early examples are the Pati-Salam model 
and the trinification model. A vast literature exists on 
this subject, but we will concentrate on post 
$AdS/CFT$ conjecture quiver gauge theory work
\cite{Butti:2005sw, Butti:2006au, Forcella:2007wk, Bertolini:2004xf, Bertolini:2005di, Smith:1993gp,Douglas:1996sw,Sardo-Infirri:1996gb,
deBoer:1996mp,Xu:1997rv,Okuyama:1998dz,Kapustin:1998fa,Sugawara:1999qp,
Feng:1999zv,He:1999xj,
Buican:2006sn,
Verlinde:2005jr,
Fiol:2000wx,Uranga:2000ck,Albertsson:2000px,Govindarajan:2000vi,Feng:2000mw,
Berenstein:2000mb,Albertsson:2001jq,Muto:2001gu,Cachazo:2001gh,
Hanany:2001py,Alvarez-Consul:2001uk,Alvarez-Consul:2001um,Kephart:2001ix,Cremades:2002te,Kim:2002fp,
Ishii:2002ee,Mukhi:2002ck,Brax:2002rz,Hollowood:2002ax,Denef:2002ru,Frenkel:2002cs,
Feng:2002kk,Lalak:2002ep,Berenstein:2002fi,Frampton:2002ta,Hollowood:2002zk,
He:2002fp,Seki:2002ti,AitBenHaddou:2003ew,Frampton:2003zt,Casero:2003gf,Dai:2003ak,
Casero:2003gr,Benhaddou:2003nd,Narayan:2003et,Mukhopadhyay:2003ky,Narayan:2003rd,
Walcher:2003rh,Belhaj:2003qa,Chiantese:2003qb,Dai:2003dy,Belhaj:2004ws,
Frampton:2004xb,Dai:2004ke,Ita:2004yn,AhlLaamara:2004yz,Robles-Llana:2004dd,
DiNapoli:2004rm,Halmagyi:2004ju,Halmagyi:2004jy,Franco:2004wp,Fucito:2004gi,
Robles-Llana:2004nq,Benvenuti:2004dw,Benvenuti:2004dy,Benvenuti:2004wx,DiNapoli:2005ma,
Billo:2005jw,Bergman:2005ba,Hanany:2005hq,Popov:2005ik,Franco:2005rj,Burrington:2005zd,
Benvenuti:2005ja,Aspinwall:2005ur,Szendroi:2005ct,Takahashi:2005qu,Zhu:2005ki,
Fujii:2005dk,Hanany:2005ss,Kajiura:2005yu,Feng:2005gw,Wijnholt:2005mp,Proudfoot:2005mz,moremodels4,
Israel:2005zp,Antebi:2005hr,Butti:2005ps,Nakayama:2005mf,Intriligator:2005aw,Belhaj:2006wh,
Burrington:2006uu,DiNapoli:2006wz,Garcia-Etxebarria:2006aq,Burrington:2006aw,
Lechtenfeld:2006wu,Burrington:2006pu,Giedt:2006dd,Herzog:2006bu,Ueda:2006jn,
Park:2006va,Butti:2006nk,
Benvenuti:2006qr,Florea:2006si,Oota:2006eg,Berenstein:2006pk,Volansky:2006wn,Zhu:2006va,
AhlLaamara:2006zj,Burrington:2007mj,Diaconescu:2007ah,Imamura:2007dc,Antebi:2007xw,
Garcia-Etxebarria:2007vh,Evslin:2007au,Jafferis:2007sg,
Butti:2007jv,Forcella:2007ps}.

Starting from $AdS_5\times S^5$ we only have an $SU(N)$ ${\cal N}=4 $ supersymmetric 
gauge theory. In order to break SUSY and generate a quiver gauge theory 
there are several options open to us. 
Orbifolds \cite{Kakushadze:2000mc,Dixon:1986jc,Dixon:1985jw,Kakushadze:1996iw}, 
conifolds \cite{Uranga:1998vf,Oh:1999sk,Klebanov:2000hb,
Hebecker:2003jt,Benvenuti:2005wi}, 
and orientifolds \cite{Kakushadze:1998yq,Kakushadze:1998hb,Fayyazuddin:1998fb,
Kakushadze:1998tz,
Kakushadze:1998tr,Kakushadze:2001bd} 
have all played a part in building quiver gauge theories.
Since our focus is quiver gauge theories in general, 
but via orbifolding of $AdS_5\times M^5$ in particular, we will not discuss 
the other options in detail but should point out that orbifolding from the eleven dimensional M theory point of view has also an active area of research
\cite{Lalak:1997ti,Faux:1999hm,Faux:2000dv,Faux:2000sp,Faux:2000mr,Doran:2001ve}. Furthermore, we are interested in orbifolds 
where the manifold $M^5$ is the five sphere $S^5$. There are 
other possible choices for $M^5$ of relevance to model building
\cite{Aldazabal:2000sa,He:2004rn} 
but we will not explore these here either. In building models from orbifolded
$AdS_5\times S^5$, it is often convenient to break the quiver gauge group to the 
trinification\cite{Glashow:1984gc} group $SU(3)^3$ 
or to the Pati-Salam\cite{PS,Blazek:2003wz,Dent:2007eu} group 
$SU(4)\times SU(2)\times SU(2)$, but there are again other possibilities, 
including more complicated intermediate groups like the 
quartification \cite{Babu:2003nw,Chen:2004jz,Demaria:2005gk,Demaria:2006uu,
Demaria:2006bd,Babu:2007fx} symmetry $SU(3)^4$
that treats quarks and leptons on an equal footing.

It is important to note that although the duality with superstrings
is a significant guide to such model building, and it is desirable
to have a string dual to give more confidence in consistency, we
shall focus on the gauge theory description in the approach to
particle phenomenology, as there are perfectly good quiver gauge
theories that have yet to be derived from string duality.

\subsection{Orbifolding}

\bigskip

The simplest superstring - gauge duality arises from the compactifiation
of a Type IIB superstring on the cleverly chosen manifold

\bigskip

\[ ~~~~~~~ AdS_5 ~~ \times ~~ S^5  \]

\bigskip

\noindent which yields an ${\cal N} = 4$ supersymmetry which is an especially
interesting gauge theory which has been intensively studied and possesses
remarkable properties of finiteness and conformal invariance for all
values of $N$ in its $SU(N)$ gauge group. By ''conformality", we shall
mean conformal invariance at high energy, also for finite $N$.

For phenomenological purposes, ${\cal N} = 4$ is too much supersymmetry.
Fortunately it is possible to breaking supersymmetries and hence
approach more nearly the real world, with less or no supersymmetry in a 
conformality theory.

By factoring out a discrete (either abelian or nonabelian) group and composing an orbifold:

\bigskip

\[ ~~~~~~S^5 / \Gamma ~~~~~~~\]

\bigskip

\noindent one may break ${\cal N} = 4$ supersymmetry to
${\cal N} = 2, ~~~~1,$ or $~~~0$. Of special interest is the ${\cal N} = 0$ case.

We may take an abelian $\Gamma = Z_p$ (non-abelian cases will also be considered 
in this review) which identifies $p$ points in a complex three dimensional space ${\cal C}_3$.

The rules for breaking the ${\cal N} = 4$ supersymmetry are:

\bigskip

\noindent If $\Gamma$ can be embedded in an $SU(2)$ of the original $SU(4)$ R-symmetry, then
\[
~~~ \Gamma \subset SU(2)~~~~\Rightarrow~~{\cal N} = 2.
\]

\noindent If $\Gamma$ can be embedded in an $SU(3)$ but not an $SU(2)$ of the original $SU(4)$ R-symmetry, then
\[
~~~ \Gamma \subset SU(3)~~~~\Rightarrow~~{\cal N} = 1.
\]

\noindent If $\Gamma$ can be embedded in the $SU(4)$ but not an $SU(3)$ of the original $SU(4)$ R-symmetry, then
\[
~~~ \Gamma \subset SU(4)~~~~\Rightarrow~~{\cal N} = 0.
\]

\bigskip

\noindent In fact to specify the embedding of $\Gamma = Z_p$ we need to identify three integers $(a_1, a_2, a_3)$:

\[
~ {\cal C}_3 :~~(X_1, X_2, X_3)~~\stackrel{Z_p}{\rightarrow}~(\alpha^{a_1} X_1, \alpha^{a_2} X_2, \alpha^{a_3}X_3)
\]

\noindent with

\[
~ \alpha = exp \left( \frac{2 \pi i}{p} \right)
\]

\noindent The $Z_p$ discrete group identifies $p$ points in ${\cal C}_3$.
The N converging D3-branes meet on all $p$ copies, giving a gauge group:
$U(N) \times U(N) \times ......\times U(N)$, $p$ times.
The matter (spin-1/2 and spin-0)
which survives is invariant
under a product of a
gauge transformation and a $Z_p$ transformation.

There is a convenient diagramatic way to find the result from
a ''quiver."   One draws $p$ points and arrows for $a_1, a_2, a_3$.

Above is an example for $Z_5$ and $a_i=(1, 3, 0)$. 

\begin{figure}
\begin{center}
\epsfxsize=4.0in
\ \epsfbox{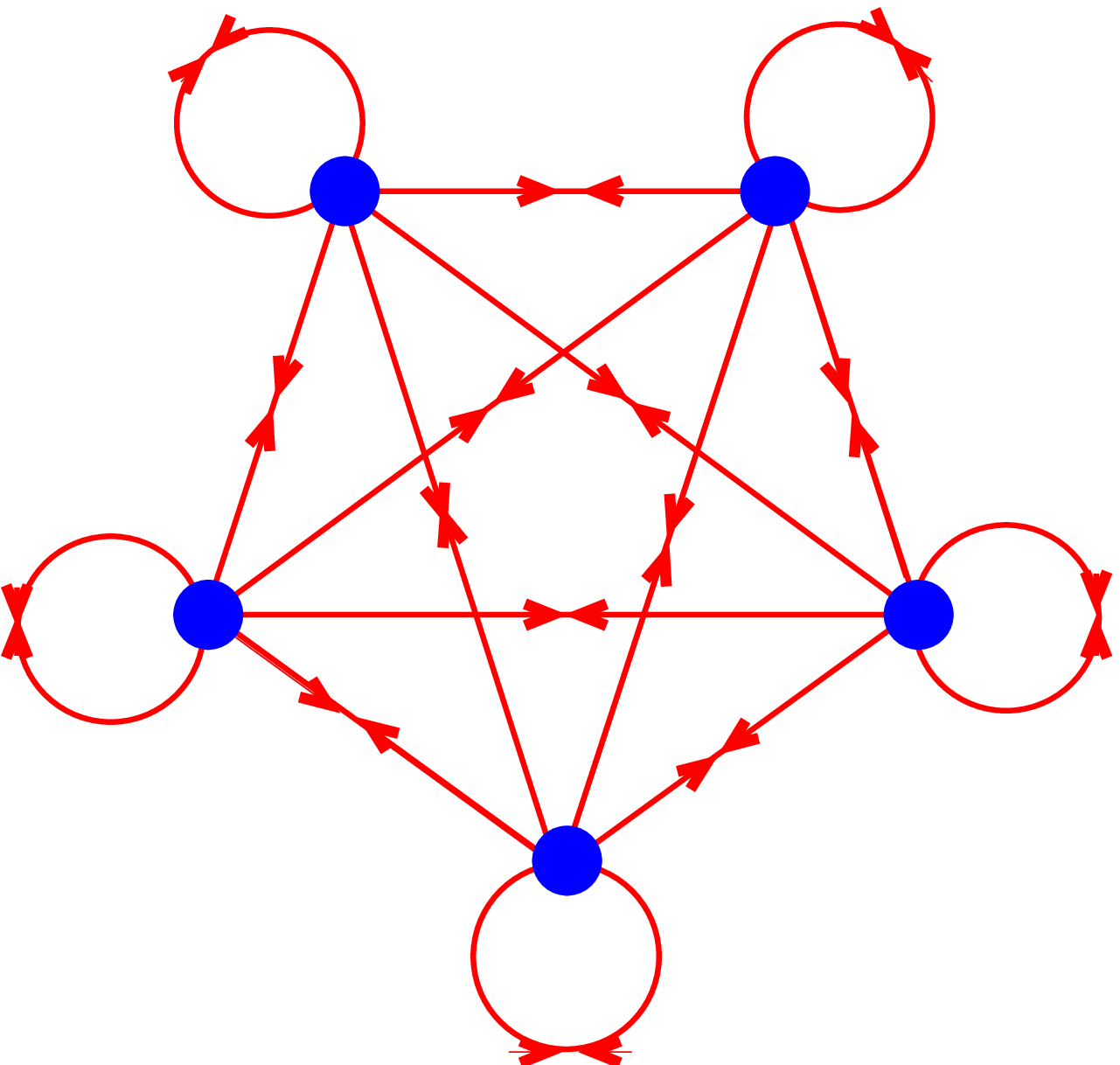}
\end{center}
\end{figure}

\bigskip

\noindent For a general case, the scalar representation contains the bifundamental
scalars

\bigskip

\[\sum_{k=1}^{3}\sum_{i = 1}^{p} (N_i, \bar{N}_{i \pm a_k})\]

\bigskip

\noindent Note that by definition
a {\it bifundamental} representation transforms as a fundamental
($N_i$) under $U(N)_i$ and simultaneously as an antifundamental
($\bar{N}_{i \pm a_k}$) under $U(N)_{i \pm a_k}$.

\bigskip

For fermions, one must first construct the {\bf 4} of R-parity $SU(4)$, isomorphic to
the isometry $SO(6)$ of the $S^5$. 
From the $a_k = (a_1, a_2, a_3)$ one constructs the 4-spinor $A_{\mu} = (A_1, A_2, A_3, A_4)$ :

\bigskip

\[ A_1 = \frac{1}{2} (a_1 + a_2 +a_3) \]

\[ A_2 = \frac{1}{2} (a_1 - a_2 -a_3) \]

\[ A_3 = \frac{1}{2} (- a_1 + a_2 - a_3) \]

\[ A_4 = \frac{1}{2} (- a_1 - a_2 +a_3) \]

\noindent These transform 
as $exp \left( \frac{2 \pi i}{p} A_{\mu} \right)$ and the 
invariants may again be derived (by a different diagram). An example 
of a fermion quiver with $p=5$ is shown above.

\begin{figure}
\begin{center}
\epsfxsize=4.0in
\ \epsfbox{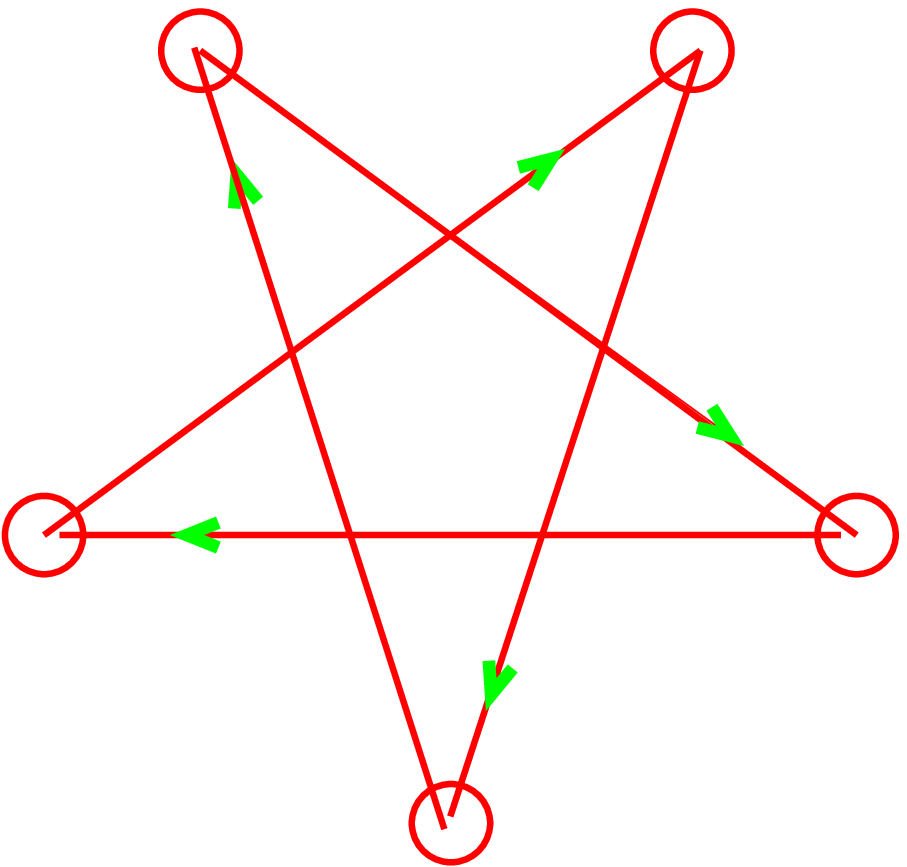}
\end{center}
\end{figure}

\bigskip

\noindent Note that these lines are oriented, as is necessary to accommodate chiral
fermions. Specifying the four $A_{\mu}$ is equivalent (there is a constraint that the
four add to zero, mod $p$) to fixing the three $a_k$ and group theoretically is more fundamental.

\bigskip

\noindent  In general,  the fermion representation contains the
bifundamentals

\bigskip

\[ \sum_{\mu = 1}^{4} \sum_{i = 1}^{p} ( N_i, \bar{N}_{i + A_{\mu}}) \]

\noindent When one of the $A_{\mu}$s is zero, it signifies a degenerate case of a bifundamental
comprised of adjoint and singlet representations of one $U(N)$.

To summarize the orbifold construction, first we select a
discrete subgroup $\Gamma $ of the $SO(6)\sim SU(4)$ isometry of $S^{5}$
with which to form the orbifold $AdS_{5}\times S^{5}/\Gamma $. As discussed above, the
replacement of $S^{5}$ by $S^{5}/\Gamma $ reduces the supersymmetry to $%
{\cal N}=$ 0, 1 or 2 from the initial ${\cal N}=4$, depending on how $\Gamma 
$ is embedded in the isometry of $S^{5}$. The cases of interest here are $%
{\cal N}=0$ and ${\cal N}=1\ $ $SUSY$ where $\Gamma $ embeds irreducibly in the $%
SU(4)$ isometry or in an $SU(3)$ subgroup of the $SU(4)$ isometry, respectively. $I.e.$, 
to achieve ${\cal N}=0$ we embed rep($\Gamma )\rightarrow {\bf 4}$ of $%
SU(4)$ as ${\bf 4}=({\bf r})$ where ${\bf r}$ is a nontrivial  
four dimensional representation of $\Gamma ;$ for ${\cal N}=1$ we
embed rep($\Gamma )\rightarrow {\bf 4}$ of $SU(4)$ as ${\bf 4}=({\bf 1},{\bf %
r})$ where ${\bf 1}\ $is the trivial 
irreducible representation
(irrep) of $\Gamma $ and ${\bf r}$ is a
nontrivial three dimensional representation of $\Gamma .$

\newpage

\section{Conformality phenomenology} 

\bigskip
\bigskip

In attempting to go beyond the standard model, one outstanding issue is
the hierarchy between GUT scale and weak scale
which is 14 orders of magnitude. Why do these two very different
scales exist?
Also, how is this hierarchy of scales stabilized
under quantum corrections?
Supersymmetry answers the second question but not the first.

The idea is to approach hierarchy problem by Conformality at a TeV Scale.
We will show how this is possible including explicit examples containing standard model states.

In some sense conformality provides more rigid constraints than supersymmetry.
It can predict additional states at TeV scale, while there can be far fewer initial 
parameters in conformality models than in SUSY models.
Conformality also provides a new approach to gauge coupling unification.
It confronts naturalness and provides cancellation of quadratic divergences.
The requirements of anomaly cancellationsi can lead
to conformality of U(1) couplings.

There is a viable dark matter candidate, and proton decay
can be consistent with experiment.

\bigskip

What is the physical intuition and picture underlying conformality?
Consider going to an energy scale higher than the weak scale,
for example at the TeV scale. Quark and lepton masses, 
QCD and weak scales small compared to TeV scale. They may
be approximated by zero. The theory is then classically
conformally invariant though not at the quantum level because
the standard model has non-vanishing renormalization group
beta functions and anomalous dimensions. So this suggests that we
add degrees of freedom to yield  a gauge field theory
with conformal invariance.  
There will be 't Hooft's naturalness since the zero mass limit 
increases symmetry to conformal symmetry.

\bigskip

We have no full understanding of how four-dimensional
conformal symmetry can be broken spontaneously so
breaking softly by relevant operators is a first step.
The theory is assumed to be given by the action:

\bigskip

\begin{equation}
S = S_0 + \int d^4x \alpha_i O_i
\end{equation}

\bigskip

\noindent where $S_0$ is the action for the conformal theory and the $O_i$ are 
operators with dimension below four ({\it i.e.} relevant)
which break conformal invariance softly.

\bigskip

\noindent The mass parameters $\alpha_i$ have mass dimension $4-\Delta_i$ where
$\Delta_i$ is the dimension of $O_i$ at the
conformal point.

\bigskip

\noindent Let $M$ be the scale set by the parameters $\alpha_i$ and
hence the scale at which conformal invariance is broken. Then for $E >> M$ the couplings
will not run while they start running for $E < M$.
To solve the hierarchy problem we assume $M$ is near the TeV scale.

\newpage

\subsection{Experimental evidence for conformality}

Consider embedding the standard model gauge group according to:

\[ SU(3) \times SU(2) \times U(1) \subset \bigotimes_i U(Nd_i) \]

Each gauge group of the SM can lie entirely in a $SU(Nd_i)$
or in a diagonal subgroup of a number thereof.

\bigskip

\noindent Only bifundamentals (including adjoints) are possible.
This implies no $(8,2), (3,3)$, etc. A conformality restriction which is new and
satisfied in Nature! The fact that the standard model has matter
fields all of which can be accommodated as bifundamentals
is expermental evidence for conformality.

\bigskip

\noindent No $U(1)$ factor can be conformal in perturbation
theory and so hypercharge is quantized
through its incorporation in a non-abelian gauge group.
This is the ``conformality'' equivalent to the GUT charge quantization
condition in {\it e.g.} $SU(5)$. It can explain the neutrality of the hydrogen atom.
While these are postdictions, the predictions of the theory are new particles,
perhaps at a low mass scale,  filling out
bifundamental representations of the gauge group that restore conformal 
invariance. The next section will begin our study of known quiver
gauge theories from orbifolded $AdS^5\times S^5$.

\newpage

\section{Tabulation of the simplest abelian quivers}

\bigskip

We consider the compactification of the type-IIB superstring
on the orbifold $AdS_5 \times S^5/\Gamma$
where $\Gamma$ is an abelian group $\Gamma = Z_p$
of order $p$ with elements ${\rm exp} \left( 2 \pi i A/p \right)$,
$0 \le A \le (p-1)$.

The resultant quiver gauge theory has ${\cal N}$
residual supersymmetries with ${\cal N} = 2,1,0$ depending
on the details of the embedding of $\Gamma$
in the $SU(4)$ group which is the isotropy
of the $S^5$. This embedding is specified
by the four integers $A_m, 1 \le m \le 4$ with

\begin{equation}
\Sigma_m A_m = 0~ {\rm mod } ~p
\nonumber
\label{SU4}
\end{equation}

\noindent which characterize
the transformation of the components of the defining
representation of $SU(4)$. We are here interested in the non-supersymmetric
case ${\cal N} = 0$ which occurs if and only if
all four $A_m$ are non-vanishing.

\bigskip

Table I.  All abelian quiver
theories with ${\cal N}=0$ from $Z_2$ to $Z_7$.

\bigskip

\newpage

\begin{tabular}{|||c||c||c|c||c|c|c||c|c|||}
\hline
\hline
\hline
& p & $A_m$ & $a_i$ & scal & scal & chir &  \\
&&&& bfds & adjs & frms & SM  \\
\hline
\hline
1 & 2 & (1111) & (000) & 0 & 6 & No  &  No \\
\hline
\hline
2 & 3 & (1122) & (001) & 2 & 4 & No  &  No \\
\hline
\hline
3 & 4 & (2222) & (000) & 0 & 6 & No  &  No \\
4 & 4 & (1133) & (002) & 2 & 4 & No  &  No \\
5 & 4 & (1223) & (011) & 4 & 2 & No  &  No \\
6 & 4 & (1111) & (222) & 6 & 0 & Yes  &  No \\
\hline
\hline
7 & 5 & (1144) & (002) & 2 & 4 & No  &  No \\
8 & 5 & (2233) & (001) & 2 & 4 & No  &  No \\
9 & 5 & (1234) & (012) & 4 & 2 & No  &  No \\
10 & 5 & (1112) & (222) & 6 & 0 & Yes  &  No \\
11 & 5 & (2224) & (111) & 6 & 0 & Yes  &  No \\
\hline
\hline
12 & 6 & (3333) & (000) & 0 & 6 & No  &  No \\
13 & 6 & (2244) & (002) & 2 & 4 & No  &  No \\
14 & 6 & (1155) & (002) & 2 & 4 & No  &  No \\
15 & 6 & (1245) & (013) & 4 & 2 & No  &  No \\
16 & 6 & (2334) & (011) & 4 & 2 & No  &  No \\
17 & 6 & (1113) & (222) & 6 & 0 & Yes  &  No \\
18 & 6 & (2235) & (112) & 6 & 0 & Yes  &  No \\
19 & 6 & (1122) & (233) & 6 & 0 & Yes  &  No \\
\hline
\hline
20 & 7 & (1166) & (002) & 2 & 4 & No  &  No \\
21 & 7 & (3344) & (001) & 2 & 4 & No  &  No \\
22 & 7 & (1256) & (013) & 4 & 2 & No  &  No \\
23 & 7 & (1346) & (023) & 4 & 2 & No  &  No \\
24 & 7 & (1355) & (113) & 6 & 0 & No  &  No \\
25 & 7 & (1114) & (222) & 6 & 0 & Yes  &  No \\
26 & 7 & (1222) & (333) & 6 & 0 & Yes  &  No \\
27 & 7 & (2444) & (111) & 6 & 0 & Yes  &  No \\
28 & 7 & (1123) & (223) & 6 & 0 & Yes  &  Yes \\
29 & 7 & (1355) & (113) & 6 & 0 & Yes  &  Yes \\
30 & 7 & (1445) & (113) & 6 & 0 & Yes  &  Yes \\
\hline
\hline
\hline
\end{tabular}

\bigskip
\bigskip

\newpage

\section{Chiral fermions}

The gauge group is $U(N)^p$. The fermions
are all in the bifundamental representations
\begin{equation}
\Sigma_{m=1}^{m=4}\Sigma_{j=1}^{j=p} (N_j, \bar{N}_{j + A_m})
\label{fermions}
\end{equation}
which are manifestly non-supersymmetric because no
fermions are in adjoint representations
of the gauge group.
Scalars appear in representations
\begin{equation}
\Sigma_{i=1}^{i=3}\Sigma_{j=1}^{i=p} (N_j, \bar{N}_{j \pm a_i})
\label{scalars}
\end{equation}
in which the six integers $(a_i, -a_i)$ characterize the
transformation of the
antisymmetric second-rank tensor representation
of $SU(4)$. The $a_i$
are given by $a_1 = (A_2+A_3), a_2= (A_3+A_1)$, and $a_3= (A_1+A_2)$.

It is possible for one or more of the $a_i$ to vanish
in which case the corresponding scalar representation
in the summation in Eq.(\ref{scalars}) is to be interpreted as an
adjoint
representation of one particular $U(N)_j$.
One may therefore
have zero, two, four or all six of the scalar
representations, in Eq.(\ref{scalars}), in such adjoints.
One purpose of the present article is to
investigate how the renormalization properties
and occurrence of quadratic divergences
depend on the distribution
of scalars into bifundamental
and adjoint representations.

Note that there is one model with all scalars in adjoints for each even
value of $p$. For general even $p$
the embedding is
$A_m=(\frac{p}{2},\frac{p}{2},\frac{p}{2},\frac{p}{2})$. This series
by itself forms the complete list of ${\cal N}=0$ abelian quivers with
all scalars in adjoints.

To be of more phenomenolgical interest the model should
contain chiral fermions. This requires that the embedding
be complex: $A_m \not\equiv -A_m$ (mod p). It will now be shown
that for
the presence of chiral fermions all scalars must be in bifundamentals.

The proof of this assertion follows by assuming the contrary,
that there is at least one adjoint arising from, say, $a_1=0$.
Therefore
$A_3=-A_2$ (mod p). But then it follows from Eq.(\ref{SU4})
that $A_1=-A_4$ (mod p). The fundamental representation of $SU(4)$
is thus real and fermions are non-chiral\footnote{This is almost
obvious
but for a complete justification, see\cite{FK03}}.

The converse also holds: If all $a_i \neq 0$ then there are chiral
fermions.
This follows since by assumption
$A_1 \neq -A_2$, $A_1 \neq -A_3$, $A_1 \neq -A_4$. Therefore
reality of the fundamental representation would require
$A_1 \equiv -A_1$ hence, since $A_1 \neq 0$, $p$ is even
and $A_1 \equiv \frac{p}{2}$; but then the other $A_m$
cannot combine to give only vector-like fermions.

It follows that:

\bigskip
\bigskip

\noindent \underline{{\it In an ${\cal N}=0$ quiver gauge theory, chiral fermions
are possible}}

\noindent \underline{{\it if and only if all scalars are in bifundamental
representaions.}}

\newpage

\vskip 1.0in

\section{Model building}

\bigskip

The next step is to examine how the framework of
quiver gauge theories can accommodate, as a sub theory,
the standard model.
This requires that the standard model gauge group
and the three families of quarks and leptons with
their correct quantum numbers be accommodated.

In such model building a stringent requirement
is that the scalar sector, prescribed by the quiver construction,
can by acquiring vacuum expection, values break
the symmetry spontaneously to the desired sub theory.
This is unlike most other model building where one
{\it chooses} the representations for the scalars
to accomplish this goal. Here the representations
for the scalars are dictated by the orbifold construction.

One useful guideline in the symmetry breaking is that
to break a semi-simple $SU(N)^n$ gauge group to
its $SU(N)$ diagonal subgroup requires at least
$(n-1)$ bifundamental scalars, connected to one another
such that each of the $n$ $SU(N)$ factors is linked 
to all of the others: it is insufficient
if the bifundamental scalars fragment 
into disconnected subsets. 

We shall describe in turn abelian orbifolds
\cite{first,WS,vafa,F2,F3}
and non abelian orbifolds
\cite{nonabelian,nonabelian1,nonabelian2,nonabelian3}
As will become clear
abelian orbifolds lead to accommodation of the standard
model in unified groups $SU(3)^n$ while non abelian
orbifolds can lead naturally to 
incorportion of the standard model in gauge groups such as
$SU(4) \times SU(2) \times SU(2)$ and generalizations.

\newpage

\subsection{Abelian model building}

We will now classify compactifications of the type $IIB$
superstring on
$AdS_{5}\times S^{5}/\Gamma $, where $\Gamma $ is an abelian  
group of order $n\leq 12$.  
Appropriate embedding of $\Gamma$ in the isometry of $S^5$
yields both $SUSY$ and non--$SUSY$ chiral models that can contain the
minimal $SUSY$ standard model or the standard model.
New non-SUSY three family models with $\Gamma=Z_8$ are introduced, 
which lead to the right Weinberg angle for TeV trinification.

We find 78 ${\cal N}=0$ chiral $Z_n$ models and discuss a few of 
phenomenological interest. For ${\cal N}=1$his results in
60 chiral $Z_n$ models, and a systematic analysis with $n<8$
yields four containing the minimal $SUSY$ standard model with three families.
One of these models extends to an infinite sequence of three-family $MSSMs$.
We also give a lower bound on the number of chiral models for all values
of $n$. For completeness, we also discuss abelian models where the orbifolding groups
are products of  cyclic groups.

\subsubsection{Abelian non-SUSY models}

When one bases models on conformal field theory gotten from the large $N$
expansion of the $AdS/CFT$ correspondence \cite{Maldacena1997}, stringy
effects can arise at an energy scale as low as a few TeV. These 
models can potentially test string theory and examples with low energy scales are known
in orbifolded $AdS_{5}\times S^{5}$. The first three-family $AdS_{5}\times S^{5}/\Gamma $ model of this
type had ${\cal N}=1$ $SUSY$ and was based on a $\Gamma=Z_{3}$ orbifold \cite
{KS}, see also \cite{Kephart:2001qu}. However, since then some of the
most studied examples have been models without supersymmetry based on both
abelian 
\cite{first,WS,vafa,F2,F3}
and non-abelian 
\cite{nonabelian,nonabelian1,nonabelian2,nonabelian3}
orbifolds of $AdS_{5}\times S^{5}$. Recently both SUSY and nonSUSY three family $Z_{12}$ 
orbifold models 
\cite{Unification,Unification2,moremodels2}
have been shown to unify at a low scale ($\sim$ 4 TeV) and to have the promise of testability. One motivation for
studying the non--$SUSY$ case is that the need for supersymmetry is less
clear as: (1) the hierarchy problem is absent or ameliorated
, (2)
the difficulties involved in breaking
the remaining ${\cal N}=1$ $SUSY$ can be avoided if the orbifolding already
results in completely broken $SUSY$, and (3) many of the
effects of $SUSY$ are still present in the theory, just hidden. For example,
the bose-fermi state count matches, RG equations preserve vanishing $\beta $
functions to some number of loops, etc. Here we concentrate on abelian
orbifolds with and without supersymmetry, where the orbifolding group $\Gamma$ has
order $n=o(\Gamma)\leq 12$. We systematically study those
cases with chiral matter ($i.e.$, in the $SUSY$ case, those with an imbalance between chiral
supermultiplets and anti-chiral supermultiplets, and in the non--$SUSY$ case with a net
imbalance between left and right handed fermions). We find all chiral models for 
$n\leq 12$. Several of these contain the standard model ($SM$)
or the minimal supersymmetric standard model ($MSSM$) with three or four families.

A summary of how orbifolded $AdS_{5}\times S^{5}$ models are
constructed was provided in Section 2. 

For ${\cal N}=0$ the fermions are given by 
$\sum_{i}{\bf 4}\otimes R_{i}$
and the scalars by 
$\sum_{i}{\bf 6}\otimes R_{i}$
where the set \{$R_{i}\}$ runs over all the irreps of $\Gamma .$ For
$\Gamma$ abelian, the irreps are all one dimensional and as a
consequence of the choice of $N$ in the $1/N$ expansion, the gauge group 
\cite{Lawrence} is $SU(N)^{n}$. In the ${\cal N}=1\ $ $SUSY$ case,
chiral supermultiples generated by this embedding are given by 
$ \sum_{i}
{\bf 4}\otimes R_{i}$
where again \{$R_{i}\}$ runs over all the 
(irreps) of $\Gamma .$ As before, for abelian $\Gamma$, the irreps
are all one dimensional and the gauge group is  $SU^{n}(N)$. Chiral
models require the {\bf 4} to be complex (${\bf 4}\neq {\bf 4}^{*})$ while a
proper embedding requires ${\bf 6}={\bf 6}^{*}$ where ${\bf 6}$=(${\bf 4}%
\otimes {\bf 4})_{antisym}$. (Even though the ${\bf 6}$ does not enter the
model in the ${\cal N}=1\ $ $SUSY$ case, mathematical consistency requires $
{\bf 6}={\bf 6}^{*}$, see \cite{FK03}.)

We now have the required background to begin building chiral models. We
choose $N=3$ throughout. This means that most of our models
will proceed to the SM or MSSM through trinification. It is also possible 
to start with larger $N$, say $N=4$ and proceed to the SM or MSSM via Pati-Salam
models. The analysis is similar, so the $N=3$ case is sufficient to gain an understanding of the techniques needed for model building, what choice of $N$ leads to the optimal model is still an open question.

 If $SU_{L}(2)$ and $U_{Y}(1)$ are embedded in
diagonal subgroups $SU^{p}(3)$ and $SU^{q}(3)$ respectively, of the initial $
SU^{n}(3)$, the ratio $\frac{\alpha _{2}}{\alpha _{Y}}$ is $\frac{p}{q},$
leading to a calculable initial value of $\theta _{W}$ with, 
$ \sin^{2}\theta _{W}=\frac{3}{3+5\left( \frac{p}{q}\right) }.$
The more standard approach is to break the initial $SU^{n}(3)$ to $%
SU_{C}(3)\otimes SU_{L}(3)\otimes SU_{R}(3)$ where $SU_{L}(3)$ and $%
SU_{R}(3) $ are embedded in diagonal subgroups $SU^{p}(3)$ and $SU^{q}(3)$
of the initial $SU^{n}(3)$. We then embed all of $SU_{L}(2)$ in $SU_{L}(3)$
but $\frac{1}{3}$ of $U_{Y}(1)$ in $SU_{L}(3)$ and the other $\frac{2}{3}$
in $SU_{R}(3)$. This modifies the $\sin ^{2}\theta _{W}$ formula to: 
$\sin ^{2}\theta _{W}=\frac{3}{3+5\left( \frac{\alpha _{2}}{\alpha _{Y}}
\right) } =\frac{3}{3+5\left( \frac{3p}{p+2q}\right) }$, which coincides 
with the previous result when $p=q$. One should use the second (standard) embedding 
when calculating $\sin ^{2}\theta _{W}$ for any of the models obtained below. A similar relation
holds for Pati--Salam type models \cite{nonabelian2} and 
their generalizations \cite{moremodels}, but this would
require investigation of models with $N\geq 4$ which are not included in this review.
Note, if $\Gamma=Z_n$ the initial ${\cal N}=0$ orbifold
model (before any symmetry breaking) is completely fixed (recall we always
are taking $N=3$) by the choice of $n$ and the embedding ${\bf 4=}%
(\alpha ^{i},\alpha ^{j},\alpha ^{k},\alpha ^{l}),$ so we define these models
by $M_{ijkl}^{n}.$ The conjugate models $M_{n-i,n-j,n-k,n-l}^{n}$ contain
the same information, so we need not study them separately.

\bigskip
\bigskip 

{\it ${\cal N}=0 $ chiral $Z_n$ models}

\bigskip
\bigskip

To get a feel for the constructions, we begin this section by 
studying the first few ${\cal N}=0 $ chiral $Z_n$ models.
Insights gained here will allow us to generalize and give results to arbitrary $n$.
First, when ${\cal N}=0$, the only allowed 
$\Gamma =Z_{2}$ orbifold  where ${\bf 4}=(\alpha ,\alpha ,\alpha
,\alpha )$ and $Z_{3}$  orbifold where ${\bf 4}=(\alpha ,\alpha ,\alpha^2
,\alpha^2 )$ have
only real representations and therefore will not yield chiral models. 
Next, for $\Gamma =Z_{4}$ the choice ${\bf 4}=(\alpha ,\alpha ,\alpha
,\alpha )$ with $N=3$ where $\alpha =e^{\frac{\pi i}{2}}$ (in what follows
we will write $\alpha =e^{\frac{2\pi i}{n}}$ for the roots of unity that
generate $Z_{n})$, yields an $SU(3)^{4}$ chiral model with the fermion content shown
in Table 2.
The scalar content of this model is given in Table 3 and a VEV for say a
(3,1,\={3},1) breaks the symmetry to $SU_{D}(3)\times SU_{2}(3)\times
SU_{4}(3)$ but renders the model vectorlike, and hence uninteresting, so we consider it no further.
(We consider only VEVs that cause symmetry breaking of the type $SU(N)\times SU(N)
\rightarrow SU_D(N).$ Other symmetry breaking patterns are possible, but for the sake of
simplicity they will not be studied here. It is clear from this and previous remarks that there are many
phenomenological avenues involving quiver gauge theories yet to be explored.)
The only other choice of embedding is a nonpartition model 
with $\Gamma = Z_4$ is ${\bf 4}=(\alpha ,\alpha ,\alpha ,\alpha ^{3})$ 
but it leads to the same
scalars with half the chiral fermions so we move on to $Z_{5}.$

There is one chiral model for $\Gamma =Z_{5}$ and it is fixed by choosing $%
{\bf 4}=(\alpha ,\alpha ,\alpha ,\alpha ^{2})$, leading to ${\bf 6}=(\alpha
^{2},\alpha ^{2},\alpha ^{2},\alpha ^{3},\alpha ^{3},\alpha ^{3})$ with real
scalars. It is straightforward to write down the particle content of this $%
M_{1112}^{5}$ model. The best one can do toward the construction of the
standard model is to give a VEV to a (3,1,\={3},1,1) to break the $SU^{5}(3)$
symmetry to $SU_{D}(3)\times SU_{2}(3)\times SU_{4}(3)\times SU_{5}(3).$ Now
a VEV for (1,3,\={3},1) completes the breaking to $SU^{3}(3),$ but the only
remaining chiral fermions are $2[(3,\bar{3},1)+(1,3,\bar{3})+(\bar{3},1,3)]$
which contains only two families.

Moving on to $\Gamma =Z_{6}$ we find two models
where, as with the previous $Z_{5}$ model, the ${\bf 4}$ is arranged so that 
${\bf 4=\ }(\alpha ^{i},\alpha ^{j},\alpha
^{k},\alpha ^{l})$ with $i+j+k+l=n.$
These have ${\bf 4}%
=(\alpha ,\alpha ,\alpha ,\alpha ^{3})$ and ${\bf 4}=(\alpha ,\alpha ,\alpha
^{2},\alpha ^{2})$ and were defined as partition
models in \cite{Kephart:2001qu} when $i$ was equal to zero.
Here we generalize and call all models satisfying $i+j+k+l=n$ partition
models. We have now introduced most of the
background and notation we need, so at this point (before completing the investigation of
the $\Gamma =Z_{6}$ models) it is useful to give a
summary (see Table 4) of all ${\cal N}=0$ chiral $Z_{n}$ models with
real ${\bf 6}$'s for $n\leq 12.$ 
We note that the $%
n=8$ partition model with ${\bf 4}=(\alpha ,\alpha ,\alpha ^{2},\alpha ^{4})$
has $\chi /N^{2}=16;$ the other four have $\chi /N^{2}=32.$ Of the nine $%
Z_{10}$ partition models$,$ 2 have $\chi /N^{2}=30$ and the other 7 have $%
\chi /N^{2}=40.$ The $Z_{12}$ partition
models derived from ${\bf 4}=(\alpha ,\alpha ,\alpha ^{4},\alpha ^{6}),$ $%
{\bf 4}=(\alpha ,\alpha ^{2},\alpha ^{3},\alpha ^{6}),$ and ${\bf 4}=(\alpha
^{2},\alpha ^{2},\alpha ^{2},\alpha ^{6})$ have $\chi /N^{2}=36;$ the others
have $\chi /N^{2}=48.$

A new class of models appears in Table 4; these are 
the double partition models. 
They have $i+j+k+l=2n$ and none are
equivalent to single partition models (if we require that $i$, $j$, $k$, and $%
l $ are all positive integers) with $i+j+k+l=n.$ The ${\cal N}=1$
nonpartition models have been classified \cite{nonabelian3}, and we
find eleven ${\cal N}=0$ examples in Table 4. While they have a self
conjugate ${\bf 6,}$ this is only a necessary condition that may be
insufficient to insure the construction of viable string theory based
models \cite{moremodels2}. 
However, as is the ${\cal N}=1$ case, the ${\cal N}=0$ 
nonpartition models may still be interesting
phenomenologically and as a testing ground for models with the potential of
broken conformal invariance.  

For $Z_{n}$ orbifold models with $n$ a prime number, only partition models
arise. The non--partition and double partition models only occur when $n$ is not
a prime number, and only a few are independent. Consider $n=12,$ here
we can write $Z_{12}=Z_{4}\times Z_{3}.$ If we write an element of this
group as $\gamma \equiv (a,b),$ where $a$ is a generator of $Z_{4}\ $and $b$
of $Z_{3},$ then $\gamma ^{2}\equiv (a^{2},b^{2}),$ $\gamma ^{3}\equiv
(a^{3},1),$ etc. The full group is generated by any one of the elements $\gamma =(a,b),$ $%
\gamma ^{5}=(a,b^{2}),$ $\gamma ^{7}=(a^{3},b),$ or $\gamma
^{11}=(a^{3},b^{2}).$ The other choices do not faithfully represent the
group. Letting $\alpha =\gamma ^{11}$ give a conjugate model, $e.g$., it
transforms $(\alpha ,\alpha ^{6},\alpha ^{8},\alpha ^{9})$ into $(\gamma
^{11},\gamma ^{6},\gamma ^{4},\gamma ^{3}),$ so this pair of double partition models are equivalent,
while letting $\alpha =\gamma ^{5}$ transforms $(\alpha ,\alpha ^{6},\alpha
^{8},\alpha ^{9})$ into the equivalent model $(\gamma ^{5},\gamma
^{6},\gamma ^{4},\gamma ^{9}),$ and $\alpha =\gamma ^{7}$ transforms $%
(\alpha ,\alpha ^{6},\alpha ^{8},\alpha ^{9})$ into the equivalent model $%
(\gamma ^{7},\gamma ^{6},\gamma ^{8},\gamma ^{3}).$ Hence a systematic use
of these operations on the non--partition and double partition models can
reduce them to the equivalence classes listed in the tables.

It is easy to prove we always have a proper embedding ($i.e$., ${\bf 6=6}%
^{*} $) for the ${\bf 4=\ }(\alpha ^{i},\alpha ^{j},\alpha ^{k},\alpha ^{l})%
{\bf \ }$ when $i+j+k+l=n$ (or $2n$). To show this note from ${\bf 6}$=(${\bf 4}%
\otimes {\bf 4})_{antisym}$ we find 

\begin{equation}
{\bf 6} =  (\alpha ^{i+j}, \alpha^{i+k}, 
\alpha^{i+l}, \alpha^{j+k}, \alpha^{j+l}, \alpha ^{k+l})
\label{six}
\end{equation}

\noindent but $(i+j) = n-k-l = - (k+l) ~~ ({\rm mod} ~~ n )$
and  $(i+k)=n-j-l=-(j+l) ~~ ({\rm mod} ~~  n)$, 
so this gives 

\begin{equation}
{\bf 6} = (\alpha ^{-(k+l)},\alpha ^{-(j+l)},\alpha
^{-(j+k)},\alpha ^{j+k},\alpha ^{j+l},\alpha ^{k+l})={\bf 6}^{*}.
\label{six2}
\end{equation}

\noindent A simple modification of this proof
also applies to the double partition models.

Now let us return to $\Gamma =Z_{6}$ where the 
partition models of interest are: (1) ${\bf 4}=(\alpha ,\alpha
,\alpha ^{2},\alpha ^{2})$ where one easily sees that VEVs for $(3,1,\bar{3}%
,1,1,1)$ and then $(1,3,\bar{3},1,1)$ lead to at most two families, while
other SSB routes lead to equal or less chirality; and (2) ${\bf 4}=(\alpha
,\alpha ,\alpha ,\alpha ^{3})$ where VEVs for $(3,1,\bar{3},1,1,1)$ followed
by a VEV for $(1,3,\bar{3},1,1)$ leads to an $SU(3)^{4}$ model containing
fermions 2[$(3,\bar{3},1,1)+(1,3,\bar{3},1)+(1,1,3,\bar{3})+(\bar{3},1,1,3)]$. 
However, there are insufficient scalars to complete the symmetry breaking
to the standard model. In fact, one cannot even achieve the trinification
spectrum.

The double partition $Z_{6}$ model ${\bf 4}=(\alpha ,\alpha ^{3},\alpha
^{4},\alpha ^{4})$ is relatively complicated, since there are 24 different
scalar representations in the spectrum, and this makes the SSB analysis
rather difficult. A number of possible 
SSB pathways were investigated \cite{moremodels3},
but  none were found that lead to the SM with at least three families.
However, since the search was not exhaustive, we cannot make a definitive
statement about this model. As stated elsewhere, the non-partition models
are difficult to interpret, if not pathological, so we have not studied the
SSB pathways for these $Z_{6}$ models.

We move on to $Z_{7}$, where there are three partition models: (1) for ${\bf %
4}=(\alpha ,\alpha ^{2},\alpha ^{2},\alpha ^{2})$, we find no SSB pathway to
the SM. There are paths to an SM with less than three families, e. g., VEVs
for $(3,1,1,\bar{3},1,1,1),$ $(1,3,1,\bar{3},1,1),$ $(3,\bar{3},1,1,1),$ and
$(1,3,\bar{3},1)$ lead to one family at the $SU^{3}(3)$ level; 
(2) for $%
{\bf 4}=(\alpha ,\alpha ,\alpha ,\alpha ^{4})$, again we find only paths to
family-deficient standard models. 
An example is where we have VEVs for 
$(3,1,\bar{3},1,1,1,1),$ $(1,3,\bar{3},1,1,1),$ $(3,1,\bar{3},1,1),$ and $%
(1,3,\bar{3},1)$,
which lead to a two-family $SU^{3}(3)$ model; 
(3) finally, ${\bf %
4}=(\alpha ,\alpha ,\alpha ^{2},\alpha ^{3})$ is the model discovered in 
\cite{F2}, where VEVs to 
$(1, 3,1,\bar{3},1,1,1)$, $(1,1,3,\bar{3},1,1)$, $(1,1,3,\bar{3},1)$
and $(1,1,3,\bar{3})$ 
lead to a three family model with the correct Weinberg angle
at the $Z$-pole, $\sin^2 \theta_W=3/13$.

For $Z_{n}$ with $n\geq 8,$ the number of representations of matter
multiplets has already grown to a degree where it makes a systematic
analysis of the models prohibitively time-consuming. It is thus helpful to
have further motivation to study particular examples or limited sets of
these models with large $n$ values. Thus we searched for examples which 
break
$SU(3)^8$ down 
to diagonal subgroups
$SU(3)^4\times SU(3)^3 \times SU(3)$, 
since this implies the right Weinberg angle for TeV trinification \cite{Glashow:1984gc},
$\sin^2 \theta_W= 
3/13$, when embedding $SU(3)_L$ and $SU(3)_R$ into the diagonal subgroups of
$SU(3)^4$ and $SU(3)$, 
respectively.
There are actually 11 different possibilities
to
break $SU(3)^8$ down 
to $SU(3)^4\times SU(3)^3 \times SU(3)$, assuming the necessary scalars 
exist. While none of these paths was successful for 
${\bf %
4}=(\alpha ,\alpha,\alpha,\alpha ^{5})$, the model ${\bf %
4}=(\alpha ,\alpha,\alpha ^{2},\alpha ^{4})$ leads to the 3 family SM.  
Assigning VEVs to 
$(3,1,\bar{3},1,1,1,1,1)$, $(3,1,1,\bar{3},1,1,1)$,
$(3,\bar{3},1,1,1,1)$, $(1,3,\bar{3},1,1)$
and $(1,3,\bar{3},1)$ breaks $SU(3)^8$ down to 
$SU(3)_{1235} \times SU(3)_{467} \times SU(3)_8$.  

Another option exists for 
${\bf 4}=(\alpha ,\alpha^4,\alpha^{5},\alpha^{6})$, when assigning VEVs to
\footnote{This SSB pathway has first been derived by Yasmin Anstruthler.}

\begin{eqnarray}
& &  (3,\bar{3},1,1,1,1,1,1), ~~ 
(3,\bar{3},1,1,1,1,1),  ~~ (3,1,1,\bar{3},1,1), \nonumber \\
& &  (1,3,\bar{3},1,1) ~~ {\rm and} ~~  (1,3,1,\bar{3})
\label{vev}
\end{eqnarray}

These models have not been 
discussed in the literature so far and have potential interesting 
phenomenology.

\subsubsection{${\cal N}=1$ chiral $Z_n$ models}

To tabulate the
possible models for each value of $n$, we first show that a
proper embedding ($i.e$., ${\bf 6=6}^{*}$) for ${\bf 4=}({\bf 1},\alpha
^{i},\alpha ^{j},\alpha ^{k})$ results when $i+j+k=n$. To do this we use the fact
that the conjugate model has $i\rightarrow i^{\prime }=n-i,$ $j\rightarrow
j^{\prime }=n-j$ and $k\rightarrow k^{\prime }=n-k.$ Summing we find $%
i^{\prime }+j^{\prime }+k^{\prime }=3n-(i+j+k)=2n.$ But from ${\bf 6}$=(${\bf 4}%
\otimes {\bf 4})_{antisym}$ we find ${\bf 6=}(\alpha ^{i},\alpha ^{j},\alpha
^{k},\alpha ^{j+k},\alpha ^{i+k},\alpha ^{i+j}),$ but $i+j=n-k=k^{\prime }.$
Likewise $i+k=j^{\prime }$ and $j+k=i^{\prime }$ so ${\bf 6=}(\alpha
^{i},\alpha ^{j},\alpha ^{k},\alpha ^{i^{\prime }},\alpha ^{j^{\prime
}},\alpha ^{k^{\prime }})$ and this is ${\bf 6}^{*}$ up to an automorphism
which is sufficient to provide a proper embedding (or to provide real scalars in the
non-SUSY models). Models with $i+j+k=n$
(we will call these partition models) are always chiral, with total
chirality (number of chiral states) $\chi =3N^{2}n$ except in the case where $n$ is even and one of $i$%
, $j$, or $k$ is $n/2$ where $\chi =2N^{2}n.$ (No more than one of $i$, $j$,
and $k$ can be $n/2$ since they sum to $n$ and are all positive.) This
immediately gives us a lower bound on the number of chiral models at fixed $%
n $. It is the number of partitions of $n$ into three non-negative
integers. There is another class of models with $i^{\prime }=k$ and $%
j^{\prime }=2j,$ and total chirality $\chi =N^{2}n;$ for example a $Z_{9}$
orbifold with ${\bf 4=}({\bf 1},\alpha ^{3},\alpha ^{3},\alpha ^{6}).$ And
there are a few other sporadically occurring cases like $M_{124}^{6}$, which
typically have reduced total chirality, $\chi <3N^{2}n$. 
Such "nonpartition'' - i.e. neither partition nor double partition - models
can fail other more subtle constraints on consistent embedding  \cite{FK03},
but we list them here because they have vanishing anomaly coefficients and
vanishing one loop $\beta$ functions, and so are still of phenomenological interest
from the gauge theory model building perspective.

\bigskip

We now list all the ${\cal N}=1$, $Z_{n}$ orbifold models up to $n=12$ along with the
total chirality of each model, (see Table 1).

A systematic
search  through $n \leq 7$ yields four models that can result a in three-family
MSSM. They are $M_{111}^{3}$, $%
M_{122}^{5}, $ $M_{123}^{6},$ and $M_{133}^{7}$. There may be many
more models with sensible phenomenology at larger $n$, and we have given
one example $M_{333}^{9}$, with particularly simple 
spontaneous symmetry breaking, that is also a member of an infinite series of
models $M_{\frac{n}{3}\frac{n}{3}\frac{n}{3}}^{n}$, which all can lead to
three-family $MSSM$s. The value of $\sin ^{2}\theta _{W}$ at $SU^n(3)$ unification
was calculated for all these three family models in \cite{Kephart:2001qu}. 
This completes the summary of 
${\cal N}=1$ chiral $Z_n$ models, so we now proceed to investigate chiral $Z_n$ models
with no remaining supersymmetry.

\subsubsection{${\cal N}=0, 1$ chiral models for abelian product groups}

\bigskip

 Now let us consider abelian orbifold groups of order $g \leq 12,$ that
are
not just $Z_{n}$. There are only four, but they will be sufficient to
teach
us how to deal with this type of orbifold. We will search for both
${\cal N}%
=1$ and ${\cal N}=0$ models. Three groups,  
$Z_{2}\times Z_{4},$ $Z_{3}\times Z_{3}$, and $Z_{2}\times Z_{2}\times Z_{3}$ 
fit our requirements. We
have dispensed with $Z_{2}\times Z_{2}\times Z_{2}$ since all its fermionic
representations are real and it cannot produce chiral models.

First for $Z_{2}\times Z_{4},$ we can write elements as ($\alpha
^{i},\beta
^{i^{\prime }})$ where $\alpha ^{2}=1,$ and $\beta ^{4}=1.$ As usual, the
supersymmetry after orbifolding is determined by the embeddings. These
are
of the form:
\[
{\bf 4}=((\alpha ^{i},\beta ^{i^{\prime }}),(\alpha ^{j},\beta
^{j^{\prime
}}),(\alpha ^{k},\beta ^{k^{\prime }}),(\alpha ^{l},\beta ^{l^{\prime
}})).
\]
If all four entries are nontrivial ${\cal N}=0$ $SUSY$ results, if one
is
trivial, then we have ${\cal N}=1.$ We can think of the $SUSY$ breaking
as a
two step process, where we first embed the $\alpha $'$s$ in the ${\bf
4}$
and then the $\beta $'$s.$ Let us proceed this way and include only the
partition, and possibly double partition models. (As we noted above, the
nonpartition models have potential pathologies.) Thus for the $\alpha
$'$s$
we must have either ${\bf 4}_{{\bf \alpha }_{{\bf 1}}}=(-1,-1,-1,-1)$ or
$%
{\bf 4}_{{\bf \alpha }_{2}}=(1,1,-1,-1).$ The ${\bf 4}_{{\bf \alpha
}_{{\bf 1%
}}}$ results in ${\cal N}=0$ $SUSY,$ while ${\bf 4}_{{\bf \alpha }_2}$
gives
${\cal N}=2.$ We do not include trivial $Z_n$ factors {\bf 4}=(1,1,1,1) in the 
discussion, since these models contain very little new information.
[Note, for any product groups $Z_n \times Z_m$, the $\alpha$'s of $Z_n$ 
must be self conjugate in the $\bf 6$, as are the $\beta$'s of $Z_m$.
Hence, the full $\bf 6$ is self conjugate since the subgroups $Z_n $ and $ Z_m$
are orthogonal. This generalizes to more complicated products $Z_n \times Z_m \times Z_p \times ... $.]

Now for the $\beta $'$s.$ These are to be combined with the $\alpha
$'$s,$
so we must consider the ${\bf 4}_{{\bf \alpha }_{{\bf 1}}}$ and ${\bf
4}_{%
{\bf \alpha }_{{\bf 2}}}$ separately. For ${\bf 4}_{{\bf \alpha }_{{\bf
1}}},
$ the inequivalent ${\bf 4}_{{\bf \beta }}$'s are ${\bf 4}_{{\bf \beta
}_{%
{\bf 1}}}=(\beta ,\beta ,\beta ,\beta )$ and ${\bf 4}_{\beta
_{2}}=(1,\beta
,\beta ,\beta ^{2}).$ [Models with ${\bf 4}=(1,1,\beta ^{2},\beta ^{2})$ are
uninteresting since they all are nonchiral.]
Both cases have ${\cal N}=0$ $SUSY$ since we were
already at ${\cal N}=0$ after the ${\bf 4}_{{\bf \alpha }_{{\bf 1}}}$
embedding. For ${\bf 4}_{{\bf \alpha }_{{\bf 2}}}$ we find five possible
inequivalent embeddings, again we can have ${\bf 4}_{\beta _{{\bf 1}%
}}=(\beta ,\beta ,\beta ,\beta )$ or ${\bf 4}_{\beta _{2}}=(1,\beta
,\beta
,\beta ^{2}),$ but now we can also have ${\bf 4}_{\beta _{3}}=(1,\beta
^{2},\beta ,\beta )$, ${\bf 4}_{\beta _{4}}=(\beta ,\beta ,1,\beta
^{2})$ and ${\bf 4}_{\beta _{5}}=(\beta^2 ,\beta ,1,\beta)$.
The embeddings ${\bf 4}_{\beta _{{\bf 1}}}$, ${\bf 4}_{\beta _{4}}$ 
and ${\bf 4}_{\beta _{5}}$ lead to
${\cal N%
}=0$ $SUSY$ while ${\bf 4}_{\beta _{2}}$ and  ${\bf 4}_{\beta _{3}}$
leave $%
{\cal N}=1$ $SUSY$ unbroken. A similar analysis can be carried out for
$Z_{3}%
\times Z_{3}$, and $Z_{2}\times Z_{2}\times Z_{3},$ with the obvious
generalization to a triple embedding for $Z_{2}\times Z_{2}\times
Z_{3}.$

For $Z_{3}\times Z_{3}$ there are five models. We can choose  ${\bf
4}_{{\bf %
\alpha }}=(1,\alpha ,\alpha ,\alpha )$ as the embedding of the first
$Z_{3}.$
Then the embedding of the second $Z_{3}$ can be ${\bf 4}_{{\bf \beta
}_{{\bf %
1}}}=(1,\beta ,\beta ,\beta ),$ ${\bf 4}_{{\bf \beta }_{{\bf 2}}}=(\beta
,1,\beta ,\beta ),$ ${\bf 4}_{{\bf \beta }_{{\bf 3}}}=(1,1,\beta ,\beta
^{2}),$ ${\bf 4}_{{\bf \beta }_{{\bf 4}}}=(\beta ,1,1,\beta ^{2}),$ or
${\bf %
4}_{{\bf \beta }_{{\bf 5}}}=(\beta ^{2},1,1,\beta ).$ The first and
third
result in ${\cal N}=1$ $SUSY$ models while the other three are ${\cal
N}=0.$

For $Z_{2}\times Z_{2}\times Z_{3}$ we find 9 chiral models. Rather than
belabor the details, we summarize all our results for $Z_{2}\times Z_{4},$
$Z_{3}%
\times Z_{3}$, and $Z_{2}\times Z_{2}\times Z_{3}$ in Table 5.

\bigskip
\bigskip

{\bf Summary}

\bigskip
\bigskip

We have now completed our task of summarizing all ${\cal N}=0$ and
${\cal N}=1$  $SUSY$ chiral models of phenomenological interest
derivable from orbifolding $AdS_{5}\times S^{5}$ with abelian
orbifold group $\Gamma $ of order
$o(\Gamma )\leq 12.$ The models fall into three classes:
partition models, double partition models, and non-partition models
as determined by how the equation $i+j+k+l=sn$ is satisfied by the
embedding where $s=1$ for
partition models, $s=2$ for double partition models and $s$ is
non--integer for non-partition models.
For $Z_{n}$ orbifolds with ${\cal N}=1$ $SUSY$, there are 53
partition models, and 7 non-partition models,
and for ${\cal N}=0$ $SUSY$, we find 54 partition, 11 double
partition, and 13 non-partition
models. The non-partition models have potential pathologies if they are
to be interpreted as coming from string theory, but they still may be of
phenomenological and technical interest, so they have been included in
our classification of $Z_{n}$ models. See also the related discussions in 
\cite{Kakushadze:2000mc} and \cite{Pickering:2001aq}.

The non--$Z_{n}$  abelian product groups of interest (we only consider
partition models here) with $g \leq 12$ are
$Z_{2}\times Z_{4}$ with five  ${\cal N}=0$ and two
${\cal N}=1$ chiral models; $Z_{3}\times Z_{3}$ with three ${\cal N}=0$ and two ${\cal N}=1$ chiral
models, and $Z_{2}\times Z_{2}\times Z_{3}$ with seven ${\cal N}=0$ and two
${\cal N}=1$ chiral models.

The relation to the SM and MSSM  have been explored in some detail for  
$Z_{n}$
models with $g \leq 7$, but we have only given a few examples with
$g >7$, and have
indicated how to build abelian orbifold models for any $g$. 
Two $Z_8$ models have been introduced, which can lead to the right
Weinberg angle, when broken down to the SM.
These results should be
useful to model builders and phenomenologists alike.

\newpage

\bigskip

\begin{tabular}{|c||c|c|c|c|}
\hline
$M_{1111}^{4}(F)$ & 1 & $\alpha $ & $\alpha ^{2}$ & $\alpha ^{3}$ \\ 
\hline\hline
$1$ &  & $\times ^{4}$ &  &  \\ \hline
$\alpha $ &  &  & $\times ^{4}$ &  \\ \hline
$\alpha ^{2}$ &  &  &  & $\times ^{4}$ \\ \hline
$\alpha ^{3}$ & $\times ^{4}$ &  &  &  \\ \hline
\end{tabular}
\\
\\
Table 2: Fermion content for the model $M_{1111}^{4}.$ The $\times ^{4}$
entry at the (1,$\alpha)$ position means the model contains $4(3,\bar{3}%
,1,1)$ of $SU^{4}(3),$ etc. Hence, the fermions in this table are 4[$(3,\bar{%
3},1,1)+(1,3,\bar{3},1)+(1,1,3,\bar{3})+(\bar{3},1,1,3)$]. Diagonal entries
do not occur in this model but, if they did, an $\times $ at say ($\alpha ^{2}$,$\alpha
^{2})$ would correspond to $(1,8+1,1,1)$, etc. See models below.

\bigskip

\bigskip

\begin{tabular}{|c||c|c|c|c|}
\hline
$M_{1111}^{4}(S)$ & 1 & $\alpha $ & $\alpha ^{2}$ & $\alpha ^{3}$ \\ 
\hline\hline
$1$ &  &  & $\times ^{6}$ &  \\ \hline
$\alpha $ &  &  &  & $\times ^{6}$ \\ \hline
$\alpha ^{2}$ & $\times ^{6}$ &  &  &  \\ \hline
$\alpha ^{3}$ &  & $\times ^{6}$ &  &  \\ \hline
\end{tabular}
\\
\\
Table 3: Scalar content of the model $M_{1111}^{4}.$

\bigskip

%%%%%%%%%%%%%%%%%%%%%%%%%%%%%%%%%%%%%%%

\newpage

\small

%%%
\begin{tabular}{|c||c|c|c|}
\hline
$n$ & {\bf 4} & $\chi /N^{2}$ & comment \\ \hline\hline
4 & $(\alpha ,\alpha ,\alpha ,\alpha )$ & 16 & $i+j+k+l=3;$ one model $%
(i=j=k=l=1)$ \\ \hline
4 & $(\alpha ,\alpha ,\alpha ,\alpha ^{3})^*$ & 8 & nonpartition model \\ 
\hline
5 & $(\alpha ^{i},\alpha ^{j},\alpha ^{k},\alpha ^{l})$ & 20 & $i+j+k+l=5;$
1 models \\ \hline
6 & $(\alpha ^{i},\alpha ^{j},\alpha ^{k},\alpha ^{l})$ & $\leq $24 & $%
i+j+k+l=6;$ 2 models \\ \hline
6 & $(\alpha ,\alpha ,\alpha ^{3},\alpha ^{5})^{*}$ & 6 
& nonpartition \\ \hline
6 & $(\alpha ,\alpha ^{2},\alpha ^{3},\alpha ^{5})^{*}$ & 6 & 
nonpartition \\ \hline
6 & $(\alpha ,\alpha ^{3},\alpha ^{4},\alpha ^{4})$ & 24 & double partition
\\ \hline
7 & $(\alpha ^{i},\alpha ^{j},\alpha ^{k},\alpha ^{l})$ & 28 & $i+j+k+l=7;$
3 models \\ \hline
8 & $(\alpha ^{i},\alpha ^{j},\alpha ^{k},\alpha ^{l})$ & $\leq $ $32$ & $%
i+j+k+l=8;$ 5 models \\ \hline
8 & $(\alpha ,\alpha ^{2},\alpha ^{3},\alpha ^{6})^{*}$ & 16 & nonpartition
\\ \hline
8 & $(\alpha ^{2},\alpha ^{2},\alpha ^{2},\alpha ^{6})^{*}$ & 16 & analog of 
$Z_{4}$ $(\alpha ,\alpha ,\alpha ,\alpha ^{3})$ model \\ \hline
8 & $(\alpha ,\alpha ^{4},\alpha ^{5},\alpha ^{6})$ & 32 & double partition
\\ \hline
9 & $(\alpha ^{i},\alpha ^{j},\alpha ^{k},\alpha ^{l})$ & 36 & $i+j+k+l=9;$
7 models \\ \hline
9 & $(\alpha ,\alpha ^{3},\alpha ^{4},\alpha ^{7})^{*}$ & 36 & nonpartition
\\ \hline
9 & $(\alpha ,\alpha ^{4},\alpha ^{6},\alpha ^{7})$ & 36 & double partition
\\ \hline
10 & $(\alpha ^{i},\alpha ^{j},\alpha ^{k},\alpha ^{l})$ & $\leq 40$ & $%
i+j+k+l=10;$ 9 models \\ \hline
10 & $(\alpha ,\alpha ^{3},\alpha ^{8},\alpha ^{8})$ & 40 & double partition
\\ \hline
10 & $(\alpha ,\alpha ^{5},\alpha ^{6},\alpha ^{8})$ & 40 & double partition
\\ \hline
11 & $(\alpha ^{i},\alpha ^{j},\alpha ^{k},\alpha ^{l})$ & 44 & $i+j+k+l=11;$
11 models \\ \hline
%%%
12 & $(\alpha ^{i},\alpha ^{j},\alpha ^{k},\alpha ^{l})$ & $\leq $ $48$ & $%
i+j+k+l=12;$ 15 models \\ \hline
12 & $(\alpha ,\alpha ^{4},\alpha ^{9},\alpha ^{10})$ & 48 & double partition
\\ \hline
12 & $(\alpha ,\alpha ^{5},\alpha ^{9},\alpha ^{9})$ & 48 & double partition
\\ \hline
12 & $(\alpha ,\alpha ^{6},\alpha ^{7},\alpha ^{10})$ & 48 & double partition
\\ \hline
12 & $(\alpha ,\alpha ^{6},\alpha ^{8},\alpha ^{9})$ & $36$ & double
partition \\ \hline
12 & $(\alpha ,\alpha ^{7},\alpha ^{8},\alpha ^{8})$ & 48 & double partition
\\ \hline
12 & $(\alpha ^{2},\alpha ^{6},\alpha ^{8},\alpha ^{8})$ & 36 & double
partition \\ \hline
12 & $(\alpha ,\alpha ,\alpha ^{5},\alpha ^{9})^*$ & 48 & nonpartition \\ 
\hline
12 & $(\alpha ,\alpha ^{3},\alpha ^{5},\alpha ^{9})^{*}$ & 24 & nonpartition
\\ \hline
12 & $(\alpha ,\alpha ^{3},\alpha ^{7},\alpha ^{11})^{*}\ $ & 24 & 
nonpartition \\ \hline
12 & $(\alpha ,\alpha ^{5},\alpha ^{5},\alpha ^{9})^{*}$ & 48 & nonpartition
\\ \hline
12 & $(\alpha ^{2},\alpha ^{2},\alpha ^{6},\alpha ^{10})^{*}$ & 12 & 
nonpartition \\ \hline
12 & $(\alpha ^{2},\alpha ^{3},\alpha ^{4},\alpha ^{9})^{*}$ & 24 & 
nonpartition \\ \hline
12 & $(\alpha ^{2},\alpha ^{4},\alpha ^{6},\alpha ^{10})^{*}$ & 24 & 
nonpartition \\ \hline
12 & $(\alpha ^{3},\alpha ^{3},\alpha ^{3},\alpha ^{9})^{*}$ & 24 & 
nonpartition \\ \hline
\end{tabular}
\\
\\

\large

Table 4. 
All chiral ${\cal N}=0,$ $Z_{n}$ orbifold models with $n\leq12.$  
The 13 non--partition models are marked with an asterisk(*).
For further explanations see text.
\newpage
%%%%%%%%%%%%%%%%%%%%%%%%%%%%%%%%%%%%%%%%%%%%%%%%%%%%%%

\bigskip

\footnotesize

\begin{tabular}{|c||c|c|c|}
\hline
$Group$ & {\bf 4} & $\chi /N^{2}$ & ${\cal N}$ \\ \hline\hline
$Z_{2}\times Z_{4}$ & $(-1,-1,-1,-1)\times (\beta ,\beta ,\beta ,\beta
)$ &
32 & $0$ \\ \hline
$Z_{2}\times Z_{4}$ & $(-1,-1,-1,-1)\times ({\bf 1},\beta ,\beta ,\beta
^{2})
$ & 16 & 0 \\ \hline
$Z_{2}\times Z_{4}$ & $(1,1,-1,-1)\times (\beta ,\beta ,\beta ,\beta )$
& 32
& $0$ \\ \hline
$Z_{2}\times Z_{4}$ & $(1,1,-1,-1)\times ({\bf 1},\beta ,\beta ,\beta
^{2})$
& 16 & $1$ \\ \hline
$Z_{2}\times Z_{4}$ & $(1,1,-1,-1)\times ({\bf 1},\beta ^{2},\beta
,\beta )$
& 16 & $1$ \\ \hline
$Z_{2}\times Z_{4}$ & $(1,1,-1,-1)\times (\beta ,\beta ,1,\beta ^{2})$ &
16
& $0$ \\ \hline
$Z_{2}\times Z_{4}$ & $(1,1,-1,-1)\times (\beta ,\beta ^{2} ,1,\beta)$ &
16
& $0$ \\ \hline
$Z_{3}\times Z_{3}$ & $({\bf 1},\alpha ,\alpha ,\alpha )\times (1,\beta
,\beta ,\beta )$ & 27 & $1$ \\ \hline
$Z_{3}\times Z_{3}$ & $({\bf 1},\alpha ,\alpha ,\alpha )\times (\beta
,1,\beta ,\beta )$ & 36 & $0$ \\ \hline
$Z_{3}\times Z_{3}$ & $({\bf 1},\alpha ,\alpha ,\alpha )\times
(1,1,\beta
,\beta ^{2})$ & 18 & 1 \\ \hline
$Z_{3}\times Z_{3}$ & $({\bf 1},\alpha ,\alpha ,\alpha )\times (\beta
,1,1,\beta ^{2})$ & 36 & 0 \\ \hline
$Z_{3}\times Z_{3}$ & $({\bf 1},\alpha ,\alpha ,\alpha )\times (\beta
^{2},1,1,\beta )$ & 36 & $0$ \\ \hline
%$Z_{2}\times Z_{2}\times Z_{3}$ & $(1,1,1,1)\times (1,1,1,1)\times
%(1,\gamma
%,\gamma ,\gamma )$ & $48$ & $1$ \\ \hline
%$Z_{2}\times Z_{2}\times Z_{3}$ & $(1,1,1,1)\times (1,1,-1,-1)\times %
%(1,\gamma ,\gamma ,\gamma )$ & $48$ & $1$ \\ \hline
%$Z_{2}\times Z_{2}\times Z_{3}$ & $(1,1,1,1)\times (1,-1,-1,1)\times %
%(1,\gamma ,\gamma ,\gamma )$ & $48$ & 1 \\ \hline
%$Z_{2}\times Z_{2}\times Z_{3}$ & $(1,1,1,1)\times (-1,-1,1,1)\times %
%(1,\gamma ,\gamma ,\gamma )$ & $48$ & 0 \\ \hline
%$Z_{2}\times Z_{2}\times Z_{3}$ & $(1,1,1,1)\times (-1,-1,-1,-1)\times %
%
%(1,\gamma ,\gamma ,\gamma )$ & $48$ & $0$ \\ \hline
$Z_{2}\times Z_{2}\times Z_{3}$ & $(1,1,-1,-1)\times (1,1,-1,-1)\times %

(1,\gamma ,\gamma ,\gamma )$ & $48$ & $1$ \\ \hline
$Z_{2}\times Z_{2}\times Z_{3}$ & $(1,1,-1,-1)\times (-1,1,1,-1)\times %

(1,\gamma ,\gamma ,\gamma )$ & $48$ & $0$ \\ \hline
$Z_{2}\times Z_{2}\times Z_{3}$ & $(1,1,-1,-1)\times (-1,-1,-1,-1)\times
(1,\gamma ,\gamma ,\gamma )$ & $48$ & 0 \\ \hline
$Z_{2}\times Z_{2}\times Z_{3}$ & $(-1,-1,1,1)\times (-1,-1,1,1)\times %

(1,\gamma ,\gamma ,\gamma )$ & $48$ & 0 \\ \hline
$Z_{2}\times Z_{2}\times Z_{3}$ & $(-1,-1,1,1)\times (-1,-1,-1,-1)\times
(1,\gamma ,\gamma ,\gamma )$ & $48$ & 0 \\ \hline
$Z_{2}\times Z_{2}\times Z_{3}$ & $(1,1,-1,-1)\times
(-1,-1,1,1)\times %
(1,\gamma ,\gamma ,\gamma )$ & $48$ & 0 \\ \hline
$Z_{2}\times Z_{2}\times Z_{3}$ & $(1,1,-1,-1)\times
(1,-1,-1,1)\times %
(1,\gamma ,\gamma ,\gamma )$ & $48$ & 1 \\ \hline
$Z_{2}\times Z_{2}\times Z_{3}$ & $(-1,1,1,-1)\times
(-1,1,-1,1)\times %
(1,\gamma ,\gamma ,\gamma )$ & $48$ & 0 \\ \hline
%$Z_{2}\times Z_{2}\times Z_{3}$ & $(-1,1,1,-1)\times
%(1,-1,-1,1)\times %
%(1,\gamma ,\gamma ,\gamma )$ & $?$ & 0 \\ \hline
$Z_{2}\times Z_{2}\times Z_{3}$ & $(-1,-1,-1,-1)\times
(-1,-1,-1,-1)\times %
(1,\gamma ,\gamma ,\gamma )$ & $48$ & 0 \\ \hline
\end{tabular}
\\
\\

\large

Table 5.: All chiral ${\cal N}=0$ and ${\cal N}=1$ $SUSY$ partition models for product
orbifolding groups $Z_{2}\times Z_{4},$ $Z_{3}\times Z_{3}$, and $Z_{2}%
\times Z_{2}\times Z_{3}$, where the embedding is nontrivial in all factors. Our notation is:
${\bf 4}=
((\alpha ^{i}),(\alpha ^{j}),(\alpha ^{k}),(\alpha ^{l}))\times
((\beta ^{i^{\prime }}),(\beta^{j^{\prime
}}),(\beta ^{k^{\prime }}),(\beta ^{l^{\prime
}}))
=((\alpha ^{i},\beta ^{i^{\prime }}),(\alpha ^{j},\beta
^{j^{\prime
}}),(\alpha ^{k},\beta ^{k^{\prime }}),(\alpha ^{l},\beta ^{l^{\prime
}}))$, etc.

\newpage

\subsubsection{Abelian SUSY models}

Orbifolded $AdS_{5}\times S^{5}$ is fertile ground for building models
which
can potentially test string theory. When one
bases the models on the conformal field theory gotten from
the large $N$ expansion of the $AdS/CFT$ correspondence \cite{Maldacena1997},
stringy effects
can
show up at the scale of a few $TeV$. The first three-family model of
this
type had ${\cal N}=1$ $SUSY$ and was based on a $Z_{3}$ orbifold 
\cite{Kachru:1998ys}. However, since
then
the most studied examples have been models without supersymmetry based
on
both abelian 
\cite{first,WS,vafa,F2,F3}
and non-abelian 
\cite{nonabelian,nonabelian1,nonabelian2,nonabelian3}
orbifolds of $AdS_{5}\times S^{5}$. Here we
return to $Z_{n}$ orbifolds with supersymmetry, and systematically study
those cases with chiral matter ($i. e.$, those with an imbalance between chiral supermultiplets and
anti-chiral
supermultiplets). We classify all cases up
to $n\leq 12$, and show that several of these contain the minimal
supersymmetric standard model ($MSSM$) with three families.

\bigskip

For details of the construcion of quiver gauge theories
from orbifolds, see Section 2.

\bigskip

The replacement of $S^{5}$ by $S^{5}/\Gamma $ reduces the
supersymmetry to
${\cal N}=$ 0, 1 or 2 from the initial ${\cal N}=4$, depending on how $%
\Gamma $ is embedded in the $SO(6) \sim SU(4)$ isometry of
$S^{5}$.
The case of interest here is ${\cal N}=1$ $SUSY$ where $\Gamma $
completely
embeds in an $SU(3)$ subgroup of the $SU(4)$ isometry. $I.e$., we embed
rep($%
\Gamma )\rightarrow {\bf 4}$ of $SU(4)$ as ${\bf 4}=({\bf 1},{\bf r})$
where
${\bf 1}\ $is the trivial irrep of $\Gamma $ and ${\bf r}$ is a
nontrivial
(but possibly reducible) three dimensional representation of $\Gamma .$
The
chiral supermultiples generated by this embedding are given by
\begin{equation}
\sum_{i}{\bf 4}\otimes R_{i}
\end{equation}
where the set $ R_{i} $ runs over all the irreps of $\Gamma .$ For our
choice, $\Gamma =Z_{n},$ the irreps are all one dimensional
and as a consequence of the choice of $N$ in the $1/N$ expansion,
the gauge group \cite{Lawrence} is $U(N)^{n}$. Chiral models require the {\bf 4} to be
complex (${\bf 4}\neq {\bf 4}^{*})$ while a proper embedding requires
${\bf 6%
}={\bf 6}^{*}$ where ${\bf 6}$=(${\bf 4}\otimes {\bf 4})_{antisym}$.,
(even
though the {\bf 6} does not enter the model). This information is
sufficient
for us to begin our investigation. We will choose $N=3$ throughout, and if we use the fact that
$SU_{L}(2)$ and $U_{Y}(1)$ are embedded in diagonal subgroups
$SU^{p}(3)$
and $SU^{q}(3)$ of the initial $SU^{n}(3)$, the ratio $\frac{\alpha
_{2}}{%
\alpha _{Y}}$ turns out to be $\frac{p}{q}.$ This implies the initial
value of $\theta_{W}$ is calculable in these models and
$\sin
^{2}\theta
_{W}$ satisfies
\begin{equation}
\sin ^{2}\theta _{W}=\frac{3}{3+5\left( \frac{p}{q}\right) }.
\end{equation}
On the other hand, a more standard approach is to break the initial $SU^{n}(3)$
to $SU_{C}(3)\otimes SU_{L}(3)\otimes SU_{R}(3)$ where $SU_{L}(3)$ and
$SU_{R}(3)$ are embedded in diagonal subgroups
$SU^{p}(3)$
and $SU^{q}(3)$ of the initial $SU^{n}(3)$. We then
embed all of $SU_{L}(2)$ in $SU_{L}(3)$ but $\frac{1}{3}$ of $U_{Y}(1)$ in $SU_{L}(3)$
and the other $\frac{2}{3}$ in $SU_{R}(3)$. This modifies the $\sin
^{2}\theta
_{W}$ formula to:
\begin{equation}
\sin ^{2}\theta _{W}=\frac{3}{3+5\left( \frac{\alpha _{2}}{\alpha _{Y}}\right) }
=\frac{3}{3+5\left( \frac{3p}{p+2q}\right) }
\end{equation}
Note, this coincides with the previous result when $p=q$. We will use the later result when calculating $\sin
^{2}\theta
_{W}$ below. A similar relation holds for Pati--Salam type models \cite{nonabelian3}.

We now go through a systematic list of low $n$ examples as we did in the ${\cal N} =1$ case
to familiarize ourselves with ${\cal N} =0$ models.
First a $Z_{2}$ orbifold has only real representations and therefore will not
yield a chiral model. (Note, although all matter is in chiral
supermultiplet, if there is a left-handed supermultiplet to match each
right-handed supermultiplet, then the model has no overall chirality,
$i.e$., it is vectorlike.)

Next, for $\Gamma =Z_{3}$ the choice ${\bf 4}=({\bf 1},\alpha ,\alpha
,\alpha )$ with $N=3$ where $\alpha =e^{\frac{2\pi i}{3}}$ (in what
follows
we will write $\alpha =e^{\frac{2\pi i}{n}}$ for $Z_{n})$, yields the
three
family trinification \cite{Glashow:1984gc} model of \cite{Kachru:1998ys}, but without sufficient scalars
to break the gauge symmetry to the MSSM. Here the initial value of
$\sin ^{2}\theta _{W}=\frac{3}{8}$, so unification at the $TeV$ scale is also problematic.
There is another chiral model for ${\bf
4}=({\bf 1}%
,\alpha ,\alpha ,\alpha ^{2})$ but it can have at most one chiral family.

$Z_{4}$ orbifolds allow only one chiral model with ${\cal N}=1$ $SUSY.$
It
is generated by ${\bf 4}=({\bf 1},\alpha ,\alpha ,\alpha ^{2})$ but can
have
at most two chiral families.

There are two chiral models for $Z_{5}$, and they are fixed by choosing
$%
{\bf 4}=({\bf 1},\alpha ,\alpha ,\alpha ^{3})$ and ${\bf 4}=({\bf
1},\alpha,\alpha ^{2},\alpha ^{2}).$ Before looking at these in detail, let us pause to define a useful notation
for
classifying models. The initial model (before any symmetry breaking) is
completely fixed (recall we always are taking $N=3$) by the choice of
$Z_{n}$
and the embedding  ${\bf 4=}({\bf 1},\alpha ^{i},\alpha ^{j},\alpha
^{k}),$
so we define the model to by $M_{ijk}^{n}.$ We immediately observe that the
conjugate model $M_{n-i,n-j,n-k}^{n}$ contains the same information, so
we
need not study it separately.

Returning now to $Z_{5},$ the two models are $M_{113}^{5}$ and
$M_{122}^{5}.$
(Other inconsistent models are eliminated by requiring ${\bf 6}={\bf
6}^{*}$
keeping the number of models limited.) We find no pattern of spontaneous
symmetry breaking ($SSB$) for $M_{113}^{5}$ that yields the $MSSM$, but $%
M_{122}^{5}$ is more interesting. The matter content of $M_{122}^{5}$ is
shown in Table 1.

For each entry, ($\times$), in the table, we have a chiral supermultiplet
in a bifundamental representation of $SU^{5}(3)$. Specifically, for an entry
at the $i^{th}$ column and $j^{th}$ row we have a bifundamental
representation of $SU_{i}(3)\times SU_{j}(3).$ We can arbitrarily assign the
fundamental representation to the rows and the anti-fundamental
representation to the columns. If $i=j$ the bifundamental is all in
$SU_{i}(3)$ and hence is a singlet plus adjoint of $SU_{i}(3)$. Hence the
complete set of chiral supermultiplets represented by Table 1 is:
\begin{eqnarray}
&&[(3,\bar{3},1,1,1)+(1,3,\bar{3},1,1)+(1,1,3,\bar{3},1)\\
&&+(1,1,1,3,\bar{3}) +(\bar{3},1,1,1,3)] \\
&&+2[(3,1,\bar{3},1,1)+(1,3,1,\bar{3},1)+(1,1,3,1,\bar{3}) \\
&&+(\bar{3} ,1,1,3,1)+(1,\bar{3},1,1,3)] \\
&&+[(1+8,1,1,1,1)+(1,1+8,1,1,1)\\
&&+(1,1,1+8,1,1)+(1,1,1,1+8,1)+(1,1,1,1,1+8)].
\end{eqnarray}

\bigskip

\begin{tabular}{|c||c|c|c|c|c|}
\hline
$M_{122}^{5}$ & 1 & $\alpha $ & $\alpha ^{2}$ & $\alpha ^{3}$ & $\alpha
^{4}$
\\ \hline\hline
$1$ & $\times $ & $\times $ & $\times \times $ &  &  \\ \hline
$\alpha $ &  & $\times $ & $\times $ & $\times \times $ &  \\ \hline
$\alpha ^{2}$ &  &  & $\times $ & $\times $ & $\times \times $ \\ \hline

$\alpha ^{3}$ & $\times \times $ &  &  & $\times $ & $\times $ \\ \hline

$\alpha ^{4}$ & $\times $ & $\times \times $ &  &  & $\times $ \\ \hline

\end{tabular}

\bigskip

Table 1: Matter content for the model $M_{122}^{5}.$ The $\times \times $ entry at the
(1,$%
\alpha ^{2})$ position means the model contains $2(3,1,\bar{3},1,1)$ of
$%
SU^{5}(3),$ etc. The diagonal entries are $(8+1,1,1,1,1)$, etc.

\bigskip

A vacuum expectation value ($VEV$) for $(3,\bar{3},1,1,1)$ breaks the
symmetry to $SU_{D}(3)\otimes SU_{3}(3)\otimes SU_{4}(3)\otimes
SU_{5}(3)$
and a further $VEV$ for $(1,3,\bar{3},1)$ breaks the
symmetry to
$SU_{D}(3)\otimes SU_{D^{\prime }}(3)\otimes SU_{5}(3).$
Identifiny $SU_{C}(3)$ with $SU_{D}(3)$, embedding $SU_{L}(2)$ in $SU_{D^{\prime }}(3)$
and $U_{Y}(1)$ partially in $SU_{5}(3)$ and partially in $SU_{D^{\prime }}(3)$ gives an
initial value of
$\sin ^{2}\theta _{W}=\frac{2}{7}=.286$,
and implies a unification scale around
$2\times 10^{7} GeV$.

The remaining
chiral multiplets are
\begin{equation}
3[(3,\bar{3},1)+(1,3,\bar{3})+(\bar{3},1,3)]
\end{equation}
We have sufficient octets to continue the symmetry breaking all the
way
to $SU(3)\otimes SU(2)\otimes U(1),$ and so arrive at the MSSM with
three
families (plus additional vector-like matter that is heavy and therefore not in
the
low energy spectrum).

 Before analyzing more models in detail, it is useful to tabulate the
possible model for each value of $n$. To this end, note we always have a
proper embedding ($i.e$., ${\bf 6=6}^{*}$) for ${\bf 4=}({\bf 1},\alpha
^{i},\alpha ^{j},\alpha ^{k})$ when $i+j+k=n$. To show this we use the
fact
that the conjugate model has $i\rightarrow i^{\prime }=n-i,$
$j\rightarrow
j^{\prime }=n-j$ and $k\rightarrow k^{\prime }=n-k.$ Summing we find $%
i^{\prime }+j^{\prime }+k^{\prime }=3n-(i+j+k)=2n.$ From ${\bf
6}$=(${\bf 4}%
\otimes {\bf 4})_{antisym}$ we find ${\bf 6=}(\alpha ^{i},\alpha
^{j},\alpha
^{k},\alpha ^{j+k},\alpha ^{i+k},\alpha ^{i+j}),$ but
$i+j=n-k=k^{\prime }.$
Likewise $i+k=j^{\prime }$ and $j+k=i^{\prime }$ so ${\bf 6=}(\alpha
^{i},\alpha ^{j},\alpha ^{k},\alpha ^{i^{\prime }},\alpha ^{j^{\prime
}},\alpha ^{k^{\prime }})$ and this is ${\bf 6}^{*}$ up to an
automorphism
which is sufficient to provide vectorlike matter in this sector in the
non-SUSY models and here provide a proper embedding. Models with
$i+j+k=n$ (we will call these partition models)
are always chiral, with total chirality $\chi =3N^{2}n$ except in the
case
where $n$ is even and one of $i$, $j$, or $k$ is $n/2$ where $\chi
=2N^{2}n.$
(No more than one of $i$, $j$, and $k$ can be $n/2$ since they add to
$n$
and are all positive.) This immediately gives us a lower bound on the
number
of chiral models at fixed $n$. It is the the number of partitions of $n$
into three non-negative integers. There is another class of models with
$%
i^{\prime }=k$ and $j^{\prime }=j^{2},$ and total chirality $\chi
=N^{2}n;$
for example a $Z_{9}$ orbifold with ${\bf 4=}({\bf 1},\alpha ^{3},\alpha
^{3},\alpha ^{6}).$ And there are a few other sporadically occurring
cases
like $M_{124}^{6}$, which typically have reduced total chirality, $\chi
<3N^{2}n$.

We now tabulate all the $Z_{n}$ orbifold models up to $n=12$ along with
the
total chirality of each model, (see Table 2).

\bigskip

\bigskip

\normalsize

\begin{tabular}{|c||c|c|c|}
\hline
$n$ & {\bf 4} & $\chi /N^{2}$ & comment \\ \hline\hline
3 & $({\bf 1},\alpha ,\alpha ,\alpha )$ & 9 & $i+j+k=3;$ one model
$(i=j=k=1)
$ \\ \hline
3 & $({\bf 1},\alpha ,\alpha ,\alpha ^{2})^{*}$ & 3 &  \\ \hline
4 & $({\bf 1},\alpha ,\alpha ,\alpha ^{2})$ & 8 & $i+j+k=4;$ one model
\\
\hline
5 & $({\bf 1},\alpha ^{i},\alpha ^{j},\alpha ^{k})$ & 15 & $i+j+k=5;$ 2 models
\\
\hline

6 & $({\bf 1},\alpha ^{i},\alpha ^{j},\alpha ^{k})$ & 12 & $i+j+k=6;$ 3
models
\\ \hline

6 & $({\bf 1},\alpha ,\alpha ^{2},\alpha ^{4})^{*}$ & 6 &  \\ \hline
6 & $({\bf 1},\alpha ^{2},\alpha ^{2},\alpha ^{4})^{*}$ & 6 &  \\ \hline

7 & $({\bf 1},\alpha ^{i},\alpha ^{j},\alpha ^{k})$ & 21 & $i+j+k=7;$ 4
models \\ \hline
8 & $({\bf 1},\alpha ^{i},\alpha ^{j},\alpha ^{k})$ & $\leq 24$ &
$i+j+k=8;$
5 models \\ \hline
9 & $({\bf 1},\alpha ^{i},\alpha ^{j},\alpha ^{k})$ & 27 & $i+j+k=9;$ 7
models \\ \hline
9 & $({\bf 1},\alpha ,\alpha ^{4},\alpha ^{7})^{*}$ & 27 &  \\ \hline
9 & $({\bf 1},\alpha ^{3},\alpha ^{3},\alpha ^{6})^{*}$ & 9 &  \\ \hline

10 & $({\bf 1},\alpha ^{i},\alpha ^{j},\alpha ^{k})$ & 30 & $i+j+k=10;$
8
models \\ \hline
11 & $({\bf 1},\alpha ^{i},\alpha ^{j},\alpha ^{k})$ & 33 & $i+j+k=11;$
10
models \\ \hline
12 & $({\bf 1},\alpha ^{i},\alpha ^{j},\alpha ^{k})$ & $\leq 36$ &
$i+j+k=12;
$ 12 models \\ \hline
12 & $({\bf 1},\alpha ^{2},\alpha ^{4},\alpha ^{8})^{*}$ & 12 &  \\
\hline
12 & $({\bf 1},\alpha ^{4},\alpha ^{4},\alpha ^{8})^{*}$ & 12 &  \\
\hline
\end{tabular}

\large

\bigskip

Table 2. All chiral $Z_{n}$ orbifold models with $n\leq 12.$ Three of
the $%
n=8$ models have $\chi /N^{2}=24;$ the other two have $\chi /N^{2}=16.$
Of
the 12 models with $i+j+k=12,$ three have models $\chi /N^{2}=24$ and
the
other nine have $\chi /N^{2}=36.$  Of the 60 models 53 are partition models, while the
remaining 7 models that do not satisfy $i+j+k=n$, are marked with an asterisk (*).

\bigskip

We have analyzed all the models for $Z_{6}$ orbifolds, and find only one
of
phenomenological interest. It is $M_{123}^{6}$ , where the matter
multiplets
are shown in Table 3.

\bigskip

\bigskip
\begin{tabular}{|c||c|c|c|c|c|c|}
\hline
$M_{123}^{6}$ & 1 & $\alpha $ & $\alpha ^{2}$ & $\alpha ^{3}$ & $\alpha
^{4}$
& $\alpha ^{5}$ \\ \hline\hline
1 & $\times $ & $\times $ & $\times $ & $\times $ &  &  \\ \hline
$\alpha $ &  & $\times $ & $\times $ & $\times $ & $\times $ &  \\
\hline
$\alpha ^{2}$ &  &  & $\times $ & $\times $ & $\times $ & $\times $ \\
\hline
$\alpha ^{3}$ & $\times $ &  &  & $\times $ & $\times $ & $\times $ \\
\hline
$\alpha ^{4}$ & $\times $ & $\times $ &  &  & $\times $ & $\times $ \\
\hline
$\alpha ^{5}$ & $\times $ & $\times $ & $\times $ &  &  & $\times $ \\
\hline
\end{tabular}

\bigskip Table 3: Chiral supermultiplets for the model $M_{123}^{6}.$

\bigskip

$VEV$s for $(3,\bar{3},1,1,1,1)$, $(1,1,3,\bar{3},1,1)$ and $(1,1,1,1,3,\bar{3})$ break the
symmetry to $SU_{12}(3)\otimes SU_{34}(3)\otimes SU_{56}(3)\ $where $%
SU_{12}(3)$ is the diagonal subgroup of $SU_{1}(3)\otimes SU_{2}(3),$
etc.,
and the remaining chirality resides in
$3[(3,\bar{3},1)+(1,3,\bar{3})+(\bar{3%
},1,3)]$. Again, we have octets of all six initial $SU(3)$s, so we can
break
to a three-family $MSSM$, but with $\sin ^{2}\theta _{W}=\frac{3}{8}$. There is no other pattern
of $SSB$ that gives three families.

\bigskip

For $Z_{7}$, we again find only one model that can break to a
three-family $%
MSSM$. It is $M_{133}^{7}$, with matter shown in Table 4.

\bigskip

\bigskip
\begin{tabular}{|c||c|c|c|c|c|c|c|}

\hline
$M_{133}^{7}$ & 1 & $\alpha $ & $\alpha ^{2}$ & $\alpha ^{3}$ & $\alpha
^{4}$
& $\alpha ^{5}$ & $\alpha ^{6}$ \\ \hline\hline
1 & $\times $ & $\times $ &  & $\times \times $ &  &  &  \\ \hline
$\alpha $ &  & $\times $ & $\times $ &  & $\times \times $ &  &  \\
\hline
$\alpha ^{2}$ &  &  & $\times $ & $\times $ &  & $\times \times $ &  \\
\hline
$\alpha ^{3}$ &  &  &  & $\times $ & $\times $ &  & $\times \times $ \\
\hline
$\alpha ^{4}$ & $\times \times $ &  &  &  & $\times $ & $\times $ &  \\
\hline
$\alpha ^{5}$ &  & $\times \times $ &  &  &  & $\times $ & $\times $ \\
\hline
$\alpha ^{6}$ & $\times $ &  & $\times \times $ &  &  &  & $\times $ \\
\hline
\end{tabular}

\large

\bigskip 

Table 4: Chiral supermultiplets for the model $M_{133}^{7}.$

\bigskip

First $VEV$s for $(3,\bar{3},1,1,1,1,1)$ and
$(1,1,3,\bar{3},1,1,1)$ breaks
the symmetry to $SU_{12}(3)\otimes SU_{34}(3)\otimes SU_{5}(3)\otimes
SU_{6}(3)\otimes SU_{7}(3)$. Then a $VEV$ for $(1,1,1,3,\bar{3})$ breaks
this to $SU_{12}(3)\otimes SU_{34}(3)\otimes SU_{5}(3)\otimes
SU_{67}(3)$,
and leaves the following multiplets chiral
\begin{eqnarray}
&&(3,\bar{3},1,1)+(1,3,\bar{3},1)+(1,1,3,\bar{3})+(\bar{3},1,1,3)\\
&&+2[(3,\bar{3}% ,1,1)+(1,3,1,\bar{3})+(\bar{3},1,1,3)]
\end{eqnarray}
 Finally, a $VEV$ for $(1,3,\bar{3},1)$ yields the $MSSM$ with three
chiral
families. Identifying $SU_{C}(3)$ with $SU_{12}(3)$ 
and embedding $SU_{L}(2)$ in $SU_{67}(3)$ and
$U_{Y}(1)$ in $SU_{345}(3)$ gives $\sin ^{2}\theta _{W}=\frac{7}{22}=.318$ and implies a
unification scale around
$ 10^{10} GeV$.

\bigskip

The $n>7$ models can be analyzed in a similar manner. The total number
of
models grows with $n$. There are also potentially interesting examples
for $%
N>4$. Although we have not made a systematic study of the models with
$n\geq %
8$, we close with a rather compact example of a three-family $MSSM$ at
$n=9$%
. The model is $M_{123}^{6}$ with matter given in Table 5.

\bigskip

\footnotesize

\begin{tabular}{|c||c|c|c|c|c|c|c|c|c|}
\hline
$M_{333}^{9}$ & 1 & $\alpha $ & $\alpha ^{2}$ & $\alpha ^{3}$ & $\alpha
^{4}$
& $\alpha ^{5}$ & $\alpha ^{6}$ & $\alpha ^{7}$ & $\alpha ^{8}$ \\
\hline\hline
1 & $\times $ &  &  & $\times \times \times $ &  &  &  &  &  \\ \hline
$\alpha $ &  & $\times $ &  &  & $\times \times \times $ &  &  &  &  \\
\hline
$\alpha ^{2}$ &  &  & $\times $ &  &  & $\times \times \times $ &  &  &
\\
\hline
$\alpha ^{3}$ &  &  &  & $\times $ &  &  & $\times \times \times $ &  &
\\
\hline
$\alpha ^{4}$ &  &  &  &  & $\times $ &  &  & $\times \times \times $ &
\\
\hline
$\alpha ^{5}$ &  &  &  &  &  & $\times $ &  &  & $\times \times \times $
\\
\hline
$\alpha ^{6}$ & $\times \times \times $ &  &  &  &  &  & $\times $ &  &
\\
\hline
$\alpha ^{7}$ &  & $\times \times \times $ &  &  &  &  &  & $\times $  &  \\
\hline
$\alpha ^{8}$ &  &  & $\times \times \times $ &  &  &  &  &  & $\times $ \\
\hline
\end{tabular}

\large

\bigskip

\bigskip

Table 5: Chiral supermultiplets for the model $M_{333}^{9}.$

\bigskip

$VEV$s for the octets of $SU_{k}(3)$, where $k=2,3,5,6,8,$ and $9$
breaks
the symmetry to $SU_{1}(3)\otimes SU_{4}(3)\otimes SU_{7}(3).$ (Each
chiral
supermultiplet of representation $R$ contains one chiral fermion
multiplet
in representation $R$, and two scalar (we need not distinguish scalars
from
pseudoscalars here) multiplets in representation $R$. Therefore, there
are
two scalar octets for each $SU_{k}(3)$. When one octet of $SU_{k}(3)$ is
given a $VEV$, gauge freedom can be used to diagonalize that $VEV$.
However,
there is not enough gauge freedom left to diagonalize the $VEV$ of the
second octet of the same $SU_{k}(3)$. Therefore $SU_{k}(3)$ can be
broken
completely by $SU_{k}(3)$s for the two octets). The chirality remaining
after this octet breaking is
$3[(3,\bar{3},1)+(1,3,\bar{3})+(\bar{3},1,3)]$.
Further symmetry breaking via single octets of $SU_{3}(3)$ and
$SU_{7}(3)$
leads us to the three-family $MSSM$, but with $\sin ^{2}\theta _{W}=\frac{3}{8}$.
Note that any model of the type
$M_{%
\frac{n}{3}\frac{n}{3}\frac{n}{3}}^{n}$ can be handled this way, and can
lead to a three-family $MSSM$. Hence this provides an infinite class of
three-family models.

We have found 60 chiral $Z_{n}$  orbifolds for $n\leq 12$. A systematic
search up through $n=7$ yields four models that can result in
three-family
minimal supersymmetric standard models. They are $M_{111}^{3}$,
$M_{122}^{5},
$ $M_{123}^{6},$ and $M_{133}^{7}$. We suspect there are many more
models
with sensible phenomenology at larger $n$, and we have pointed out one
example $M_{333}^{9}$, which is particularly simple in its spontaneous
symmetry breaking, and is also a member of an infinite series of models
$M_{%
\frac{n}{3}\frac{n}{3}\frac{n}{3}}^{n}$, which all can lead to
three-family
$MSSM$s.
Orbifolded $AdS/CFT$ models hold great promise for testing string theory not far above the
the $TeV$ scale, and they have inspired models \cite{Aldazabal:2000sa} with phenomenology ranging from light magnetic
monopoles \cite{Kephart:2001ix} to an anomalous muon magnetic moment\cite{Kephart:2001iu}. They have also provided a check on higher loop $\beta$
functions \cite{Pickering:2001aq}, and raised interesting cosmological questions \cite{Kakushadze:2000mc}.

\newpage

\subsection{Non abelian model building}

We shall present what we believe is the minimal three-family $AdS/CFT$ model
compactified on a nonabelian orbifold $S^{5}/(Q\times Z_{3})$.
Nontrivial irreps of the discrete nonabelian group
$Q\times Z_{3}$ are identified with the $4$ of $SU(4)$ $R$
symmetry to break all supersymmetries, and the scalar content of the
model is sufficient to break the gauge symmetry to the standard model.
According to the conformality hypothesis the progenitor $SU(4)^3 \times SU(2)^{12}$
theory becomes conformally invariant at an infra-red fixed point of the
renormalization group.

\subsubsection{Non-Abelian non-SUSY models}

\bigskip

For construction of quiver gauge theories from orbifolds,
please refer back again to Section 2.

\bigskip

It was conjectured in \cite{first} that 
at least a subset of the resultant nonsupersymmetric
${\cal N} = 0$ theories are conformal even for finite $N$. Some
first steps to check this idea were made in \cite{WS}.
Model-building based on abelian
$\Gamma$ was studied further in \cite{vafa,F2,F3}, arriving in \cite{F3}
at an $SU(3)^7$ model based on $\Gamma = Z_7$
which has three families of chiral fermions,
a correct value for ${\rm sin}^2 \theta$ and a conformal scale $\sim
10$~~TeV.

The case of non-abelian orbifolds bases on non-abelian $\Gamma$ is
somewhat more mathematically
sophisticated. However, we shall show here that it can be handled
equally
as systematically as the abelian case and leads to richer structures
and interesting results.

We consider all non-abelian discrete groups of order $g < 32$. These
are described in detail in \cite{books,books2,books3,books4,FK,nonabelian2}. 
There are exactly 45 such
non-abelian groups. Because the gauge group arrived at
by this construction\cite{vafa}
is $\otimes_i U(Nd_i)$ where $d_i$ are the dimensions of the
irreducible representations of $\Gamma$, one can expect to arrive
at models such as the Pati-Salam
$SU(4) \times SU(2) \times SU(2)$ type\cite{PS}
by choosing $N = 2$ and combining two singlets
and a doublet in the {\bf 4}
of $SU(4)$. Indeed we shall show that such an
accommodation of the standard model
is possible by using a non-abelian $\Gamma$.

The procedures for building a model within such a conformality approach
are:
(1) Choose $\Gamma$; (2) Choose a proper
embedding $\Gamma \subset SU(4)$ by assigning
the components of the {\bf 4} of $SU(4)$ to irreps of $\Gamma$,
while at the same time ensuring that the {\bf 6} of $SU(4)$ is real;
(3) Choose $N$, in the gauge group $\otimes_i SU(Nd_i)$. (4) Analyse the
patterns of spontaneous symmetry breaking.

In the present study we shall choose $N = 2$ and aim
at the gauge group $SU(4) \times SU(2) \times SU(2)$.
To obtain chiral fermions, it is necessary\cite{FK03}
that the {\bf 4} of $SU(4)$
be complex ${\bf 4} \neq {\bf 4}^*$. Actually this condition is not
quite
sufficient to ensure chirality in the present case because
of the pseudoreality of $SU(2)$. We must ensure that the {\bf 4} is
not just pseudoreal.

This last condition means that many of our 45 candidates for $\Gamma$
do not lead to chiral fermions. For example, $\Gamma = Q_{2n} \subset
SU(2)$
has irreps of appropriate dimensionalities for our purpose
but it will not sustain chiral fermions under $SU(4)\times SU(2) \times
SU(2)$
because these
irreps are all, like $SU(2)$, pseudoreal.\footnote{Note that
were we using $N \geq 3$
then a pseudoreal {\bf 4} would give chiral fermions.}
Applying the rule that {\bf 4} must be
neither real nor pseudoreal
leaves a total of only 19 possible non-abelian discrete groups of
order $g \leq 31$. The smallest group which avoids pseudoreality
has order $g = 16$ but gives only two families. The
technical details
of our systematic search will be postponed
to a future publication. Here we shall present only the
simplest interesting non-abelian case
which has $g = 24$ and gives three chiral families in a
Pati-Salam-type model\cite{PS}.

Before proceeding to the details of
the specific $g = 24$ case, it is worth reminding the reader that
the Conformal Field Theory (CFT) that it exemplifies should
be free of all divergences,
even logarithmic ones, if the conformality conjecture is correct,
and be completely finite. Further
the theory is originating from a superstring theory
in a higher-dimension (ten) and contains gravity
\cite{antoniadis,antoniadis2,antoniadis3,antoniadis4,antoniadis5,
antoniadis6,antoniadis7,antoniadis8,antoniadis9,antoniadis10,
antoniadis11,antoniadis12,antoniadis13,antoniadis14,antoniadis15,
antoniadis16}
by compactification of the
higher-dimensional graviton already contained in that superstring
theory. In
the CFT as we derive it, gravity is absent because we have not kept
these graviton modes - of course, their influence on high-energy
physics experiments is generally completely negligible unless the
compactification
scale is ``large''\cite{antoniadis,antoniadis16}; here we shall neglect the effects
of
gravity.

To motivate our model it is instructive to comment on the choice of
$\Gamma$ and
on the
choice of embedding.

If we embed only four singlets of $\Gamma$ in the {\bf 4} of $SU(4)$
then this
has the effect of abelianizing $\Gamma$ and the gauge group obtained in
the
chiral sector of the
theory is $SU(N)^q$. These cases can be interesting but have already
been
studied\cite{vafa,F2}.
Thus, we require at least one irrep of $\Gamma$ to have $d_i \geq 2$ in
the
embedding.

The only $\Gamma$ of order $g \leq 31$ with a {\bf 4} is $Z_5
\tilde{\times}
Z_4$ and this embedding
leads to a non-chiral theory. This leaves only embeddings with two
singlets and
a doublet,
a triplet and a singlet or two doublets.

The third of these choices leads to richer structures for
low order $\Gamma$. Concentrating on them shows that
of the chiral models possible, those from groups of low order
result in an insufficient number (below three) of
chiral families.

The first group that can lead to exactly three families occurs at
order $g = 24$ and is $\Gamma = Z_3 \times Q$ where $Q (\equiv Q_4)$
is the group of unit quarternions which is the smallest dicyclic group
$Q_{2n}$.

There are several potential models due to the different choices for the
{\bf 4}
of $SU(4)$ but only the case {\bf 4} = $(1\alpha, 1^{'}, 2\alpha)$ leads
to three families so let
us describe this in some detail:

Since $Q \times Z_3$ is a direct product group, we can write the irreps
as $R_i
\otimes \alpha^{a}$
where $R_i$ is a $Q$ irrep and $\alpha^{a}$ is a $Z_3$ irrep. We write
$Q$
irreps as $1,~1^{'},~1^{''},~
1^{'''},~2$ while the irreps of $Z_3$ are all singlets which we call
$\alpha, \alpha^2, \alpha^3 = 1$.
Thus $Q \times Z_3$ has fiveteen irreps in all and the gauge group will
be of Pati-Salam type for $N = 2$.

If we wish to break all supersymmetry, the {\bf 4} may not contain the
trivial
singlet of
$\Gamma$.
Due to permutational symmetry among the singlets
it is sufficiently general to choose {\bf 4} =
$(1\alpha^{a_1},~1^{'}\alpha^{a_2},~2\alpha^{a_3})$
with $a_1 \neq 0$.

To fix the $a_i$ we note that the scalar sector of the theory which is
generated
by the
{\bf 6} of $SU(4)$ can be used as a constraint since the {\bf 6} is
required to be real. This leads to
$a_1 + a_2 = - 2a_3 ({\rm mod}~3)$. Up to permutations in the chiral
fermion
sector the most
general choice is $a_1 = a_3= +1$ and $a_2 = 0$. Hence our choice of
embedding
is
\begin{equation}
{\bf 4} = (1\alpha,~1^{'},~2\alpha)
\label{embed}
\end{equation}
with
\begin{equation}
{\bf 6} = (1^{'}\alpha,~2\alpha,~2\alpha^{2},~1^{'}\alpha^{2})
\label{Six}
\end{equation}
which is real as required.

We are now in a position to summarize the particle content of the
theory. The
fermions are given by
\begin{equation}
\sum_I~{\bf 4}\times R_I
\end{equation}
where the $R_I$ are all the irreps of $\Gamma = Q \times Z_3$. This is:
\[
\sum_{i=1}^{3} [(2_{1}\alpha^{i},2_{2}\alpha^{i})
+(2_{3}\alpha^{i},2_{4}\alpha^{i})+(2_{2}\alpha^{i},2_{1}\alpha^{i})
+(2_{4}\alpha^{i},2_{3}\alpha^{i})+(4\alpha^{i},\overline{4}\alpha^{i})] 
\]

\begin{equation}
+ \sum_{i=1}^{3} \sum_{a=1}^{4} [(2_{a}\alpha^{i},
2_{a}\alpha^{i+1})+(2_{a}\alpha^{i},4\alpha^{i+1})
+ (\bar{4}\alpha^{i},2_{a}\alpha^{i+1})]
\label{Fermions}
\end{equation} 

 It is convenient to represent the chiral portions of these in a given
diagram (see Figure 1).

The scalars are given by
\begin{equation}
\sum_I~{\bf 6}\times R_I
\end{equation}
and are:
\begin{equation}
\sum_{i=1}^{3} \sum_{j=1(j\neq i)}^{3} 
[(2_{1}\alpha^{i},2_{2}
\alpha^{j})+(2_{2}\alpha^{i}, 2_{1}\alpha^{j})+(2_{3}\alpha^{i},
2_{4}\alpha^{j})+(2_{4}\alpha^{i},2_{3}\alpha^{j})
\end{equation}
\begin{equation}
+(2_{2}\alpha^{i},2_{1}\alpha^{i})+(2_{4}\alpha^{i},2_{3}\alpha^{i})] 
\end{equation}
\begin{equation}
+ \sum_{i=1}^{3} \sum_{j=1(j\neq i)}^{3} 
\{ \sum_{a=1}^{4}[(2_{a}\alpha^{i},4\alpha^{j})
+\bar{(4}\alpha^{i},2_{a}\alpha^{j} )]
+(4\alpha^{i}, \bar{4}\alpha ^{i}) \}
\label{Scalars}
\end{equation}
which is easily checked to be real.

The gauge group $SU(4)^3 \times SU(2)^{12}$ with chiral fermions of
Eq.(\ref{Fermions}) and scalars of Eq.(\ref{Scalars}) 
is expected to acquire confromal invariance
at an infra-red fixed point of the  renormalization group, as discussed in \cite{first}.

To begin our examination of the symmetry breaking we first
observe
that if we break the three $SU(4)$s to the totally diagonal $SU(4)$,
then chirality in the
fermionic sector is lost. To avoid this we break $SU_{1}(4)$ completely
and then break $SU_{\alpha }(4)\times SU_{\alpha ^{2}}(4)$ to
its diagonal subgroup $SU_{D}(4).$ The first of these steps can be
achieved with VEVs of the form $[(4_{1},2_{b}\alpha ^{k})+h.c.]$ where
we
are free to choose $b$, but $k$ must be $1$ or $2$ since there
are no $(4_{1},2_{b}\alpha ^{k=0})$ scalars. The second step requires an
$SU_{D}(4)$ singlet VEV from
($\overline{4}_{\alpha }$,4$_{\alpha^{2}})$ and/or
(4$_{\alpha }$, $\overline{4}_{\alpha ^{2}})$. Once we
make a choice for $b$ (we take $b=4$), the remaining chiral fermions
are,
in an intuitive notation:

\bigskip

\noindent $\ \sum_{a=1}^{3}\left[ (2_{a}\alpha \
,1,4_{D})+(1,2_{a}\alpha ^{-1},\overline{4_{D}})\right] $

\bigskip

\noindent which has the same content as as a three family Pati-Salam
model,
though with a separate $SU_{L}(2)\times SU_{R}(2)$ per family.

To further reduce the symmetry we must arrange to break to a single
$SU_{L}(2)$ and a single $SU_{R}(2).$ This is achieved by modifying step
one where $SU_{1}(4)$ was broken. Consider the block diagonal
decomposition of $SU_{1}(4)$ into
$SU_{1L}(2) \times SU_{1R}(2).$ The representations
$(2_{a}\alpha ,4_{1})$ and $(2_{a}\alpha ^{-1},4_{1})$ then decompose as
$$(2_{a}\alpha ,4_{1})\rightarrow (2_{a}\alpha ,2,1)+(2_{a}\alpha ,1,2)$$
and 
$$(2_{a}\alpha ^{-1},4_{1})\rightarrow (2_{a}\alpha
^{-1},,2,1)+(2_{a}\alpha ^{-1},1,2).$$ Now if we give $VEVs$ of equal
magnitude to the $(2_{a}\alpha ,,2,1),$ $a=1,2,3$, and equal magnitude
$VEVs$ to the $(2_{a}\alpha ^{-1},1,2)$ $a=1,2,3,$ we break
$SU_{1L}(2) \times \prod\limits_{a=1}^{3}SU(2_{a}\alpha )$ to a
single $SU_{L}(2)$ and we break
$SU_{1R}(2) \times \prod\limits_{a=1}^{3}SU(2_{a}\alpha )$ to a
single $SU_{R}(2).$  Finally, $VEVs$ for $(2_{4}\alpha ,2,1)\ $and
$(2_{4}\alpha ,1,2)$ as well as $(2_{4}\alpha ^{-1},2,1)\ $and
$(2_{4}\alpha ^{-1},1,2)$ insures that both $SU(2_{4}\alpha )$ and
$SU(2_{4}\alpha ^{-1})$ are broken and that only three families remain
chiral. The final set of chiral fermions is then
$3[(2,1,4)+(1,2,\bar{4})]$ with gauge symmetry
$SU_{L}(2) \times SU_{R}(2) \times SU_{D}(4).$

 To achieve the final reduction to the standard model, an adjoint VEV\
from
($\overline{4}_{\alpha }$,4$_{\alpha ^{2}})$ and/or
(4$_{\alpha }$,$\overline{4}_{\alpha ^{2}})$
is used to break $SU_{D}(4)$ to the
$SU(3)\times U(1),$ and a right handed doublet is used to break
$SU_{R}(2).$

While this completes our analysis of symmetry breaking, it is worthwhile
noting the degree of constraint imposed on the symmetry and particle
content of a model as the number of irreps $N_{R}$ of the discrete group
$\Gamma $ associated with the choice of orbifold changes. The number of
guage groups grows linearly in $N_{R}$, the number of scalar irreps
grows roughly quadratically with $N_{R}$, and the chiral fermion content
is highly $\Gamma $ dependent. If we require the minimal $\Gamma $ that
is large enough for the model generated to contain the fermions of the
standard model and have sufficient scalars to break the symmetry to that
of the standard model, then $\Gamma = Q \times Z_{3}$ appears to
be that minimal choice\cite{nonabelian,nonabelian2}.

Although a decade ago the chances of testing string theory seemed at
best remote, recent progress has given us hope that such tests may
indeed be possible in AdS/CFTs. The model provided here demonstrates the
standard model can be accomodated in these theories and suggests the
possibility of a rich spectrum of new physics just around the TeV
corner.

\newpage

\subsection{Non-Abelian Groups with order $g\leq 31$}

From any good textbook on finite groups\cite{books,books2,books3,books4} 
we may find a tabulation of
the number of finite groups as a function of the order g, the number of
elements in the group. Up to order 31 there is a total of 93 different
finite groups of which slightly over one half (48) are abelian.

Amongst finite groups, the non-abelian examples have the advantage
of non-singlet irreducible representations which can be used to inter-relate
families. Which such group to select is based on simplicity: the minimum
order and most economical use of representations.

Let us first dispense with the abelian groups. These are all made up from
the basic unit $Z_p$, the order p group formed from the $p^{th}$ roots
of unity. It is important to note that the product $Z_p\times Z_q$ is identical
to $Z_{pq}$ if and only if p and q have no common prime factor.

If we write the prime factorization of g as:
\begin{equation}
g = \prod_{i}p_i^{k_i}
\end{equation}
where the product is over primes, it follows that the number
$N_a(g)$ of inequivalent abelian groups of order g is given by:
\begin{equation}
N_a(g) = \prod_{k_i}P(k_i)
\end{equation}
where $P(x)$ is the number of unordered partitions of $x$.
For example, for order $g = 144 = 2^43^2$ the value would be
$N_a(144) = P(4)P(2) = 5\times2 = 10$. For $g\leq31$ it is simple
to evaluate $N_a(g)$ by inspection. $N_a(g) = 1$ unless g contains
a nontrivial power ($k_i\geq2$) of a prime. These exceptions are:
$N_a(g = 4,9,12,18,20,25,28) = 2; N_a(8,24,27) = 3$; and $N_a(16) = 5$.
This confirms that:
\begin{equation}
\sum_{g = 1}^{31}N_a(g) = 48
\end{equation}
We do not consider the abelian cases further in this section.\\

Of the nonabelian finite groups, the best known are perhaps the
permutation groups $S_N$ (with $N \geq 3$) of order $N!$
The smallest non-abelian finite group is $S_3$ ($\equiv D_3$),
the symmetry of an equilateral triangle with respect to all
rotations in a three dimensional sense. This group initiates two
infinite series, the $S_N$ and the $D_N$. Both have elementary
geometrical significance since the symmetric permutation group
$S_N$ is the symmetry of the N-plex in N dimensions while the dihedral group
$D_N$ is the symmetry of the planar N-agon in 3 dimensions.
As a family symmetry, the $S_N$ series becomes uninteresting rapidly
as the order and the dimensions of the representions increase. Only $S_3$
and $S_4$ are of any interest as symmetries associated with the particle
spectrum\cite{Pak}, also the order (number of elements) of the $S_N$ groups
grow factorially with N. The order of the dihedral groups increase only
linearly with N and their irreducible representations are all one- and
two- dimensional. This is reminiscent of the representations of the
electroweak $SU(2)_L$ used in Nature.

Each $D_N$ is a subgroup of $O(3)$ and has a counterpart double dihedral (also known as dicyclic)
group $Q_{2N}$, of order $4N$, which is a subgroup of the double covering
$SU(2)$ of $O(3)$.

With only the use of $D_N$, $Q_{2N}$, $S_N$ and the tetrahedral group T ( of
order
12, the even permutations subgroup of $S_4$ ) we find 32 of the 45
nonabelian groups up to order 31, either as simple groups or as
products of simple nonabelian groups with abelian groups:
(Note that $D_6 \simeq Z_2 \times D_3, D_{10} \simeq Z_2 \times D_5$ and $
D_{14} \simeq Z_2 \times D_7$ ) Some of these groups are firmiliar from crystalography and chemistry, but the 
nonabelian groups that do not embed in in $SU(2)$ are less likely to have seen wide usage.

\begin{center}

\begin{tabular}{||c||c||}   \hline
g & \\    \hline
$6$  & $D_3 \equiv S_3$\\  \hline
$8$ & $ D_4 , Q = Q_4 $\\    \hline
$10$& $D_5$\\   \hline
$12$&  $D_6, Q_6, T$ \\ \hline
$14$& $D_7$\\  \hline
$16$& $D_8, Q_8, Z_2 \times D_4, Z_2 \times Q$\\  \hline
$18$& $D_9, Z_3 \times D_3$\\  \hline
$20$& $D_{10}, Q_{10}$ \\  \hline
$22$& $D_{11}$\\  \hline
$24$& $D_{12}, Q_{12}, Z_2 \times D_6, Z_2 \times Q_6, Z_2 \times T$,\\  \hline
 & $Z_3 \times D_4, Z_3 \times Q, Z_4 \times D_3, S_4$\\  \hline
$26$& $D_{13}$\\  \hline
$28$& $D_{14}, Q_{14}$ \\  \hline
$30$& $D_{15}, D_5 \times Z_3, D_3 \times Z_5$\\  \hline
\end{tabular}

\end{center}

\bigskip
\bigskip

There remain thirteen others formed by twisted products of abelian factors.
Only certain such twistings are permissable, namely (completing all $g \leq 31$
)

$$\begin{tabular}{||c||c||}   \hline
g & \\    \hline
$16$  & $Z_2 \tilde{\times} Z_8$ (two, excluding $D_8$), $Z_4 \tilde{\times}
Z_4, Z_2 \tilde{\times}(Z_2 \times Z_4)$
(two)\\  \hline
$18$ & $Z_2 \tilde{\times} (Z_3 \times Z_3)$\\    \hline
$20$&  $Z_4 \tilde{\times} Z_5$ \\   \hline
$21$&  $Z_3 \tilde{\times} Z_7$ \\    \hline
$24$&  $Z_3 \tilde{\times} Q, Z_3 \tilde{\times} Z_8, Z_3 \tilde{\times} D_4$
\\  \hline
$27$&  $ Z_9 \tilde{\times} Z_3, Z_3 \tilde{\times} (Z_3 \times Z_3)$ \\
\hline
\end{tabular}$$

It can be shown that these thirteen exhaust the classification of {\it all}
inequivalent finite groups up to order thirty-one\cite{books4}.

Of the 45 nonabelian groups, the dihedrals ($D_N$) and double dihedrals
($Q_{2N}$), of order 2N and 4N respectively,
form the simplest sequences. In particular, they fall into subgroups of
$O(3)$ and $SU(2)$ respectively,
the two simplest nonabelian continuous groups.

For the $D_N$ and $Q_{2N}$, the multiplication tables, as derivable from the
character tables,
are simple to express in general. $D_N$, for odd N, has two singlet
representations $1,1^{'}$ and $m = (N-1)/2$
doublets $2_{(j)}$ ($1 \leq j \leq m$). The multiplication rules are:

\begin{equation}
1^{'}\times 1^{'} = 1 ; ~~~1^{'}\times 2_{(j)} = 2_{(j)}
\end{equation}
\begin{equation}
2_{(i)}\times 2_{(j)} = \delta_{ij} (1 + 1^{'}) + 2_{(min[i+j,N-i-j])}
+ (1 - \delta_{ij}) 2_{(|i - j|)}
\end{equation}
\noindent

For even N, $D_N$ has four singlets $1, 1^{'},1^{''},1^{'''}$ and $(m - 1)$
doublets
$2_{(j)}$ ($ 1 \leq j \leq m - 1$)where $m = N/2$ with multiplication rules:

\begin{equation}
1^{'}\times 1^{'} = 1^{''} \times 1^{''} = 1^{'''} \times 1^{'''} = 1
\end{equation}
\begin{equation}
1^{'} \times 1^{''} = 1^{'''}; 1^{''} \times 1^{'''} = 1^{'}; 1^{'''} \times
1^{'} = 1^{''}
\end{equation}
\begin{equation}
1^{'}\times 2_{(j)} = 2_{(j)}
\end{equation}
\begin{equation}
1^{''}\times 2_{(j)} = 1^{'''} \times 2_{(j)} = 2_{(m-j)}
\end{equation}
\begin{equation}
2_{(j)} \times 2_{(k)} = 2_{|j-k|} + 2_{(min[j+k,N-j-k])}
\end{equation}

\noindent
(if $k \neq j, (m - j)$)

\begin{equation}
2_{(j)} \times 2_{(j)} = 2 _{(min[2j,N-2j])} + 1 + 1^{'}
\end{equation}

\noindent
(if $j \neq m/2$ )

\begin{equation}
2_{(j)} \times 2_{(m - j)} = 2_{|m - 2j|} + 1^{''} + 1^{'''}
\end{equation}

\noindent
(if $j \neq m/2 $)

\begin{equation}
2_{m/2} \times 2_{m/2} = 1 + 1^{'} + 1^{''} + 1^{'''}
\end{equation}

\noindent
This last rule is possible only if m is even and hence if N is divisible by {\it
four}.\\

For $Q_{2N}$, there are four singlets $1$, $1^{'}$ ,$1^{''}$, $1^{'''}$ and
$(N - 1)$ doublets $2_{(j)}$ ($ 1 \leq j \leq (N-1) $).
The singlets have the multiplication rules:

\begin{equation}
1 \times 1 = 1^{'} \times 1^{'} = 1
\end{equation}
\begin{equation}
1^{''} \times 1^{''} = 1^{'''} \times 1^{'''} = 1^{'}
\end{equation}
\begin{equation}
 1^{'} \times 1^{''} = 1^{'''} ; 1^{'''} \times 1^{'} = 1^{''}
\end{equation}

\noindent
for $N = (2k + 1)$ but are identical to those for $D_N$ when N = 2k.

The products involving the $2_{(j)}$ are identical to those given
for $D_N$ (N even) above.

This completes the multiplication rules for 19 of the 45 groups. 
As they are not available in the literature, and somewhat tedious to work out, we have 
provided the complete multiplication tables for all the nonabelian groups
with order $g \leq 31$ in \cite{nonabelian2}.

\bigskip

\newpage

\bigskip
\bigskip
\bigskip

{\it Mathematical Theorem:}

\underline{A Pseudoreal $4$ of $SU(4)$ Cannot Yield Chiral Fermions.}

\bigskip

In \cite{FK03} it was proved that if the embedding in $SU(4)$ is such that
the ${\bf 4}${\bf \ }is real: ${\bf 4}={\bf 4}^{{\bf *}}$, then the
resultant fermions are always non-chiral. It was implied there that the
converse holds, that if ${\bf 4}$ is complex, ${\bf 4}={\bf 4}^{{\bf *}}$ ,
then the resulting fermions are necessarily chiral. Actually for $\Gamma $ $%
\subset $ $SU(2)$ one encounters the intermediate possibility that the {\bf 4%
} is {\it pseudoreal}. In the present section we shall show that if ${\bf 4}$
is pseudoreal then the resultant fermions are necessarily non-chiral. The
converse now holds: if the ${\bf 4}$ is neither real nor pseudoreal then the
resutant fermions are chiral.

\bigskip

For $\Gamma \subset SU(2)$ it is important that the embedding be contained
within the chain $\Gamma \subset SU(2)\subset SU(4)$ otherwise the embedding
is not a consistent one. One way to see the inconsistency is to check the
reality of the ${\bf 6}=({\bf 4}\otimes {\bf 4)}_{antisymmetric}$. If ${\bf 6%
}\neq {\bf 6}^{{\bf *}}${\bf \ }then the embedding is clearly improper. To
avoid this inconsistency it is sufficient to include in the ${\bf 4}$ of $%
SU(4)$ only complete irreducible representations of $SU(2)$.

\bigskip

An explicit example will best illustrate this propriety constraint on
embeddings. Let us consider $\Gamma =Q_{6}$, the dicyclic group of order $%
g=12$. This group has six inequivalent irreducible representations: $%
1,1^{\prime },1^{\prime \prime },1^{\prime \prime \prime },2_{1},2_{2}$. The
1, $1^{\prime }$, 2$_{1}$ are real. The $1^{\prime \prime }$ and $1^{\prime
\prime \prime }$ are a complex conjugate pair, The $2_{2}$ is pseudoreal. To
embed $\Gamma =Q_{6}\subset SU(4)$ we must choose from the special
combinations which are complete irreducible representations of $SU(2)$
namely 1, $2=2_{2}$, $3=1^{\prime }+2_{1}$ and $4=1^{\prime \prime
}+1^{\prime \prime \prime }+2_{2}$. In this way the embedding either makes
the ${\bf 4}$ of $SU(4)$ real {\it e.g}. $4=1+1^{\prime }+2_{1}\ $and the
theorem of \cite{FK03} applies, and non-chirality results, or the ${\bf 4}$
is pseudoreal {\it e.g}. $4=2_{2}+2_{2}$. In this case one can check that
the embedding is consistent because $({\bf 4}\otimes {\bf 4)}%
_{antisymmetric} $ is real. But it is equally easy to check that the product
of this pseudoreal ${\bf 4}$ with the complete set of irreducible
representations of $Q_{6}$ is again real and that the resultant fermions are
non-chiral.

The lesson is:

{\it To obtain chiral fermions from compactification on }${\it AdS}_{{\it 5}%
}\times S_{5}/\Gamma ${\it , the embedding of }${\it \Gamma }${\it \ in }$%
SU(4)${\it \ must be such that the 4 of }$SU(4)${\it \ is neither real nor
pseudoreal}.

Other cases where the {\bf 4} of $SU(4)$ is complex, but not a complete irrep of $SU(2)$ coming from $Q_6$, and where the  {\bf 6} of $SU(4)$ is real, are improper embeddings, but form perfectly good quiver gauge theories (not necessarily derivable from $AdS/CFT$). These theories may well be of phenomenological interest and are the analogs of the non-partition models related to the abelian orbifolded models. In the following we do not include these improper non-abelian orbifold model in our analysis, but one should keep them in mind as phenomenological viable gauge theories.

We will now enumerate all the possible minimal $N$,  {\cal N = 0} chiral models for non-abelian orbifolded $AdS_5\times S^5$ and then briefly consider breaking the resulting quiver gauge theory to the standard model. Our analysis concentrates on symmetry breaking via reduction to diagonal subgroups through either the Pati-Salam group $SU(4)\times SU(2)\times SU(2)$ or the trinification gauge group $SU(3)\times SU(3)\times SU(3)$ before reaching the standard model. While our list of proper ciral models is complete, we study only a limited number of possible patters on symmetry breaking, so our discussion should be used as a guide rather than a compendium. There are still many unexplored avenues for model building, both for abelian and non-abelian orbifolds.

\bigskip
\bigskip

\newpage

\bigskip
\bigskip
\bigskip

{\it Chiral Fermions for all nonabelian $g\leq 31$}

Looking at the full list of non-abelian discrete groups
of order $g \leq 31$ as given explicitly in \cite{FK} we see that
of the 45 such groups 32 are simple groups or semi-direct products thereof; these
32 are listed in the Table on page 4691 of \cite{FK}, and  reproduced in section 6.3 above.
The remaining 13 are formed as semi-direct product groups (SDPGs)
and are listed in the Table on page 4692 of \cite{FK} and in section 6.3. We shall
follow closely this classification.

\bigskip

Using the pseudoreality considerations of the previous section, we can pare
down the full list of 45 to only 19 which include 13 SDPGs. The
lowest order nonabelian group $\Gamma$ which can lead to
chiral fermions is $g = 16$. The only possible orders
$g \leq 31$ are the seven values:
$g = 16(5[5 SDPGs]),~~ 18(2[1 SDPG])$,

\noindent $20(1[1 SDPG]), ~~ 21(1[1 SDPG]), ~~ 24(6[3 SDPGs])$

\noindent $ 27(2[2 SDPGs]), ~~{\rm and} ~~ 30(2[0 SDPG])$. 

\bigskip

\noindent In parenthesis are the number of groups
at order $g$ and the number of these that are SDPGS is in square brackets;
they add to (19[13 SDPGs]). We shall proceed with the analysis 
systematically, in progressively increasing magnitude of $g$.

\bigskip
\bigskip

\newpage

\bigskip
\bigskip
\subsection{${\cal N}=0$ non-Abelian models}
\underline{{\bf g = 16.}}

\bigskip

\noindent The non-pseudoreal groups number five, and all are SDPGs. In the
notation of Thomas and Wood\cite{books}, which we shall follow for
definiteness, they are:
$16/8,9,10,11,13$. We now treat these in the order they are
enumerated by Thomas and Wood. Again, the relevant multiplication tables are collected in Appendix A of \cite{nonabelian2}.

\bigskip

\noindent \underline{Group 16/8; also designated $(Z_4 \times Z_2) \tilde{\times} Z_2$}.

\bigskip

\noindent This group has eight singlets $1_1, 1_2, ......, 1_8$ and two doublets
$2_1$ and $2_2$. In the embedding of 16/8 in $SU(4)$ we must avoid the singlet
$1_1$ otherwise there will be a residual supersymmetry with ${\cal N} \geq 1$.
Consider the embedding defined by ${\bf 4} = (2_1, 2_1)$. To find the 
surviving chiral fermions we need to product the ${\bf 4}$ with all ten of the irreps
of 16/8. This results in the table:

\bigskip

\begin{tabular}{||c||c|c|c|c|c|c|c|c||c|c||}
\hline
 & $1_1$ & $1_2$ & $1_3$ & $1_4$ & $1_5$ & $1_6$ & $1_7$ & $1_8$ & $2_1$ & $2_2$ \\
\hline\hline
$1_1$&&&&&&&&&$\times\times$& \\
\hline
$1_2$&&&&&&&&&$\times\times$& \\
\hline
$1_3$&&&&&&&&&$\times\times$& \\
\hline
$1_4$&&&&&&&&&$\times\times$& \\
\hline
$1_5$&&&&&&&&&&$\times\times$ \\
\hline
$1_6$&&&&&&&&&&$\times\times$ \\
\hline
$1_7$&&&&&&&&&&$\times\times$ \\
\hline
$1_8$&&&&&&&&&&$\times\times$ \\
\hline
\hline
$2_1$&&&&&$\times\times$&$\times\times$&$\times\times$&$\times\times$&& \\
\hline
$2_2$&$\times\times$&$\times\times$&$\times\times$&$\times\times$&&&&&& \\
\hline
\hline
\end{tabular}

\bigskip

If we choose $N = 2$, the gauge group is $SU(2)^8 \times SU(4)^2$, and the entries in the table correspond to bifundamental representations of this group (e.g., the entry nearest
the top right corner at the position ($1_1$, $2_1$) is the representation $2(2,1,1,1,1,1,1,1;\bar{4},1))$. If
we identify the diagonal subgroup of the first four SU(2)s as $SU(2)_L$,
of the second four as $SU(2)_R$ and of the two SU(4) as color SU(4)
the result is non-chiral due to the lack of symmetry about the main diagonal of the above table.

On the other hand, if we identify ${\bf 4_1}$ with ${\bf \bar{4}_2}$ there
are potentially eight chiral families:
\begin{equation}
8[(2, 1, 4) + (1, 2, \bar{4})]
\end{equation}
under $SU(2)_L \times SU(2)_R \times SU(4)$. This is the maximum total chirality for this orbifold, 
but as we will see below, the allowed chiral at any stage is as usual determined by 
spontaneous symmetry breaking (SSB) generated in the scalar sector.
In this section we give the maximum chirality for each orbifold, in the next section we
study SSB
for those models with sufficient chirality too accomodate at least three families.

\bigskip

Because $2_1=2_2^*$ form a complex conjugate pair, the embedding ${\bf 4} = (2_1, 2_2)$
is pseudoreal ${\bf 4} \equiv {\bf 4^*}$ and the fermions are non-chiral as
easily confirmed by the resultant table:

\bigskip

\begin{tabular}{||c||c|c|c|c|c|c|c|c||c|c||}
\hline
 & $1_1$ & $1_2$ & $1_3$ & $1_4$ & $1_5$ & $1_6$ & $1_7$ & $1_8$ & $2_1$ & $2_2$ \\
\hline\hline
$1_1$&&&&&&&&&$\times$&$\times$ \\
\hline
$1_2$&&&&&&&&&$\times$&$\times$ \\
\hline
$1_3$&&&&&&&&&$\times$&$\times$ \\
\hline
$1_4$&&&&&&&&&$\times$&$\times$ \\
\hline
$1_5$&&&&&&&&&$\times$&$\times$ \\
\hline
$1_6$&&&&&&&&&$\times$&$\times$ \\
\hline
$1_7$&&&&&&&&&$\times$&$\times$ \\
\hline
$1_8$&&&&&&&&&$\times$&$\times$ \\
\hline
\hline
$2_1$&$\times$&$\times$&$\times$&$\times$&$\times$&$\times$&$\times$&$\times$&& \\
\hline
$2_2$&$\times$&$\times$&$\times$&$\times$&$\times$&$\times$&$\times$&$\times$& & \\
\hline
\hline
\end{tabular}

\bigskip

\noindent For this embedding, the result is non-chiral for either of the
cases ${\bf 4_1} \equiv {\bf 4_2}$ or
${\bf 4_1} \equiv {\bf \bar{4}_2}$.
(In the future, we shall not even consider such trivially real or pseudo-real non-chiral embeddings).

\bigskip

\noindent Finally, for 16/8, consider the embedding
${\bf 4} = (1_2,1_5, 2_1)$. (In general there will be many equivalent embeddings. We will give
one member of each equivalence class. Cases that are obviously nonchiral (vectorlike)
will, in general, be ignored, except for a few instructive examples at orders 16 and 18.) The table is now:

\bigskip

\begin{tabular}{||c||c|c|c|c|c|c|c|c||c|c||}
\hline
 & $1_1$ & $1_2$ & $1_3$ & $1_4$ & $1_5$ & $1_6$ & $1_7$ & $1_8$ & $2_1$ & $2_2$ \\
\hline\hline
$1_1$&&$\times$&&&$\times$&&&&$\times$& \\
\hline
$1_2$&$\times$&&&&&$\times$&&&$\times$& \\
\hline
$1_3$&&&&$\times$&&&$\times$&&$\times$& \\
\hline
$1_4$&&&$\times$&&&&&$\times$&$\times$& \\
\hline
$1_5$&$\times$&&&&&&$\times$&&&$\times$ \\
\hline
$1_6$&&$\times$&&&&$\times$&&&&$\times$ \\
\hline
$1_7$&&&$\times$&&&&&$\times$&&$\times$ \\
\hline
$1_8$&&&&$\times$&&&$\times$&&&$\times$ \\
\hline
\hline
$2_1$&&&&&$\times$&$\times$&$\times$&$\times$&$\times$&$\times$ \\
\hline
$2_2$&$\times$&$\times$&$\times$&$\times$&&&&&$\times$&$\times$ \\
\hline
\hline
\end{tabular}

\bigskip

\noindent which is chiral.

\bigskip
\bigskip

\noindent These examples of embeddings for $\Gamma = 16/8$ show clearly how the number of
chiral families depends critically on the choice of embedding $\Gamma \subset SU(4)$. To
actual achieve a model with a viable phenomenologically, we must study
the possible routes through SSB for each chiral model. This 
we postpone until we have found all models of potential 
interest.

\bigskip

\noindent \underline{Group 16/9; also designated $[(Z_4 \times Z_2) \tilde{\times} Z_2]^{'}$}

\bigskip

\noindent This group has irreps which comprise eight singlets $1_1,..., 1_8$
and two doublets $2_1, 2_2$. With the embedding ${\bf 4} = (2_1, 2_2)$
and using the multiplication table from Appendix A of \cite{nonabelian2} we arrive at the table of fermion bilinears:

\bigskip

\begin{tabular}{||c||c|c|c|c|c|c|c|c||c|c||}
\hline
 & $1_1$ & $1_2$ & $1_3$ & $1_4$ & $1_5$ & $1_6$ & $1_7$ & $1_8$ & $2_1$ & $2_2$ \\
\hline\hline
$1_1$&&&&&&&&&$\times\times$& \\
\hline
$1_2$&&&&&&&&&&$\times\times$ \\
\hline
$1_3$&&&&&&&&&$\times\times$& \\
\hline
$1_4$&&&&&&&&&&$\times\times$ \\
\hline
$1_5$&&&&&&&&&$\times\times$& \\
\hline
$1_6$&&&&&&&&&&$\times\times$ \\
\hline
$1_7$&&&&&&&&&$\times\times$& \\
\hline
$1_8$&&&&&&&&&&$\times\times$ \\
\hline
\hline
$2_1$&$\times\times$&&$\times\times$&&$\times\times$&&$\times\times$&&& \\
\hline
$2_2$&&$\times\times$&&$\times\times$&&$\times\times$&&$\times\times$&& \\
\hline
\hline
\end{tabular}

\bigskip
\bigskip

This is non-chiral and has no families. This was the only potentially chiral embedding. In what follows,
nonchiral models will not be displayed, however, as the unification scale can be rather low in AdS/CFT
models, it would also be interesting to investigate vectorlike models of this class.

\bigskip

\noindent \underline{Group 16/10; also designated $Z_4 \tilde{\times} Z_4$}

\bigskip

The multiplication table is identical to that for 16/9, as mentioned in 
Appendix A of \cite{nonabelian2}; thus the model building for 16/10 is also identical to 16/9
and merits no additional discussion.

\bigskip

\noindent \underline{Group 16/11; also designated $Z_8 \tilde{\times} Z_2$}

\bigskip

Again there are eight singlets and two doublets. 
The singlets $1_{1,3,5,7}$ are real while the other
singlets fall into two conjugate pairs: $1_2 = 1_4^*$
and $1_6 = 1_8^*$. The doublets are complex: $2_1 = 2_2^*$.

The multiplication table includes the products:
$1_{1,3,5,7} \times 2_{1,2} = 2_{1,2}$ 
and
$1_{2,4,6,8} \times 2_{1,2} = 2_{2,1}$.
Also $2_1 \times 2_1 = 2_2 \times 2_2 = 1_2 + 1_4 + 1_6 + 1_8$,
while $2_1 \times 2_2 = 1_1 + 1_3 + 1_5 + 1_7$. 

This means that there are no interesting (legitimate and chiral)
embeddings of the type 1+1+2 or 2+2. 

Upon choosing $N=3$, the most chiral possibility is the embedding ${\bf 4} = (1_2, 1_2, 1_2, 1_2)$
which leads to the fermions in the following table. In this table,
$(\times)^6 \equiv (\times\times\times\times\times\times)$.

\bigskip

\small

\begin{tabular}{||c||c|c|c|c|c|c|c|c||c|c||}
\hline
 & $1_1$ & $1_2$ & $1_3$ & $1_4$ & $1_5$ & $1_6$ & $1_7$ & $1_8$ & $2_1$ & $2_2$ \\
\hline\hline
$1_1$&&$(\times)^6$&&&&&&&& \\
\hline
$1_2$&&&$(\times)^6$&&&&&&& \\
\hline
$1_3$&&&&$(\times)^6$&&&&&& \\
\hline
$1_4$&$(\times)^6$&&&&&&&&& \\
\hline
$1_5$&&&&&&$(\times)^6$&&&& \\
\hline
$1_6$&&&&&&&$(\times)^6$&&& \\
\hline
$1_7$&&&&&&&&$(\times)^6$&& \\
\hline
$1_8$&&&&&$(\times)^6$&&&&& \\
\hline
\hline
$2_1$&&&&&&&&&&$(\times)^6$ \\
\hline
$2_2$&&&&&&&&&$(\times)^6$& \\
\hline
\hline
\end{tabular}

\large

\bigskip
\bigskip

\noindent This gives rise to 
twelve chiral families if we identify $N=3$
and $3_1=3_4=3_5=3_8$, $3_2=3_6$ and $3_3=3_7$.
Under $SU(3)^3$ the chiral fermions are:

\begin{equation}
12[(3, \bar{3}, 1) + (1, 3, \bar{3}) + (\bar{3}, 3, 1)]
\end{equation}

\noindent together with real non-chiral representations. 
When we discuss spontaneous 
symmetry breaking, we will see if this type of unification is possible.

\noindent With the different embedding ${\bf 4} = (1_2, 1_2, 1_2, 1_4)$
the model changes to a less chiral but still interesting fermion
configuration:

\bigskip

\scriptsize

\begin{tabular}{||c||c|c|c|c|c|c|c|c||c|c||}
\hline
 & $1_1$ & $1_2$ & $1_3$ & $1_4$ & $1_5$ & $1_6$ & $1_7$ & $1_8$ & $2_1$ & $2_2$ \\
\hline\hline
$1_1$&&$\times\times\times$&&$\times$&&&&&& \\
\hline
$1_2$&$\times$&&$\times\times\times$&&&&&&& \\
\hline
$1_3$&&$\times$&&$\times\times\times$&&&&&& \\
\hline
$1_4$&$\times\times\times$&&$\times$&&&&&&& \\
\hline
$1_5$&&&&&&$\times\times\times$&&$\times$&& \\
\hline
$1_6$&&&&&$\times$&&$\times\times\times$&&& \\
\hline
$1_7$&&&&&&$\times$&&$\times\times\times$&& \\
\hline
$1_8$&&&&&$\times\times\times$&&$\times$&&& \\
\hline
\hline
$2_1$&&&&&&&&&&$\times\times\times\times$ \\
\hline
$2_2$&&&&&&&&&$\times\times\times\times$& \\
\hline
\hline
\end{tabular}

\large

\bigskip
\bigskip

\noindent If we can identify
$SU(3)'s$ as $3_1 \equiv 3_4 \equiv 3_5 \equiv 3_8$,
$3_2 \equiv 3_6$ and $3_3 \equiv 3_7$ this embedding give just four chiral families:

\begin{equation}
4[(3, \bar{3}, 1) + (1, 3, \bar{3}) + (\bar{3}, 3, 1)]
\end{equation}
under $SU(3)^3$ together with real representations.

\bigskip

\noindent To check consistency, we have verified that real and legitimate
embeddings for 16/11 like: ${\bf 4} = (1_3, 1_3, 1_3, 1_3)$ and
${\bf 4} = (2_1, 2_2)$ give no chiral fermions.

\bigskip
\bigskip

\newpage

\bigskip

\noindent \underline{Group 16/13; also designated $[Z_8 \tilde{\times} Z_2]^{''}$}

\bigskip
\bigskip

\noindent Of the five non-pseudoreal $g =16$ nonabelian $\Gamma$, 16/13
is unique in having only four inequivalent singlets $1_1, 1_2, 1_3, 1_4$
but three doublets $2_1, 2_2, 2_3$.

All the four singlet are real $1_i = 1_i^*$. The three doublets comprise
a conjugate complex pair $2_1= 2_3^* \neq 2_1^*$ and the real $2_2 = 2_2^*$.

With the embedding ${\bf 4} = (1_3, 1_4, 2_1)$ the resultant model has a chiral 
fermion quiver corresponding to the Table:

\bigskip

\begin{tabular}{||c||c|c|c|c||c|c|c||}
\hline
 & $1_1$ & $1_2$ & $1_3$ & $1_4$ & $2_1$ & $2_2$ & $2_3$ \\
\hline\hline
$1_1$&&&$\times$&$\times$&$\times$&& \\
\hline
$1_2$&&&$\times$&$\times$&&&$\times$ \\
\hline
$1_3$&$\times$&$\times$&&&&&$\times$ \\
\hline
$1_4$&$\times$&$\times$&&&$\times$&& \\
\hline
\hline
$2_1$&&$\times$&$\times$&&$\times$&$\times$&$\times$ \\
\hline
$2_2$&&&&&$\times$&$\times\times$&$\times$ \\
\hline
$2_3$&$\times$&&&$\times$&$\times$&$\times$&$\times$ \\
\hline
\end{tabular}

\bigskip
\bigskip

\noindent Here and in the following, unless otherwise specified, we let $N=2$. If we identify $SU(2)_L$ with the diagonal subgroup of the first
and fourth $SU(2)$s, and $SU(2)_R$ with the diagonal subgroup of the
2nd and 3rd, then there are four chiral families if we embed
${\bf 4_1} \equiv {\bf \bar{4}_3}$ and break $SU(4)_2$ completely.

\bigskip

\noindent Again consider 16/13 but now with
${\bf 4} = (2_1, 2_1)$. The table becomes:

\bigskip
\bigskip

\begin{tabular}{||c||c|c|c|c||c|c|c||}
\hline
 & $1_1$ & $1_2$ & $1_3$ & $1_4$ & $2_1$ & $2_2$ & $2_3$ \\
\hline\hline
$1_1$&&&&&$\times\times$&& \\
\hline
$1_2$&&&&&&&$\times\times$ \\
\hline
$1_3$&&&&&&&$\times\times$ \\
\hline
$1_4$&&&&&$\times\times$&& \\
\hline
\hline
$2_1$&&$\times\times$&$\times\times$&&&$\times\times$& \\
\hline
$2_2$&&&&&$\times\times$&&$\times\times$ \\
\hline
$2_3$&$\times\times$&&&$\times\times$&&$\times\times$& \\
\hline
\end{tabular}

\bigskip
\bigskip

\noindent With ${\bf 4_1} \equiv {\bf \bar{4}_3}$ there are eight
chiral families.

\bigskip
\bigskip

A similar result occurs, of course, for ${\bf 4} = (2_3, 2_3)$. But the
embedding ${\bf 4} = (2_1, 2_3)$ is non-chiral, leading to the symmetric
fermion quiver/table:

\bigskip
\bigskip

\begin{tabular}{||c||c|c|c|c||c|c|c||}
\hline
 & $1_1$ & $1_2$ & $1_3$ & $1_4$ & $2_1$ & $2_2$ & $2_3$ \\
\hline\hline
$1_1$&&&&&$\times$&&$\times$ \\
\hline
$1_2$&&&&&$\times$&&$\times$ \\
\hline
$1_3$&&&&&$\times$&&$\times$ \\
\hline
$1_4$&&&&&$\times$&&$\times$ \\
\hline
\hline
$2_1$&$\times$&$\times$&$\times$&$\times$&&$\times\times$& \\
\hline
$2_2$&&&&&$\times\times$&&$\times\times$ \\
\hline
$2_3$&$\times$&$\times$&$\times$&$\times$&&$\times\times$& \\
\hline
\end{tabular}

\bigskip
\bigskip

\noindent This arrangement is manifestly non-chiral because of the
symmetry of the table. Even though $2_1$ and $2_3$ are complex,  
$2_1={2_3^*}$, so $4^*=(2_1,2_3)^*=(2_3,2_1)$. We can rotate this
within $SU(4)$ to $4=(2_1,2_3)$. Therefore, the $4$ is pseudoreal 
and the fermions are vectorlike as expected.

\bigskip
\bigskip

\noindent The embedding ${\bf 4} = (2_2, 2_2)$ in 16/13 gives rise to
the table:

\bigskip
\bigskip

\begin{tabular}{||c||c|c|c|c||c|c|c||}
\hline
 & $1_1$ & $1_2$ & $1_3$ & $1_4$ & $2_1$ & $2_2$ & $2_3$ \\
\hline\hline
$1_1$&&&&&&$\times\times$& \\
\hline
$1_2$&&&&&&$\times\times$& \\
\hline
$1_3$&&&&&&$\times\times$& \\
\hline
$1_4$&&&&&&$\times\times$& \\
\hline
\hline
$2_1$&&&&&$\times\times$&&$\times\times$ \\
\hline
$2_2$&$\times\times$&$\times\times$&$\times\times$&$\times\times$&&& \\
\hline
$2_3$&&&&&$\times\times$&&$\times\times$ \\
\hline
\end{tabular}

\bigskip
\bigskip

\noindent This embedding leads to no chirality and zero families.

\bigskip

\newpage

\bigskip

\noindent Finally, the embedding ${\bf 4} = (2_1, 2_2)$ of 16/13 leads to the intermediate
situation:

\bigskip
\bigskip

\begin{tabular}{||c||c|c|c|c||c|c|c||}
\hline
 & $1_1$ & $1_2$ & $1_3$ & $1_4$ & $2_1$ & $2_2$ & $2_3$ \\
\hline\hline
$1_1$&&&&&$\times$&$\times$& \\
\hline
$1_2$&&&&&&$\times$&$\times$ \\
\hline
$1_3$&&&&&&$\times$&$\times$ \\
\hline
$1_4$&&&&&$\times$&$\times$& \\
\hline
\hline
$2_1$&&$\times$&$\times$&&$\times$&$\times$&$\times$ \\
\hline
$2_2$&$\times$&$\times$&$\times$&$\times$&$\times$&&$\times$ \\
\hline
$2_3$&$\times$&&&$\times$&$\times$&$\times$&$\times$ \\
\hline
\end{tabular}

\bigskip
\bigskip

\noindent This give rise to four chiral families with the identification
${\bf 4_1} \equiv {\bf \bar{4}_3}$.

\bigskip
\bigskip

To summarize the ``double doublet'' embeddings ${\bf 4} = (2_i, 2_j)$ of 16/13:
for the equivalent embeddings 
(i, j) = (1, 1) or (3, 3), there are up to eight chiral families;
for the other mutually equivalent cases
(i, j) = (1, 2), (3, 2), (2, 3), or (2, 1) there are up to
four chiral families and finally for the pseudoreal
cases (i, j) = (1, 3), (3, 1) and the real case (2, 2) there are, because
of the mathematical theorem (and as we have now easily verified by direct calculation)
no chiral fermions.

\bigskip
\bigskip
\bigskip

\newpage

\underline{{\bf g = 18.}}

\bigskip

\noindent Here the non-pseudoreal groups number two, and one is an SDPG. In the
notation of Thomas and Wood\cite{books}
they are:
$18/3,5$. So we now treat these in the order they are
enumerated by Thomas and Wood.

\bigskip

\noindent \underline{Group 18/3; also designated $D_3 \times Z_3$}

\bigskip

\noindent This group has irreps which fall into six singlets
$1, 1^{'}, 1\alpha, 1^{'}\alpha, 1\alpha^2, 1^{'}\alpha^2$ and three
doublets $2, 2\alpha, 2\alpha^2$. Using the $D_3$ multiplication
table from appendix A of \cite{nonabelian2} the embedding ${\bf 4} = (1\alpha, 1^{'}, 2\alpha)$
yields the table:

\bigskip
\bigskip

\bigskip
\bigskip

\begin{tabular}{||c||c|c|c||c|c|c||c|c|c||}
\hline
 & $1$ & $1^{'}$ & $2$ & $1\alpha$ & $1^{'}\alpha$ & $2\alpha$ & $1\alpha^2$ & $1^{'}\alpha^2$ & $ 2\alpha^2$ \\
\hline\hline
$1$&&$\times$&&$\times$&&$\times$&&& \\
\hline
$1^{'}$&$\times$&&&&$\times$&$\times$&&& \\
\hline
$2$&&&$\times$&$\times$&$\times$&$\times\times$&&& \\
\hline\hline
$1\alpha$&&&&&$\times$&&$\times$&&$\times$ \\
\hline
$1^{'}\alpha$&&&&$\times$&&&&$\times$&$\times$ \\
\hline
$2\alpha$&&&&&&$\times$&$\times$&$\times$&$\times\times$ \\
\hline\hline
$1\alpha^2$&$\times$&&$\times$&&&&&$\times$& \\
\hline
$1^{'}\alpha^2$&&$\times$&$\times$&&&&$\times$&& \\
\hline
$2\alpha^2$ & $\times$&$\times$&$\times\times$&&&&&&$\times$ \\
\hline
\end{tabular}

\bigskip
\bigskip

\noindent Identifying $SU(2)_{L,R}$ with the diagonal subgroups
of respectively $SU(2)_3 \times SU(2)_4$ and
$SU(2)_5 \times SU(2)_6$ gives rise to two chiral families
when it is assumed that $SU(2)_{1,2}$ and $SU(4)_{1,2}$
are broken.

\bigskip
\bigskip
\bigskip

\newpage

\noindent \underline{Group 18/5; also designated $(Z_3 \times Z_3) \tilde{\times} Z_2$}

\bigskip
\bigskip
\bigskip

\noindent This group has two singlets $1, 1^{'}$ and four doublets
$2_1, 2_2, 2_3, 2_4$. Using the multiplication table from \cite{nonabelian2}
we compute the models corresponding to the three inequivalent embeddings
${\bf 4} = (1^{'}, 1^{'}, 2_1)$,
${\bf 4} = (2_1, 2_1)$
and
${\bf 4} = (2_1, 2_2)$.

\bigskip

\noindent For ${\bf 4} = (1^{'}, 1^{'}, 2_1)$ the table is:

\bigskip
\bigskip
\bigskip

\begin{tabular}{||c||c|c||c|c|c|c||}
\hline
 & $1$ & $1^{'}$ & $2_1$ & $2_2$ & $2_3$ & $2_4$  \\
\hline\hline
$1$&&$\times\times$&$\times$&&& \\
\hline
$1^{'}$&$\times\times$&&$\times$&&& \\
\hline\hline
$2_1$&$\times$&$\times$&$\times\times\times$&&& \\
\hline
$2_2$&&&&$\times\times$&$\times$&$\times$ \\
\hline
$2_3$&&&&$\times$&$\times\times$&$\times$\\
\hline
$2_4$&&&&$\times$&$\times$&$\times\times$ \\
\hline
\end{tabular}

\bigskip
\bigskip

\noindent This model is manifestly non-chiral due to the symmetry
of the table.

\bigskip
\bigskip

\newpage   

\bigskip

\noindent For ${\bf 4} = (2_1, 2_1)$ the table is:

\bigskip
\bigskip
\bigskip

\begin{tabular}{||c||c|c||c|c|c|c||}
\hline
 & $1$ & $1^{'}$ & $2_1$ & $2_2$ & $2_3$ & $2_4$  \\
\hline\hline
$1$&&&$\times\times$&&& \\
\hline
$1^{'}$&&&$\times\times$&&& \\
\hline\hline
$2_1$&$\times\times$&$\times\times$&$\times\times$&&& \\
\hline
$2_2$&&&&&$\times\times$&$\times\times$ \\
\hline
$2_3$&&&&$\times\times$&&$\times\times$\\
\hline
$2_4$&&&&$\times\times$&$\times\times$& \\
\hline
\end{tabular}

\bigskip
\bigskip

\noindent This model is also manifestly non-chiral due to the symmetry
of the table.

\bigskip
\bigskip

\bigskip

\noindent For ${\bf 4} = (2_1, 2_2)$ the table is:

\bigskip
\bigskip
\bigskip

\begin{tabular}{||c||c|c||c|c|c|c||}
\hline
 & $1$ & $1^{'}$ & $2_1$ & $2_2$ & $2_3$ & $2_4$  \\
\hline\hline
$1$&&&$\times$&$\times$&& \\
\hline
$1^{'}$&&&$\times$&$\times$&& \\
\hline\hline
$2_1$&$\times$&$\times$&$\times$&&$\times$&$\times$ \\
\hline
$2_2$&$\times$&$\times$&&$\times$&$\times$&$\times$ \\
\hline
$2_3$&&&$\times$&$\times$&&$\times\times$\\
\hline
$2_4$&&&$\times$&$\times$&$\times\times$& \\
\hline
\end{tabular}

\bigskip
\bigskip

\noindent Again, this model is manifestly non-chiral.
18/5 does not lend itself to chirality.
This is easy to understand when one realizes that all
of the irreducible representations of 18/5
are individually either real or pseudoreal \cite{books} making a complex embedding of {\bf 4} impossible.

\bigskip
\bigskip

\newpage

\underline{{\bf g = 20.}}

\bigskip

\noindent One non-pseudoreal group, an SDPG. In the
notation of Thomas and Wood\cite{books} it is $20/5$.

\bigskip

\noindent \underline{Group 20/5; also designated $Z_5 \tilde{\times} Z_4$}   

\bigskip
\bigskip
\bigskip

\noindent The group has four singlets $1_1, 1_2, 1_3, 1_4$ and a $4$.
The singlets $1_1, 1_3$ are real and the other two form a
complex conjugate pair $1_2 = 1_4^*$.
The {\bf 6} which is the antisymmetric product
{\bf 6} = $(4 \times 4)_a$ must be real for a legitimate embedding.
The two inequivalent choices, bearing in mind the multiplication table
provided in the Appendix of \cite{nonabelian2} are ${\bf 4} = (1_2, 1_2, 1_2, 1_2)$ 
and ${\bf 4} = (1_2, 1_2, 1_2, 1_4)$.

\bigskip

The first ${\bf 4} = (1_2, 1_2, 1_2, 1_2)$ yields the chiral fermions
in the following table:

\bigskip
\bigskip

\begin{tabular}{||c||c|c|c|c||c||}
\hline
 & $1_1$ & $1_2$ & $1_3$ & $1_4$ & $4$  \\
\hline\hline
$1_1$&&$\times\times\times\times$&&& \\
\hline
$1_2$&&&$\times\times\times\times$&& \\
\hline
$1_3$&&&&$\times\times\times\times$& \\
\hline
$1_4$&$\times\times\times\times$&&&& \\
\hline\hline
$4$&&&&&$\times\times\times\times\times$ \\
\hline
\end{tabular}

\bigskip
\bigskip

\noindent Putting $N=3$ this embedding gives four chiral families
when we identify $SU(3)_3 \equiv SU(3)_4$ and drop
real representations, giving:

\begin{equation}
4[ (3, \bar{3}, 1) + (1, 3, \bar{3}) + (\bar{3}, 1, 3)]
\end{equation}

\noindent under $SU(3) \times SU(3) \times SU(3)$. The chiral sector of
this 20/5 nonabelian orbifold coincides with a $Z_4$ orbifold.

\bigskip

\bigskip

\begin{tabular}{||c||c|c|c|c||c||}
\hline
 & $1_1$ & $1_2$ & $1_3$ & $1_4$ & $4$  \\
\hline\hline
$1_1$&&$\times\times\times$&&$\times$& \\
\hline
$1_2$&$\times$&&$\times\times\times$&& \\
\hline
$1_3$&&$\times$&&$\times\times\times$& \\
\hline
$1_4$&$\times\times\times$&&$\times$&& \\
\hline\hline
$4$&&&&&$\times\times\times\times\times$ \\
\hline
\end{tabular}

\bigskip

\bigskip
\bigskip

\noindent Identifying $SU(3)_3 \equiv SU(3)_4$ as before for $N=3$
this is less chiral and gives rise to just two chiral
families.

\begin{equation}
4[ (3, \bar{3}, 1) + (1, 3, \bar{3}) + (\bar{3}, 1, 3)]
\end{equation}

\noindent under $SU(3) \times SU(3) \times SU(3)$.

\bigskip
\bigskip

\newpage

\bigskip
\bigskip

\underline{{\bf g = 21.}}

\bigskip

\noindent One non-pseudoreal group, an SDPG. In the
notation of Thomas and Wood\cite{books} it is: $21/2$.

\bigskip

\noindent \underline{Group 21/2; also designated $Z_7 \tilde{\times} Z_3$}

\bigskip
\bigskip
\bigskip

This group has irreps which comprise three singlets $1_1, 1_2, 1_3$ and
two triplets $3_1, 3_2$. With the embedding
${\bf 4} = (1_2, 3_1)$ (recall that $1_1$ must be avoided to obtain
${\cal N} = 0$), the resultant fermions are given by:

\bigskip
\bigskip

\bigskip

\begin{tabular}{||c||c|c|c||c|c||}
\hline
 & $1_1$ & $1_2$ & $1_3$ & $3_1$ & $3_2$  \\
\hline\hline
$1_1$&&$\times$&&$\times$& \\
\hline
$1_2$&&&$\times$&$\times$& \\
\hline
$1_3$&$\times$&&&$\times$& \\
\hline\hline
$3_1$&&&&$\times\times$&$\times\times$ \\
\hline
$3_2$&$\times$&$\times$&$\times$&$\times$&$\times\times$ \\
\hline
\end{tabular}

\bigskip
\bigskip

Putting $N = 2$, the gauge group is $SU(2)^3 \times SU(6)^2$. 
Clearly the model is chiral as seen in the asymmetry of the table.
For example, put $SU(2)_L \equiv SU(2)_1$,
$SU(2)_R \equiv SU(2)_2$, break $SU(2)_3$ entirely and break $SU(6)^2$ such that
${\bf 6_1} \rightarrow {\bf 4}, {\bf 6_2} \rightarrow {\bf \bar{4}}$,
to find two chiral families.

\newpage

\underline{{\bf g = 24.}}

\bigskip

\noindent The non-pseudoreal groups number six, and three are SDPGs. In the
notation of Thomas and Wood\cite{books}
they are: $24/7,8,9,13,14,15$.
We now treat these in the order they are
enumerated by Thomas and Wood.

\bigskip
\bigskip

\noindent \underline{Group 24/7; also designated $D_4 \times Z_3$}

\bigskip
\bigskip

\noindent This has twelve singlets $1_1\alpha^i, 1_2\alpha^i, 1_3\alpha^i, 1_4\alpha^i$ 
(i = 0 - 2)
and three doublets $2\alpha^i$ (i = 0 - 2); here $\alpha = exp (i\pi/3)$.
The embedding ${\bf 4} = (1_1\alpha, 1_2, 2\alpha)$
was studied in detail in \cite{nonabelian} 
where it was shown how it can lead to precisely three
chiral families in the standard model.

\bigskip

For completeness we include the table for the chiral fermions (it was
presented in a different equivalent way in \cite{nonabelian}):

\bigskip
\bigskip

\scriptsize

\begin{tabular}{||c||c|c|c|c|c||c|c|c|c|c||c|c|c|c|c||}
\hline
 & $1_1$ & $1_2$ & $1_3$ & $1_4$ & $2$ &$1_1\alpha$& $1_2\alpha$ & $1_3\alpha$ & $1_4\alpha$ & $2\alpha$ 
& $1_1\alpha^2$ & $1_2\alpha^2$ & $1_3\alpha^2$ & $1_4\alpha^2$& $2\alpha^2$  \\
\hline\hline
$1_1$&&$\times$&&&&$\times$&&&&$\times$&&&&& \\
\hline
$1_2$&$\times$&&&&&&$\times$&&&$\times$&&&&& \\
\hline
$1_3$&&&&$\times$&&&&$\times$&&$\times$&&&&& \\
\hline
$1_4$&&&$\times$&&&&&&$\times$&$\times$&&&&& \\
\hline
$2$&&&&&$\times$&$\times$&$\times$&$\times$&$\times$&$\times$&&&&& \\
\hline\hline
$1_1\alpha$&&&&&&&$\times$&&&&$\times$&&&&$\times$ \\
\hline
$1_2\alpha$&&&&&&$\times$&&&&&&$\times$&&&$\times$  \\
\hline
$1_3\alpha$&&&&&&&&&$\times$&&&&$\times$&&$\times$ \\
\hline
$1_4\alpha$ &&&&&&&&$\times$&&&&&&$\times$&$\times$ \\
\hline
$2\alpha$ &&&&&&&&&&$\times$&$\times$&$\times$&$\times$&$\times$&$\times$ \\
\hline\hline
$1_1\alpha^2$&$\times$&&&&$\times$&&&&&&&$\times$&&&  \\
\hline
$1_2\alpha^2$&&$\times$&&&$\times$&&&&&&$\times$&&&&\\
\hline
$1_3\alpha^2$&&&$\times$&&$\times$&&&&&&&&&$\times$& \\
\hline
$1_4\alpha^2$&&&&$\times$&$\times$&&&&&&&&$\times$&&\\
\hline
$2\alpha^2$&$\times$&$\times$&$\times$&$\times$&$\times$&&&&&&&&&&$\times$  \\
\hline\hline
\end{tabular}

\large

\bigskip
\bigskip

\bigskip
\bigskip

\noindent By identifying $SU(4)$ with the diagonal subgroup of
$SU(4)_{2,3}$, breaking $SU(4)_1$ to $SU(2)_L^{'}\times
SU(2)_R^{'}$, then identifying
$SU(2)_L$ with the diagonal subgroup
of $SU(2)_{6,7,8}$ and $SU(2)_L^{'}$
and $SU(2)_R$ with the diagonal subgroup of
$SU(2)_{10,11,12}$ and $SU(2)_R^{'}$
then leads to a three-family model as explained in \cite{nonabelian}.

\bigskip

It is convenient to represent the chiral fermions in a quiver diagram.
This model is especially interesting because, uniquely among
the large number of models examined in this study, the prescribed scalars are 
sufficient to break the gauge symmetry to that of the standard model.

\bigskip
\bigskip

\noindent \underline{Group 24/8; also designated $Q \times Z_3$}

\bigskip

\noindent The multiplication tables of $D_4$ and $Q$ and hence the
multiplication tables of 24/7 and 24/8 are identical. Model building
for 24/8 is therefore the same as 24/7 and merits
no additional discussion.

\bigskip
\bigskip 

\noindent \underline{Group 24/9; also designated $D_3 \times Z_4$}

\bigskip

This group generates one of the richest sets of chiral model in the class of models discussed
in this paper. The group has as irreps eight singlets $(1_1\alpha^j, 1_2\alpha^j)$ 
and four doublets $2\alpha^j$ (j = 0, 1, 2, 3),
where $\alpha = exp (i \pi / 4)$.

\bigskip

\noindent The embedding ${\bf 4} = (1_1 \alpha^{a_1}, 1_2\alpha^{a_2}, 2 \alpha^{a_3})$
must satisfy $a_1 \neq 0$ (for ${\cal N} = 0$) and
$a_1 + a_2 = -2a_3 $ (mod 4) (to ensure reality of ${\bf 6} = ({\bf 4} \times {\bf 4})_a$). 
There are several interesting possibilities including:
$(1_1\alpha, 1_2\alpha, 2\alpha))$,
$(1_1\alpha, 1_2\alpha^3, 2\alpha^2))$,
$(1_1\alpha^2, 1_2, 2\alpha^3))$,
$(1_1\alpha^2, 1_2, 2\alpha))$, and
$(1_1\alpha^2, 1_2\alpha^2, 2))$. The third and fourth cases are equivalent as can be seen by letting $\alpha$ go to ${\alpha}^{-1}$, 
and the last case has only real fermions since ${\alpha}^{2}=-1$.

\bigskip

\bigskip

The fermions for ${\bf 4=(1}_{{\bf 1}}{\bf \alpha }^{{\bf 2}}{\bf
,1}%
_{2}{\bf \alpha }^{{\bf 2}},{\bf 2)}$ are vectorlike.

\bigskip
\bigskip

Moving on to 24/9 with ${\bf 4=(1}_{{\bf 1}}{\bf \alpha ,{\bf
1}_{2}\alpha }%
^{{\bf 3}},{\bf 2{\bf \alpha }^{{\bf 2}})}$ we find the fermions are
chiral
and fall into the irrep:

\bigskip

\begin{tabular}{|c||c|c|c|c|c|c|c|c|c|c|c|c|}
\hline
 & $1_{1}$ & $1_{2}$ & 2 & $1_{1}\alpha $ & $1_{2}\alpha $ &
$2\alpha $ & $1_{1}\alpha^{2}$ & $1_{2}\alpha^{2}$ & $2\alpha ^{2}$ 
&$1_{1}\alpha^{3}$ & $1_{2}\alpha^{3}$ & $2\alpha^{3}$ \\ \hline\hline
$1_{1}$ &  &  &  & $\times $ &  &  &  &  & $\times $ &  & $\times $ &
\\
\hline
$1_{2}$ &  &  &  &  & $\times $ &  &  &  & $\times $ & $\times $ &  &
\\
\hline
2 &  &  &  &  &  & $\times $ & $\times $ & $\times $ & $\times $ &  &  &
$%
\times $ \\ \hline\hline
$1_{1}\alpha $ &  & $\times $ &  &  &  &  & $\times $ &  &  &  &  &
$\times$
\\ \hline
$1_{2}\alpha $ & $\times $ &  &  &  &  &  &  & $\times $ &  &  &  &
$\times $
\\ \hline
2$\alpha $ &  &  & $\times $ &  &  &  &  &  & $\times $ & $\times $ & $%

\times $ & $\times $ \\ \hline\hline
$1_{1}\alpha^{2}$ &  &  & $\times $ &  & $\times $ &  &  &  &  &
$\times $
&  &  \\ \hline
$1_{2}\alpha^{2}$ &  &  & $\times $ & $\times $ &  &  &  &  &  &  &
$\times
$ &  \\ \hline
2$\alpha^{2}$ & $\times $ & $\times $ & $\times $ &  &  & $\times $ &
&  &
&  &  & $\times $ \\ \hline\hline
$1_{1}\alpha^{3}$ & $\times $ &  &  &  &  & $\times $ &  & $\times $ &
&
&  &  \\ \hline
$1_{1}\alpha^{3}$ &  & $\times $ &  &  &  & $\times $ & $\times $ &  &
&
&  &  \\ \hline
$2\alpha^{3}$ &  &  & $\times $ & $\times $ & $\times $ & $\times $ &
&  &
$\times $ &  &  &  \\ \hline\hline
\end{tabular}

\bigskip
\bigskip

\noindent With the embedding ${\bf 4} = (1_1\alpha, 1_2\alpha, 2\alpha)$ the
chiral fermions are:

\bigskip
\bigskip

\small

\begin{tabular}{|c||c|c|c|c|c|c|c|c|c|c|c|c|}
\hline
& $1_{1}$&$1_{2}$ & 2 & $1_{1}\alpha $ & $1_{2}\alpha $ &
$2\alpha $ & $1_{1}\alpha^{2}$ & $1_{2}\alpha^{2}$ & $2\alpha^{2}$ &
$1_{1}\alpha^{3}$ & $1_{2}\alpha^{3}$ & $2\alpha^{3}$ \\ 

\hline\hline

1$_{1}$ &  &  &  & $\times $ & $\times$  &$\times$  &  &  & &  & &\\

\hline

1$_{2}$ &  &  &  & $\times $ & $\times$  &$\times$  &  &  & &  & &\\

\hline

$2$ &  &  &  & $\times $ & $\times$  &$\times\times\times$  &  &  & &  & &\\

\hline\hline

$1_{1}\alpha$ &  & &  &  &  &  & $\times $ &$\times$  &$\times$  &  &  &\\ \hline
$1_{2}\alpha$ &  & &  &  &  &  & $\times $ &$\times$  &$\times$  &  &  &\\ \hline
$2\alpha$ &  & &  &  &  &  & $\times $ &$\times$  & $\times\times\times$  &  &  &\\ 
\hline\hline

$1_{1}\alpha ^{2}$ &  &  & &  & &  &  &  &  &$\times $& $\times$ & $\times$ \\ \hline
$1_{2}\alpha ^{2}$ &  &  & &  & &  &  &  &  &$\times $& $\times$ & $\times$ \\ \hline
$2\alpha ^{2}$ &  &  & &  & &  &  &  &  &$\times $& $\times$ & $\times\times\times$ \\ 
\hline\hline
$1_{1}\alpha^{3}$&$\times$&$\times$&$\times$&&&&&&&&&  \\ \hline
$1_{2}\alpha^{3}$&$\times$&$\times$&$\times$&&&&&&&&&  \\ \hline
$2\alpha^{3}$&$\times$&$\times$&$\times\times\times$&&&&&&&&&  \\

\hline\hline
\end{tabular}

\large

\bigskip
\bigskip

\noindent Identifying $SU(2)_L$ with the diagonal subgroup of $SU(2)_{1,2,3,4}$,
$SU(2)_R$ with the diagonal subgroup of $SU(2)_{5,6,7,8}$
and the ${\bf 4}$ of $SU(4)$ with the ${\bf 4}$ of $SU(4)_{2,3}$
and the ${\bf \bar{4}}$ of $SU(4)_{1,4}$ leads to eight chiral families.

\bigskip
\bigskip

\noindent Taking the embedding ${\bf 4} = (1_1\alpha^2, 1_2, 2\alpha)$
gives as chiral fermions:

\bigskip
\bigskip

\bigskip

\normalsize

\begin{tabular}{||c||c|c|c||c|c|c||c|c|c||c|c|c||}
\hline
 & $1_1$ & $1_2$ & $2$ & $1_1\alpha$ & $1_2\alpha$& $2\alpha$ & $1_1\alpha^2$ & $1_2\alpha^2$ 
&$2\alpha^2$&$1_1\alpha^3$&$1_2\alpha^3$&$2\alpha^3$  \\
\hline\hline
$1_1$&&$\times$&&&&$\times$&$\times$&&&&& \\
\hline
$1_2$&$\times$&&&&&$\times$&&$\times$&&& & \\
\hline
$2$&&&$\times$&$\times$&$\times$&$\times$& &&$\times$&&&\\
\hline\hline
$1_1\alpha$&&&&&$\times$&&&&$\times$ &$\times$&&\\
\hline
$1_2\alpha$&&&&$\times$&&&&&$\times$&&$\times$& \\
\hline
$2\alpha$&&&&&&$\times$&$\times$&$\times$&$\times$&&&$\times$ \\
\hline\hline
$1_1\alpha^2$ &$\times$&&&&&&&$\times$&&&&$\times$ \\
\hline
$1_2\alpha^2$ &&$\times$& &&& &$\times$&& &&&$\times$ \\
\hline
$2\alpha^2$ &&&$\times$ &&& &&&$\times$ &$\times$ &$\times$& $\times$ \\
\hline\hline
$1_1\alpha^3$ &&&$\times$ &$\times$&& &&& &&$\times$& \\
\hline
$1_2\alpha^3$ &&&$\times$ &&$\times$& &&& &$\times$&& \\
\hline
$2\alpha^3$ &$\times$&$\times$&$\times$ &&&$\times$ &&& &&&$\times$ \\
\hline
\end{tabular}

\large

\bigskip
\bigskip

\noindent We identify $SU(2)_L$, $SU(2)_R$ with the diagonal subgroups
of $SU(2)_{1,2}$ and $SU(2)_{3,4}$, respectively
and break completely $SU(2)_{5,6,7,8}$. The generalized
color embedding ${\bf 4} \equiv {\bf 4_1} \equiv {\bf 4_2} \equiv {\bf \bar{4}_3} \equiv {\bf \bar{4}_4}$
leads to four chiral families. This can be reduced to three families
by further symmetry breaking.

\bigskip

An even more interesting embedding for 24/9 is to set ${\bf 4} = (2\alpha,2\alpha)$
which gives a real ${\bf 6}$ as required (since $\alpha^2=-1$ is real).
The table for fermions is:

\bigskip
\bigskip

\small

\begin{tabular}{||c||c|c|c||c|c|c||c|c|c||c|c|c||}
\hline
 & $1_1$ & $1_2$ & $2$ & $1_1\alpha$ & $1_2\alpha$& $2\alpha$ & $1_1\alpha^2$ & $1_2\alpha^2$ 
&$2\alpha^2$&$1_1\alpha^3$&$1_2\alpha^3$&$2\alpha^3$  \\
\hline\hline
$1_1$&&&&&&$\times\times$&&&&&& \\
\hline
$1_2$&&&&&&$\times\times$&&&&& & \\
\hline
$2$&&&&$\times\times$&$\times\times$&$\times\times$&&&&&&\\
\hline\hline
$1_1\alpha$&&&&&&&&&$\times\times$&&&\\
\hline
$1_2\alpha$&&&&&&&&&$\times\times$&&& \\
\hline
$2\alpha$&&&&&&&$\times\times$&$\times\times$&$\times\times$&&& \\
\hline\hline
$1_1\alpha^2$&&&&&&&&&&&&$\times\times$ \\
\hline
$1_2\alpha^2$&&&&&&&&&&&&$\times\times$ \\
\hline
$2\alpha^2$&&&&&&&&&&$\times\times$&$\times\times$&$\times\times$ \\
\hline\hline
$1_1\alpha^3$&&&$\times\times$&&&&&&&&& \\
\hline
$1_2\alpha^3$&&&$\times\times$&&&&&&&&& \\
\hline
$2\alpha^3$&$\times\times$&$\times\times$&$\times\times$&&&&&&&&& \\
\hline
\end{tabular}

\large

\bigskip
\bigskip
Identifying $SU(2)_L$ with the diagonal subgroup of $SU(2)_{1,3,5,7}$,
$SU(2)_R$ with the diagonal subgroup of $SU(2)_{2,4,6,8}$, 
breaking $SU(4)_{1,3}$ and keeping the unbroken $SU(4)$
which is the diagonal subgroup of $SU(4)_{2,4}$ gives rise to 
eight chiral families:
\begin{equation}
8[(2, 1, \bar{4}) + (1, 2, 4)]
\end{equation}

The possibility of achieving the relevant symmetry breaking will be
examined below.

\bigskip

\newpage

\bigskip

\noindent \underline{Group 24/13; also designated $Q \tilde{\times} Z_3$}

\bigskip

This group has three singlets $1_1, 1_2, 1_3$, three doublets
$2_1, 2_2, 2_3$ and one triplet $3$. For $N = 2$ the gauge group is
therefore $SU(2)^3 \times SU(4)^3 \times SU(6)$.

\bigskip

\noindent With the embedding ${\bf 4} = (2_1, 2_2)$ the chiral fermions
are:

\bigskip
\bigskip

\bigskip

\begin{tabular}{||c||c|c|c||c|c|c||c||}
\hline
 & $1_1$ & $1_2$ & $1_3$ & $2_1$ & $2_2$& $2_3$ & $3$  \\
\hline\hline
$1_1$&&&&$\times$&$\times$&& \\
\hline
$1_2$&&&&&$\times$&$\times$& \\
\hline
$1_3$&&&&$\times$&&$\times$& \\
\hline\hline
$2_1$&$\times$&$\times$&&&&&$\times\times$ \\
\hline
$2_2$&&$\times$&$\times$&&&&$\times\times$ \\
\hline
$2_3$&$\times$&&$\times$&&&&$\times\times$ \\
\hline\hline
$3$ &&&&$\times\times$&$\times\times$&$\times\times$& \\
\hline
\end{tabular}

\bigskip
\bigskip

\noindent If we identify $SU(2)_L \equiv SU(2)_3$,
$SU(2)_R \equiv SU(2)_2$, and break $SU(2)_1$ there
are two chiral families for
${\bf 4} \equiv {\bf 4_1} \equiv {\bf \bar{4}_2} \equiv {\bf \bar{4}_3}$.

\bigskip

\newpage

\bigskip

\noindent If, instead, we embed ${\bf 4} = (2_2, 2_3)$ the
fermions fall according to the following table:

\bigskip
\bigskip
\bigskip

\begin{tabular}{||c||c|c|c||c|c|c||c||}
\hline
 & $1_1$ & $1_2$ & $1_3$ & $2_1$ & $2_2$& $2_3$ & $3$  \\
\hline\hline
$1_1$&&&&&$\times$&$\times$& \\
\hline
$1_2$&&&&$\times$&&$\times$& \\
\hline
$1_3$&&&&$\times$&$\times$&& \\
\hline\hline
$2_1$&&$\times$&$\times$&&&&$\times\times$ \\
\hline
$2_2$&$\times$&&$\times$&&&&$\times\times$ \\
\hline
$2_3$&$\times$&$\times$&&&&&$\times\times$ \\
\hline\hline
$3$ &&&&$\times\times$&$\times\times$&$\times\times$& \\
\hline
\end{tabular}

\bigskip
\bigskip

\noindent This model is manifestly non-chiral because of the
total symmetry of the table.

\bigskip
\bigskip

\newpage

\noindent \underline{Group 24/14; also designated $Z_8 \tilde{\times} Z_3$}

\bigskip
\bigskip

\noindent There are eight singlets and four doublets, with multiplication
table as in \cite{nonabelian2}. With the embedding ${\bf 4} = (2_2, 2_4)$ one
arrives at the fermions:

\bigskip
\bigskip

\begin{tabular}{||c||c|c|c|c|c|c|c|c||c|c|c|c||}
\hline
 & $1_1$ & $1_2$ & $1_3$ & $1_4$ & $1_5$& $1_6$ & $1_7$ & $1_8$ & $2_1$ & $2_2$ & $2_3$ & $2_4$ \\
\hline\hline
$1_1$&&&&&&&&&&$\times$&&$\times$ \\
\hline
$1_2$&&&&&&&&&$\times$&&$\times$& \\
\hline
$1_3$&&&&&&&&&&$\times$&&$\times$ \\
\hline
$1_4$&&&&&&&&&$\times$&&$\times$& \\
\hline
$1_5$&&&&&&&&&&$\times$&&$\times$ \\
\hline
$1_6$&&&&&&&&&$\times$&&$\times$& \\
\hline
$1_7$ &&&&&&&&&&$\times$&&$\times$ \\
\hline
$1_8$ &&&&&&&&&$\times$&&$\times$&  \\
\hline\hline
$2_1$ && $\times$ && $\times$ && $\times$ && $\times$ && $\times$ && $\times$ \\
\hline
$2_2$ & $\times$ && $\times$ && $\times$ && $\times$ && $\times$ && $\times$ & \\
\hline
$2_3$ && $\times$ && $\times$ && $\times$ && $\times$ && $\times$ && $\times$ \\
\hline
$2_4$ & $\times$ && $\times$ && $\times$ && $\times$ && $\times$ && $\times$ & \\
\hline
\end{tabular}

\bigskip
\bigskip

\noindent This arrangement has zero families.

\bigskip

\bigskip

\bigskip

A chiral embedding is ${\bf 4} = (2_1, 2_2)$ giving rise to the fermions:

\bigskip
\bigskip

\begin{tabular}{||c||c|c|c|c|c|c|c|c||c|c|c|c||}
\hline
 & $1_1$ & $1_2$ & $1_3$ & $1_4$ & $1_5$& $1_6$ & $1_7$ & $1_8$ & $2_1$ & $2_2$ & $2_3$ & $2_4$ \\
\hline\hline
$1_1$&&&&&&&&&$\times$&$\times$&& \\
\hline
$1_2$&&&&&&&&&&$\times$&$\times$& \\
\hline
$1_3$&&&&&&&&&&&$\times$&$\times$ \\
\hline
$1_4$&&&&&&&&&$\times$&&&$\times$ \\
\hline
$1_5$&&&&&&&&&$\times$&$\times$&& \\
\hline
$1_6$&&&&&&&&&&$\times$&$\times$& \\
\hline
$1_7$ &&&&&&&&&&&$\times$&$\times$ \\
\hline
$1_8$ &&&&&&&&&$\times$&&&$\times$  \\
\hline\hline
$2_1$&$\times$&$\times$&&&$\times$&$\times$&&&$\times$&$\times$&& \\
\hline
$2_2$&&$\times$&$\times$&&&$\times$&$\times$&&&$\times$&$\times$& \\
\hline
$2_3$ &&&$\times$&$\times$&&&$\times$&$\times$&&&$\times$&$\times$ \\
\hline
$2_4$&$\times$&&&$\times$&$\times$&&&$\times$&$\times$&&&$\times$ \\
\hline
\end{tabular}

\bigskip
\bigskip

If we identify $SU(2)_L$ as the diagonal subgroup
of $SU(2)_{1,2,5,6}$ and $SU(2)_R$ as the diagonal
subgroup of $SU(2)_{3,4,7,8}$, then identify the {\bf 4} of $SU(4)$
with the ${\bf 4}$ of $SU(4)_{2,3}$
and the ${\bf \bar{4}}$ of $SU(4)_{1,4}$, this
model has eight chiral families under
$SU(2)_L \times SU(2)_R \times SU(4)$.

\bigskip
\bigskip

\noindent \underline{Group 24/15; also designated $D_4 \tilde{\times} Z_3$}

\bigskip
\bigskip
\bigskip

\noindent The group 24/15 has nine inequivalent irreducible representations,
four singlets and five doublets.

\bigskip
\bigskip

\noindent With the embedding ${\bf 4} = (2_3, 2_5)$, the fermion table is:

\bigskip
\bigskip

\begin{tabular}{||c||c|c|c|c||c|c|c|c|c||}
\hline
 & $1_1$ & $1_2$ & $1_3$ & $1_4$ & $2_1$ & $2_2$ & $2_3$ & $2_4$ & $2_5$ \\
\hline\hline
$1_1$&&&&&&&$\times$&&$\times$ \\
\hline
$1_2$&&&&&&&&$\times$&$\times$ \\
\hline
$1_3$&&&&&&&$\times$&&$\times$ \\
\hline
$1_4$&&&&&&&&$\times$&$\times$ \\
\hline\hline
$2_1$&&&&&&&$\times$&$\times\times$&$\times$ \\
\hline
$2_2$&&&&&&&$\times\times$&$\times$&$\times$ \\
\hline
$2_3$ &&$\times$&&$\times$&$\times\times$&$\times$&&& \\
\hline
$2_4$ &$\times$&&$\times$&&$\times$&$\times\times$&&&  \\
\hline
$2_5$ &$\times$& $\times$ &$\times$& $\times$ &$\times$& $\times$ &&&  \\
\hline
\end{tabular}

\bigskip
\bigskip

\noindent Identifying $SU(2)_L \equiv SU(2)_{1,3}$ and
$SU(2)_R \equiv SU(2)_{2,4}$ gives rise to two chiral families.

\bigskip

\bigskip

\newpage

\bigskip

Another chiral embedding is ${\bf 4} = (1_2, 1_3, 2_3)$ which
gives the chiral fermions:

\bigskip
\bigskip

\begin{tabular}{||c||c|c|c|c||c|c|c|c|c||}
\hline
 & $1_1$ & $1_2$ & $1_3$ & $1_4$ & $2_1$ & $2_2$ & $2_3$ & $2_4$ & $2_5$ \\
\hline\hline
$1_1$&&$\times$&$\times$&&&&$\times$&& \\
\hline
$1_2$&$\times$&&&$\times$&&&&$\times$& \\
\hline
$1_3$&$\times$&&&$\times$&&&$\times$&& \\
\hline
$1_4$&&$\times$&$\times$&&&&&$\times$& \\
\hline\hline
$2_1$&&&&&$\times$&$\times$&&$\times$&$\times$ \\
\hline
$2_2$&&&&&$\times$&$\times$&$\times$&&$\times$ \\
\hline
$2_3$&&$\times$&&$\times$&$\times$&&$\times$&$\times$& \\
\hline
$2_4$&$\times$&&$\times$&&&$\times$&$\times$&$\times$&  \\
\hline
$2_5$&&&&&$\times$&$\times$&&&$\times\times$  \\
\hline
\end{tabular}

\bigskip

\noindent Identifying $SU(2)_L$ with the diagonal subgroup
of $1_1$ and $1_3$, $SU(2)_R$ with $1_2$ and $1_4$, and then identifying 
$2_3={\bf 4}$ and $2_4 = {\bf \bar{4}}$ and finally breaking the other
three $SU(4)$'s gives rise
to six chiral families.

\bigskip

\newpage

\bigskip

\noindent As an alternative 24/15 model we can embed
${\bf 4} = (2_3, 2_3)$ and obtain:

\bigskip
\bigskip

\begin{tabular}{||c||c|c|c|c||c|c|c|c|c||}
\hline
 & $1_1$ & $1_2$ & $1_3$ & $1_4$ & $2_1$ & $2_2$ & $2_3$ & $2_4$ & $2_5$ \\
\hline\hline
$1_1$&&&&&&&$\times\times$&& \\
\hline
$1_2$&&&&&&&&$\times\times$& \\
\hline
$1_3$&&&&&&&$\times\times$&& \\
\hline
$1_4$&&&&&&&&$\times\times$& \\
\hline\hline
$2_1$&&&&&&&&$\times\times$&$\times\times$ \\
\hline
$2_2$&&&&&&&$\times\times$&&$\times\times$ \\
\hline
$2_3$ &&$\times\times$&&$\times\times$&$\times\times$&&&& \\
\hline
$2_4$ &$\times\times$&&$\times\times$&&&$\times\times$&&&  \\
\hline
$2_5$ &&&&&$\times\times$& $\times\times$ &&&  \\
\hline
\end{tabular}

\bigskip
\bigskip

\noindent With $SU(2)_L$, $SU(2)_R$ as diagonal subgroups of
$SU(2)_1 \times SU(2)_3$ and
$SU(2)_2 \times SU(2)_4$ respectively,
and breaking completely $SU(4)_4$,
this leads to four chiral families.

\bigskip
\bigskip

\newpage

\underline{{\bf g = 27.}}

\bigskip

\noindent The non-pseudoreal groups number two and both are SDPGs. In the
notation of Thomas and Wood\cite{books}
they are: $27/4,5$.
So we now treat these in the order they are
enumerated by Thomas and Wood.

\bigskip

\noindent \underline{Group 27/4; also designated $Z_9 \tilde{\times} Z_3$}

\bigskip

\noindent 27/4 has nine singlet $1_1, ...., 1_9$ and two triplet $3_1, 3_2$
irreducible representations.

\noindent 
We may choose the embedding $4 = (1_2, 3_1)$. The chiral fermions are:

\bigskip

\begin{tabular}{||c||c|c|c|c|c|c|c|c|c||c|c||}
\hline
& $1_1$ & $1_2$ & $1_3$ & $1_4$ & $1_5$ & $1_6$ & $1_7$ & $1_8$ & $1_9$ & $3_1$ & $3_2$ \\
\hline\hline
$1_1$&&$\times$&&&&&&&&$\times$& \\
\hline
$1_2$&&&$\times$&&&&&&&$\times$& \\
\hline
$1_3$&$\times$&&&&&&&&&$\times$& \\
\hline
$1_4$&&&&&$\times$&&&&&$\times$& \\
\hline
$1_5$&&&&&&$\times$&&&&$\times$& \\
\hline
$1_6$&&&&$\times$&&&&&&$\times$& \\
\hline
$1_7$ &&&&&&&&$\times$&&$\times$& \\
\hline
$1_8$ &&&&&&&&&$\times$&$\times$&  \\
\hline
$1_9$ &&&&&&&$\times$&&&$\times$&\\
\hline\hline
$3_1$ &&&&&&&&&&$\times$&$\times\times\times$ \\
\hline
$3_2$ &$\times$&$\times$&$\times$&$\times$&$\times$&$\times$&$\times$&
$\times$&$\times$&&$\times$ \\
\hline
\hline
\end{tabular}

\bigskip
\bigskip

\noindent Putting $N=2$, the gauge group is $SU(2)^9 \times SU(6)_1 \times SU(6)_2$
and the chiral fermions are, from the above table:

\begin{equation}
( \sum_{i=1}^{i=9} 2_i, \bar{6}_1 ) + (6_1, \bar{6}_1 + 3 (\bar{6_2}) )
+ (6_2, \sum_{i=1}^{i=9} 2_i ) + (6_2, \bar{6_2} )
\end{equation}

\noindent Though asymmetric in representations, this result is anomaly-free with respect to both $SU(6)_1$ and
$SU(6)_2$.

\bigskip
\bigskip

\bigskip
\bigskip
\noindent \underline{Group 27/5; also designated $(Z_3 \times Z_3) \tilde{\times} Z_3$}

\bigskip

\noindent The multiplication tables, and hence the model-building, are identical
for 27/4 and 27/5. The group 27/5 merits no further separate discussion.

\bigskip
\bigskip
\bigskip

\newpage

\underline{{\bf g = 30.}}

\bigskip

\noindent The non-pseudoreal groups number two, and neither is an SDPG. In the
notation of Thomas and Wood\cite{books}, they are: $30/2,3$.
We treat these in the order they are
enumerated by Thomas and Wood.

\bigskip

\noindent \underline{Group 30/2; also designated $D_5 \times Z_3$}

\bigskip

\noindent 30/2 has six singlets $1\alpha^i, 1^{'}\alpha^i$ and six
doublets $2\alpha^i, 2^{'}\alpha^i$ with $\alpha = exp(i \pi/3)$ and 
i = 0, 1, 2.

\bigskip

Choosing ${\bf 4} = (1\alpha, 1^{'}, 2\alpha)$ yields as fermions:

\bigskip
\bigskip

\small

\begin{tabular}{||c||c|c|c|c||c|c|c|c||c|c|c|c||}
\hline
 & $1$ & $1^{'}$ & $2$ & $2^{'}\alpha$ & $1\alpha$& $1^{'}\alpha$ & $2\alpha$ 
& $2^{'}\alpha$ & $1\alpha^2$ & $1^{'}\alpha^2$ & $2\alpha^2$ & $2^{'}\alpha^2$ \\
\hline\hline
$1$&&$\times$&&&$\times$&&$\times$&&&&& \\
\hline
$1^{'}$&$\times$&&&&&$\times$&$\times$&&&&& \\
\hline
$2$&&&$\times$&&$\times$&$\times$&$\times$&$\times$&&&& \\
\hline
$2^{'}$&&&&$\times$&&&$\times$&$\times\times$&&&& \\
\hline\hline
$1\alpha$&&&&&&$\times$&&&$\times$&&$\times$& \\
\hline
$1^{'}\alpha$&&&&&$\times$&&&&&$\times$&$\times$& \\
\hline
$2\alpha$&&&&&&&$\times$&&$\times$&$\times$&$\times$&$\times$ \\
\hline
$2^{'}\alpha$&&&&&&&&$\times$&&&$\times$&$\times\times$  \\
\hline\hline
$1\alpha^2$&$\times$&&$\times$&&&&&&&$\times$&& \\
\hline
$1^{'}\alpha^2$ & &$\times$ & $\times$ &&&& && $\times$ &&& \\
\hline
$2\alpha^2$ &$\times$& $\times$ &$\times$& $\times$ && && & && $\times$ & \\
\hline
$2^{'}\alpha^2$ & && $\times$ &$\times\times$&  &&  && & && $\times$  \\
\hline
\end{tabular}

\bigskip
\bigskip

\large

\noindent Identify $SU(2)_L$ with the diagonal subgroup of
$SU(2)_1 \times SU(2)_2$ (associated with $1, 1^{'}$) and
$SU(2)_R$ with the diagonal subgroup of
$SU(2)_5 \times SU(2)_6$ (associated with $1\alpha^2, 1^{'}\alpha^2$);
break the $SU(4)$s associated with $2, 2\alpha^2$ to arrive at
two chiral families.

\bigskip
\bigskip

\newpage

\noindent \underline{Group 30/3; also designated $D_3 \times Z_5$}

\bigskip
\bigskip
\bigskip

\noindent This group has irreps which comprise ten singlets
and five doublets and yields, for $N = 2$, the gauge group
$SU(2)^{10} \times SU(4)^5$.

\bigskip
\bigskip

\noindent As we have encountered for groups $D_3 \times Z_p$
(with $g = 6p$) the embedding
${\bf 4} = (1\alpha^{a_1}, 1^{'}\alpha^{a_2}, 2\alpha^{a_3})$
must satisfy $a_1 + a_2 = - 2 a_3$ (mod p) for consistency,
as well as $a_1 \neq 0$ to ensure ${\cal N} = 0$.

\bigskip

\noindent There are several interesting such examples, one of which
is ${\bf 4} = (1\alpha, 1^{'}, 2\alpha^2)$ which gives as fermions:

\bigskip
\bigskip

\scriptsize

\begin{tabular}{||c||c|c|c||c|c|c||c|c|c||c|c|c||c|c|c|}
\hline
 & $1$ & $1^{'}$ & $2$ & $1\alpha$ & $1^{'}\alpha$& $2\alpha$ & $1\alpha^2$ & $1^{'}\alpha^2$ & $2\alpha^2$ 
& $1\alpha^3$ & $1^{'}\alpha^3$ & $2\alpha^3$ & $1\alpha^4$& $1^{'}\alpha^4$ & $2\alpha^4$ \\
\hline\hline
$1$&&$\times$&&$\times$&&&&&$\times$&&&&&& \\
\hline
$1^{'}$&$\times$&&&&$\times$&&&&$\times$&&&&&& \\
\hline
$2$&&&$\times$&&&$\times$&$\times$&$\times$&$\times$&&&&&& \\
\hline\hline
$1\alpha$&&&&&$\times$&&$\times$&&&&&$\times$&&& \\
\hline
$1^{'}\alpha$&&&&$\times$&&&&$\times$&&&&$\times$&&& \\
\hline
$2\alpha$&&&&&&$\times$&&&$\times$&$\times$&$\times$&$\times$&&& \\
\hline\hline
$1\alpha^2$&&&&&&&&$\times$&&$\times$&&&&&$\times$  \\
\hline
$1^{'}\alpha^2$&&&&&&&$\times$&&&&$\times$& &&&$\times$ \\
\hline
$2\alpha^2$ & && &&&& && $\times$ &&&$\times$&$\times$&$\times$&$\times$ \\
\hline\hline
$1\alpha^3$ && &$\times$&&& && &&& $\times$&& $\times$ && \\
\hline
$1^{'}\alpha^3$ & && $\times$ &&  &&  && & $\times$&&&& $\times$&  \\
\hline
$2\alpha^3$&$\times$&$\times$&$\times$ &&& &&& &&&$\times$ &&&$\times$ \\
\hline\hline
$1\alpha^4$ &$\times$&& &&&$\times$ &&& &&& &&$\times$& \\
\hline
$1^{'}\alpha^4$ &&$\times$& &&&$\times$ &&& &&& &$\times$&& \\
\hline
$2\alpha^4$ &&&$\times$ &$\times$&$\times$&$\times$ &&& &&& &&&$\times$ \\
\hline
\end{tabular}

\large

\bigskip
\bigskip
\bigskip

\noindent In an obvious notation, the chiral fermions are:

\begin{eqnarray}
&&(2_1 + 2_2, \bar{4}_3 + 4_4) +
(2_3 + 2_4, \bar{4}_4 + 4_5) +
(2_5 + 2_6, \bar{4}_5 + 4_1) \nonumber \\ 
&& + (2_7 + 2_8, \bar{4}_1 + 4_2) + 
(2_9 + 2_{10}, \bar{4}_2 + 4_3)
\label{30/3}
\end{eqnarray}

\bigskip

\noindent By identifying, for example (there are equivalent cyclic permutations)
$SU(2)_L$ as the diagonal subgroup of
$SU(2)_1 \times SU(2)_2 \times SU(2)_7 \times SU(2)_8$,
$SU(2)_R$ as the diagonal subgroup
of
$SU(2)_5 \times SU(2)_6 \times SU(2)_9 \times SU(2)_{10}$,
generalized color $SU(4)$ as the diagonal subgroup of
$SU(4)_1 \times SU(4)_3$,
and breaking completely $SU(4)_{2,4,5}$
give rise to four chiral families.

\bigskip
\bigskip

\noindent We can examine the infinite series $D_3 \times Z_p$ for
$p \geq 3$ (as necessary for non-pseudoreality). The order is $g = 6p$.
By generalizing the above discussions of 18/3 ($D_3 \times Z_3$),
24/9 ($D_3 \times Z_4$) and 30/3 ($D_3 \times Z_5$) we find
that with the same type of embedding one arrives at a maximal
number of 2[p] chiral families where [x] is the largest integer
not greater than x.
For example, with p = 3, 4, 5, 6, 7, 8, 9, 10,....
one obtains 2, 4, 4, 6, 6, 8, 8, 10.... chiral families resspectively.
This is an example of accessing the more difficult nonabelian $\Gamma$
with $g \geq 32$
at least for orders $g = 6p \geq 36$.

\bigskip
\bigskip

That completes the analysis of the occurrence of chiral fermions for $\Gamma$
with $g \leq 31$. For the cases where there are $ \geq 3$ chiral families, it remains
to check whether the spectrum of complex scalars is sufficient to allow
spontaneous symmetry breaking to the Standard Model gauge group.

\bigskip
\bigskip
\bigskip

\newpage

\bigskip
\bigskip
\bigskip

\subsection{The Scalar Sector}

In order to carry out the spontaneous symmetry breaking (SSB) in the
chiral
models we found in the last section, we must first extract the scalar
sector
from eq. (5), where the $6$ is gotten from the embedding of $({\bf
4}\times
{\bf 4})_{A}$ which in turn follows from the embedding of the ${\bf
4}$. We
only consider models of phenomenological interest, i.e., those which
potentially have three or more families, but preferably three. With
this
perspective in mind we first collect the models, they are:

\newpage

\underline{16/8 with ${\bf 4}=({\bf 2}_{1},{\bf 2}_{1})$} and $\chi = 2^8$ with
$N=2$.

\underline{16/8 with ${\bf 4}=({\bf 1}_{2},{\bf 1}_{5},{\bf 2}_{1})$} and
$\chi = 2^7$
with $N=2$.

\underline{16/11 with ${\bf 4}=({\bf 1}_{2},{\bf 1}_{2},{\bf 1}_{2},{\bf 1}_{2})$} and
$\chi = 432$
with $N=3$.

\underline{16/11 with ${\bf 4}=({\bf 1}_{2},{\bf 1}_{2},{\bf 1}_{2},{\bf 1}_{4})$} and
$\chi = 216$
with $N=3$.

\underline{16/13 with ${\bf 4}=({\bf 1}_{3},{\bf 1}_{4},{\bf 2}_{1})$} and
$\chi = 2^6$
with $N=2$.

\underline{16/13 with ${\bf 4}=({\bf 2}_{1},{\bf 2}_{2})$} and $\chi = 2^6$ with
$N=2$

\underline{16/13 with ${\bf 4}=({\bf 2}_{1},{\bf 2}_{1})$} and $\chi = 2^7$ with
$N=2$.

\underline{18/3 with ${\bf 4}=({\bf 1}\alpha ,{\bf 1}^{\prime },{\bf 2}\alpha )$}
and $\chi = 192$ with $N=2$.

\underline{20/5 with ${\bf 4}=({\bf 1}_{2},{\bf 1}_{2},{\bf 1}_{2},{\bf 1}_{2})$}
and $\chi = 144$ with $N=3$.

\underline{20/5 with ${\bf 4}=({\bf 1}_{2},{\bf 1}_{2},{\bf 1}_{2},{\bf 1}_{4})$}
and $\chi=72$ with $N=3$.

\underline{21/2 with ${\bf 4}=({\bf 1}_{2},{\bf 3}_{1})$} and $\chi = 108$ with
$N=2$.

\underline{24/7 with ${\bf 4}=({\bf 1}\alpha ,{\bf 1}^{\prime },{\bf 2}\alpha )$} 
and $\chi = 240$ with $N=2$.

\underline{24/9 with ${\bf 4}=({\bf 1}_{1}\alpha ,{\bf 1}_{2}\alpha^3 ,{\bf 2}\alpha^2)$} 
and $\chi = 320$ with $N=2$.

\underline{24/9 with ${\bf 4}=({\bf 1}_{1}\alpha ,{\bf 1}_{2}\alpha ,{\bf 2}\alpha)$}
and $\chi = 320$ with $N=2$.

\underline{24/9 with ${\bf 4}=({\bf 1}_{1}\alpha^2 ,{\bf 1}_{2}, {\bf 2}\alpha)$}
and $\chi = 192$ with $N=2$.

\underline{24/9 with ${\bf 4}=({\bf 2}\alpha ,{\bf 2}\alpha )$} and $\chi = 384$
with $ N=2$.

\underline{24/13 with ${\bf 4}=({\bf 2}_{1},{\bf 2}_{2})$} and $\chi = 48$ with
$N=2$.

\underline{24/14 with ${\bf 4}=({\bf 2}_{1},{\bf 2}_{2})$} and $\chi = 192$ with
$N=2$.

\underline{24/15 with ${\bf 4}=({\bf 1}_{2},{\bf 1}_{3},{\bf 2}_{3})$} and $\chi = 2^7$ with
$N=2$.

\underline{24/15 with ${\bf 4}=({\bf 2}_{3},{\bf 2}_{5})$} and $\chi = 2^7$ with
$N=2$.

\underline{24/15 with ${\bf 4}=({\bf 2}_{3},{\bf 2}_{3})$} and $\chi = 2^8$ with
$N=2$.

\underline{27/4 with ${\bf 4}=({\bf 1}_{2},{\bf 3}_{1})$} and $\chi = 324$ with
$N=2$.

\underline{30/2 with ${\bf 4}=({\bf 1}\alpha ,{\bf 1}^{\prime },{\bf 2}\alpha )$}
and $\chi = 336$ with $N=2$.

\underline{30/3 with ${\bf 4}=({\bf 1}\alpha ,{\bf 1}^{\prime },{\bf 2}\alpha^{2})$}
and $\chi = 320$ with $N=2$.

\bigskip
where $\chi$ counts chirality, see below.
\newpage

First we consider \underline{16/8 with ${\bf 4}=({\bf 1}_{2},{\bf 1}_{2},{\bf
2}_{1})$},
where we have included this example to demonstrate improper embedding.
This
representation is complex and would be expected to lead to chiral
fermions,
but ${\bf 6}=({\bf 4}\times {\bf 4})_{A}={\bf 1}_{1}+2({\bf 2}_{1}+{\bf
2}
_{1})+({\bf 1}_{5}+{\bf 1}_{6}+{\bf 1}_{7}+{\bf 1}_{8})_{A}$ is complex
(for
any choice of singlet in the last parenthetical expression), and
therefore
the embedding ${\bf 4}=({\bf 1}_{2},{\bf 1}_{2},{\bf 2}_{1})$ is
improper
and we need not consider this or other such models further.

\bigskip

Let us define the chirality measure $\chi $ of a model as the number of
chiral fermion states. 
This variable applies to any irreps and provides
a somewhat finer measure of chirality than the number of families. 
As spontaneous symmetry breaking (SSB) proceeds,
$%
\chi $ decreased (except under unusual circumstances). For instance, the
standard model and minimal $SU(5)$ both have $\chi =45$ initially. By
the
time the symmetry is broken to $SU(3)\times U_{EM}(1)$, $\chi =3$ since
the
neutrino's cannot acquire mass due to global $B-L$ symmetry. On the
other
hand, three family $SO(10)$ and $E_{6}$ models start with $\chi =48$ and
$%
\chi =81$ respectively, but both break to $\chi =0.$

In model building with $AdS/CFT$s we are faced with a number of choices. if
we
require the initial model be chiral before SSB, then we need $\chi \geq
45$ initially. However, since the scale of SSB  $M_{AdS}$ in these models
can be relatively low (few 10s of $TeV$), vector like models are more
appealing
than usual, and we could allow an initial $\chi =0$ without resorting to
incredibly detailed fine tunings. Our prejudice is to still require a
chiral
model with $\chi \geq 45$ initially in order to gain some control in
model
building, but we want to make it clear that, even though
we have not displayed them explicitly,   
 the entire class of
vectorlike
model based on the nonabelian orbifold classification given here would
be
worthy of detailed study. There are also models (chiral or vectorlike)
that
break from $G_{AdS}$ to $SU(3)\times U_{EM}(1)$ but without going
through $SU(3)\times SU(2)\times U(1)$ directly. As $M_{AdS}$ may be not
far
above $M_{Z}$, there may be models in this class that could be in
agreement
with current data, but again we restrict most of our discussion to
chiral
models that break through the standard model. What is encouraging is the
fact that orbifold AdS/CFTs provide such a wealth of potentially
interesting
models.

\bigskip
\bigskip

\newpage

\bigskip
\bigskip

We now begin a relatively  systematic study of the scalar sectors of the chiral models.

\underline{16/8 with ${\bf 4}=({\bf 2}_{1},{\bf 2}_{1})$}. Here ${\bf 6}=\ 3({\bf
1}%
_{5})+{\bf 1}_{6}+{\bf 1}_{7}+{\bf 1}_{8}\ \ $ which is real so the
embedding is proper and the scalar sector is:

\bigskip

\small

\begin{tabular}{|c||c|c|c|c|c|c|c|c|c|c|}
\hline
$\otimes $ & 1$_{1}$ & 1$_{2}$ & 1$_{3}$ & 1$_{4}$ & 1$_{5}$ & 1$_{6}$ &
1$%
_{7}$ & 1$_{8}$ & 2 & 2$^{\prime }$ \\ \hline\hline
1$_{1}$ &  &  &  &  & $\times \times \times $ & $\times $ & $\times $ &
$%
\times $ &  &  \\ \hline
1$_{2}$ &  &  &  &  & $\times $ & $\times $ & $\times $ & $\times $ &
&  \\
\hline
1$_{3}$ &  &  &  &  & $\times $ & $\times $ & $\times $ & $\times $ &
&  \\
\hline
1$_{4}$ &  &  &  &  & $\times $ & $\times $ & $\times $ & $\times \times

\times $ &  &  \\ \hline
1$_{5}$ & $\times \times \times $ & $\times $ & $\times $ & $\times $ &
&
&  &  &  &  \\ \hline
1$_{6}$ & $\times $ & $\times $ & $\times $ & $\times $ &  &  &  &  &
&  \\
\hline
1$_{7}$ & $\times $ & $\times $ & $\times $ & $\times $ &  &  &  &  &
&  \\
\hline
1$_{8}$ & $\times $ & $\times $ & $\times $ & $\times \times \times $ &
&
&  &  &  &  \\ \hline
2 &  &  &  &  &  &  &  &  &  &
\begin{tabular}{c}
$\times \times \times $ \\
$\times \times \times $%
\end{tabular}
\\ \hline
2$^{\prime }$ &  &  &  &  &  &  &  &  &
\begin{tabular}{c}
$\times \times \times $ \\
$\times \times \times $%
\end{tabular}
&  \\ \hline
\end{tabular}

\large

\bigskip
\bigskip

\bigskip

\noindent \underline{16/8 with ${\bf 4}=({\bf 1}_{2},{\bf 1}_{4+i},{\bf 2}_{1})$} and ${\bf
6}=(%
{\bf 1}_{x(i)},{\bf 2},{\bf 2}^{{\bf \prime }}{\bf ,({\bf 1}_{5}+{\bf
1}_{6}+%
{\bf 1}_{7}+{\bf 1}_{8})}_{{\bf A}}{\bf )}$ where $x=6,5,8,$ or 7 for $
i=1,2,3,4$. The fermionic sectors of these models are identical up to
permutation, but there are two potential types of scalar sectors, depending on
whether
${\bf 1}_{x(i)}$ is the same as or different from the antisymmetric
product $
({\bf 2}_{1}\times {\bf 2}_{1})_{A}$ . Let us relabel the singlets so
$({\bf
2}_{1}\times {\bf 2}_{1})_{A}={\bf {\bf 1}_{6},}$ and then choose ${\bf
1}%
_{x(i)}$ to be either ${\bf {\bf 1}_{5}}$ or ${\bf {\bf 1}_{6}}$. Now the
two inequivalent scalar sectors (In this instance, it is easier to analyse both models and show that neither 
phenomenology is interesting, rather than untangle the correct antisymmetric 
singlet in $({\bf 2}_{1}\times {\bf 2}_{1})_{A}$. See the next section.) are:

\small

\begin{tabular}{|c||c|c|c|c|c|c|c|c|c|c|}
\hline
$\otimes $ & 1$_{1}$ & 1$_{2}$ & 1$_{3}$ & 1$_{4}$ & 1$_{5}$ & 1$_{6}$ &
1$
_{7}$ & 1$_{8}$ & 2 & 2$^{\prime }$ \\ \hline\hline
1$_{1}$ &  &  &  &  & $\times (5)$ & (6) &  &  & $\times $ & $\times $
\\
\hline
1$_{2}$ &  &  &  &  & (6) & $\times (5)$ &  &  & $\times $ & $\times $
\\
\hline
1$_{3}$ &  &  &  &  &  &  & $\times (5)$ & (6) & $\times $ & $\times $
\\
\hline
1$_{4}$ &  &  &  &  &  &  & (6) & $\times (5)$ & $\times $ & $\times $
\\
\hline
1$_{5}$ & $\times (5)$ & (6) &  &  &  &  &  &  & $\times $ & $\times $
\\
\hline
1$_{6}$ & (6) & $\times (5)$ &  &  &  &  &  &  & $\times $ & $\times $
\\
\hline
1$_{7}$ &  &  & $\times (5)$ & (6) &  &  &  &  & $\times $ & $\times $
\\
\hline
1$_{8}$ &  &  & (6) & $\times (5)$ &  &  &  &  & $\times $ & $\times $
\\
\hline
2 & $\times $ & $\times $ & $\times $ & $\times $ & $\times $ & $\times
$ & $%
\times $ & $\times $ &  & $\times \times $ \\ \hline
2$^{\prime }$ & $\times $ & $\times $ & $\times $ & $\times $ & $\times
$ & $%
\times $ & $\times $ & $\times $ & $\times \times $ &  \\ \hline
\end{tabular}

\large

\bigskip

\bigskip 

\noindent where(5) is replaced by an $"\times "$ and (6) by a blank if
${\bf 1%
}_{x(i)}={\bf {\bf 1}_{5}}$ and {\it vice versa} if ${\bf 1}_{x(i)}={\bf {\bf
1}%
_{6}.}$

\bigskip

\newpage

\bigskip

For \underline{16/11 with ${\bf 4}=({\bf 1}_{2},{\bf 1}_{2},{\bf 1}_{2},{\bf 1}_{2})$}
and ${\bf 6}=({\bf 1}_{3},{\bf 1}_{3},{\bf 1}_{3},{\bf 1}_{3},{\bf
1}_{3},{\bf 1}%
_{3}) $ the scalars are:

\bigskip

\small
\begin{tabular}{|c||c|c|c|c|c|c|c|c|c|c|}
\hline
$\otimes $ & 1$_{1}$ & 1$_{2}$ & 1$_{3}$ & 1$_{4}$ & 1$_{5}$ & 1$_{6}$ &
1$%
_{7}$ & 1$_{8}$ & 2 & 2$^{\prime }$ \\ \hline\hline
1$_{1}$ &  &  & ($\times )^{6}$ &  &  &  &  &  &  &  \\ \hline
1$_{2}$ &  & ($\times )^{6}$ &  &  &  &  &  &  &  &  \\ \hline
1$_{3}$ & ($\times )^{6}$ &  &  &  &  &  &  &  &  &  \\ \hline
1$_{4}$ &  &  &  & ($\times )^{6}$ &  &  &  &  &  &  \\ \hline
1$_{5}$ &  &  &  &  &  &  & ($\times )^{6}$ &  &  &  \\ \hline
1$_{6}$ &  &  &  &  &  & ($\times )^{6}$ &  &  &  &  \\ \hline
1$_{7}$ &  &  &  &  & ($\times )^{6}$ &  &  &  &  &  \\ \hline
1$_{8}$ &  &  &  &  &  &  &  & ($\times )^{6}$ &  &  \\ \hline
2 &  &  &  &  &  &  &  &  & ($\times )^{6}$ &  \\ \hline
2$^{\prime }$ &  &  &  &  &  &  &  &  &  & ($\times )^{6}$ \\ \hline
\end{tabular}

\large

\bigskip
\bigskip

\newpage

\bigskip
\bigskip
Similarly, for
\underline{16/11 with ${\bf 4}=({\bf 1}_{2},{\bf 1}_{2},{\bf 1}_{2},{\bf 1}_{4})$}
and ${\bf 6}=({\bf 1}_{1},{\bf 1}_{1},{\bf 1}_{1},{\bf 1}_{3},{\bf
1}_{3},{\bf 1}%
_{3}{\bf \ )}$
we find the scalars:
\bigskip

\bigskip
\begin{tabular}{|c||c|c|c|c|c|c|c|c|c|c|}
\hline
$\otimes $ & 1$_{1}$ & 1$_{2}$ & 1$_{3}$ & 1$_{4}$ & 1$_{5}$ & 1$_{6}$ &
1$%
_{7}$ & 1$_{8}$ & 2 & 2$^{\prime }$ \\ \hline\hline
1$_{1}$ & ($\times )^{3}$ &  & ($\times )^{3}$ &  &  &  &  &  &  &  \\
\hline
1$_{2}$ &  & ($\times )^{6}$ &  &  &  &  &  &  &  &  \\ \hline
1$_{3}$ & ($\times )^{3}$ &  & ($\times )^{3}$ &  &  &  &  &  &  &  \\
\hline
1$_{4}$ &  &  &  & ($\times )^{6}$ &  &  &  &  &  &  \\ \hline
1$_{5}$ &  &  &  &  & ($\times )^{3}$ &  & ($\times )^{3}$ &  &  &  \\
\hline
1$_{6}$ &  &  &  &  &  & ($\times )^{6}$ &  &  &  &  \\ \hline
1$_{7}$ &  &  &  &  & ($\times )^{3}$ &  & ($\times )^{3}$ &  &  &  \\
\hline
1$_{8}$ &  &  &  &  &  &  &  & ($\times )^{6}$ &  &  \\ \hline
2 &  &  &  &  &  &  &  &  & ($\times )^{6}$ &  \\ \hline
2$^{\prime }$ &  &  &  &  &  &  &  &  &  & ($\times )^{6}$ \\ \hline
\end{tabular}

\bigskip
Next, \underline{$16/13$ with ${\bf 4=({\bf 1}_{3},{\bf 1}_{4},2}_{1}{\bf )}$}
and ${\bf6}=({\bf 1}_{2},{\bf 1}_{c},{\bf 2}_{1},{\bf 2}_{3})$, 
where ${\bf 1}_{c}= ({\bf 2}_{1} \times {\bf 2}_{1})_{{\bf A}}$ 
so we have ${\bf 1}_{c}$ is
either
${\bf 1}_{2}$ or ${\bf 1}_{3},$ yields

\bigskip

\bigskip

\bigskip
\begin{tabular}{|c||c|c|c|c|c|c|c|}
\hline
$\otimes $ & 1$_{1}$ & 1$_{2}$ & 1$_{3}$ & 1$_{4}$ & 2$_{1}$ & 2$_{2}$ &
2$%
_{3}$ \\ \hline\hline
1$_{1}$ &  & $\times (2)$ & (3) &  & $\times $ &  & $\times $ \\ \hline
1$_{2}$ & $\times (2)$ &  &  & (3) & $\times $ &  & $\times $ \\ \hline
1$_{3}$ & (3) &  &  & $\times (2)$ & $\times $ &  & $\times $ \\ \hline
1$_{4}$ &  & (3) & $\times (2)$ &  & $\times $ &  & $\times $ \\ \hline
2$_{1}$ & $\times $ & $\times $ & $\times $ & $\times $ &  & $\times
\times $
& $\times \times $ \\ \hline
2$_{2}$ &  &  &  &  & $\times \times $ & $\times \times $ & $\times
\times $
\\ \hline
2$_{3}$ & $\times $ & $\times $ & $\times $ & $\times $ & $\times \times
$ &
$\times \times $ &  \\ \hline
\end{tabular}

\bigskip

\bigskip

\bigskip

Next, \underline{$16/13$ with ${\bf 4=(2}_{1}{\bf ,2}_{2}{\bf )}$} and ${\bf 6}=({\bf
1}_{a},%
{\bf 1}_{b},{\bf 2}_{1},{\bf 2}_{3})$, where ${\bf 1}_{a}={\bf
(2}_{1}{\bf
\times 2}_{1}{\bf )}_{{\bf A}}=(1_2+1_3+2_2)_{{\bf A}}$ and ${\bf 1}_{b}={\bf (2}_{2}{\bf \times
2}%
_{2}{\bf )}_{{\bf A}}+(1_1+1_2+1_3+1_4)_{{\bf A}}$ gives

\bigskip

\bigskip

\bigskip

\begin{tabular}{|c||c|c|c|c|c|c|c|}
\hline
$\otimes $ & 1$_{1}$ & 1$_{2}$ & 1$_{3}$ & 1$_{4}$ & 2$_{1}$ & 2$_{2}$ &
2$%
_{3}$ \\ \hline\hline
1$_{1}$ & (1) & (2) & (3) & (4) & $\times $ &  & $\times $ \\ \hline
1$_{2}$ & (2) & (1) & (4) & (3) & $\times $ &  & $\times $ \\ \hline
1$_{3}$ & (3) & (4) & (1) & (2) & $\times $ &  & $\times $ \\ \hline
1$_{4}$ & (4) & (3) & (2) & (1) & $\times $ &  & $\times $ \\ \hline
2$_{1}$ & $\times $ & $\times $ & $\times $ & $\times $ & (1)(4) &
$\times
\times $ & (2)(3) \\ \hline
2$_{2}$ &  &  &  &  & $\times \times $ & $
\begin{array}{l}
(1)(2) \\
(3)(4)
\end{array}
$ & $\times \times $ \\ \hline
2$_{3}$ & $\times $ & $\times $ & $\times $ & $\times $ & (2)(3) &
$\times
\times $ & (1)(4) \\ \hline
\end{tabular}

\bigskip

\noindent where again we insert $\times $s at the locations in parenthesis when the
singlets
are chosen properly from the antisymmetric products of the doublets.
There are three inequivalent choices, either (i) put $\times \times$ at
location (2), or (ii) put an $\times $ at (2) and one at (3), or (iii)
put $%
\times $ at (2) and $\times $ at (1). All other choices lead to
equivalent
models. Thus, without the full detailed knowledge of the antisymmetric products,
we
can still reduce the analysis to the consideration of these three cases.

\bigskip

\newpage

\bigskip

\bigskip

Finally, we have \underline{$16/13$ with ${\bf 4=(2}_{1}{\bf ,2}_{1}{\bf )}$} and ${\bf 6}=({\bf
1}_{2},%
{\bf 1}_{2},{\bf 1}_{2},{\bf 1}_{3},{\bf 2}_{2})$ (which is equivalent
to $%
{\bf 6}=({\bf 1}_{2},{\bf 1}_{3},{\bf 1}_{3},{\bf 1}_{3},{\bf 2}_{2})$
for
SSB up to a relabeling of irreps).

\bigskip

\bigskip

\bigskip
\begin{tabular}{|c||c|c|c|c|c|c|c|}
\hline
$\otimes $ & 1$_{1}$ & 1$_{2}$ & 1$_{3}$ & 1$_{4}$ & 2$_{1}$ & 2$_{2}$ &
2$%
_{3}$ \\ \hline\hline
1$_{1}$ &  & $\times \times \times $ & $\times $ &  &  & $\times $ &  \\

\hline
1$_{2}$ & $\times \times \times $ &  &  &  &  & $\times $ &  \\ \hline
1$_{3}$ & $\times $ &  &  & $\times \times \times $ &  & $\times $ &  \\

\hline
1$_{4}$ &  &  & $\times \times \times $ &  &  & $\times $ &  \\ \hline
2$_{1}$ &  &  &  &  & $\times $ &  & $
\begin{array}{l}
\times \times  \\
\times \times \times
\end{array}
$ \\ \hline
2$_{2}$ & $\times $ & $\times $ & $\times $ & $\times $ &  & $
\begin{array}{l}
\times \times  \\
\times \times
\end{array}
$ &  \\ \hline
2$_{3}$ &  &  &  &  & $
\begin{array}{l}
\times \times  \\
\times \times \times
\end{array}
$ &  & $\times $ \\ \hline
\end{tabular}

\bigskip

\newpage

\bigskip
\bigskip

Moving on to $o(\Gamma)=18$ we have 
\underline{18/3 with $4=(1^{\prime }\alpha ,1^{\prime },2\alpha )$} and
$6=(1^{\prime
}\alpha ,2\alpha ,2\alpha ^{2},1^{\prime }\alpha ^{2})$ with scalars

\bigskip

\begin{tabular}{|c||c|c|c|c|c|c|c|c|c|}
\hline
$\otimes $ & 1 & 1$^{\prime }$ & 2 & 1$\alpha $ & 1$^{\prime }\alpha $ &
2$%
\alpha $ & 1$\alpha ^{2}$ & 1$^{\prime }\alpha ^{2}$ & 2$\alpha ^{2}$ \\

\hline\hline
1 &  &  &  &  & $\times $ & $\times $ &  & $\times $ & $\times $ \\
\hline
1$^{\prime }$ &  &  &  & $\times $ &  & $\times $ & $\times $ &  &
$\times $
\\ \hline
2 &  &  &  & $\times $ & $\times $ & $\times \times $ & $\times $ &
$\times $
& $\times \times $ \\ \hline
1$\alpha $ &  & $\times $ & $\times $ &  &  &  &  & $\times $ & $\times
$ \\
\hline
1$^{\prime }\alpha $ & $\times $ &  & $\times $ &  &  &  & $\times $ &
& $%
\times $ \\ \hline
2$\alpha $ & $\times $ & $\times $ & $\times \times $ &  &  &  & $\times
$ &
$\times $ & $\times \times $ \\ \hline
1$\alpha ^{2}$ &  & $\times $ & $\times $ &  & $\times $ & $\times $ &
&  &
\\ \hline
1$^{\prime }\alpha ^{2}$ & $\times $ &  & $\times $ & $\times $ &  &
$\times
$ &  &  &  \\ \hline
2$\alpha ^{2}$ & $\times $ & $\times $ & $\times \times $ & $\times $ &
$%
\times $ & $\times \times $ &  &  &  \\ \hline
\end{tabular}

\bigskip

\newpage

\bigskip

\bigskip

The two $o(\Gamma)=20$ models are 
\underline{20/5 with ${\bf 4}=({\bf 1}_{2},{\bf 1}_{2},{\bf 1}_{2},{\bf 1}_{2})$}
and ${\bf 6}=({\bf 1}_{3},{\bf 1}_{3},{\bf 1}_{3},{\bf 1}_{3},{\bf
1}_{3},{\bf 1}%
_{3}{\bf \ )}$ which is very much like the 16/11 model because of the similar
embedding, with scalars

\bigskip

\begin{tabular}{|c||c|c|c|c|c|}
\hline
$\otimes $ & 1$_{1}$ & 1$_{2}$ & 1$_{3}$ & 1$_{4}$ & 4 \\ \hline\hline
1$_{1}$ &  &  & ($\times )^{6}$ &  &  \\ \hline
1$_{2}\ $ &  &  &  & ($\times )^{6}$ &  \\ \hline
1$_{3}$ & ($\times )^{6}$ &  &  &  &  \\ \hline
1$_{4}$ &  & ($\times )^{6}$ &  &  &  \\ \hline
4 &  &  &  &  &  \\ \hline
\end{tabular}

\bigskip

But a VEV for any of these renders the entire fermion sector vectorlike.

\bigskip

\newpage

\bigskip
\bigskip

The second model has \underline{20/5 with ${\bf 4}=({\bf 1}_{2},{\bf 1}_{2},{\bf 1}_{2},{\bf
1}_{4})$}
and ${\bf 6}=({\bf 1}_{1},{\bf 1}_{1},{\bf 1}_{1},{\bf 1}_{3},{\bf
1}_{3},%
{\bf 1}_{3}{\bf \ )}$ with scalars:

\bigskip

\bigskip

\begin{tabular}{|c||c|c|c|c|c|}
\hline
$\otimes $ & 1$_{1}$ & 1$_{2}$ & 1$_{3}$ & 1$_{4}$ & 4 \\ \hline\hline
1$_{1}$ & $\times \times \times $ &  & $\times \times \times $ &  &  \\
\hline
1$_{2}\ $ &  & $\times \times \times $ &  & $\times \times \times $ &
\\
\hline
1$_{3}$ & $\times \times \times $ &  & $\times \times \times $ &  &  \\
\hline
1$_{4}$ &  & $\times \times \times $ &  & $\times \times \times $ &  \\
\hline
4 &  &  &  &  & ($\times $)$^{6}$ \\ \hline
\end{tabular}

\bigskip

\newpage

\bigskip
\bigskip
At $o(\Gamma)=21$ we have a single model.
\underline{21/2 with ${\bf 4}=({\bf 1}_{2},{\bf 3}_{1})$} and ${\bf 6}%
=\ {\bf 3}_{1}+{\bf 3}_{2}\ $ with
$N=2$.
(All other embeddings of the ${\bf 4}$ with chiral fermions and ${\cal N}
=0$
SUSY are permutations and therefore equivalent to this model.). Here
${\bf 6}$ is real so the embedding is proper
and
the scalar sector is:

\bigskip

\begin{tabular}{|c||c|c|c|c|c|}
\hline
$\otimes $ & 1$_{1}$ & 1$_{2}$ & 1$_{3}$ & 3$_{1}$ & 3$_{2}$ \\
\hline\hline
1$_{1}$ &  &  &  & $\times $ & $\times $ \\ \hline
1$_{2}\ $ &  &  &  & $\times $ & $\times $ \\ \hline
1$_{3}$ &  &  &  & $\times $ & $\times $ \\ \hline
3$_{1}$ & $\times $ & $\times $ & $\times $ & $\times \times $ & $\times

\times \times $ \\ \hline
3$_{2}$ & $\times $ & $\times $ & $\times $ & $\times \times \times $ &
$%
\times \times $ \\ \hline
\end{tabular}

\bigskip

\bigskip

\newpage
The rich scalar sectors at $o(\Gamma)=24$ offer a number varied model building opportunities
\underline{24/7 or equivalently 24/8}

\noindent (since they have isomorphic irrep product
tables)
with ${\bf 4}=({\bf 1}_{1}, {\bf \alpha_1}_{2},{\bf 2}\alpha )$ and
${\bf 6}=({\bf 1}_{2}{\bf \alpha ,1}_{2}\alpha ^{2},{\bf 2}\alpha ,{\bf
2}%
\alpha ^{2})$ give the scalars

\bigskip

\bigskip

\small
\begin{tabular}{|c||c|c|c|c|c|c|c|c|c|c|c|c|c|c|c|}
\hline
$\otimes $ & 1$_{1}$ & 1$_{2}$ & 1$_{3}$ & 1$_{4}$ & 2 & 1$_{1}\alpha $
& 1$%
_{2}\alpha $ & 1$_{3}\alpha $ & 1$_{4}\alpha $ & 2$\alpha $ &
1$_{1}\alpha
^{2}$ & 1$_{2}\alpha ^{2}$ & 1$_{3}\alpha ^{2}$ & 1$_{4}\alpha ^{2}$ &
2$%
\alpha ^{2}$ \\ \hline\hline
1$_{1}$ &  &  &  &  &  &  & $\times $ &  &  & $\times $ &  & $\times $
&  &
& $\times $ \\ \hline
1$_{2}$ &  &  &  &  &  & $\times $ &  &  &  & $\times $ & $\times $ &
&  &
& $\times $ \\ \hline
1$_{3}$ &  &  &  &  &  &  &  &  & $\times $ & $\times $ &  &  &  &
$\times $
& $\times $ \\ \hline
1$_{4}$ &  &  &  &  &  &  &  & $\times $ &  & $\times $ &  &  & $\times
$ &
& $\times $ \\ \hline
2 &  &  &  &  &  & $\times $ & $\times $ & $\times $ & $\times $ &
$\times $
& $\times $ & $\times $ & $\times $ & $\times $ & $\times $ \\ \hline
1$_{1}\alpha $ &  & $\times $ &  &  & $\times $ &  &  &  &  &  &  &
$\times $
&  &  & $\times $ \\ \hline
1$_{2}\alpha $ & $\times $ &  &  &  & $\times $ &  &  &  &  &  & $\times
$ &
&  &  & $\times $ \\ \hline
1$_{3}\alpha $ &  &  &  & $\times $ & $\times $ &  &  &  &  &  &  &  &
& $%
\times $ & $\times $ \\ \hline
1$_{4}\alpha $ &  &  & $\times $ &  & $\times $ &  &  &  &  &  &  &  &
$%
\times $ &  & $\times $ \\ \hline
2$\alpha $ & $\times $ & $\times $ & $\times $ & $\times $ & $\times $
&  &
&  &  &  & $\times $ & $\times $ & $\times $ & $\times $ & $\times $ \\
\hline
1$_{1}\alpha ^{2}$ &  & $\times $ &  &  & $\times $ &  & $\times $ &  &
& $%
\times $ &  &  &  &  &  \\ \hline
1$_{2}\alpha ^{2}$ & $\times $ &  &  &  & $\times $ & $\times $ &  &  &
& $%
\times $ &  &  &  &  &  \\ \hline
1$_{3}\alpha ^{2}$ &  &  &  & $\times $ & $\times $ &  &  &  & $\times $
& $%
\times $ &  &  &  &  &  \\ \hline
1$_{4}\alpha ^{2}$ &  &  & $\times $ &  & $\times $ &  &  & $\times $ &
& $%
\times $ &  &  &  &  &  \\ \hline
2$\alpha ^{2}$ & $\times $ & $\times $ & $\times $ & $\times $ & $\times
$ &
$\times $ & $\times $ & $\times $ & $\times $ & $\times $ &  &  &  &  &
\\
\hline
\end{tabular}

\large

\bigskip

\newpage

\bigskip

The scalars for \underline{24/9 with ${\bf 4=(1}_{{\bf 1}}{\bf \alpha ,{\bf
1}%
_{2}\alpha }^{{\bf 3}},{\bf 2{\bf \alpha }^{{\bf 2}})}$} and ${\bf
6}=({\bf 1}%
_{2}{\bf ,1}_{2},{\bf 2}\alpha ,{\bf 2}\alpha ^{3})$ are:

\bigskip

\small

\begin{tabular}{|c||c|c|c|c|c|c|c|c|c|c|c|c|}
\hline
$\otimes $ & 1$_{1}$ & 1$_{2}$ & 2 & 1$_{1}\alpha $ & 1$_{2}\alpha $ &
2$%
\alpha $ & 1$_{1}\alpha ^{2}$ & 1$_{2}\alpha ^{2}$ & 2$\alpha ^{2}$ &
1$%
_{1}\alpha ^{3}$ & 1$_{2}\alpha ^{3}$ & 2$\alpha ^{3}$ \\ \hline\hline
1$_{1}$ &  & $\times \times $ &  &  &  & $\times $ &  &  &  &  &  &
$\times $
\\ \hline
1$_{2}$ & $\times \times $ &  &  &  &  & $\times $ &  &  &  &  &  &
$\times $
\\ \hline
2 &  &  & $\times \times $ & $\times $ & $\times $ & $\times $ &  &  &
& $%
\times $ & $\times $ & $\times $ \\ \hline
1$_{1}\alpha $ &  &  & $\times $ &  & $\times \times $ &  &  &  &
$\times $
&  &  &  \\ \hline
1$_{2}\alpha $ &  &  & $\times $ & $\times \times $ &  &  &  &  &
$\times $
&  &  &  \\ \hline
2$\alpha $ & $\times $ & $\times $ & $\times $ &  &  & $\times \times $
& $%
\times $ & $\times $ & $\times $ &  &  &  \\ \hline
1$_{1}\alpha ^{2}$ &  &  &  &  &  & $\times $ &  & $\times \times $ &
&  &
& $\times $ \\ \hline
1$_{2}\alpha ^{2}$ &  &  &  &  &  & $\times $ & $\times \times $ &  &
&  &
& $\times $ \\ \hline
2$\alpha ^{2}$ &  &  &  & $\times $ & $\times $ & $\times $ &  &  &
$\times
\times $ & $\times $ & $\times $ & $\times $ \\ \hline
1$_{1}\alpha ^{3}$ &  &  & $\times $ &  &  &  &  &  & $\times $ &  &
$\times
\times $ &  \\ \hline
1$_{1}\alpha ^{3}$ &  &  & $\times $ &  &  &  &  &  & $\times $ &
$\times
\times $ &  &  \\ \hline
2$\alpha ^{3}$ & $\times $ & $\times $ & $\times $ &  &  &  & $\times $
& $%
\times $ & $\times $ &  &  & $\times \times $ \\ \hline
\end{tabular}

\large

\bigskip

\newpage

\bigskip

\noindent The scalars for \underline{24/9 with 
${\bf 4= ({1}_{1}\alpha , 1_{2}\alpha, 2\alpha)}$} 
and 
${\bf 6}=({\bf 1_2\alpha, 2,1_2\alpha , 2 \alpha^2})$
are:
\bigskip

\bigskip

\bigskip

\small

\begin{tabular}{|c||c|c|c|c|c|c|c|c|c|c|c|c|}
\hline
$\otimes $ & 1$_{1}$ & 1$_{2}$ & 2 & 1$_{1}\alpha $ & 1$_{2}\alpha $ &
2$%
\alpha $ & 1$_{1}\alpha ^{2}$ & 1$_{2}\alpha ^{2}$ & 2$\alpha ^{2}$ &
1$%
_{1}\alpha ^{3}$ & 1$_{2}\alpha ^{3}$ & 2$\alpha ^{3}$ \\ \hline\hline
1$_{1}$ &  &  &  &  &  &  &  & $\times \times $ & $\times \times $ &  &
&
\\ \hline
1$_{2}$ &  &  &  &  &  &  & $\times \times $ &  & $\times \times $ &  &
&
\\ \hline
2 &  &  &  &  &  &  & $\times \times $ & $\times \times $ & $
\begin{array}{l}
\times \times  \\
\times \times
\end{array}
$ &  &  &  \\ \hline
1$_{1}\alpha $ &  &  &  &  &  &  &  &  &  &  & $\times \times $ &
$\times
\times $ \\ \hline
1$_{2}\alpha $ &  &  &  &  &  &  &  &  &  & $\times \times $ &  &
$\times
\times $ \\ \hline
2$\alpha $ &  &  &  &  &  &  &  &  &  & $\times \times $ & $\times
\times $
& $
\begin{array}{l}
\times \times  \\
\times \times
\end{array}
$ \\ \hline
1$_{1}\alpha ^{2}$ &  & $\times \times $ & $\times \times $ &  &  &  &
&  &
&  &  &  \\ \hline
1$_{2}\alpha ^{2}$ & $\times \times $ &  & $\times \times $ &  &  &  &
&  &
&  &  &  \\ \hline
2$\alpha ^{2}$ & $\times \times $ & $\times \times $ & $
\begin{array}{l}
\times \times  \\
\times \times
\end{array}
$ &  &  &  &  &  &  &  &  &  \\ \hline
1$_{1}\alpha ^{3}$ &  &  &  &  & $\times \times $ & $\times \times $ &
&  &
&  &  &  \\ \hline
1$_{1}\alpha ^{3}$ &  &  &  & $\times \times $ &  & $\times \times $ &
&  &
&  &  &  \\ \hline
2$\alpha ^{3}$ &  &  &  & $\times \times $ & $\times \times $ & $
\begin{array}{l}
\times \times  \\
\times \times
\end{array}
$ &  &  &  &  &  &  \\ \hline
\end{tabular}

\large

\bigskip

\newpage

\bigskip

For \underline{24/9 with ${\bf 4}=({\bf 1}_{1}\alpha ,{\bf 1}_{2},{\bf 2}\alpha )$} and
   ${\bf 6}=({\bf 1}_{2}\alpha ^{2},{\bf
2}\alpha
,{\bf 2}\alpha ^{-1},{\bf 1}_{2}\alpha ^{-2})$ where $\alpha ^{4}=1,$
the
scalar sector is:

\bigskip

\small

\begin{tabular}{|c||c|c|c|c|c|c|c|c|c|c|c|c|}
\hline
$\otimes $ & 1$_{1}$ & 1$_{2}$ & 2 & 1$_{1}\alpha $ & 1$_{2}\alpha $ &
2$%
\alpha $ & 1$_{1}\alpha ^{2}$ & 1$_{2}\alpha ^{2}$ & 2$\alpha ^{2}$ &
1$%
_{1}\alpha ^{3}$ & 1$_{2}\alpha ^{3}$ & 2$\alpha ^{3}$ \\ \hline\hline
1$_{1}$ &  &  &  &  &  & $\times $ &  & $\times \times $ &  &  &  &
$\times $
\\ \hline
1$_{2}$ &  &  &  &  &  & $\times $ & $\times \times $ &  &  &  &  &
$\times $
\\ \hline
2 &  &  &  & $\times $ & $\times $ & $\times $ &  &  & $\times \times $
& $%
\times $ & $\times $ & $\times $ \\ \hline
1$_{1}\alpha $ &  &  & $\times $ &  &  &  &  &  & $\times $ &  & $\times

\times $ &  \\ \hline
1$_{2}\alpha $ &  &  & $\times $ &  &  &  &  &  & $\times $ & $\times
\times
$ &  &  \\ \hline
2$\alpha $ & $\times $ & $\times $ & $\times $ &  &  &  & $\times $ & $%

\times $ & $\times $ &  &  & $\times \times $ \\ \hline
1$_{1}\alpha ^{2}$ &  & $\times \times $ &  &  &  & $\times $ &  &  &
&  &
& $\times $ \\ \hline
1$_{2}\alpha ^{2}$ & $\times \times $ &  &  &  &  & $\times $ &  &  &
&  &
& $\times $ \\ \hline
2$\alpha ^{2}$ &  &  & $\times \times $ & $\times $ & $\times $ &
$\times $
&  &  &  & $\times $ & $\times $ & $\times $ \\ \hline
1$_{1}\alpha ^{3}$ &  &  & $\times $ &  & $\times \times $ &  &  &  & $%

\times $ &  &  &  \\ \hline
1$_{2}\alpha ^{3}$ &  &  & $\times $ & $\times \times $ &  &  &  &  & $%

\times $ &  &  &  \\ \hline
2$\alpha ^{3}$ & $\times $ & $\times $ & $\times $ &  &  & $\times
\times $
& $\times $ & $\times $ & $\times $ &  &  &  \\ \hline
\end{tabular}

\large

\bigskip

\newpage

\bigskip

For \underline{24/9 with ${\bf 4}=({\bf 2}\alpha ,{\bf 2}\alpha )$} 
where ${\bf 6}=3({\bf 1}_{2}\alpha^{2})+{\bf 1}_{1}
\alpha^{2}+{\bf 2}\alpha^{2},$the scalars are:

\bigskip

\scriptsize

\begin{tabular}{|c||c|c|c|c|c|c|c|c|c|c|c|c|}
\hline
$\otimes $ & 1$_{1}$ & 1$_{2}$ & 2 & 1$_{1}\alpha $ & 1$_{2}\alpha $ &
2$%
\alpha $ & 1$_{1}\alpha ^{2}$ & 1$_{2}\alpha ^{2}$ & 2$\alpha ^{2}$ &
1$%
_{1}\alpha ^{3}$ & 1$_{2}\alpha^{3}$ & 2$\alpha^{3}$ \\ \hline\hline
1$_{1}$ &  &  &  &  &  &  & $\times $ & $\times \times \times $ &
$\times
\times $ &  &  &  \\ \hline
1$_{2}$ &  &  &  &  &  &  & $\times \times \times $ & $\times $ &
$\times
\times $ &  &  &  \\ \hline
2 &  &  &  &  &  &  & $\times \times $ & $\times \times $ &
\begin{tabular}{c}
$\times \times \times $ \\
$\times \times \times $%
\end{tabular}
&  &  &  \\ \hline
1$_{1}\alpha $ &  &  &  &  &  &  &  &  &  & $\times $ & $\times \times
\times $ & $\times \times $ \\ \hline
1$_{2}\alpha $ &  &  &  &  &  &  &  &  &  & $\times \times \times $ & $%

\times $ & $\times \times $ \\ \hline
2$\alpha $ &  &  &  &  &  &  &  &  &  & $\times \times $ & $\times
\times $
&
\begin{tabular}{c}
$\times \times \times $ \\
$\times \times \times $%
\end{tabular}
\\ \hline
1$_{1}\alpha ^{2}$ & $\times $ & $\times \times \times $ & $\times
\times $
&  &  &  &  &  &  &  &  &  \\ \hline
1$_{2}\alpha ^{2}$ & $\times \times \times $ & $\times $ & $\times
\times $
&  &  &  &  &  &  &  &  &  \\ \hline
2$\alpha ^{2}$ & $\times \times $ & $\times \times $ &
\begin{tabular}{c}
$\times \times \times $ \\
$\times \times \times $%
\end{tabular}
&  &  &  &  &  &  &  &  &  \\ \hline
1$_{1}\alpha ^{3}$ &  &  &  & $\times $ & $\times \times \times $ &
$\times
\times $ &  &  &  &  &  &  \\ \hline
1$_{2}\alpha ^{3}$ &  &  &  & $\times \times \times $ & $\times $ &
$\times
\times $ &  &  &  &  &  &  \\ \hline
2$\alpha ^{3}$ &  &  &  & $\times \times $ & $\times \times $ &
\begin{tabular}{c}
$\times \times \times $ \\
$\times \times \times $%
\end{tabular}
&  &  &  &  &  &  \\ \hline
\end{tabular}

\large

\bigskip

\newpage

\bigskip

The next example of interest is \underline{24/13 with ${\bf 4}=({\bf 2}_{1},{\bf
2}%
_{2})\ $} and ${\bf 6}=\ \ {\bf 1}_{1}+{\bf 1}_{2}+{\bf 1}_{3}+{\bf
3}$
with scalars:

\bigskip

\begin{tabular}{|c||c|c|c|c|c|c|c|}
\hline
$\otimes $ & 1$_{1}$ & 1$_{2}$ & 1$_{3}$ & 2$_{1}$ & 2$_{2}$ & 2$_{3}$ &
3
\\ \hline\hline
1$_{1}$ & $\times $ & $\times $ & $\times $ &  &  &  & $\times $ \\
\hline
1$_{2}$ & $\times $ & $\times $ & $\times $ &  &  &  & $\times $ \\
\hline
1$_{3}$ & $\times $ & $\times $ & $\times $ &  &  &  & $\times $ \\
\hline
2$_{1}$ &  &  &  & $\times \times $ & $\times \times $ & $\times \times
$ &
\\ \hline
2$_{2}$ &  &  &  & $\times \times $ & $\times \times $ & $\times \times
$ &
\\ \hline
2$_{3}$ &  &  &  & $\times \times $ & $\times \times $ & $\times \times
$ &
\\ \hline
3 & $\times $ & $\times $ & $\times $ &  &  &  & $\times \times $ \\
\hline
\end{tabular}

\bigskip

\bigskip

There are two ineqivalent models for the group \underline{24/15}, they are  
\underline{${\bf 4}=({\bf 1}_{2},{\bf 1}_{3},{\bf 2}_{3})$} where ${\bf 6}={\bf
1}%
_{4}+{\bf 1}_{2[4]}+{\bf 2}_{3}+{\bf 2}_{4}$ and the scalars are:

\bigskip

\begin{tabular}{|c||c|c|c|c|c|c|c|c|c|}
\hline
$\otimes $ & 1$_{1}$ & 1$_{2}$ & 1$_{3}$ & 1$_{4}$ & 2$_{1}$ & 2$_{2}$ &
2$%
_{3}$ & 2$_{4}$ & 2$_{5}$ \\ \hline\hline
1$_{1}$ &  &  &  & $\times \times $ &  &  & $\times $ & $\times $ &  \\
\hline
1$_{2}$ &  &  & $\times \times $ &  &  &  & $\times $ & $\times $ &  \\
\hline
1$_{3}$ &  & $\times \times $ &  &  &  &  & $\times $ & $\times $ &  \\
\hline
1$_{4}$ & $\times \times $ &  &  &  &  &  & $\times $ & $\times $ &  \\
\hline
2$_{1}$ &  &  &  &  &  & $\times \times $ & $\times $ & $\times $ &
$\times
\times $ \\ \hline
2$_{2}$ &  &  &  &  & $\times \times $ &  & $\times $ & $\times $ &
$\times
\times $ \\ \hline
2$_{3}$ & $\times $ & $\times $ & $\times $ & $\times $ & $\times $ & $%

\times $ &  & $\times \times $ &  \\ \hline
2$_{4}$ & $\times $ & $\times $ & $\times $ & $\times $ & $\times $ & $%

\times $ & $\times \times $ &  &  \\ \hline
2$_{5}$ &  &  &  &  & $\times \times $ & $\times \times $ &  &  &
$\times
\times $ \\ \hline
\end{tabular}

\bigskip

if (${\bf 2}_{3}\times {\bf 2}_{3})_{A}={\bf 1}_{4}$ but if it is ${\bf
1}%
_{2}$ then the top $4\times 4$ changes to:

\bigskip

\begin{tabular}{||c|c|c|c|}
\hline\hline
& $\times $ &  & $\times $ \\ \hline
$\times $ &  & $\times $ &  \\ \hline
& $\times $ &  & $\times $ \\ \hline
$\times $ &  & $\times $ &  \\ \hline
\end{tabular}

\bigskip

\newpage

\bigskip
\bigskip

The other \underline{24/15} case has \underline{${\bf 4}=({\bf 2}_{3},{\bf 2}_{3})$} 
where ${\bf 6}=3({\bf 1}_{2})+{\bf 1}_{4}+{\bf 2}_{1}$ and the scalars
are (this time swapping ${\bf 1}_{2}\ $and ${\bf 1}_{4}$ gives equivalent
models):

\bigskip

\small

\begin{tabular}{|c||c|c|c|c|c|c|c|c|c|}
\hline
$\otimes $ & 1$_{1}$ & 1$_{2}$ & 1$_{3}$ & 1$_{4}$ & 2$_{1}$ & 2$_{2}$ &
2$%
_{3}$ & 2$_{4}$ & 2$_{5}$ \\ \hline\hline
1$_{1}$ &  & $\times \times \times $ &  & $\times $ & $\times $ &  &  &
&
\\ \hline
1$_{2}$ & $\times \times \times $ &  & $\times $ &  &  & $\times $ &  &
&
\\ \hline
1$_{3}$ &  & $\times $ &  & $\times \times \times $ & $\times $ &  &  &
&
\\ \hline
1$_{4}$ & $\times $ &  & $\times \times \times $ &  &  &  &  &  &  \\
\hline
2$_{1}$ & $\times $ &  & $\times $ &  & $\times $ &
\begin{tabular}{c}
$\times \times $ \\
$\times \times $%
\end{tabular}
&  &  &  \\ \hline
2$_{2}$ &  & $\times $ &  &  &
\begin{tabular}{c}
$\times \times $ \\
$\times \times $%
\end{tabular}
& $\times $ &  &  &  \\ \hline
2$_{3}$ &  &  &  &  &  &  &  &
\begin{tabular}{c}
$\times \times \times $ \\
$\times \times $%
\end{tabular}
& $\times $ \\ \hline
2$_{4}$ &  &  &  &  &  &  &
\begin{tabular}{c}
$\times \times \times $ \\
$\times \times $%
\end{tabular}
&  & $\times $ \\ \hline
2$_{5}$ &  &  &  &  &  &  & $\times $ & $\times $ &
\begin{tabular}{c}
$\times \times $ \\
$\times \times $%
\end{tabular}
\\ \hline
\end{tabular}

\large

\bigskip

\newpage

\bigskip
\bigskip

The only $o(\Gamma)=27$ model to evaluate is \underline{27/4 with ${\bf 4}=({\bf 1}_{2},
{\bf3}_{1})$},
 where ${\bf 6}=\ {\bf 3}_{1}+{\bf 3}_{2}$ is real, and where the scalar
sector is:

\bigskip

\begin{tabular}{|c||c|c|c|c|c|c|c|c|c|c|c|}
\hline
$\otimes $ & 1$_{1}$ & 1$_{2}$ & 1$_{3}$ & 1$_{4}$ & 1$_{5}$ & 1$_{6}$ &
1$%
_{7}$ & 1$_{8}$ & 1$_{9}$ & 3$_{1}$ & 3$_{2}$ \\ \hline\hline
1$_{1}$ &  &  &  &  &  &  &  &  &  & $\times $ & $\times $ \\ \hline
1$_{2}$ &  &  &  &  &  &  &  &  &  & $\times $ & $\times $ \\ \hline
1$_{3}$ &  &  &  &  &  &  &  &  &  & $\times $ & $\times $ \\ \hline
1$_{4}$ &  &  &  &  &  &  &  &  &  & $\times $ & $\times $ \\ \hline
1$_{5}$ &  &  &  &  &  &  &  &  &  & $\times $ & $\times $ \\ \hline
1$_{6}$ &  &  &  &  &  &  &  &  &  & $\times $ & $\times $ \\ \hline
1$_{7}$ &  &  &  &  &  &  &  &  &  & $\times $ & $\times $ \\ \hline
1$_{8}$ &  &  &  &  &  &  &  &  &  & $\times $ & $\times $ \\ \hline
1$_{9}$ &  &  &  &  &  &  &  &  &  & $\times $ & $\times $ \\ \hline
3$_{1}$ & $\times $ & $\times $ & $\times $ & $\times $ & $\times $ & $%

\times $ & $\times $ & $\times $ & $\times $ &  & $\times \times \times
$ \\
\hline
3$_{2}$ & $\times $ & $\times $ & $\times $ & $\times $ & $\times $ & $%

\times $ & $\times $ & $\times $ & $\times $ & $\times \times \times $
&  \\
\hline
\end{tabular}

\bigskip

\newpage

\bigskip

And finally, at order 30 we have for \underline{30/2 with  
${\bf 4}=({\bf 1}\alpha ,{\bf 1}^{\prime },{\bf 2}\alpha )$}
 and  ${\bf 6}=({\bf 1}^{\prime }\alpha +{\bf 2}\alpha +{\bf 2}\alpha
^{-1}+%
{\bf 1}^{\prime }\alpha ^{-1})$ where $\alpha ^{3}=1,$ a model with scalar sector:

\bigskip

\bigskip
\begin{tabular}{|c||c|c|c|c|c|c|c|c|c|c|c|c|}
\hline
$\otimes $ & 1 & 1$^{\prime }$ & 2 & $2^{\prime }$ & 1$\alpha $ &
1$^{\prime
}\alpha $ & 2$\alpha $ & $2^{\prime }\alpha $ & 1$\alpha ^{2}$ &
1$^{\prime
}\alpha ^{2}$ & 2$\alpha ^{2}$ & $2^{\prime }\alpha ^{2}$ \\
\hline\hline
1 &  &  &  &  &  & $\times $ & $\times $ &  &  & $\times $ & $\times $
&  \\
\hline
1$^{\prime }$ &  &  &  &  & $\times $ &  & $\times $ &  & $\times $ &  &
$%
\times $ &  \\ \hline
$2$ &  &  &  &  & $\times $ & $\times $ & $\times $ & $\times $ &
$\times $
& $\times $ & $\times $ & $\times $ \\ \hline
$2^{\prime }$ &  &  &  &  &  &  & $\times $ & $\times \times $ &  &  &
$%
\times $ & $\times \times $ \\ \hline
1$\alpha $ &  & $\times $ & $\times $ &  &  &  &  &  &  & $\times $ & $%

\times $ &  \\ \hline
1$^{\prime }\alpha $ & $\times $ &  & $\times $ &  &  &  &  &  & $\times
$ &
& $\times $ &  \\ \hline
2$\alpha $ & $\times $ & $\times $ & $\times $ & $\times $ &  &  &  &  &
$%
\times $ & $\times $ & $\times $ & $\times $ \\ \hline
$2^{\prime }\alpha $ &  &  & $\times $ & $\times \times $ &  &  &  &  &
&
& $\times $ & $\times \times $ \\ \hline
1$\alpha ^{2}$ &  & $\times $ & $\times $ &  &  & $\times $ & $\times $
&  &
&  &  &  \\ \hline
1$^{\prime }\alpha ^{2}$ & $\times $ &  & $\times $ &  & $\times $ &  &
$%
\times $ &  &  &  &  &  \\ \hline
2$\alpha ^{2}$ & $\times $ & $\times $ & $\times $ & $\times $ & $\times
$ &
$\times $ & $\times $ & $\times $ &  &  &  &  \\ \hline
$2^{\prime }\alpha ^{2}$ &  &  & $\times $ & $\times \times $ &  &  & $%

\times $ & $\times \times $ &  &  &  &  \\ \hline
\end{tabular}

\bigskip

\newpage

\bigskip
\bigskip

The other possibility at order 30 is 
\underline{30/3 with  ${\bf 4}=({\bf 1}\alpha ,
{\bf 1}^{\prime },{\bf 2}\alpha^{2})$}
 where ${\bf 6}={\bf 1}^{\prime }\alpha +{\bf
2}\alpha
^{2}+{\bf 2}\alpha ^{3}+{\bf 1}^{\prime }\alpha ^{4}$ and $\alpha
^{5}=1,$
where the scalars are:

\bigskip

\bigskip

\footnotesize

\begin{tabular}{|c||c|c|c|c|c|c|c|c|c|c|c|c|c|c|c|}
\hline
$\otimes $ & 1 & 1$^{\prime }$ & 2 & 1$\alpha $ & 1$^{\prime }\alpha $ &
2$%
\alpha $ & 1$\alpha ^{2}$ & 1$^{\prime }\alpha ^{2}$ & 2$\alpha ^{2}$ &
1$%
\alpha ^{3}$ & 1$^{\prime }\alpha ^{3}$ & 2$\alpha ^{3}$ & 1$\alpha
^{4}$ & 1%
$^{\prime }\alpha ^{4}$ & 2$\alpha ^{4}$ \\ \hline\hline
1 &  &  &  &  & $\times $ &  &  &  & $\times $ &  &  & $\times $ &  & $%

\times $ &  \\ \hline
1$^{\prime }$ &  &  &  & $\times $ &  &  &  &  & $\times $ &  &  &
$\times $
& $\times $ &  &  \\ \hline
2 &  &  &  &  &  & $\times $ & $\times $ & $\times $ & $\times $ &
$\times $
& $\times $ & $\times $ &  &  & $\times $ \\ \hline
1$\alpha $ &  & $\times $ &  &  &  &  &  & $\times $ &  &  &  & $\times
$ &
&  & $\times $ \\ \hline
1$^{\prime }\alpha $ & $\times $ &  &  &  &  &  & $\times $ &  &  &  &
& $%
\times $ &  &  & $\times $ \\ \hline
2$\alpha $ &  &  & $\times $ &  &  &  &  &  & $\times $ & $\times $ & $%

\times $ & $\times $ & $\times $ & $\times $ & $\times $ \\ \hline
1$\alpha ^{2}$ &  &  & $\times $ &  & $\times $ &  &  &  &  &  & $\times
$ &
&  &  & $\times $ \\ \hline
1$^{\prime }\alpha ^{2}$ &  &  & $\times $ & $\times $ &  &  &  &  &  &
$%
\times $ &  &  &  &  & $\times $ \\ \hline
2$\alpha ^{2}$ & $\times $ & $\times $ & $\times $ &  &  & $\times $ &
&  &
&  &  & $\times $ & $\times $ & $\times $ & $\times $ \\ \hline
1$\alpha ^{3}$ &  &  & $\times $ &  &  & $\times $ &  & $\times $ &  &
&  &
&  & $\times $ &  \\ \hline
1$^{\prime }\alpha ^{3}$ &  &  & $\times $ &  &  & $\times $ & $\times $
&
&  &  &  &  & $\times $ &  &  \\ \hline
2$\alpha ^{3}$ & $\times $ & $\times $ & $\times $ & $\times $ & $\times
$ &
$\times $ &  &  & $\times $ &  &  &  &  &  & $\times $ \\ \hline
1$\alpha ^{4}$ &  & $\times $ &  &  &  & $\times $ &  &  & $\times $ &
& $%
\times $ &  &  &  &  \\ \hline
1$^{\prime }\alpha ^{4}$ & $\times $ &  &  &  &  & $\times $ &  &  &
$\times
$ & $\times $ &  &  &  &  &  \\ \hline
2$\alpha ^{4}$ &  &  & $\times $ & $\times $ & $\times $ & $\times $ &
$%
\times $ & $\times $ & $\times $ &  &  & $\times $ &  &  &  \\ \hline
\end{tabular}

\bigskip
\bigskip
\bigskip

\large

This concludes the enumeration of the scalar sectors of the properly embedded models. We will now investigate a number of symmetry breaking scenarios for these models.

\large

\newpage

\subsection{ Spontaneous Symmetry Breaking}

\bigskip
\bigskip

We are now in a position to carry out the spontaneous symmetry breaking for the models with 
fermions and scalars
given in the previous sections. 
We restrict ourselves to chiral models with the potential of
at least three families
($\chi \geq 45$) and for the most part consider only models with $N = 2$, although we have 
included two $N = 3$
models. Again, we move progressively through the models of increasing order of $\Gamma$.
The model is completely fixed by $\Gamma$, the 
embedding of $\bf{4}$ in $\Gamma$, and 
the choice of $N$. 

\bigskip

It is worth mentioning explicitly that no attempt is
made to stationarize the scalar potential in any of
our models. We are tacitly assuming that there are
sufficient parameters in the potential to arrange
a minimum with the scalar VEVs displayed. It
would be an interesting research project to study this
in specific cases and will generally require
a computer analysis.

\bigskip

The first
relevant model is:

\underline{16/8 with ${\bf 4=(2_{1}{\bf ,}2_{1})}$ and $N=2$}

The chiral fermions are 

$2[(2,1,1,1,1,1,1,1;4,1)  +(1,1,1,1,2,1,1,1;1,4) \\ 
+(1,2,1,1,1,1,1,1;4,1)  +(1,1,1,1,1,2,1,1;1,4) \\
+(1,1,2,1,1,1,1,1;4,1)  +(1,1,1,1,1,1,2,1;1,4) \\
+(1,1,1,2,1,1,1,1;4,1)  +(1,1,1,1,1,1,1,2;1,4) \\
+(2,1,1,1,1,1,1,1;1,\bar{4})  +(1,1,1,1,2,1,1,1;\bar{4},1) \\
+(1,2,1,1,1,1,1,1;1,\bar{4})  +(1,1,1,1,1,2,1,1;\bar{4},1) \\
+(1,1,2,1,1,1,1,1;1,\bar{4})  +(1,1,1,1,1,1,2,1;\bar{4},1) \\
+(1,1,1,2,1,1,1,1;1,\bar{4}) + (1,1,1,1,1,1,1,2; \bar{4}, 1)]$ 

\noindent and $\chi = 2^8$. From the table of scalars
for this model, we find that if we break $SU(4) \times SU(4)$ 
to the diagonal $SU_{D}(4)$,
then the model becomes vectorlike.

All scalars that are nontrivial in the $SU(4)$s are of the form 
$(1,1,1,1,1,1,1,1;4,\bar{4})+h.c.$, 
and a VEV for any one can be rotated such that the
unbroken
symmetry is $SU_{D}(4)$. All other scalars are $SU_{i}(2)\times
SU_{j}(2)$
bilinears, hence we cannot break to a Pati-Salam (PS) model or any
standard
type chiral model with this strategy.

As mentioned above, symmetry breaking via non-diagonal subgroups is another possibility, but we will not investigate such models here. They usually lead to somewhat more complicated patterns of spontaneous symmetry breaking (SSB) and while they are more cumbersome to analyze, they do offer further model building opportunities.

\bigskip
\bigskip

We continue with
\underline{16/8 with  ${\bf 4=(1_{2}{\bf ,}1_{4+i}{\bf ,}2_{1})}$ and $N=2,$}
\noindent where ${\bf
6=(1_{x(i)}%
{\bf ,}2_{1},2_{2,}({\bf 1}_{5}{\bf ,1}_{6},{\bf 1}}_{{\bf 7}}
{\bf ,1}_{8})_{{\bf A}}{\bf )}$ with $x=6,5,8,7$ for
$i=1,2,3,4.$
This model has only half the initial chirality of the previous
model ($%
\chi =2^{7})$, and the fermions are given  above if the overall factor
of 2
is removed. As above, we need to break one $SU(4)$, either will do. We
choose $SU_{2}(4).$ For the scalars shown, we can do this with, say, $%
(1,1,1,2,1,1,1,1;1,\bar{4})$ and $(1,1,1,1,1,1,1,2;1,4)$ VEVs. The
remaining
chiral fermion sector is 

$
(2,1,1,1,1,1;4)  +(1,1,1,2,1,1;\bar{4})  +(1,2,1,1,1,1;4) \\
 +(1,1,1,1,2,1;\bar{4}) +(1,1,2,1,1,1;4)  +(1,1,1,1,1,2;\bar{4})$

for
$G=\prod\limits_{k}SU_{k}(2)\times SU(4),$ with k=1,2,3,5,6,7.

There are only $SU_{i}(2)\times SU_{j}(2)$ bilinear scalars of the form$
(2_{i},2_{j})$ where $i=1,2,$ or $3$ and $j=4,5, $or $6$, whose VEVs reduce
chirality
further, so we cannot reach a three-family P-S model.

Note: what one would need is bilinears that allow us to break
$SU_{1}(2)%
\times SU_{2}(2)\times SU_{3}(2)$ to a diagonal subgroup $SU_{L}(2)$,
and
similarly for $SU_{4}(2)\times SU_{5}(2)\times SU_{6}(2)$ to
$SU_{R}(2)$.
This would then have been a three-family P-S model, however, such scalars do not exist in the model.

\bigskip

\bigskip

\bigskip
Next consider
\underline{16/11 with ${\bf 4=({\bf 1}_{2}{\bf ,1}_{2},{\bf 1}_{2}{\bf ,1}_{2})}$
and $%
N=3$}.
This model is highly chiral, with $\chi =432$,
and the chiral fermions are 

$6[(3,\bar{3},1,1,1,1,1,1;1,1)  +(1,1,1,1,3,\bar{3},1,1;1,1) \\  
+(1,3,\bar{3},1,1,1,1,1;1,1)  +(1,1,1,1,1,3,\bar{3},1;1,1) \\
+(1,1,3,\bar{3},1,1,1,1;1,1)  +(1,1,1,1,1,1,3,\bar{3};1,1) \\ 
+(\bar{3},1,1,3,1,1,1,1;1,1)  +(1,1,1,1,\bar{3},1,1,3;1,1)
].$

\noindent We can ignore the $SU(6)\times SU(6)$  sector, since it can be broken
completely without affecting the chirality. If we then give VEVs to 
$(1,1,1,8,1,1,1,1)$ and $(1,1,1,1,1,1,1,8)$ representations of
$SU(3)^8$,
we arrive at $6[(3,\bar{3},1)+(1,3,\bar{3})+(1,1,3)+(\bar{3},1,1)]$ in
the
$SU_{i+1}(3)\times $ $SU_{i+2}(3)\times SU_{i+3}(3)$ sector for both
$i=0$
and $i=1$. The $i=0$ sector can be broken completely with
$(1,1,1,1,8,1)$%
-type VEVs plus $(1,1,1,3,1,\bar{3})$-type VEVs. The
remaining fermions falling nearly into six $E_{6}\longrightarrow
SU(3)\times %
SU(3)\times SU(3)$-type families. While close to containing the standard model, this model is still
unsuccessful.

\bigskip

\bigskip

\bigskip 
\underline{16/11 with ${\bf 4=({\bf 1}_{2}{\bf ,1}_{2},{\bf 1}_{2}{\bf
,1}_{4})%
}$ and $N=3$}
The chiral fermion sector is exactly half the previous case. Again we
break $%
SU(6)\times SU(6)$ completely. Then breaking
$\prod\limits_{j=4}^{8}SU_{j}(3)
$ completely with $SU_{j}(3)$ octet VEVs gives us finally a chiral
fermion
sector $3[(3,\bar{3},1)+(1,3,\bar{3})+(1,1,3)+(\bar{3},1,1)]$. 
This is
tantalizingly close to a three-family trinification model, but still lacks
the requisite $(\bar{3},1,3)$ fermions.

\bigskip

\bigskip

{\bf 16/13}: There are three potential models for this group.

Consider first the case with 

\underline{${\bf 4} =(2_{1},2_{1})$ and $N=2$. }

Here ${\bf 6}=(1_{2},1_{2},1_{2},1_{3},2_{2})$ and
the chiral fermions are 

\bigskip

2[$(2,1,1,1;4,1,1)+(1,2,1,1;1,1,4)+(1,1,2,1;1,1,4)+ \\
(1,1,1,2;4,1,1)+(2,1,1,1;1,1,\bar{4}) +(1,2,1,1;\bar{4},1,1)\\
+(1,1,2,1;\bar{4},1,1)+(1,1,1,2;1,1,\bar{4})$]

\bigskip

VEVs of the form
%TCIMACRO{\TEXTsymbol{<}}
%BeginExpansion
\mbox{$<$}%
%EndExpansion
$4_{2},\bar{4}_{2}>$ etc., can break  $SU_{2}(4)$ completely (this group
is
irrelevant,
since there are no chiral fermions with $SU_{2}(4)$ quantum numbers).
VEVs for ($4_{1},\bar{4}_{3})$ scalars then breaks $SU_{1}(4)\times
SU_{3}(4)
$ to $SU_{D}(4)$, such that the fermions become vectorlike. On the other
hand, VEVs for (2$_{4}$,$4_{2})+h.c.$ reduces the chiral sector to 

\bigskip

2[$%
(2,1,1;1,4)+(1,2,1;4,1)+(1,1,2;4,1)+2(1,1,1;1,4) \\$
$ +(2,1,1;\bar{4},1) 
+2(1,1,1;\bar{4},1)+(1,2,1;1,\bar{4})+(1,1,2;1,\bar{4})]$

and then a VEV for (2$_{3}$,$4_{2})+h.c$ reduces this farther to $%
2[(2,1;1,4)+(1,2;4,1)+(1,2;1,\bar{4})+(2,1;\bar{4},1)].$

As above a VEV for ($4_{1},\bar{4}_{3})$ scalars would render the model
vectorlike, while just breaking $SU_{3}(4)$ would give a one-family
model.
However, this needs VEVs for (2$_{1}$,$2_{4})$ and (2$_{2}$,$2_{3})$,
but no
scalars of this type exist in the model. We conclude the model has no
Pati-Salam type phenomology.

\bigskip

\bigskip

Consider next 

\underline{{\bf 16/13} with ${\bf 4}=({\bf 2}_{1}{\bf ,2}_{2})$ and $N=2$.}

This time ${\bf
6}$ is as given in Section 6.5, but undetermined up to the identification of
antisymmetric
singlets in ({\bf 2}$_{i}\times {\bf 2}_{i})_{A}$ with $i=1,2$. The
chiral
fermions are as in the ${\bf 4=(2_{1},2_{1})}$ case, but with the
overall
factor of 2 deleted. A useful strategy is to do a generic spontaneous
symmetry breaking analysis to try to obtain a realistic Pati-Salam type
phenomenology and then, if successful, one asks if the scalars to carry
out
the breaking are in the model. As above, $SU_{2}(4)$ is irrelevant and
so
can be ignored. If we identify $SU_{1}(2)\times SU_{4}(2)$ with
$SU_{L}(2)$
and $SU_{2}(2)\times SU_{3}(2)$ with $SU_{R}(2)$, we find 2$%
[(2,1;1,4)+(1,2;4,1)+(1,2;1,\bar{4})+(2,1;\bar{4},1)]$. Now breaking one
of
the remaining $SU(4)s$ completely gives two families, and this is the
best
one can do. Hence independent of what scalars are available, there is no
chance to get a model with three or more families.

The remaining 16/13 case is:

\underline{ ${\bf 4=(1_{3},1_{4},2_{1})}$ with $N=2$.}

Now ${\bf 6=(1_{2},2_{1},2_{3,}1}_{{\bf c}}{\bf )}$. 
but the chiral fermions are in the
same representations as the previous model, and so we can immediately
conclude on general grounds that there is no viable phenomenology for
this case.

\bigskip

{\bf \bigskip 18/3}

\bigskip

Now consider

\underline{18/3 with  ${\bf 4=(1\alpha ,1}^{{\bf \prime }}{\bf
;,2\alpha )}
$ and $N=3$.} 
This model has $\chi =192$ and chiral fermions 
$(2,1,1,1,1,1;1,4,1)+(1,2,1,1,1,1;1,4,1)+(1,1,2,1,1,1;\bar{4},1,1) \\
+(1,1,1,2,1,1;\bar{4},1,1)+(1,1,2,1,1,1;1,1,4)+(1,1,1,2,1,1;1,1,4) \\
+(1,1,1,1,2,1;1,\bar{4},1)+(1,1,1,1,1,2;1,\bar{4},1)+(1,1,1,1,2,1;4,1,1) \\
+(1,1,1,1,1,2;4,1,1)+(2,1,1,1,1,1;1,1,\bar{4})+(1,2,1,1,1,1;1,1,\bar{4}) \\
+2[(1,1,1,1,1,1;\bar{4},4,1)+
(1,1,1,1,1,1;1,\bar{4},4)+(1,1,1,1,1,1;4,1,\bar{4})]$. 
Breaking $SU^{6}(2)$ to a single diagonal $SU(2)$with all
six (2$_{i}$,2$_{j}$) type VEVs of $SU_{i}(2)\times SU_{j}(2)$, and
then
further VEVs of the type (2;4,1,1), (2;1,4,1), and (2;1,1,4) 
to break the $SU(4)$s to $SU(3)$s leads to
the
set of remaining chiral fermions:

2[$(3,\bar{3},1)+(1,3,\bar{3})+(\bar{3},1,3)].$ 

So this route leads to two families.

\bigskip
\bigskip

\noindent If instead we seek a Pati-Salam model, there are several spontaneous
symmetry breaking routes we need to investigate. If we break with
(1,1,1,1,1,1;$\bar{4},4,1)$ scalars to $SU^{6}(2)\times SU_{D}(4)\times
SU_{3}(4)$ we find the fermions remaining chiral are 

$%
(2,1,1,1,1,1;4,1)+(1,2,1,1,1,1;4,1)+ (1,1,2,1,1,1;1,4)+(1,1,1,2,1,1;1,4) \\
+(1,1,2,1,1,1;\bar{4},1)
+(1,1,1,2,1,1;\bar{4},1) +(2,1,1,1,1,1;1,\bar{4})+(1,2,1,1,1,1;1,\bar{4}).$

Now breaking with a 
($4_1,\bar{4}_{3})$ or ($4_2,\bar{4}_{3})$ VEV would render the model vectorlike, so we avoid
this and insted give
VEVs to (2$_{5}$,4$_{1}$) and (2$_{6}$,4$_{1}$) to break $SU_{D}(4)$ to
$SU^{\prime }(2)$. However, this yields at most two families.

We must try another route. If we avoid ($\bar{4},4)$ type VEVs and give
VEVs
only to ($2,4)$ type scalars, we can proceed as follows:
%TCIMACRO{\TEXTsymbol{<}}
%BeginExpansion
\mbox{$<$}%
%EndExpansion
2$_{1}$,4$_{2}>,$
%TCIMACRO{\TEXTsymbol{<}}
%BeginExpansion
\mbox{$<$}%
%EndExpansion
2$_{2}$,4$_{2}>,$
%TCIMACRO{\TEXTsymbol{<}}
%BeginExpansion
\mbox{$<$}%
%EndExpansion
2$_{3}$,\={4}$_{1}>$ and
%TCIMACRO{\TEXTsymbol{<}}
%BeginExpansion
\mbox{$<$}%
%EndExpansion
2$_{4}$,\={4}$_{1}>$ VEVs break $SU^{6}(2)\times SU^{3}(4)$ to
$SU_{5}(2)\times %
SU_{6}(2)\times SU^{\prime }(2)\times SU^{\prime ^{{}}\prime }(2)\times
SU(4)
$.
Some fermions remain chiral but they are insufficient to construct
families. We conclude that this model will not provide viable
phenomenology.

\bigskip

\underline{20/5 with ${\bf 4=(1_{2},1_{2},1_{2},1_{2})}$ and $N=3$}

The chiral $SU^{4}(3)$ fermions are 4[$(3,\bar{3},1,1)+$ $(1,3,\bar{3}%
,1)+(1,1,3,\bar{3})+(\bar{3},1,1,3)].$ (The $SU(6)$ does not
participate; it
will be ignored.) The only scalars are in representations $(3,1,\bar{3},1)$
+h.c. and $(1,3,1,\bar{3})$+h.c. A
VEV to, say, the first of these, would break $SU_{1}(3)\times SU_{3}(3)$
to
a diagonal $SU_{D}(3)$, and the fermions would become 4[$(3,\bar{3},1)+$
$(%
\bar{3},3,1)+(3,1,\bar{3})+(\bar{3},1,3)]$ under $SU_{D}(3)\times
SU_{2}(3)%
\times SU_{4}(3)$, which is
vectorlike. Hence any allowed VEVs immediately renders the model
vectorlike.

We get no farther with ${\bf 4=(1_{2},1_{2},1_{2},1_{2})}$ and N = 3,
where $%
{\bf 6}={\bf (1_{3},1_{3},1_{3},1_{1},1_{1},1_{1})}$, since this model
has
only half the chirality content of the previous case, and
again VEVs will render it vectorlike.

\bigskip

\underline{21/2 with ${\bf 4}=({\bf 1}_{2}{\bf ,3}_{1})$ and $N=2$.} Now ${\bf
6}=({\bf 3}%
_{1}{\bf ,3}_{2}).$ (Other embeddings of the {\bf 4} with n=0 SUSY are
permutation of the representations of this model and are therefore all
equivalent.) The fermions have $\chi =108$ and are $%
(2,1,1;6,1)+(1,2,1;6,1)+(1,1,2;6,1)$
$+(2,1,1;1,\bar{6})+(1,2,1;1,\bar{6}%
)+(1,1,2;1,\bar{6})+(1,1,1;\bar{6},6).$ A VEV for a ($\bar{6},6)$ scalar
renders the model vectorlike. Our only other option is to give (2,6)
type
VEVs. A
%TCIMACRO{\TEXTsymbol{<}}
%BeginExpansion
\mbox{$<$}%
%EndExpansion
2,1,1;6,1%
%TCIMACRO{\TEXTsymbol{>} }
%BeginExpansion
\mbox{$>$}%
%EndExpansion
 breaks the gauge group to $SU_{2}(4)\times SU_{3}(2)\times SU(5)\times
SU(6)$
with chiral fermions 2$(1,1;5,1)+(1,2;5,1)+(2,1;5,1)$ $+(1,1;1,\bar{6}%
)+(2,1;1,\bar{6})+(1,2;1,\bar{6})+(1,1,1;\bar{5},6).$ There is
insufficient
fermion content for a three family Pati-Salam model if we identify
$SU_{2}(4)%
\times SU_{3}(2)$ with $SU_{L}(4)\times SU_{R}(2).$ Our only other
choice is
to get one of these $SU(2)$s from $SU(5)\times SU(6).$ For instance a
%TCIMACRO{\TEXTsymbol{<}}
%BeginExpansion
\mbox{$<$}%
%EndExpansion
2$_{2}$,5%
%TCIMACRO{\TEXTsymbol{>} }
%BeginExpansion
\mbox{$>$}%
%EndExpansion
VEV breaks the gauge group to $SU_{3}(2)\times SU(4)\times SU(6)$ but
the
remaing chiral fermions are 4$(1,4,1)+(2,4,1)$
$+3(1,1,\bar{6})+(2,1,\bar{6}
)+(1,2;1,\bar{6})+(1,1,6)+(1,\bar{4},6).$ We can not identify $SU(4)$
with $%
SU_{PS}(4)$, so this group can only be in $SU(6).$ Breaking $SU(6)$
with an
adjoint to $SU(2)\times SU(4)$ leaves us with $SU(2)\times SU(4)\times
SU(2)%
\times SU(4)$ fermions that are again insufficient for a three family
Pati-Salam model.

\bigskip

\bigskip

\underline{24/7 with ${\bf 4}=({\bf 1}\alpha ,{\bf 1}^{\prime },{\bf 2}\alpha )$ for $N=2$}

This model, the only successful one in the present rather broad yet 
still not comprehensive search, 
has been discussed in detail in \cite{nonabelian}.

The original gauge group at the conformality scale is $SU(4)^3 \times SU(2)^{12}$ with
chiral fermions as given in Section 6.4 and complex scalars as stated in Section 6.5 above.

If we break the three $SU(4)$s to a single diagonal $SU(4)$ subgroup, chirality is
lost. To avoid this we break $SU(4)_1$ completely and then break
$SU(4)_{\alpha} \times SU(4)_{\alpha^2}$ to its diagonal
subgroup $SU(4)_D$. The appropriate VEVs are available as $[(4_1, 2_b \alpha^k) + h.c.]$
with $b$ arbitrary but $k=1$ or $k=2$. The second step requires an $SU(4)_D$ singlet VEV
from $(\bar{4}_{\alpha}, 4_{\alpha^2})$ and/or $(4_{\alpha}, \bar{4}_{\alpha^2})$.
Once a choice is made for $b$ (we take $b=4$), the remaining fermions are, in an intuitive notation,:
\begin{equation}
\sum_{\alpha = 1}^{\alpha = 3} [ (2_{\alpha}\alpha, 1, 4_D) + (1, 2_{\alpha}\alpha^{-1}, \bar{4}_D)]
\label{24/7}
\end{equation}
which has the same content as a three family Pati-Salam model, though
with a separate $SU(2)_L \times SU(2)_R$ per family.

To further reduce the symmetry we must arrange to break to a single $SU(2)_L$
and a single $SU(2)_R$.
This is achieved by modifying step one where $SU(4)_1$ was broken. Consider
the block diagonal decomposition of $SU(4)_1$ into $SU(2)_{1L} \times SU(2)_{1R}$.
The representations $(2_{\alpha}\alpha, 4_1)$ and $(2_{\alpha}\alpha^{-1}, 4_1)$ 
decompose as $(2_{\alpha}\alpha, 4_1) \rightarrow (2_{\alpha}\alpha, 2, 1) + (2_{\alpha}\alpha,
1, 2)$ and $(2_{\alpha}\alpha^{-1}, 4_1) \rightarrow (2_{\alpha}\alpha^{-1}, 2, 1) + (2_{\alpha}^{-1},
1, 2)$.
Now if we give VEVs of equal magnitude to the
$(2_a\alpha, 2, 1), a = 1, 2, 3$
and equal magnitude VEVs to the
$(2_a\alpha^{-1}, 1, 2), a = 1, 2, 3$,
we break 
$SU(2)_{1L} \times \Pi_{a=1}^{a=3}SU(2_a\alpha)$
to a single
$SU(2)_L$
and we break
$SU(2)_{1R} \times \Pi_{a=1}^{a=3}SU(2_a\alpha^{-1})$
to a single
$SU(2)_R$.
Finally, VEVs for $(2_4\alpha, 2, 1)$
and $(2_4\alpha, 1, 2)$ as well as
$(2_4\alpha^{-1}, 2, 1)$ and $(2_4\alpha^{-1}, 1, 2)$
ensure that both $SU(2_4\alpha)$ and $SU(2_4\alpha^{-1})$
are broken and that only three families remain chiral.
The final set of chiral fermions is then
$3[(2, 1, 4) + (1, 2, \bar{4})]$ with gauge symmetry
$SU(2)_L \times SU(2)_R \times SU(4)_D$.

To achieve the final reduction to the standard model, an adjoint VEV
from $(\bar{4}_{\alpha}, 4_{\alpha^2})$ and/or $(4_{\alpha}, \bar{4}_{\alpha^2})$
is used to break $SU(4)_D$ to 
$SU(3) \times U(1)$, and
a right-handed doublet is used to break $SU(2)_R$.

\bigskip
\bigskip  

\underline{24/9 with ${\bf 4}=({\bf 1}_{1}\alpha ,{\bf 1}_{2}\alpha^3 ,{\bf 2}\alpha^2
)$ for $N=2$}

The original gauge group at the conformality scale is $SU(4)^4 \times SU(2)^{8}$ with
chiral fermions as given in Section 6.4 and complex scalars as stated in Section 6.5 above.

Achievement of chiral families under the Pati-Salam subgroup
$SU(4) \times SU(2)_L \times SU(2)_R$
requires the identifications
$SU(2)_{1_1} = SU(2)_{1_2} = SU(2)_{1_1\alpha} = SU(2)_{1_2\alpha} = SU(2)_L$;
$SU(2)_{1_1\alpha^3} = SU(2)_{1_2\alpha^2} = SU(2)_{1_1\alpha^3} = SU(2)_{1_2\alpha^3} = SU(2)_R$;
while, for example,
$SU(4)_{2} = SU(4)_{2\alpha} = {\bf \bar{4}}$ of $SU(4)$;
$SU(4)_{2\alpha^2} = SU(4)_{2\alpha^3} = {\bf 4}$ of $SU(4)$
where by this simplified notation we imply diagonal subgroups.

But the scalars tabulated for this case in Section 6.5
are insufficient to allow this pattern of spontaneous symmetry breaking,
and hence no interesting model emerges.

\bigskip
\bigskip

\underline{24/9 with ${\bf 4}=({\bf 1}_{1}\alpha ,{\bf 1}_{2}\alpha ,{\bf 2}\alpha
)$ for $N=2$}

The original gauge group at the conformality scale is $SU(4)^4 \times SU(2)^{8}$ with
chiral fermions as given in Section 6.4 and complex scalars as stated in Section 6.5 above.

Achievement of chiral families under the Pati-Salam subgroup
$SU(4) \times SU(2)_L \times SU(2)_R$
requires the identifications
$SU(2)_{1_1} = SU(2)_{1_2} = SU(2)_{1_1\alpha} = SU(2)_{1_2\alpha} = SU(2)_L$;
$SU(2)_{1_1\alpha^3} = SU(2)_{1_2\alpha^2} = SU(2)_{1_1\alpha^3} = SU(2)_{1_2\alpha^3} = SU(2)_R$;
while, for example,
$SU(4)_{2} = SU(4)_{2\alpha^3} = {\bf \bar{4}}$ of $SU(4)$;
$SU(4)_{2\alpha} = SU(4)_{2\alpha^2} = {\bf 4}$ of $SU(4)$
where again by this simplified notation we imply diagonal subgroups.

But the scalars tabulated for this case in Section 6.5
are insufficient to allow this pattern of spontaneous symmetry breaking,
and hence no interesting model emerges.

\bigskip 

\bigskip

\underline{24/9 with ${\bf 4}=({\bf 1}_{1}\alpha^2 ,{\bf 1}_{2}, {\bf 2}\alpha
)$ for $N=2$}

The original gauge group at the conformality scale is $SU(4)^4 \times SU(2)^{8}$ with
chiral fermions as given in Section 6.4 and complex scalars as stated in Section 6.5 above.

Achievement of chiral families under the Pati-Salam subgroup
$SU(4) \times SU(2)_L \times SU(2)_R$
requires the identifications
$SU(2)_{1_1} = SU(2)_{1_2} = SU(2)_{1_1\alpha} = SU(2)_{1_2\alpha} = SU(2)_L$;
$SU(2)_{1_1\alpha^3} = SU(2)_{1_2\alpha^2} = SU(2)_{1_1\alpha^3} = SU(2)_{1_2\alpha^3} = SU(2)_R$;
while, for example,
$SU(4)_{2} = SU(4)_{2\alpha^3} = {\bf \bar{4}}$ of $SU(4)$;
$SU(4)_{2\alpha} = SU(4)_{2\alpha^2} = {\bf 4}$ of $SU(4)$
where by this simplified notation we imply diagonal subgroups.

But again, as in the previous two embeddings, the scalars tabulated for this case in Section 6.5
are insufficient to allow this pattern of spontaneous symmetry breaking,
and hence no interesting model emerges.

\bigskip
\bigskip

\newpage

\bigskip
The final model to consider for this group is

\underline{24/9 with ${\bf 4}=({\bf 2}\alpha ,{\bf 2}\alpha )$ for $N=2$}

The original gauge group at the conformality scale is $SU(4)^4 \times SU(2)^{8}$ with
chiral fermions as given in Section 6.4 and complex scalars as stated in Section 6.5 above.

Achievement of chiral families under the Pati-Salam subgroup
$SU(4) \times SU(2)_L \times SU(2)_R$
requires the identifications
$SU(2)_{1_1} = SU(2)_{1_1\alpha} = SU(2)_{1_1\alpha^2} = SU(2)_{1_1\alpha^3} = SU(2)_L$;
$SU(2)_{1_2\alpha} = SU(2)_{1_2\alpha} = SU(2)_{1_2\alpha^2} = SU(2)_{1_2\alpha^3} = SU(2)_R$;
while, for example,
$SU(4)_{2\alpha} = SU(4)_{2\alpha^3} = {\bf 4}$ of $SU(4)$
where by this simplified notation we imply diagonal subgroups, and
$SU(4)_{2}$ and $SU(4)_{2\alpha^2}$ are broken.

But the scalars tabulated for this case in Section 6.5
are insufficient to allow this pattern of spontaneous symmetry breaking,
and hence no interesting model emerges.

\bigskip
\bigskip
Moving on we next consider 
\underline{24/13 with ${\bf 4}=({\bf 2}_{1},{\bf 2}_{2})$ for $N=2$}

The original gauge group at the conformality scale is $SU(6) \times SU(4)^3 \times SU(2)^3$ with
chiral fermions as given in Section 6.4 and complex scalars as stated in Section 6.5 above.

\bigskip

According to the analysis in Section 6.4 this orbifold permits only two chiral
families and is therefore not of phenomenological interest.

\bigskip
\bigskip

\underline{24/14 with ${\bf 4}=({\bf 2}_{1},{\bf 2}_{2})$ for $N=2$}

The original gauge group at the conformality scale is $SU(4)^4 \times SU(2)^8$ with
chiral fermions as given in Section 6.4 and complex scalars as stated in Section 6.5 above.

Achievement of chiral families under the Pati-Salam subgroup
$SU(4) \times SU(2)_L \times SU(2)_R$
requires the identifications
$SU(2)_{1_1} = SU(2)_{1_2} = SU(2)_{1_5} = SU(2)_{1_6} = SU(2)_L$;
$SU(2)_{1_3} = SU(2)_{1_4} = SU(2)_{1_5} = SU(2)_{1_6} = SU(2)_R$;
while, for example,
$SU(4)_{2_2} = SU(4)_{2_3} = {\bf 4}$ of $SU(4)$;
$SU(4)_{2_1} = SU(4)_{2_4} = {\bf \bar{4}}$ of $SU(4)$
where by this simplified notation we imply diagonal subgroups.

But the scalars tabulated for this case in Section 6.5
are insufficient to allow this pattern of spontaneous symmetry breaking,
and hence no interesting model emerges.

%WHERE ARE THE SCALARS FOR THIS GROUP?

\bigskip
\bigskip

\underline{24/15 with ${\bf 4}=({\bf 1}_{2},{\bf 1}_{3},{\bf 2}_{3})$ for $N=2$}

The original gauge group at the conformality scale is $SU(4)^5 \times SU(2)^4$ with
chiral fermions as given in Section 6.4 and complex scalars as stated in Section 6.5 above.

Achievement of chiral families under the Pati-Salam subgroup
$SU(4) \times SU(2)_L \times SU(2)_R$
requires the identifications
$SU(2)_{1_1} = SU(2)_{1_3} = SU(2)_L$;
$SU(2)_{1_2} = SU(2)_{1_4} = SU(2)_R$;
while, for example,
$SU(4)_{2_3} = SU(4)_{2_4} = {\bf 4}$ of $SU(4)$,
where by this simplified notation we imply diagonal subgroups.

But again the scalars tabulated for this case in Section 6.5
are insufficient to allow this pattern of spontaneous symmetry breaking,
and hence no interesting model emerges.

\bigskip
\bigskip

\underline{24/15 with ${\bf 4}=({\bf 2}_{3},{\bf 2}_{5})$ for $N=2$}

The original gauge group at the conformality scale is $SU(4)^5 \times SU(2)^4$ with
chiral fermions as given in Section 6.4 and complex scalars as stated in Section 6.5 above.

\bigskip

According to the analysis in Section 6.4 this orbifold permits only two chiral
families and is hence not phenomenologically interesting.

\bigskip
\bigskip

\underline{24/15 with ${\bf 4}=({\bf 2}_{3},{\bf 2}_{3})$ for $N=2$}

The original gauge group at the conformality scale is $SU(4)^5 \times SU(2)^4$ with
chiral fermions as given in Section 6.4 and complex scalars as stated in Section 6.5 above.

Achievement of chiral families under the Pati-Salam subgroup
$SU(4) \times SU(2)_L \times SU(2)_R$
requires the identifications
$SU(2)_{1_1} = SU(2)_{1_3} = SU(2)_L$;
$SU(2)_{1_2} = SU(2)_{1_4} = SU(2)_R$;
while, for example,
$SU(4)_{2_3} = SU(4)_{2_4} = {\bf 4}$ of $SU(4)$
where by this simplified notation we imply diagonal subgroups.

But the scalars tabulated for this case in Section 6.5
are insufficient to allow this pattern of spontaneous symmetry breaking,
and hence no interesting model emerges.

\bigskip

\bigskip
Continuing on to $o(\Gamma)=27$ we have

\underline{27/4 with ${\bf 4}=({\bf 1}_{2}{\bf ,3}_{1})$ with $N=2.$}

Here ${\bf 6}=({\bf
3}_{1}{\bf %
,3}_{2})$ and the chiral fermions are given by Equation 49 and all scalars are of type
$%
(2_{i},\bar{6}_{1}),(2_{i},6_{2})$ or $(6_{1},\bar{6}_{2})\ $for
$i=1,2,...,9
$.. A VEV for $(6_{1},\bar{6}_{2})+h.c.$ scalar breaks
$SU_{1}(6)\times SU_{2}(6)$ to $SU_{D}(6)$, and the model becomes
vectorlike. Hence we must break only with (2,6) type scalars if there is
any
hope of a viable model. We give VEVs to $(2_{i},6_{1})$ scalars for $%
i=1,2,...,5$  to break $SU_{1}(6)$ completely, and VEVs to
$(2_{j},6_{2})$
for $j=6,7$ to break $SU_{2}(6)$ to $SU(4)$. Then the remaining unbroken
gauge group is $SU_{8}(2)\times SU_{9}(2)\times SU(4)$ with  
fermions
$(2,1,4)+(1,2,4)+4(1,1,\bar{4})$, which are chiral but not of the correct
form. 

A more successful variation is obtained with $(2_{i},6_{1})$
scalars VEVs for $i=1,2,3$ and $4$ to break the gauge group to
$SU_{5}(2)\times %
SU_{6}(2)\times SU_{7}(2)\times SU_{8}(2)\times SU_{9}(2)\times
SU^{\prime
}(2)\times SU(6)$ and thenVEVs for $(2_{5},6_{2})$ and $(2_{6},6_{2})$
to
break to $SU_{7}(2)\times SU_{8}(2)\times SU_{9}(2)\times SU^{\prime
}(2)%
\times SU(4)$ which has chiral fermions

\noindent $(2,1,1,1,4)+(1,2,1,1,4)+(1,1,2,1,4)+3(1,1,1,2,4)$. 

\noindent If we could break
$%
SU_{7}(2)\times SU_{8}(2)\times SU_{9}(2)$ to a diagonal
$SU(2)$ subgroup, we
would have a three-family Pati-Salam model. However, the scalars to
accomplish this are not in the spectrum. If we could give VEVs to $%
(2_{i},6_{1})$ scalars for $i=7,8,9$ to break $SU_{7}(2)\times
SU_{8}(2)%
\times SU_{9}(2)$ to a $U_{Y}(1)$ without disturbing the $SU^{\prime
}(2)$
subgroup of $SU_{1}(6),$ and a further $(2_{j},6_{2})$ VEV, say $%
(2_{1},6_{2}),$ to break $SU(4)$ to $SU_{C}(3),$ then we would have a true
three family standard (i.e., $U_{Y}(1)\times SU_{EW}(2)\times $
$SU_{C}(3))$%
model upon identifying $SU^{\prime }(2)$ with $SU_{EW}(2)$.

\bigskip
 
Finally at $o(\Gamma)=30$ we have
\underline{30/2 with  ${\bf 4=(1}_{{\bf 1}}
{\bf \alpha ,1_{2}},{\bf 2\alpha)}$
and $N=2$}.

\bigskip
\bigskip

Here ${\bf 6}=({\bf 1}_{2}{\bf \alpha ,1}_{2}{\bf \alpha ^{2}},{\bf
2\alpha },%
{\bf 2\alpha ^{2}})$, and the gauge group is $SU^{6}(2)\times
SU^{6}(4).$ This
group has chiral fermions:

 $(2,1,1,1,1,1;1,1,4,1,1,1)
+(1,2,1,1,1,1;1,1,4,1,1,1) \\
+(1,1,2,1,1,1;\bar{4},1,1,1,1,1)
+(1,1,1,2,1,1;\bar{4},1,1,1,1,1)$ \\

$+(1,1,1,1,1,1;1,\bar{4},4,1,1,1)
+(1,1,1,1,1,1;\bar{4},1,4,1,1,1) \\
+2(1,1,1,1,1,1;1,\bar{4},1,4,1,1)
+(1,1,1,1,1,1;\ \bar{4},1,1,4,1,1)$ \\

+$($1$,1,2,1,1,1;1,1,1,1,4,1)+(1,1,1,2,1,1;1,1,1,1,4,1) \\
+(1,1,1,1,2,1;1,1,\bar{4},1,1,1)
+(1,1,1,1,1,2;1,1,\bar{4},1,1,1)$ \\

$+(1,1,1,1,1,1;1,1,1,\bar{4},4,1)
+(1,1,1,1,1,1;1,1,\bar{4},1,4,1) \\
+2(1,1,1,1,1,1;1,1,1,\bar{4},1,4)+(1,1,1,1,1,1;1,1,\bar{4},1,1,4)$ \\

+$(1,1,1,1,2,1;4,1,1,1,1,1)+(1,1,1,1,1,2;4,1,1,1,1,1) \\
+(2,1,1,1,1,1;1,1,1,1,\bar{4},1)
+(1,2,1,1,1,1;1,1,1,1,\bar{4},1)$ \\
$+(1,1,1,1,1,1;4,1,1,1,\bar{4},1)+(1,1,1,1,1,1;4,1,1,1,1,\bar{4}) \\
+2(1,1,1,1,1,1;1,4,1,1,1,\bar{4})+(1,1,1,1,1,1;1,4,1,1,\bar{4},1)$

The spontaneous symmetry breaking analysis for this model is quite
unwieldy,
but for the most part can be carried out systematically.  For example,
breaking with 

\noindent $(1,1,1,1,1,1;1,\bar{4},1,4,1,1),$ $(1,1,1,1,1,1;\bar{4}
,1,1,4,1,1),$ \\

$(1,1,1,1,1,1;4,1,1,1,\bar{4},1)$ and
$(1,1,1,1,1,1;1,1,\bar{4} ,1,1,4)$ 

\noindent VEVs reduces $SU^{6}(4)$ to $SU_{1}(4)\times SU_{D}(4),$ with
fermions remaining chiral in representations: 

$%
(2,1,1,1,1,1;1,4)+(1,2,1,1,1,1;1,4)
+(1,1,2,1,1,1;\bar{4},1)+(1,1,1,2,1,1;\bar{4},1)$

$+($1$,1,2,1,1,1;1,4)+(1,1,1,2,1,1;1,4)
+(1,1,1,1,2,1;1,\bar{4})+(1,1,1,1,1,2;1,\bar{4})$

+$(1,1,1,1,2,1;4,1)+(1,1,1,1,1,2;4,1)+(2,1,1,1,1,1;1,\bar{4})
+(1,2,1,1,1,1;1,\bar{4}).$ 

Now (1,1,1,1,2,1;4,1) and (1,1,1,1,1,2;4,1) VEVs break
$SU_{5}(2)\times SU_{6}(2)\times SU_{1}(4)$ 
to $SU^{\prime }(2)$ with fermions
remaining chiral in the representations:

$(2,1,1,1;4)+(1,2,1,1;4)+($1$,1,2,1;4)
+(1,1,1,2;4) \\
+2(1,1,1,1;\bar{4})+2(1,1,1,1;\bar{4})+(2,1,1,1;\bar{4})
+(1,2,1,1;\bar{4})$ 

\noindent which is already insufficient to provide three
normal
families. Other analyses of spontaneous symmetry breaking toward
constructing a Pati-Salam model starting with this 30/2 model are
similarly
unsucessful. 

An alternative is to seek a trinification model. To
this
end, consider only the $SU^{6}(4)$ fermion
sector

$+(1,\bar{4},4,1,1,1)+(\bar{4},1,4,1,1,1)
+2(1,\bar{4},1,4,1,1) \\
+(\bar{4},1,1,4,1,1)+(1,1,1,\bar{4},4,1)
+(1,1,\bar{4},1,4,1)$

$+2(1,1,1,\bar{4},1,4)+(1,1,\bar{4},1,1,4)
+(4,1,1,1,\bar{4},1) \\
+2(4,1,1,1,1,\bar{4})
+2(1,4,1,1,1,\bar{4})+(1,4,1,1,\bar{4},1)$

\noindent Identifying $SU_{1}(4)$ with $SU_{2}(4),$ $SU_{3}(4)$ with $SU_{4}(4)$
and $SU_{5}(4)$ with $SU_{6}(4)$ would lead to five families of the form
$5[(\bar{4},4,1)+(1,\bar{4},4)+(4,1,\bar{4})],$ however 
there are no scalars of
the
type needed to carry this out.

This analysis is not exhaustive and there may be models where
$SU_{L}(2),$ $SU_{R}(2)$ or both are contained in $SU^{6}(4)$. 
Since we are starting
with
a group of rank 24, and seek the standard model of rank 4 or a unified
model
thereof of rank 5 or 6, and since there are 66 Higgs representations in
the
theory, the spontaneous symmetry breaking possibilities are rather
complex.
The $N=3$ case is obviously even more complicated, with initial rank 42,
and
one could try to automate the search for phenomenological models,
although
we have not attempted to do so.

\bigskip

\underline{30/3 with ${\bf 4=(1}_{{\bf 1}}{\bf \alpha ,1_{2}},{\bf 2\alpha }^{{\bf
2}}%
{\bf )}$ and $N=2$.} We now have ${\bf 6}=({\bf 1}_{2}{\bf \alpha ,1}_{2}{\bf \alpha
^{4}},{\bf %
2\alpha }^{{\bf 3}},{\bf 2\alpha ^{2}})$ where ${\bf \alpha ^{5}=1.}$

The chiral  $SU^{10}(2)\times SU^{5}(4)$fermions are

(1$^{10}$;$\bar{4},4,1,1,1)+$(1$^{10}$;$\bar{4},1,4,1,1)+
$(1$^{4},2,1^{5}$;$\bar{4},1,1,1,1) \\
+$(1$^{5},2,1^{4}$;$\bar{4},1,1,1,1)$$ 
$+(1$^{10}$;1,$\bar{4},4,1,1)+
$(1$^{10}$;1,$\bar{4},1,4,1)\\
+$(1$^{6},2,1^{3}$;$1,\bar{4},1,1,1)
+$(1$^{7},2,1^{2}$;$1,\bar{4},1,1,1)$$ 
$+(1$^{10}$;$1,1,\bar{4},4,1) \\
+$(1$^{10}$;1,1,$\bar{4},1,4)+$(1$^{8},2,1^{1}$;
$1,1,\bar{4},1,1)+$(1$^{9},2$;$1,1,\bar{4},1,1)$$ \\
$+(1$^{10}$;$1,1,1,\bar{4},4)+$(1$^{10}$;$4,1,1,\bar{4},1)
+$(2,1$^{9}$;$1,1,1,\bar{4},1) \\
+$(1$^{1},2,1^{8}$;$1,1,1,\bar{4},1)$$ 
$+(1$^{10}$;$4,1,1,1,\bar{4})+$(1$^{10}$;$1,4,1,1,\bar{4})+$($1^{2},2,1^{7}$;
$ 1,1,1,1,\bar{4}) \\
+$(1$^{3},2,1^{6}$;$1,1,1,1,\bar{4})$

\bigskip

\noindent Consider the bifundamentals only. VEVs for 
($1,1,1,\bar{4},4)\ $and (1,$\bar{4},4,1,1)$ scalars reduce the chiral fermion
sector to $2[(\bar{4},4,1)+(1,\bar{4},4)+(4,1,\bar{4})]$ which provides
at
most a two family model.

If instead we try to construct a Pati-Salam model, and note that there
are
20 (2;4) type fermions, and that we need six appropriate ones of these
for
three families, we must take care in the spontaneous symmetry breaking
to
preserve this much chirality. If we (i), break  $SU_{2}(4)\times $
$SU_{4}(4)%
\times SU_{5}(4)$ completely and (ii) $SU_{1}(4)\times SU_{3}(4)$ to $%
SU_{PS}(4)$ while (iii) equating $SU_{5}(2),$ $SU_{6}(2),$ $SU_{9}(2)\
$and
(iv) equating $SU_{10}(2)$ with $SU_{L}(2)$, and $SU_{1}(2),$
$SU_{2}(2),$ $%
SU_{7}(2)\ $and $SU_{8}(2)$ with $SU_{R}(2),$ and (v) breaking
$SU_{3}(2)%
\times SU_{4}(2)$ completely, we would be left with a 4 family
Pati--Salam
Fmodel. Can we do this?
(ii) is accomplished with (a) (1$^{10}$;$\bar{4},1,4,1,1),$ then (i)
requires (b) (1$^{10}$;1,$\bar{4},1,4,1)$ and (c)
(1$^{10}$;1,$\bar{4},1,1,4)
$to get a $SU_{D}(4)$. Breaking this to nothing, assuming VEVs (a) and
(b)
allow no freedom to rotate the (c) VEV to diagonal form. Now, at this
point,
we are stymied, as there are insufficient (2$_{i}$,2$_{j}$)
representations
of $SU_{i}(2)\times $ $SU_{j}(2)$ to accomplish (v).

Finally, one can imagine that there exist models with either $SU_{L}(2)$
or $%
SU_{R}(2)$ or both coming from $SU^{5}(4)$, but we see not obvious way
to
cary this out, while on the other hand since there are 60 Higgs
representations we are unable to categorically eliminate this
possibility.

\newpage

\bigskip
\bigskip
\bigskip

{\bf Summary}

We have shown how $AdS/CFT$ duality leads to a large
class of models which can provide interesting extensions of the standard model of
particle phenomenology. The naturally occurring ${\cal N} = 4$ extended
supersymmetry was completely broken to
${\cal N} = 0$ by choice of orbifolds $S^5/\Gamma$ such that 
$\Gamma \not\subset SU(3)$.

\bigskip

In the present work, we studied systematically all such non-abelian $\Gamma$
with order $g \leq 31$. We have seen how chiral fermions require that the
embedding of $\Gamma$ be neither real nor pseudoreal. This reduces dramatically
the number of possibilities to obtain chiral fermions.
Nevertheless, many candidates for models which contain the
chiral fermions of the three-family standard model were found.

\bigskip
\bigskip

However, the requirement that the spontaneous symmetry breaking down to the correct gauge symmetry
of the standard model be permitted by the prescribed scalar representations
eliminates most of the surviving models. We found only one allowed model
based on the $\Gamma = 24/7$ orbifold. We had initially expected to find more examples in our search.
The moral for model-building is interesting. Without the rigid framework of string duality 
the scalar sector would be arbitrarily chosen to 
permit the required spontaneous symmetry breaking. This is the normal practice in the standard model,
in grand unification, in supersymmetry and so on.
With string duality, the scalar sector is prescribed by the construction
and only in one very special case does it permit the required symmetry breaking.

\bigskip

This leads us to give more credence to the $\Gamma = 24/7$ example that does work
and to encourage its further study to check whether it can have any connection
to the real world.

Note that we have usually taken the minimal choice for $N$, either 2 or 3, and have only considered SSB through diagonal subgroup chains. Both these constraints can obviously be relaxed to generate more model building opportunities, but clearly this is an open ended process as is choosing an orbifolding group. If properly motivated, then one of these directions may be productive. We stress that we are providing a guide to the process of model building, and not a compendium of carefully selected models.

\newpage

\section{Unification} 

\bigskip

Going through the abelian quiver gauge theories, we may
search for a model that contains as a sub theory
the states of the standard model.

The scalar sector is dictated by the construction so
that one does not have the usual model building luxury
of choosing those scalar represntations which will
most conveniently break the symmetry spontaneously
in the manner desired. The constraint that the
scalars present can break the gauge symmetry to the
standard model is an exceptionally strong constraint 
which is, in general, not satisfied.

For $p \le 6$ there is no solution. Only when we 
arrive at $p = 7$ there are viable models\cite{F2,F3}. 
Actually three different $p=7$ quiver diagrams can give:
3 chiral families, adequate scalars to spontaneously 
break $U(3)^7 \rightarrow SU(3) \times SU(2) \times U(1)$
and
${\rm sin}^2 \theta_W = 3/13 = 0.231$

\bigskip

The possible embeddings of $\Gamma = Z_7$ in SU(4) are:

7A. ~~$(\alpha, \alpha, \alpha, \alpha^4)$

\bigskip

7B. ~~$(\alpha, \alpha, \alpha^2, \alpha^3)^*$ ~~ C-H-H-H-W-H-W

\bigskip

7C. ~~$(\alpha, \alpha^2, \alpha^2, \alpha^2)$

\bigskip

7D. ~~$(\alpha, \alpha^3, \alpha^5, \alpha^5)^*$ ~~ C-H-W-H-H-H-W

\bigskip

7E. ~~$(\alpha, \alpha^4, \alpha^4, \alpha^5)^*$ ~~ C-H-W-W-H-H-H

\bigskip

7F. ~~$(\alpha^2, \alpha^4, \alpha^4, \alpha^4)$

\bigskip

The three marked $^*$ are those which possess the
desired properties,
The corresponding scalar quivers are indicated in the following figures.

\newpage

\begin{figure}
7B ~~ 4 = (1, 1, 2, 3) ~ 6 = (2, 3, 3, -3, -3, -2)
\begin{center}
\epsfxsize=2.0in
\ \epsfbox{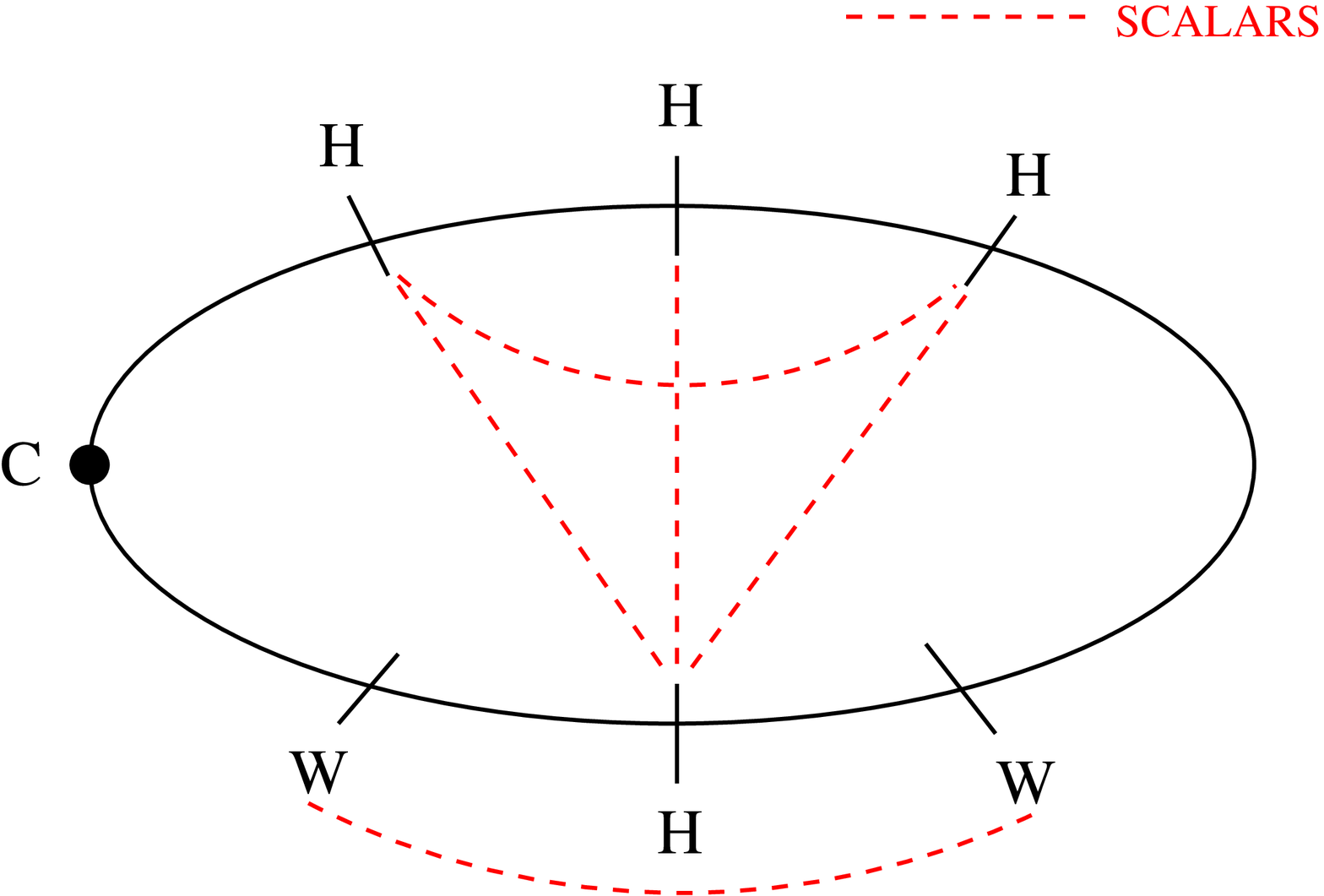}
\end{center}
\end{figure}
\begin{figure}
7D ~~ 4 = (1, 3, 5, 5) ~ 6 = (1, 1, 3, -3, -1, -1)
\begin{center}
\epsfxsize=2.0in
\ \epsfbox{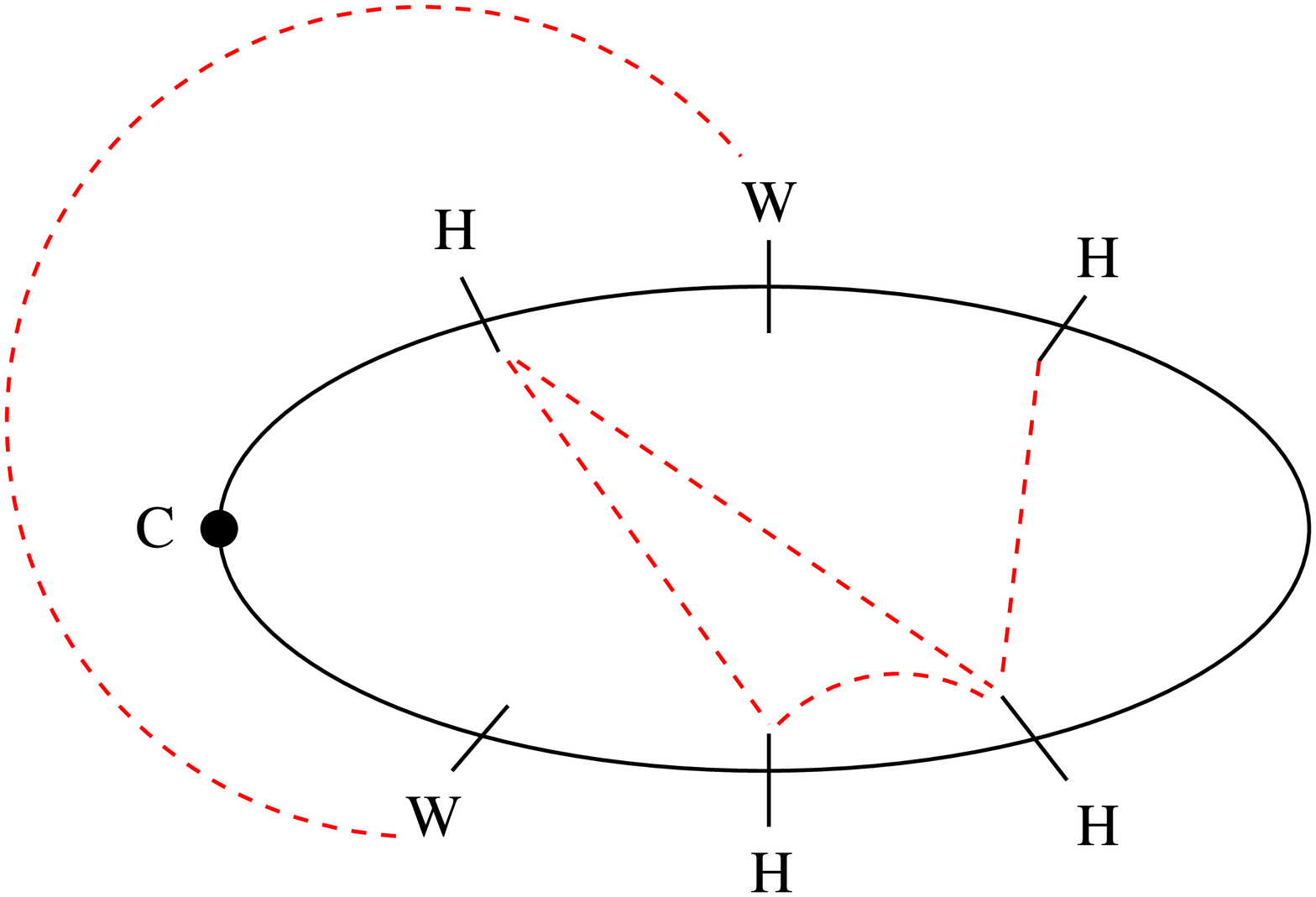}
\end{center}
\end{figure}
\begin{figure}
7E ~~ 4 = (1, 4, 4, 5) ~ 6 = (1, 2, 2, -2, -2, -1)
\begin{center}
\epsfxsize=2.0in
\ \epsfbox{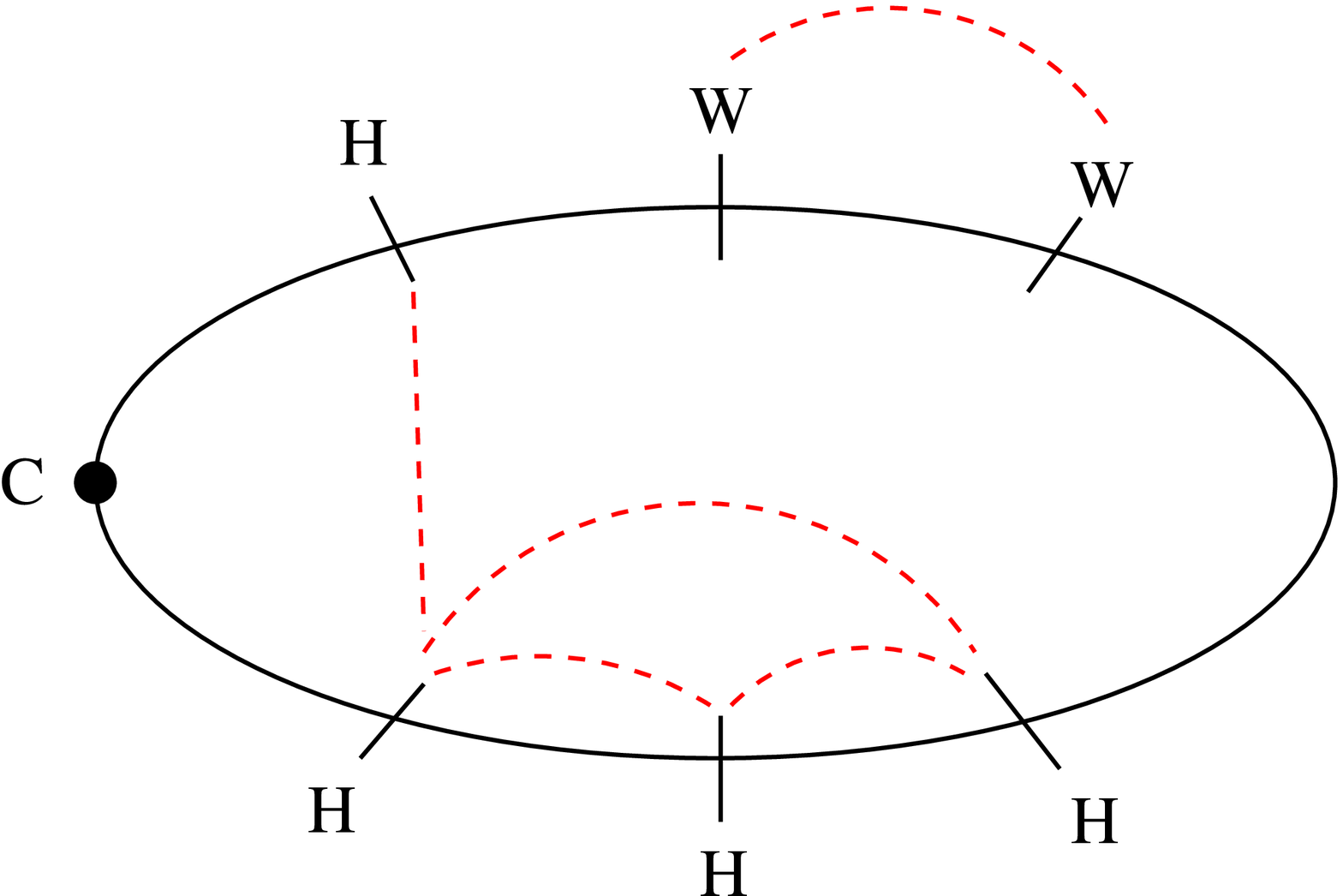}
\end{center}
\end{figure}

So the simplest abelian orbifold conformal extension of the standard model has
$U(3)^7 \rightarrow SU(3)^3$ trinification $\rightarrow (321)_{SM}$.
In these cases we have $\alpha_2$ and $\alpha_1$ related correctly for
low energy. But $\alpha_3(M) \simeq 0.07$ suggesting a conformal scale $M \geq 10$ TeV
 - too high for detection at the L.H.C. The simple cases are not, in any cases,
grand unified and far more interesting example uses $p=12$ and
accommodates a TeV scale grand unification of a new type.
The above merely illustrates how all the standard model states
can be accommodated in abelian quiver theories.

\bigskip
\bigskip

\newpage

\subsection{Grand unification at 4 TeV}

\bigskip

A grand unified model \cite{Unification} which introduces the 4 TeV scale is:
obtained by taking as orbifold $S^5/Z_{12}$ with embedding of $Z_{12}$
in the SU(4) R-parity  specified by
{\bf 4} $\equiv \alpha^{A_1}, \alpha^{A_2}, \alpha^{A_3}, \alpha^{A_4})$
and $A_{\mu} = (1, 2, 3, 6)$.

Firstly, this accommodates the scalars necessary to spontaneously
break to the SM.

\bigskip

\noindent As a bonus, the dodecagonal quiver predicts three chiral families
(see next page).

\bigskip

\noindent
In this teravolt unification of $SU(3)$, $SU(2)$ and $U(1)$ at 4 TeV,
we shall assume that all non standard model states have masses
satisfying $M \leq 4 TeV$ and that the couplings of $SU(3)^{12}$
are scale invariant at higher energies. As will be discussed in
Section 8, cancellation of quadratic divergences requires that
the $U(1)$ charged states also be at teravolt scales. A
mechanism by which $U(1)$ couplings can (surprisingly) become
scale invariant will be discussed in Section 9.

\newpage

\begin{figure}
\begin{center}
\epsfxsize=4.0in
\ \epsfbox{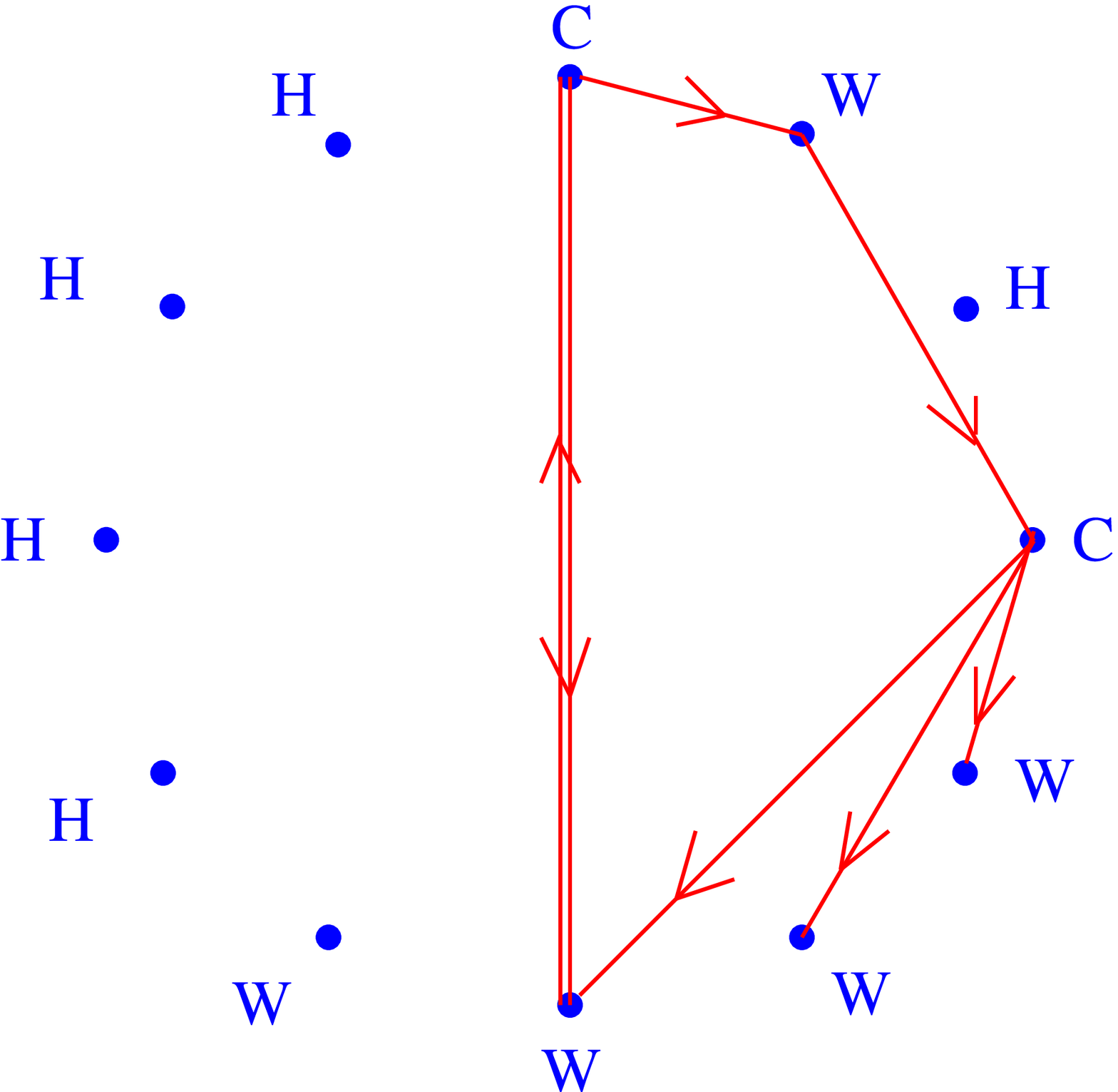}
\end{center}
\end{figure}

\bigskip

\noindent $A_{\mu} = (1, 2, 3, 6)$

\bigskip

\noindent $SU(3)_C \times SU(3)_H \times SU(3)_H$

\bigskip

\noindent $5(3, \bar{3}, 1) + 2 (\bar{3}, 3, 1)$

\newpage

This grand unification works very precisely for the following
bottom up reason. If we take the minimal standard model and
run the couplings up in energy we find that the electroweak
mixing angle $\sin^2 \theta_W(\mu)$ which has the value
$\sim 0.231$ at scale $\mu=M_Z$ becomes 1/4 at a scale
$\mu \sim 4$ TeV, actually nearer to 3.8 TeV. Another ratio
$R(\mu) = \alpha_3(\mu)/\alpha_2(\mu)$ has a value 
larger than 3 at $\mu = M_Z$ then decreases through 3
at 400 GeV, becoming 2 at 140 TeV then 1 at the GUT scale
$\sim 10^{16}$ GeV. It goes through the simple rational
fraction $R(\mu) = 5/2$ at a value $\mu=3.8$ TeV within
one percent of the scale where $\sin^2 \theta(\mu) = 1/4$.

As a result the unification is perfect within errors as illustrated
in the figure where all additional states are taken to be at
$M_U = 4$ TeV.

\begin{figure}
\begin{center}
\epsfxsize=6.0in
\ \epsfbox{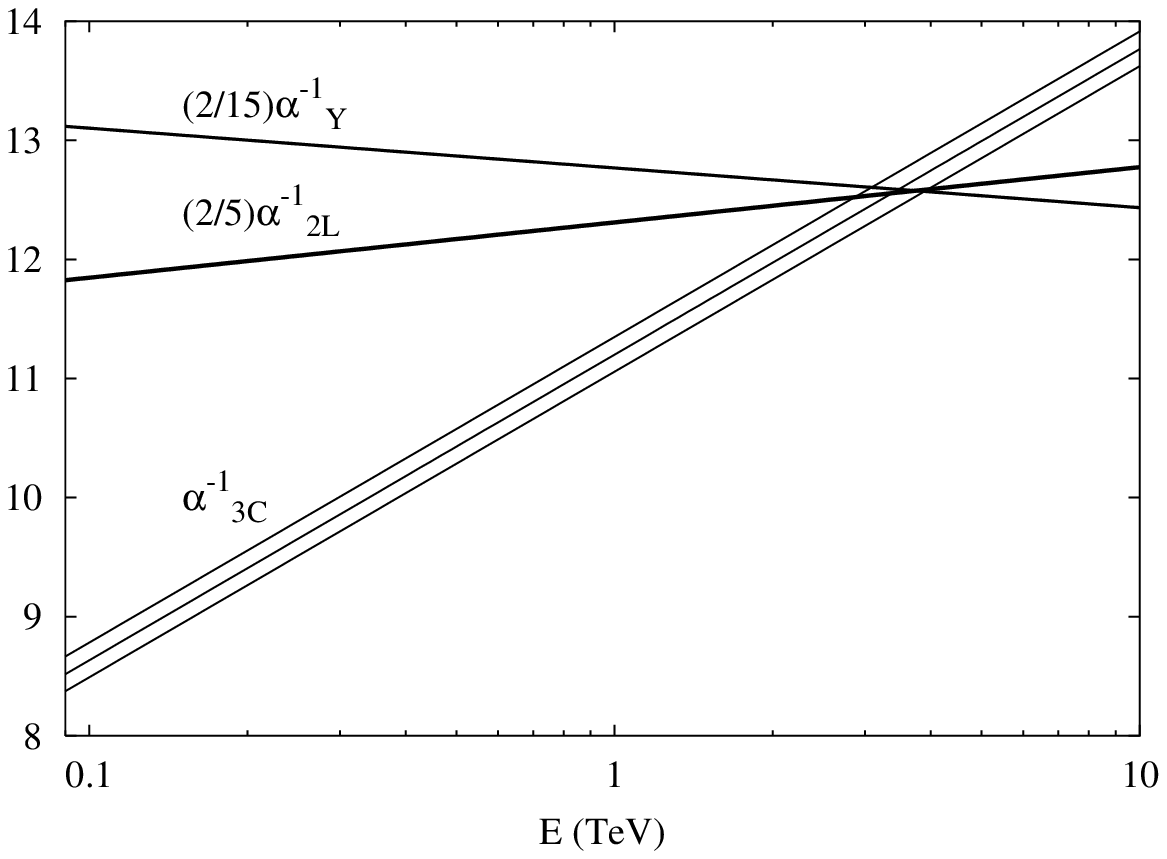}
\end{center}
\end{figure}

\newpage

When the spectrum of additional states is altered there are threshold
corrections which may spoil the perfect unification, as in
more conventional grand unification. This example has "CH" 
fermions at 2 TeV, see \cite{Unification2}.  

\begin{figure}
\begin{center}
\epsfxsize=6.0in
\ \epsfbox{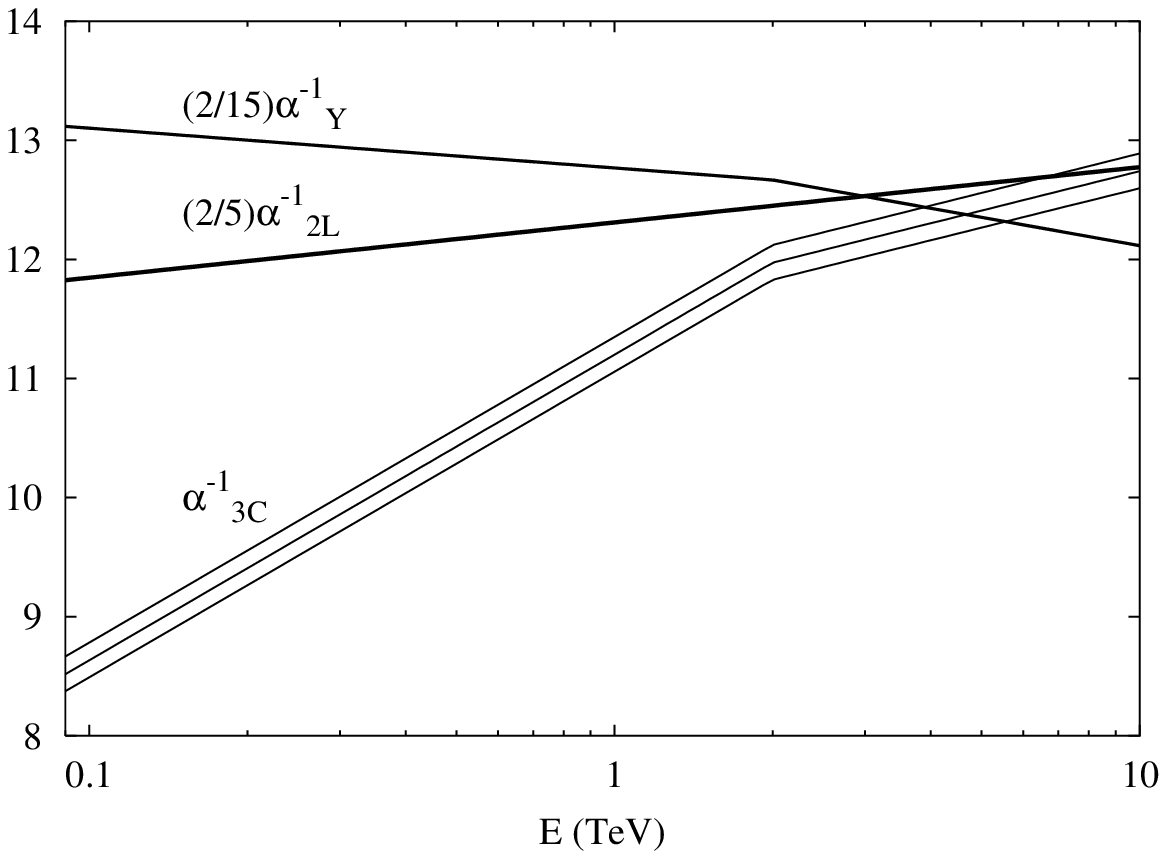}
\end{center}
\end{figure}

\bigskip

\newpage

On the other hand, perfect unification can easily be maintained
even with significant threshold corrections as indicated in
this example which has "CW" fermions at 2 TeV, see \cite{Unification2}.

\begin{figure}
\begin{center}
\epsfxsize=6.0in
\ \epsfbox{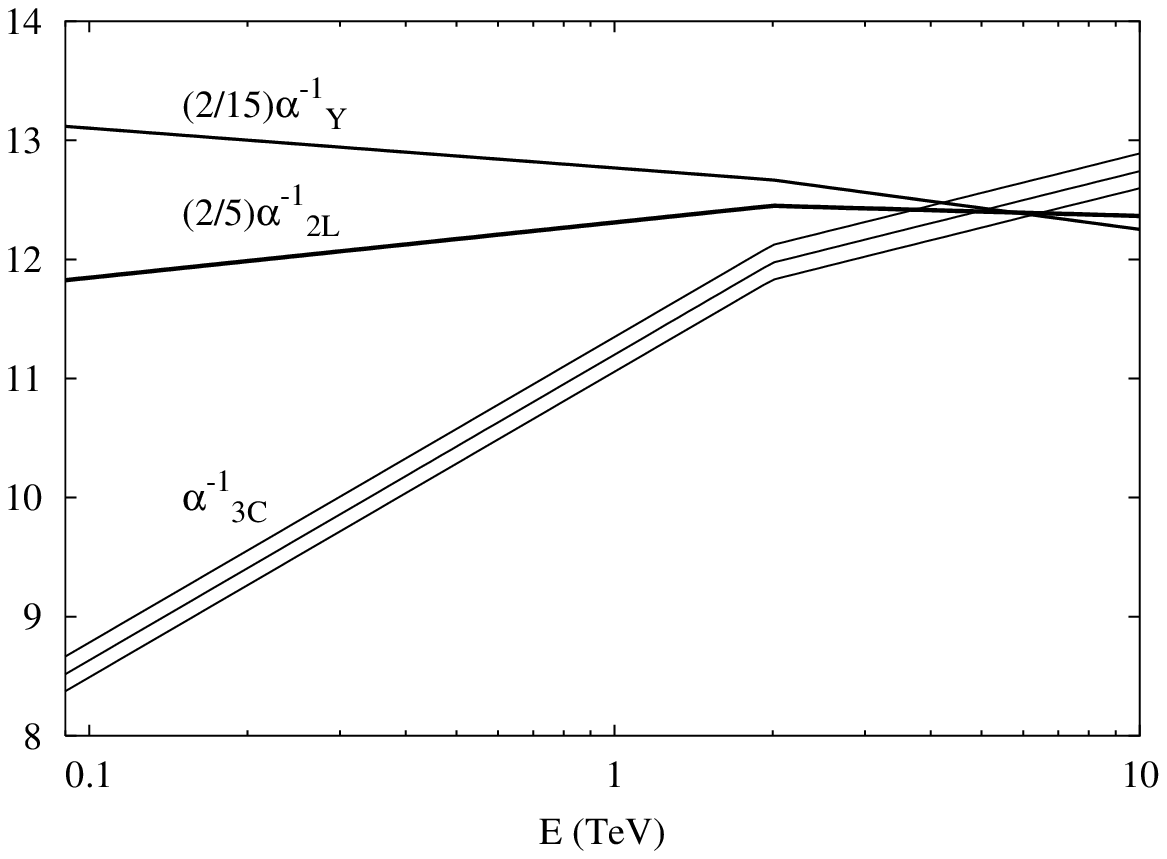}
\end{center}
\end{figure}

\bigskip

\newpage

\section{Quadratic divergences}

\bigskip

The lagrangian for the nonsupersymmetric $Z_p$ theory can be written in
a convenient
notation which accommodates simultaneously both adjoint and
bifundamental scalars as

\small

\begin{eqnarray}
{\cal L} & = &
-\frac{1}{4} F_{\mu\nu; r,r}^{ab}F_{\mu\nu; r,r}^{ba}
+i \bar{\lambda}_{r + A_4, r}^{ab} \gamma^{\mu} D_{\mu} \lambda_{r,
r+A_4}^{ba}
+ 2 D_{\mu} \Phi_{r+a_i, r}^{ab \dagger} D_{\mu} \Phi_{r, r+a_i}^{ba}
+i \bar{\Psi}_{r+A_m, r}^{ab} \gamma^{\mu} D_{\mu} \Psi_{r, r+A_m}^{ba}
\nonumber \\
&  &
- 2 i g
\left[ \bar{\Psi}_{r, r+A_i}^{ab} P_L \lambda_{r + A_i, r + A_i +
A_4}^{bc}
\Phi_{r + A_i+A_4, r}^{\dagger ca}
- \bar{\Psi}_{r, r+A_i}^{ab} P_L \Phi_{r + A_i, r - A_4}^{\dagger bc}
\lambda_{r - A_4, r}^{ca}
\right] \nonumber \\
& & -   \sqrt{2} i g \epsilon_{ijk}
\left[
\bar{\Psi}_{r, r + A_i}^{ab} P_L \Psi_{r +A_i, r + A_i + A_j}^{bc}
\Phi_{r -A_k - A_4, r}^{ca}
-
\bar{\Psi}_{r, r + A_i}^{ab} P_L \Phi_{r +A_i, r + A_i + A_k +
A_4}^{bc} \Psi_{r - A_j, r}^{ca}
\right] \nonumber \\
& & - g^2 \left(
\Phi_{r, r + a_i}^{ab} \Phi_{r+a_i,r}^{\dagger bc}
-
\Phi_{r, r - a_i}^{\dagger ab} \Phi_{r-a_i,r}^{bc}
\right)
\left(
\Phi_{r, r + a_j}^{cd} \Phi_{r+a_j,r}^{\dagger da}
-
\Phi_{r, r - a_j}^{\dagger cd} \Phi_{r- a_j,r}^{da}
\right)  \nonumber \\
& & + 4 g^2
\left(
\Phi_{r, r+a_i}^{ab}\Phi_{r+a_i, r+a_i+a_j}^{bc}
\Phi_{r+a_i+a_j,r+a_j}^{\dagger cd}\Phi_{r+a_j,r}^{\dagger da}
-
\Phi_{r, r+a_i}^{ab}\Phi_{r+a_i, r+a_i+a_j}^{bc}
\Phi_{r+a_i+a_j,r+a_i}^{\dagger cd}\Phi_{r+a_i, r}^{\dagger da}
\right)
\nonumber
\label{N=0L}
\end{eqnarray}

\large

\noindent where $\mu, \nu = 0, 1, 2, 3$ are lorentz indices; $a, b, c, d = 1$ to
$N$ are $U(N)^p$
group labels; $r = 1$ to $p$ labels the node of the quiver diagram
;$a_i ~~ (i = \{1, 2, 3\}) $ label the first three of the {\bf 6} of
SU(4);
$A_m ~~ (m = \{1, 2, 3 ,4\}) = (A_i, A_4)$ label the {\bf 4} of SU(4).
By definition
$A_4$ denotes an arbitrarily-chosen fermion ($\lambda$) associated with
the gauge boson,
similarly to the notation in the ${\cal N} = 1$ supersymmetric case.

\bigskip
\bigskip
\bigskip

Let us first consider the quadratic divergence question in the mother
${\cal N} = 4$ theory. The ${\cal N}=4$ lagrangian is like
Eq.(\ref{N=0L})
but since there is only one node all those subscripts become
unnecessary
so the form is simply

\begin{eqnarray}
{\cal L} & = &
-\frac{1}{4} F_{\mu\nu}^{ab}F_{\mu\nu}^{ba}
+i \bar{\lambda}^{ab} \gamma^{\mu} D_{\mu} \lambda^{ba}
+ 2 D_{\mu} \Phi_{i}^{ab \dagger} D_{\mu} \Phi_{i}^{ba}
+i \bar{\Psi}_{m}^{ab} \gamma^{\mu} D_{\mu} \Psi_{m}^{ba} \nonumber \\
&  &
- 2 i g
\left[\bar{\Psi}_{i}^{ab} P_L \lambda^{bc}
\Phi_{i, r}^{\dagger ca}
- \bar{\Psi}_{i}^{ab} P_L \Phi_{i}^{bc}\lambda^{ca}
\right] \nonumber \\
& & - \sqrt{2} i g \epsilon_{ijk}
\left[
\bar{\Psi}_{i}^{ab} P_L \Psi_{j}^{bc} \Phi_{k}^{\dagger ca}
-
\bar{\Psi}_{i}^{ab} P_L \Phi_{j}^{bc} \Psi_{k}^{ca}
\right] \nonumber \\
& & - g^2 \left(
\Phi_{i}^{ab} \Phi_{i}^{\dagger bc}
-
\Phi_{i}^{\dagger ab} \Phi_{i}^{bc}
\right)
\left(
\Phi_{j}^{cd} \Phi_{j}^{\dagger da}
-
\Phi_{j}^{\dagger cd} \Phi_{j}^{da}
\right)  \nonumber \\
& & + 4 g^2
\left(
\Phi_{i}^{ab}\Phi_{j}^{bc}
\Phi_{i}^{\dagger cd}\Phi_{j}^{\dagger da}
-
\Phi_{i}^{ab}\Phi_{j}^{bc}
\Phi_{j}^{\dagger cd}\Phi_{i}^{\dagger da}
\right)
\nonumber
\label{N=4L}
\end{eqnarray}

\newpage

\bigskip

All ${\cal N} = 4$ scalars are in adjoints and the scalar propagator
has one-loop quadratic divergences coming potentially from
three scalar self-energy diagrams:
(a) the gauge loop (one quartic vertex);
(b) the fermion loop (two trilinear vertices);
and (c) the scalar loop (one quartic vertex).

For ${\cal N} = 4$ the respective contributions
of (a, b, c) are computable from
Eq.(\ref{N=4L}) as proportional to $g^2N (1, -4, 3)$ which cancel
exactly.

\bigskip
\bigskip
\bigskip

The ${\cal N} = 0$ results for the scalar self-energies (a, b, c)
are computable from the lagrangian of Eq.(\ref{N=0L}).
Fortunately, the calculation was already done in the literature.
The result is amazing! The quadratic divergences cancel
if and only if x = 3, exactly the same ``if and only if"
as to have chiral fermions.
It is pleasing that one can independently confirm
the results directly from the interactions
in Eq.(\ref{N=0L}) To give just one explicit
example, in the contributions to
diagram (c) from the last term in Eq.(\ref{N=0L}), the
1/N corrections arise from a contraction of $\Phi$ with
$\Phi^{\dagger}$
when all the four color superscripts are distinct
and there is consequently no sum
over color in the loop. For this case, examination of the node
subscripts then
confirms proportionality to the kronecker delta, $\delta_{0, a_i}$.
If and only if all $a_i \neq 0$, all the other terms
in Eq.(\ref{N=0L}) do not lead to 1/N corrections
to the ${\cal N}=4$.

We find it remarkable\cite{quadratic} and encouraging that the condition
for naturalness, absence of one loop quadratic divergences,
coincides with the necessary and sufficient condition
for presence of chiral fermions. If the conditions
had conflicted, as seemed {\it a priori} possible,
the approach would have had less phenomenological interest.

\newpage

\subsection{Is there a global symmetry?}

\bigskip

It is an old notion that ${\cal N}=4$ supersymmetric gauge theory
is germane to the generalization of the standard model of particle
phenomenology. The ${\cal N}=4$ theory has remarkable properties
which include ultra violet finiteness and conformal invariance.
Nevertheless, one striking feature of the standard model is the
presence of chiral fermions which excludes both ${\cal N}=4$ and
${\cal N}=2$ extended supersymmetries. Also, although the situation
might change, the absence of any experimental
support even for ${\cal N}=1$ supersymmetry is striking.

Shortly after the discovery
of supersymmetric theories, Haag, Lopuszanski and Sohnius
generalized this to show in a 1975 publication \cite{Haag1975}
that under certain assumptions, supersymmetry is the only possibility.

No-go theorems can be useful because
they provide a set of assumptions some or all of which must be violated in order
to make progress.
Here, we shall suggest \cite{global} that the way around
the generalized no-go theorem of Haag, Lopuszanki and Sohnius is to
relax their assumption that
the generators of the symmetry commute with gauge transformations.
In particular, for a $U(N)^n$ quiver gauge theory, we suggest
fermionic generators which transform as bi-bifundamentals under $U(N)^n$.

\bigskip

\noindent {\it ${\cal N}=4$ supersymmetry}

\bigskip

\noindent Here we collect briefly some well-known \cite{BSS}
facts, for convenience.

\bigskip

\noindent The action for ${\cal N}=4$ Yang-Mills can be written
\begin{eqnarray}
S & = &
\int d^4x \left[ -\frac{1}{4} F_{\mu\nu a}F^{\mu\nu a}
+ \frac{1}{2} D_{\mu} \Phi_{ij}^{a} D^{\mu} \Phi_{ij}^{a}
+i \bar{\chi}^{a} \gamma.D L \chi^{a} \right.  \nonumber \\
&  &
\left.   - \frac{1}{2} i g f_{abc} \left( \bar{\tilde{\chi}}^{ai} L \chi^{jb}
\Phi_{ij}^{c}
- \bar{\chi}_{i}^{a} R \tilde{\chi}_{j}^{b} \Phi^{ijc} \right)  \right.  \nonumber \\
& & \left.  - \frac{1}{4} g^2 \left(
f_{abc} \Phi_{ij}^{b} \Phi_{kl}^{c} \right)
\left(
f_{ade} \Phi^{ijd} \Phi^{kle} \right)
\right]
\label{N=4S}
\end{eqnarray}
where $\mu,\nu = 0,  1,  2,  3$; $i, j, k, l = 1, 2,  3$; $L=\frac{1}{2} (1+\gamma_5)$,
$R=\frac{1}{2}(1-\gamma_5)$; and $\tilde{\chi}_i = C\bar{\chi}^{iT}$ with $C$ the charge
conjugation operator.

\bigskip

\noindent The action (\ref{N=4S}) is invariant under the ${\cal N}=4$ supersymmetry
\cite{BSS}

\small

\begin{eqnarray}
\delta A_{\mu}^{a} & =  & i \left(\bar{\alpha}_i \gamma_{\mu} L \chi^{ia}
- \bar{\chi}_i^a \gamma_{\mu} L \alpha^i \right).  \nonumber  \\
\delta \Phi_{ij}^{a} & = &  i \left( \bar{\alpha}_j R \tilde{\chi}_i^a
- \bar{\alpha}_i R \tilde{\chi}_i^a
+ \epsilon_{ijkl} \bar{\alpha}^k L \chi^{la}  \right).   \nonumber  \\
\delta L \chi^{ia} & = &
\sigma_{\mu\nu} F^{\mu\nu a} L \alpha^i
- \gamma.D \Phi^{ij a} R \tilde{\alpha}_j
+ \frac{1}{2} g f_{abc} \phi_b^{ik} \Phi_{kj}^{c} L \alpha^{j}   \nonumber   \\
\delta R \tilde{\chi}_i^a & = &
\sigma_{\mu\nu} F^{\mu\nu a} R \tilde{\alpha}_i
+ \gamma.D \Phi_{ij}^a L \alpha^i
+ \frac{1}{2} g f_{abc} \Phi^b_{ik} \Phi^{kj}_{c} L \tilde{\alpha}_{j}.
\label{susy}
\end{eqnarray}

\large

\noindent where $\alpha^i$ transforms as a ${\bf 4}$ and $\bar{\alpha}_i$
as a ${\bf \bar{4}}$ under an internal $SU(4)$ symmetry.

\bigskip

\noindent The group indices $a,b,c$ run over the dimension of the gauge group
$a, b, c = 1,.....,d_G$. For $G = SU(N)~~ {\rm or} ~~ U(N)$, $d_G = (N^2-1) ~~ {\rm or} ~~
N^2$ respectively. Note that the infintesimal supersymmetry parameter $\alpha^i$
is singlet under the gauge group $G$. This assumption will be
relaxed for misaligned supersymmetry in the next section.

\bigskip

\noindent {\it Misaligned supersymmetric gauge field theory (MSGFT)}

\bigskip

The name is taken from \cite{Dienes}
where string models without supersymmetry were studied, particularly
the supertrace conditions necessary for cancellation of
ultra violet divergences.

The nonsupersymmetric quiver gauge
theories here discussed
satisfy such supertrace conditions
if all scalars are in bifundamentals
so the name ``misaligned" supersymmetric gauge
field theory (MSGFT)
is appropriate.
In \cite{Dienes}, however, no explicit field transformation
underlying misaligned supersymmetry was given
and my aim here is to suggest how this may be
accomplished.

\bigskip

A specific MSGFT model is defined by several integers, namely $N$ (the number of coalescing
parallel D3 branes in AdS/CFT, also the $N$ in the gauge group
$U(N)^n$), $n$ (defining the abelian orbifold group $Z_n$,
also the $n$ in the gauge group $U(N)^n$); and
three integers $A_1, A_2, A_3$ which specify the embedding
$Z_n \subset SU(4)$ where $SU(4)$ is the internal symmetry of the ${\cal N}=4$
case corresponding to replacing the orbifold by a manifold.
Note that the fourth integer $A_4$ defining the transformation of the {\bf 4}
of $SU(4)$ is not independent because $A_4 = - A_1 - A_2 - A_3$ (mod $n$).
In summary, MSGFT  models 
are specified by five integers $\{N, n, A_1, A_2, A_3\}$.

\bigskip

The action for such a MSGFT in the present notation

\small

\begin{eqnarray}
S & = &
\int d^4x \left[ -\frac{1}{4} F_{\mu\nu a;r,r}F_{r,r}^{\mu\nu a}
+ \frac{1}{2} D_{\mu} \Phi_{ij;r+a_i,r}^{a} D^{\mu} \Phi_{ij;r,r+a_i}^{a}
+i \bar{\chi}_{r+A_m,r}^{a} \gamma.D L \chi_{r,r+A_m}^{a} \right.  \nonumber \\
&  &
\left.   - \frac{1}{2} i g f_{abc} \left( \bar{\tilde{\chi}}_{r,r+A_m}^{ai} L \chi_{r+A_m,r+A_m+A_n}^{jb}
\Phi_{ij;r+A_m+A_n,r}^{c} \right. \right. \nonumber  \\
&  & - \left.  \left. \bar{\chi}_{i;r,r+A_m}^{a} R \tilde{\chi}_{j;r+A_m,r-A_n}^{b} \Phi_{r-A_n,r}^{ijc}
\right)  \right.  \nonumber \\
& & \left.  - \frac{1}{4} g^2 \left(
f_{abc} \Phi_{ij;r,r+a_i}^{b} \Phi_{kl;r+a_i,r+a_i+a_j}^{c} \right)
\left(
f_{ade} \Phi_{r+a_i+a_j,r+a_j}^{ijd} \Phi_{r+a_j,r}^{kle} \right)
\right]
\label{MSGFTaction}
\end{eqnarray}

\large

\noindent in which the $a_i$ are defined by $a_i = A_2+A_3, a_2=A_3+A_1, a_3=A_1+A_2$;
the subscript $r=1, 2, .... n$ is a node label; when the two node
superscripts are equal it is an adjoint plus
singlet of that $U(N)_r$; when the two subscripts are unequal it is a bifundamental
and the two gauge labels transform under different $U(N)$ gauge groups.

\bigskip

Now we address the question of what variation of the
fields in the action (\ref{MSGFTaction}) will leave it invariant.
Given the field content, the infinitesimal fermionic
parameters must transform under the $U(N)^n$ gauge group.
As a generalization of equations (\ref{susy}), we suggest

\small

\begin{eqnarray}
& & \delta \left( A^{(p)}_{\mu} \right)^{\beta_p}_{\alpha_p} \nonumber \\
& =  & i \left( [\bar{\alpha}_i]^{\beta_p,\gamma_p}
_{\alpha_p,\delta_{p+A_m}} \gamma_{\mu} L \left( \chi^{i(p,p+A_m)} \right)^{\delta_{p+A_m}}_{\gamma_p}
- \left( \bar{\chi}_i^{(p-A_m,p)} \right)^{\delta_p}_{\gamma_{p-A_m}} \gamma_{\mu} L
[\alpha^i]^{\alpha_p,\gamma_{p-A_m}}_{\alpha_p,\delta_p} \right) \nonumber  \\
& & \delta \left( \Phi_{ij}^{(p,p+a_i)}\right)^{\alpha_{p+a_i}}_{\alpha_p} \nonumber \\
& = &
i \left( [\bar{\alpha}_j]^{\alpha_{p+a_i},\beta_{p}}_{\alpha_p,\beta_{p+A_m}} R
\left( \tilde{\chi}_i^{(p,p+A_m)} \right)_{\beta_{p}}^{\beta_{p+A_m}}
- [\bar{\alpha}_i]^{\alpha_{p+a_i},\beta_{p}}_{\alpha_p,\beta_{p+A_m}} R
\left(\tilde{\chi}_j^{(p,p+A_m)} \right)_{\beta_{p}}^{\beta_{p+A_m}}  \right. \nonumber \\
& & \left. + \epsilon_{ijkl} [\bar{\alpha}^k]^{\alpha_{p+a_i},\beta_p}_{\alpha_p,\beta_{p+A_m}} L
\left( \chi^{l (p,p+A_m)} \right)_{\beta_p}^{\beta_{p+A_m}}  \right). \nonumber  \\
& & \delta \left( L \chi^{i(p,p+A_m)} \right)_{\alpha_p}^{\alpha_{p+A_m}} 
\nonumber \\
& = &
\sigma_{\mu\nu} \left( F^{\mu\nu (p)} \right) _{\beta_p}^{\gamma_p}
L [\alpha^i]_{\alpha_p \gamma_p}^{\alpha_{p+A_m} \beta_p} \nonumber \\
& & - \gamma.D \left( \Phi^{ij (p,p+a_i)} \right)_{\beta_p}^{\beta_{p+a_i}}  R
[\tilde{\alpha}_j]_{\alpha_p \beta_{p+a_i}}^{\alpha_{p+A_m} \beta_p}  \nonumber  \\
& & + \frac{1}{2} g \epsilon_{\alpha_{p+a_i}\beta_{p+a_i}\gamma_{p+a_i}}
\epsilon^{\beta_p\gamma_p\delta_p}
\left( \phi^{ik (p,p+a_i)} \right)_{\gamma_p}^{\beta_{p+a_i}}
\left( \Phi_{kj}^{(p,p+a_i)} \right)_{\delta_p}^{\gamma_{p+a_i}}
L [\alpha^{j}]_{\alpha_p\beta_p}^{\alpha_{p+A_m}\alpha_{p+a_i}} \nonumber \\
& & \delta \left( R \tilde{\chi}_{i}^{(p-A_m,p)} \right)_{\alpha_{p-A_m}}^{\alpha_{p}} 
\nonumber  \\
& = &
\sigma_{\mu\nu} \left( F^{\mu\nu (p)} \right) _{\beta_p}^{\gamma_p}
R [\tilde{\alpha_i}]_{\alpha_{p-A_m} \gamma_p}^{\alpha_{p} \beta_p} \nonumber \\
& & + \gamma.D \left( \Phi_{ij}^{(p-a_i,p)} \right)_{\beta_{p-a_i}}^{\beta_{p}}  L
[\alpha^j]_{\alpha_{p-A_m} \beta_{p}}^{\alpha_{p} \beta_{p-a_i}}  \nonumber  \\
& & + \frac{1}{2} g \epsilon_{\alpha_{p+a_i}\beta_{p+a_i}\gamma_{p+a_i}}
\epsilon^{\beta_p\gamma_p\delta_p}
\left( \phi_{ik}^{(p,p+a_i)} \right)_{\gamma_p}^{\beta_{p+a_i}}
\left( \Phi^{kj (p,p+a_i)} \right)_{\delta_p}^{\gamma_{p+a_i}}
R [\tilde{\alpha_{j}}]_{\alpha_{p-A_m}\beta_p}^{\alpha_{p}\alpha_{p+a_i}}
\label{misalignedsusy}
\end{eqnarray}

\large

\bigskip

\noindent The equations (\ref{misalignedsusy}) are written so that they reduce
to the ${\cal N}=4$ equations (\ref{susy}) when the internal $U(N)^n$ dependence of the fermionic
generators is removed and are written such that the transformation properties
under the gauge group $U(N)^n$ are consistent for each term in the
field transformations (\ref{misalignedsusy}).

In the limit $A_m=a_i=0$ and $n=1$, the bifundamentals become adjoints and
the couplings in the transformations Eq.(\ref{misalignedsusy})
reduce to those in Eq.(\ref{susy}); this requirement excludes
further (symmetric) cubic couplings in Eq. (\ref{misalignedsusy}).

We see that the infinitesimal generators n Eq. (\ref{misalignedsusy})
must generically be outer products
of two bifundamentals under $U(N)^n$ although in all
terms of (\ref{misalignedsusy}) this reduces
to an outer product of one adjoint with one bifundamental.
In the transformation of the $\chi^i$ fields
I have for definiteness specialized
to the case $N=3$ in generalizing the structure constants $f_{abc}$
of (\ref{susy}) for adjoint representations to the antisymmetric tensors
$\epsilon_{\alpha\beta\gamma}$ in (\ref{misalignedsusy}) for
bifundamental representations; for
general $N$ one can form
a unique antisymmetric cubic invariant from bifundamentals writable in two equivalent forms
\begin{equation}
f_{abc} (\lambda^a)^i_{i^{'}} (\lambda^b)^j_{j^{'}} (\lambda^c)^k_{k^{'}}
\Phi^{i^{'}}_i \Phi^{j^{'}}_j \Phi^{k^{'}}_k ~~~ {\rm or} ~~~
\epsilon^{ijklmn...xyz}\epsilon_{i'j'k'lmn...xyz} \Phi_i^{i^{'}} \Phi_j^{j^{'}}
\Phi_k^{k^{'}}
\end{equation}

\bigskip

\noindent {\it Discussion.}

\bigskip

A first issue concerns the no-go theorems of
\cite{Haag1975}. There is no problem with the earlier paper 
which did not consider fermionic generators and the
generalized no-go theorem implicitly
assumes that the fermionic generators are singlets
under the gauge group; since this assumption is
violated in misaligned supersymmetry, the no-go
theorem is inapplicable.

\bigskip

There remain a number of questions to be explored:
Does variation \cite{global} under the field transformations (\ref{misalignedsusy})
really provide an exact symmetry of the action (\ref{MSGFTaction})?
Do the generators form a closed algebra and the transformations
a group?
What are the representations of this group? The quiver diagram must
form a representation but it may be reducible. It would be interesting to
know the irreducible representations.
Do MSGFT share properties of supersymmetric gauge theories
such as non renormalization theorems?
Can a MSGFT be conformally invariant?

\newpage

\section{U(1) factors}

\bigskip
\bigskip

The lagrangian for the nonsupersymmetric $Z_n$ theory can be written in
a convenient
notation which accommodates simultaneously both adjoint and
bifundamental scalars as mentioned before.

As also mentioned above we shall restrict attention to models
where all scalars are in bifundamentals which
requires all $a_i$ to be non zero. Recall that
$a_1=A_2+A_3$, $a_2=A_3+A_1$; $a_3=A_1+A_2$.

\bigskip

\bigskip
\bigskip
\bigskip

The lagrangian is classically $U(N)^p$ gauge
invariant. There are, however, triangle anomalies of the
$U(1)_p U(1)^2_q$ and $U(1)_p SU(N)_q^2$ types. Making gauge transformations under
the $U(1)_r$ (r = 1,2,...,n) with gauge parameters $\Lambda_r$
leads to a variation

\begin{equation}
\delta {\cal L} = - \frac{g^2}{4\pi^2}\Sigma_{p=1}^{p=n} A_{pq} F_{\mu\nu}^{(p)}
\tilde{F}^{(p) \mu\nu} \Lambda_q
\label{Apq}
\end{equation}
which defines an $n \times n$ matrix $A_{pq}$ which is given by

\begin{equation}
A_{pq} = {\rm Tr} (Q_p Q_q^2)
\label{chiraltrace}
\end{equation}
where the trace is over all chiral fermion links and $Q_r$ is the
charge of the bifundamental under $U(1)_r$. We shall adopt the sign
convention that ${\bf N}$ has $Q=+1$ and ${\bf N^{*}}$ has $Q=-1$.

\bigskip
\bigskip
\bigskip

It is straightforward to write $A_{pq}$ in terms of
Kronecker deltas because the content of chiral fermions is

\begin{equation}
\Sigma_{m=1}^{m=4} \Sigma_{r=1}^{r=n} ({\bf N}_r, {\bf N^{*}}_{r+A_{m}})
\end{equation}
This implies that the antisymmetric matrix $A_{pq}$ is explicitly

\begin{equation}
A_{pq} = - A_{qp} = \Sigma_{m=1}^{m=4} \left( \delta_{p, q-A_{m}} -
\delta_{p, q+A_{m}} \right)
\label{ApqKronecker}
\end{equation}

\bigskip

Now we are ready to construct ${\cal L}_{comp}^{(1)}$, the compensatory
term\cite{DiNapoli:2006wz}. Under the $U(1)_r$ gauge transformations with
gauge parameters $\Lambda_r$ we require that

\begin{eqnarray}
\delta {\cal L}_{comp}^{(1)} & = &   - \delta {\cal L} \nonumber \\
& = & + \frac{g^2}{4\pi^2}\Sigma_{p=1}^{p=n} A_{pq} F_{\mu\nu}^{(p)}
\tilde{F}^{(p) \mu\nu} \Lambda_q
\label{compensatory}
\end{eqnarray}
To accomplish this property, we
construct a compensatory term in the form
\begin{equation}
{\cal L}_{comp}^{(1)} = \frac{g^2}{4 \pi} \Sigma_{p=1}^{p=n}
\Sigma_{k} B_{pk} ~{\rm Im}~ {\rm Tr} ~{\rm ln}
\left( \frac{\Phi_k}{v} \right) F_{\mu\nu}^{(p)} \tilde{F}^{(p) \mu\nu}
\label{compensatory2}
\end{equation}
where $\Sigma_{k}$ runs over scalar links.
To see
that ${\cal L}_{comr}^{(1)}$ of Eq.(\ref{compensatory2}) has
$SU(N)^n$.
invariance
rewrite Tr ln $\equiv$ exp det and note that the $SU(N)$
matrices have unit determinant.

\bigskip
\bigskip
\bigskip

We note {\it en passant} that one cannot take the
$v \rightarrow 0$ limit in Eq.(\ref{compensatory2}); the chiral anomaly
enforces a breaking of conformal invariance.

We define a matrix $C_{kq}$ by

\begin{equation}
\delta \left( \Sigma_{p=1}^{p=n} \Sigma_k~ {\rm Im}
~{\rm Tr}~ {\rm ln} \left( \frac{\Phi_k}{v} \right) \right)
= \Sigma_{q=1}^{q=n} C_{kq} \Lambda_q
\label{Ckq}
\end{equation}
whereupon Eq.(\ref{compensatory}) will be satisfied if
the matrix $B_{pk}$ satisfies $A=BC$. The inversion $B=AC^{-1}$ is
non trivial because $C$ is singular but $C_{kq}$ can be written in terms
of Kronecker deltas by noting that the content of complex scalar fields in the model
implies that the matrix $C_{kq}$ must be of the form
\begin{equation}
C_{kq} = 3 \delta_{kq} - \Sigma_{i} \delta_{k+a_{i},q}
\label{CkqKronecker}
\end{equation}

\bigskip

\newpage

\bigskip

\subsection{Conformality of U(1) couplings.}

\bigskip
$U(1)$ factors are a major concern in quiver gauge theories.
In the absence of the compensatory term, the two independent $U(N)^n$
gauge couplings $g_N$ for SU(N) and $g_1$ for U(1) are taken to be
equal $g_N(\mu_0) = g_1(\mu_0)$ at a chosen scale, {\it e.g.} $\mu_0$=4 TeV
to enable cancellation of quadratic
divergences. Note that the $n$ SU(N) couplings
$g_N^{(p)}$ are equal by the overall $Z_n$ symmetry,
as are the $n$ U(1) couplings $g_1^{(p)}$, $1 \le p \le n$.

As one evolves to higher scales $\mu > \mu_0$, the renormalization
group beta function $\beta_N$ for SU(N)
vanishes $\beta_N =0$ at least at one-loop level so
the $g_N(\mu)$ can behave independent of the scale as expected by conformality.
On the other hand, the beta function $\beta_1$ for
U(1)
is positive definite in the unadorned theory, given at one loop by
\begin{equation}
b_1 = \frac{11N}{48\pi^2}
\label{b1}
\end{equation}
where N is the number of colors.

\bigskip
\bigskip
\bigskip

The corresponding coupling satisfies
\begin{equation}
\frac{1}{\alpha_1(\mu)} = \frac{1}{\alpha_1(M)} + 8\pi b_1 {\rm ln} \left( \frac{M}{\mu}
\right)
\end{equation}
so the Landau pole, putting $\alpha(\mu)=0.1$ and $N=3$, occurs at
\begin{equation}
\frac{M}{\mu} = {\rm exp} \left[ \frac{20 \pi}{11} \right] \simeq 302
\end{equation}
so for $\mu = 4$ TeV, $M \sim 1200$ TeV. The coupling becomes ``strong''
$\alpha(\mu) = 1$ at
\begin{equation}
\frac{M}{\mu} = {\rm exp} \left[ \frac{18 \pi}{11} \right] \simeq 171
\end{equation}
or $M \sim 680$ TeV.

We may therefore ask whether the new term ${\cal L}_{comp}$
in the lagrangian, necessary for anomaly cancellation, can
solve \cite{DiNapoli:2006wz} this problem for conformality?

\bigskip

\newpage

\bigskip

Since the scale $v$ breaks conformal invariance,
the matter fields acquire mass, so the one-loop
diagram \footnote{The usual one-loop $\beta-$function is
of order $h^2$ regarded as an expansion in Planck's constant:
four propagators each $\sim h$ and two vertices each $\sim h^{-1}$
(c.f. \cite{Nambu}). 
The diagram considered
is also $\sim h^2$ since it has three propagators,
one quantum vertex $\sim h$ and an additional $h^{-2}$ associated with
$\Delta m^2_{kk'}$.} has a logarithmic divergence proportional to

\begin{equation}
\int \frac{d^4p}{v^2} \left[ \frac{1}{(p^2-m_k^2)} - \frac{1}{(p^2-m_{k'}^2)}
\right]
\sim - \frac{ \Delta m^2_{kk'}}{v^2} {\rm ln} \left( \frac{\Lambda}{v} \right)
\end{equation}
the sign of which depends on $\delta m^2_{kk'} = (m_k^2 - m_{k'}^2)$.

\bigskip
\bigskip
\bigskip

\noindent To achieve conformality of U(1), a constraint must be imposed
on the mass spectrum of matter bifundamentals, {\it viz}
\begin{equation}
\Delta m^2_{kk'} \propto v^2 \left( \frac{11N}{48\pi^2} \right)
\end{equation}
\noindent with a proportionality constant of order one which depends on the choice
of model, the $n$ of $Z_n$ and the values chosen for $A_m, m=1,2,3$. This
signals how conformal invariance must be broken at
the TeV scale in order that it can be restored at high energy;
it is interesting that such a constraint arises
in connection with an anomaly cancellation mechanism
which necessarily breaks conformal symmetry.

\bigskip
\bigskip

\newpage

\section{Dark matter candidate} 

\bigskip

In the nonsupersymmetric quiver gauge theories, the
gauge group, for abelian orbifold $AdS_5 \times S^5/Z_n$
is $U(N)^n$. In phenomenological application $N=3$
and $n$ reduces eventually after symmetry breaking
to $n=3$ as in trinification.
The chiral fermions are then in the representation of $SU(3)^3$:
\begin{equation}
(3, 3^*, 1) + (3^*, 1, 3) + (1, 3, 3^*)
\end{equation}
This is as in the {\bf 27} of $E_6$ where the particles
break down in to the following representations of the
$SU(3) \times SU(2) \times U(1)$ standard model group:
\begin{equation}
Q, ~~~~~ u^c, ~~~~ d^c, ~~~~ L ~~~~ e^c ~~~~~ N^c
\end{equation}
transforming as
\begin{equation}
(3,2), ~~ (3^*,1), ~~ (3^*,1), ~~ (1,2), ~~ (1,1), ~~ (1,1) \\
\end{equation}
in a {\bf 16} of the $SO(10)$ subgroup.

\bigskip

\noindent In addition there are the states
\begin{equation}
h, ~~~~~ h*, ~~~~~ E, ~~~~~ E*
\end{equation}
transforming as
\begin{equation}
(3, 1), ~~ (3^*, 1), ~~ (2,1), ~~ (2,1)
\end{equation}
in a {\bf 10} of $SO(10)$
and finally
\begin{equation}
S
\end{equation}
transforming as the singlet
\begin{equation}
(1, 1)
\end{equation}

It is natural to define\cite{LCP}
a $Z_2$ symmetry $R$ which commutes
with the $SO(10)$ subgroup of $E_6 \rightarrow O(10) \times U(1)$

It is natural to define a $Z_2$ symmetry $R$ which commutes
with the $SO(10)$ subgroup of $E_6 \rightarrow O(10) \times U(1)$
such that $R=+1$ for the first {\bf 16} of states.
Then it is mandated that $R=-1$ for the {\bf 10} and {\bf 1}
of SO(10) because the following Yukawa couplings must
be present to generate mass for the fermions:
\begin{equation}
16_f 16_f 10_s, ~~~~16_f 10_f 16_s, ~~~~ 10_f 10_f 1_s, ~~~~
10_f 1_f 1_s
\end{equation}
\noindent which require $R=+1$ for $10_s, 1_s$ and $R=-1$ for $16_s$.

\bigskip

\noindent Contribution of the Lightest Conformality Particle (LCP) to the cosmological energy density:

\bigskip

The LCP act as cold dark matter WIMPs, and the calculation of the
resultant energy density follows a well-known path. Here
we follow the procedure in a recent technical book by Mukhanov.

The LCP decouple at temperature $T_*$, considerably less than their
mass $M_{LCP}$; we define $x_* = M_{LCP}/T_*$. Let the
annihilation cross-section of the LCP
at decoupling be $\sigma_*$. Then the dark matter density
$\Omega_m$, relative to the critical density,
\begin{equation}
\Omega_m h_{75}^2 = \frac{\tilde{g}_*^{1/2}}{g_*} x_*^{3/2} \left(
\frac{3 \times 10^{-38} cm^2}{\sigma_*} \right)
\label{OmegaM}
\end{equation}
where $h_{75}$ is the Hubble constant in units
of $75km/s/Mpc$.
$g_* = (g_b + \frac{7}{8}g_f)$ is the effective number of
degrees of freedom (dof) at freeze-out for all particles
which later convert their energy into photons;
and $\tilde{g}_*$ is the number
of dof which are relativistic at $T_*$.

\bigskip
\bigskip

\noindent The LCP \cite{LCP}
is a viable candidate for a
cold dark matter particle which can be produced at the LHC.
The distinction from other candidates
will require establishment of the $U(3)^3$ gauge bosons, extending the 3-2-1 standard model
and the discovery that the LCP is in a bifundamental
representation thereof.
To confirm that the LCP is the dark matter particle would, however,
require direct detection of dark matter.

\newpage

\section{Proton decay}

\bigskip

Here we address the issue of proton decay.
Under $SU(3)_C \times SU(3) \times SU(3)$, the quarks and leptons
families each appear as in trinification\cite{Glashow:1984gc}
in the representations
\begin{equation}
[(3, 3^*, 1) + (1, 3, 3^*) + (3^*, 1, 3)]
\end{equation}
for which the baryon numbers are respectively $B=+1/3, 0, -1/3$
for quarks, leptons and antiquarks.

The gauge bosons cannot transform quarks
into leptons or vice versa because of the factoring out of the
color $SU(3)_C$ group. So unlike in $SU(5)$(\cite{GG,GG2}) proton
decay is absent in the gauge sector.

The scalars are likewise in 27's according to
\begin{equation}
[(3, 3^*, 1) + (1, 3, 3^*) + (3^*, 1, 3)] + c.c.
\end{equation}
although here, unlike for the fermions the complex conjugate
representations must be included.

\noindent Fermion masses arise in trinification \cite{Glashow:1984gc}
from Yukawa couplings of the form
\begin{equation}
(3, 3^*, 1)_q (3^*, 1, 3)_q (1, 3, 3^*)_{\phi}
\end{equation}
for the quarks and
\begin{equation}
(1, 3, 3^*)_f (1, 3, 3^*)_f (1, 3, 3^*)_{\phi}
\end{equation}
for the leptons. Because these are two independent couplings,
trinification has the feature of giving no relationship between
quark and lepton masses.

There are additional Yukawa couplings possible which would violate
baryon number $B$ and cause catastrophically rapid proton decay
with a TeV unification scale. Therefore these must be forbidden,
as is achieved \cite{proton} by assigning a generalization of baryon
number $B=+1/3, 0 -1/3$ respectively to the three representation
listed.
\begin{equation}
(3, 3^*, 1)_{\phi} ~~~~~~~~~~~~~~~~ B=1/3
\end{equation}
\begin{equation}
(1, 3, 3^*)_{\phi} ~~~~~~~~~~~~~~~  B=0
\end{equation}
\begin{equation}
(3^*, 1, 3)_{\phi} ~~~~~~~~~~~~~~~ B=-1/3
\end{equation}
Such assignmnents are very natural.

In phenomenological analysis of trinification\cite{Pak},
however, such a procedure is avoided in order to be able
to obtain acceptable
quark and lepton mass matrices. With teravolt unification
such departure from the simplest case
is not possible as proton decay must be totally forbidden.

However, the quark and lepton masses in conformality will
acquire contributions in relevant operators $m_i\psi\psi$
from breaking of conformal symmetry. The pattern of these
masses cannot yet be calculated, although a constraint on
the pattern of the corresponding scalar masses has been
suggested in \cite{quadratic} based on the cancellation of
quadratic divergences.

In grand unification based on conformality, therefore,
it is expected that proton decay will be completely
absent (See however \cite{Dent:2007ed}.) in the non-gravitational theory. Gravitational
effects may eventually destabilize the proton but with
lifetime $\sim 10^{50} y$ far beyond any forseeable experiment.

Equally or more important is the realization that consistency
of the conformality with proton decay dictates that the
quark and lepton masses receive significant contributions
from the four-dimensional conformal symmetry breaking, not only
from the Yukawa coupling to Higgs as universally assumed in previous
studies of grand unification.

In the standard model,
while the $Z^0, W^{\pm}$ masses are accurately
predicted by a Higgs mechanism, there is no
similar statement about fermion masses.
The conclusion is that quark and lepton masses arise
principally as relevant operators
\footnote{One important question is how such mass
terms are induced above the electroweak scale. One possibility is via four fermion
operators.
} arising from breaking
of four-dimensional conformal invariance; a corollary is
that the couplings of the standard Higgs scalar to
quarks and leptons depart from the values usually assumed.

\newpage

\section{Conclusions} 

\bigskip

\noindent It has been established that
conformality
can provide (i) naturalness without one-loop
quadratic divergence for the scalar mass and
anomaly cancellation;
(ii) precise unification of the coupling constants;
and (iii) a viable dark matter candidate.
It remains for experiment to check that quiver gauge
theories with gauge group $U(3)^3$ or $U(3)^n$ with $n \geq 4$
are actually employed by Nature.

For completeness, we should note that possible problems
with ${\cal N} = 0$ orbifolds have been pointed out
both in one-loop calculations in field 
theory\cite{AdamsSilverstein} and from studies of tachyonic
instability in the ancestral string theory\cite{AdamsPolchinskiSilverstein}.

One technical point worth another mention
is that while the $U(1)$ anomalies
discussed in Section 9 are cancelled in sting theory, in the gauge theory
we have shown a different description of such cancellation which
has the advantage of suggesting how $U(1)$ gauge couplings may
be conformally invariant at high energies. This is important because
in string theory, except for special linear combinations, all
such $U(1)$ factors acquire mass by the St\"{u}ckelberg mechanism
while in the gauge field theory 
as discussed in Section 8
the cancellation of quadratic
divergences and solution of the hierarchy problem for ${\cal N} = 0$ 
require they be instead at the teravolt scale.
 
We have described how phenomenology of conformality has striking resonances
with the standard model, as we have described optimistically
as experimental evidence in its favor.
We have described how 4 TeV Unification predicts three families and new particles
around 4 TeV accessible to experiment (LHC).

It is encouraging that the scalar propagator in these theories has no
quadratic divergence if and only if there are chiral fermions.
Anomaly cancellation in the effective lagrangian
has been tied to the conformality of U(1) gauge couplings.
 
A dark matter candidate (LCP = Lightest Conformality Particle) may be
produced at the Large Hadron Collider (LHC), then directly detected from the cosmos. Study
of proton decay leads to the conclusion that quark and lepton
masses arise not from the Yukawa couplings of the standard
model but from operators induced in breaking of the
four-dimensional conformal invariance. This implies that
the Higgs couplings differ from those usually assumed
in the unadorned standard model. This is yet another
prediction from conformality to be tested
when the Higgs scalar couplings and decay products
are examined at the LHC in the near
future.

\newpage

\begin{center}

{\bf Acknowledgements}

\end{center}

\bigskip
TWK thanks the Aspen Center for Physics for hospitality while this work was in progress.
This work was supported in part by
the U.S. Department of Energy under Grants 
No. DE-FG02-06ER41418 (PHF) and DE-FG05-85ER40226 (TWK).

\newpage

\noindent \underline{{\bf Note Added}}

\bigskip

\noindent When the speculation was originally made in \cite{first} that
non supersymmetric ${\cal N} = 0$ abelian orbifolded theories
with gauge group $U(N)^n$ become conformally
invariant ({\it conformality})
at high energy for finite $N$ it was believed that 
in the limit $N \rightarrow \infty$ conformality
is correct, as suggested by the papers
\cite{Vafa1,Vafa2} which provided a {\it ``proof"}
that in the large N limit the ${\cal N} = 0$
theory coincided to all orders of perturbation theory
with the ancestor ${\cal N} = 4$ theory which is
known to be conformally invariant.

\bigskip

\noindent For ${\cal N}=4$ gauge theory conformality is known
\cite{Mandelstam1983,Mandelstam19832,Mandelstam19833}
not only for the limit $N \rightarrow \infty$ but also
for finite $N$. It was therefore natural in \cite{first}
to conjecture the same for ${\cal N} \leq 2$ including
for the nonsupersymmetric case ${\cal N} = 0$.

\bigskip
{
\noindent More recent work by Dymarsky, Klebanov and Roiban
\cite{DKR1,DKR2} shows that the {\it ``proof"} in
\cite{Vafa1,Vafa2} is false because there are double trace
operators of the bifundamental scalars with the
general form $Tr(\Phi\Phi)Tr(\Phi\Phi)$ which
are forbidden by ${\cal N}=4$ symmetry from appearing
in the ancestor theory but which can and do appear as counterterms
in the ${\cal N} = 0$ theory. The renormalization group
$\beta-$functions for these double trace couplings do
not vanish generically in the $N \rightarrow \infty$ limit so
the conformality of \cite{first}
requires more than cancellation of $1/N$ corrections.

\bigskip

\noindent The important papers \cite{DKR1,DKR2} study this phenomenon 
at the one-loop level. The impact on the all orders speculation
of \cite{first} about ${\cal N} = 0$ and finite $N$
merits further investigation.


\newpage

\begin{thebibliography}{99}

\bibitem{Wilson1971}
K.G. Wilson, Phys. Rev. {\bf D3,} 1818 (1971).

\bibitem{Wilson2004}
K.G. Wilson, Nucl. Phys. Proc. Suppl. {\bf 140,} 3 (2005).
{\tt hep-lat/0412043}.

\bibitem{Glashow1961}
S.L. Glashow, Nucl. Phys. {\bf 22,} 579 (1961). 

\bibitem{Weinberg1967}
S. Weinberg, Phys. Rev. Lett. {\bf 19,} 1264 (1967).

\bibitem{Salam1968}
A. Salam, in Svartholm: Elementary Particle Theory, Proceedings
of the Nobel Symposium
held 1968 at Lerum, Sweden, Stockholm (1968) page 367-377.

\bibitem{Wess1974}
J. Wess and B. Zumino, Nucl. Phys. {\bf B70,} 39 (1974).

\bibitem{Haag1975}
R. Haag, J.T. Lopuszansi and M. Sohnius, Nucl. Phys. {\bf B88,} 257 (1975).

\bibitem{Haag19752}
S. Coleman and J.E. Mandula, Phys. Rev. {\bf 159,} 1251 (1967).

\bibitem{Woit2006}
P. Woit, {\it Not Even Wrong}. Basic Books (2006).

\bibitem{Mandelstam1983}
S. Mandelstam, Nucl. Phys. {\bf B213,} 149 (1983).

\bibitem{Mandelstam19832}
P. Howe and K. Stelle, Phys. Lett. {\bf B137} 135 (1984);

\bibitem{Mandelstam19833}
L. Brink, Talk at Johns Hopkins Workshop
on Current Problems in High Energy Particle Theory, Bad Honnef, Germany (1983).

\bibitem{Maldacena1997}
J.M. Maldacena, Adv. Theor. Math. Phys. {\bf 2,} 231 (1998).
{\tt hep-th/9711200}.

\bibitem{first}
P.H. Frampton, Phys.Rev. {\bf D60,} 041901 (1999). {\tt hep-th/9812117}.

\bibitem{Smith:1993gp}
  F.~D.~Smith,
  %``Sets and C**n: quivers and A-D-E: Triality: generalized supersymmetry: and
  %D(4) - D(5) - E(6),''
{\tt hep-th/9306011}.

%\cite{Douglas:1996sw}
\bibitem{Douglas:1996sw}
  M.~R.~Douglas and G.~W.~Moore,
  %``D-branes, Quivers, and ALE Instantons,''
  {\tt hep-th/9603167}.

%\cite{Sardo-Infirri:1996gb}
\bibitem{Sardo-Infirri:1996gb}
  A.~V.~Sardo-Infirri,
  %``Resolutions of Orbifold Singularities and Flows on the McKay Quiver,''
  {\tt alg-geom/9610005}.

%\cite{deBoer:1996mp}
\bibitem{deBoer:1996mp}
  J.~de Boer, K.~Hori, H.~Ooguri and Y.~Oz,
  %``Mirror symmetry in three-dimensional gauge theories, quivers and
  %D-branes,''
  Nucl.\ Phys.\  B {\bf 493} (1997) 101
  {\tt hep-th/9611063}.

%\cite{Xu:1997rv}
\bibitem{Xu:1997rv}
  F.~Xu,
  %``A note on quivers with symmetries,''
  {\tt q-alg/9707003}.
  %%CITATION = Q-ALG/9707003;%%

%\cite{Okuyama:1998dz}
\bibitem{Okuyama:1998dz}
  K.~Okuyama and Y.~Sugawara,
  %``Fractional strings in (p,q) 5-brane and quiver matrix string theory,''
  JHEP {\bf 9808} (1998) 002
  {\tt hep-th/9806001}.
  %%CITATION = JHEPA,9808,002;%%

%\cite{Kapustin:1998fa}
\bibitem{Kapustin:1998fa}
  A.~Kapustin,
  %``D(n) quivers from branes,''
  JHEP {\bf 9812} (1998) 015
  {\tt hep-th/9806238}.
  %%CITATION = JHEPA,9812,015;%%

%\cite{Sugawara:1999qp}
\bibitem{Sugawara:1999qp}
  Y.~Sugawara,
  %``N = (0,4) quiver SCFT(2) and supergravity on AdS(3) x S(2),''
  JHEP {\bf 9906} (1999) 035
  {\tt hep-th/9903120}.
  %%CITATION = JHEPA,9906,035;%%

%\cite{Feng:1999zv}
\bibitem{Feng:1999zv}
  B.~Feng, A.~Hanany and Y.~H.~He,
  %``Z-D brane box models and non-chiral dihedral quivers,''
  {\tt hep-th/9909125}.
  %%CITATION = HEP-TH/9909125;%%

%\cite{He:1999xj}
\bibitem{He:1999xj}
  Y.~H.~He,
  %``Some remarks on the finitude of quiver theories,''
  {\tt hep-th/9911114}.
  %%CITATION = HEP-TH/9911114;%%
  
\bibitem{Buican:2006sn}
M.~Buican, D.~Malyshev, D.~R.~Morrison, M.~Wijnholt and H.~Verlinde,
%``D-branes at singularities, compactification, and hypercharge,''
JHEP {\bf 0701} (2007) 107 {\tt hep-th/0610007}.
  
%\cite{Verlinde:2005jr}
\bibitem{Verlinde:2005jr}  
H.~Verlinde and M.~Wijnholt,
%``Building the standard model on a D3-brane,''
JHEP {\bf 0701} (2007) 106  {\tt hep-th/0508089}.

%\cite{Fiol:2000wx}
\bibitem{Fiol:2000wx}
  B.~Fiol and M.~Marino,
  %``BPS states and algebras from quivers,''
  JHEP {\bf 0007} (2000) 031
  {\tt hep-th/0006189}.
  %%CITATION = JHEPA,0007,031;%%

%\cite{Uranga:2000ck}
\bibitem{Uranga:2000ck}
  A.~M.~Uranga,
  %``From quiver diagrams to particle physics,''
  {\tt hep-th/0007173}.
  %%CITATION = HEP-TH/0007173;%%

%\cite{Albertsson:2000px}
\bibitem{Albertsson:2000px}
  C.~Albertsson, B.~Brinne, U.~Lindstrom, M.~Rocek and R.~von Unge,
  %``ADE-quiver theories and mirror symmetry,''
  Nucl.\ Phys.\ Proc.\ Suppl.\  {\bf 102} (2001) 3
  {\tt hep-th/0103084}.
  %%CITATION = NUPHZ,102,3;%%

%\cite{Govindarajan:2000vi}
\bibitem{Govindarajan:2000vi}
  S.~Govindarajan and T.~Jayaraman,
  %``D-branes, exceptional sheaves and quivers on Calabi-Yau manifolds: From
  %Mukai to McKay,''
  Nucl.\ Phys.\  B {\bf 600} (2001) 457
  {\tt hep-th/0010196}.
  %%CITATION = NUPHA,B600,457;%%

%\cite{Feng:2000mw}
\bibitem{Feng:2000mw}
  B.~Feng, A.~Hanany, Y.~H.~He and N.~Prezas,
  %``Discrete torsion, covering groups and quiver diagrams,''
  JHEP {\bf 0104} (2001) 037
  {\tt hep-th/0011192}.
  %%CITATION = JHEPA,0104,037;%%

%\cite{Berenstein:2000mb}
\bibitem{Berenstein:2000mb}
  D.~Berenstein, V.~Jejjala and R.~G.~Leigh,
  %``D-branes on singularities: New quivers from old,''
  Phys.\ Rev.\  D {\bf 64} (2001) 046011
  {\tt hep-th/0012050}.
  %%CITATION = PHRVA,D64,046011;%%

%\cite{Albertsson:2001jq}
\bibitem{Albertsson:2001jq}
  C.~Albertsson, B.~Brinne, U.~Lindstrom and R.~von Unge,
  %``E(8) quiver gauge theory and mirror symmetry,''
  JHEP {\bf 0105} (2001) 021
  {\tt hep-th/0102038}.
  %%CITATION = JHEPA,0105,021;%%

\bibitem{Kephart:2001ix}
T.~W.~Kephart and Q.~Shafi, Phys. Lett. {\bf B520,} 313 (2001).
%``Family unification, exotic states and magnetic monopoles,''
{\tt hep-ph/0105237}.


%\cite{Muto:2001gu}
\bibitem{Muto:2001gu}
  T.~Muto and T.~Tani,
  %``Stability of quiver representations and topology change,''
  JHEP {\bf 0109} (2001) 008
  {\tt hep-th/0107217}.
  %%CITATION = JHEPA,0109,008;%%

%\cite{Cachazo:2001gh}
\bibitem{Cachazo:2001gh}
  F.~Cachazo, S.~Katz and C.~Vafa,
  %``Geometric transitions and N = 1 quiver theories,''
  {\tt hep-th/0108120}.
  %%CITATION = HEP-TH/0108120;%%

%\cite{Hanany:2001py}
\bibitem{Hanany:2001py}
  A.~Hanany and A.~Iqbal,
  %``Quiver theories from D6-branes via mirror symmetry,''
  JHEP {\bf 0204} (2002) 009
  {\tt hep-th/0108137}.
  %%CITATION = JHEPA,0204,009;%%

%\cite{Alvarez-Consul:2001uk}
\bibitem{Alvarez-Consul:2001uk}
  L.~Alvarez-Consul and O.~Garcia-Prada,
  %``Dimensional reduction and quiver bundles,''
  arXiv:math.dg/0112160.
  %%CITATION = MATH.DG/0112160;%%

%\cite{Alvarez-Consul:2001um}
\bibitem{Alvarez-Consul:2001um}
  L.~Alvarez-Consul and O.~Garcia-Prada,
  %``Hitchin-Kobayashi correspondence, quivers, and vortices,''
  Commun.\ Math.\ Phys.\  {\bf 238} (2003) 1
  [arXiv:math.dg/0112161].
  %%CITATION = CMPHA,238,1;%%

%\cite{Cremades:2002te}
\bibitem{Cremades:2002te}
  D.~Cremades, L.~E.~Ibanez and F.~Marchesano,
  %``SUSY quivers, intersecting branes and the modest hierarchy problem,''
  JHEP {\bf 0207} (2002) 009
  {\tt hep-th/0201205}.
  %%CITATION = JHEPA,0207,009;%%

%\cite{Kim:2002fp}
\bibitem{Kim:2002fp}
  N.~W.~Kim, A.~Pankiewicz, S.~J.~Rey and S.~Theisen,
  %``Superstring on pp-wave orbifold from large-N quiver gauge theory,''
  Eur.\ Phys.\ J.\  C {\bf 25} (2002) 327
  {\tt hep-th/0203080}.
  %%CITATION = EPHJA,C25,327;%%

%\cite{Ishii:2002ee}
\bibitem{Ishii:2002ee}
  A.~Ishii,
  %``Representation Moduli Of The Mckay Quiver For Finite Abelian Subgroups Of
  %Sl(3,C),''
%\href{http://www.slac.stanford.edu/spires/find/hep/www?irn=6198775}{SPIRES entry}
{\it Prepared for School on Geometry and String Theory, Cambridge, England, 24 Mar - 20 Apr 2002}

%\cite{Mukhi:2002ck}
\bibitem{Mukhi:2002ck}
  S.~Mukhi, M.~Rangamani and E.~P.~Verlinde,
  %``Strings from quivers, membranes from moose,''
  JHEP {\bf 0205} (2002) 023
  {\tt hep-th/0204147}.
  %%CITATION = JHEPA,0205,023;%%

%\cite{Brax:2002rz}
\bibitem{Brax:2002rz}
  P.~Brax, A.~Falkowski, Z.~Lalak and S.~Pokorski,
  %``Custodial supersymmetry in non-supersymmetric quiver theories,''
  Phys.\ Lett.\  B {\bf 538} (2002) 426
  {\tt hep-th/0204195}.
  %%CITATION = PHLTA,B538,426;%%

%\cite{Hollowood:2002ax}
\bibitem{Hollowood:2002ax}
  T.~J.~Hollowood and S.~P.~Kumar,
  %``World sheet instantons via the Myers effect and N = 1* quiver
  %superpotentials,''
  JHEP {\bf 0210} (2002) 077
  {\tt hep-th/0206051}.
  %%CITATION = JHEPA,0210,077;%%

%\cite{Denef:2002ru}
\bibitem{Denef:2002ru}
  F.~Denef,
  %``Quantum quivers and Hall/hole halos,''
  JHEP {\bf 0210} (2002) 023
  {\tt hep-th/0206072}.
  %%CITATION = JHEPA,0210,023;%%

%\cite{Frenkel:2002cs}
\bibitem{Frenkel:2002cs}
  I.~Frenkel, A.~Malkin and M.~Vybornov,
  %``Quiver varieties, affine Lie algebras, algebras of BPS states, and
  %semicanonical basis,''
  arXiv:math-ph/0206012.
  %%CITATION = MATH-PH/0206012;%%

%\cite{Feng:2002kk}
\bibitem{Feng:2002kk}
  B.~Feng, A.~Hanany, Y.~H.~He and A.~Iqbal,
  %``Quiver theories, soliton spectra and Picard-Lefschetz transformations,''
  JHEP {\bf 0302} (2003) 056
  {\tt hep-th/0206152}.
  %%CITATION = JHEPA,0302,056;%%

%\cite{Lalak:2002ep}
\bibitem{Lalak:2002ep}
  Z.~Lalak,
  %``Custodial supersymmetry in non-supersymmetric quivers,''
%\href{http://www.slac.stanford.edu/spires/find/hep/www?irn=5576784}{SPIRES entry}
{\it Prepared for 10th International Conference on Supersymmetry and Unification of Fundamental Interactions (SUSY02), Hamburg, Germany, 17-23
Jun 2002}

%\cite{Berenstein:2002fi}
\bibitem{Berenstein:2002fi}
  D.~Berenstein and M.~R.~Douglas,
  %``Seiberg duality for quiver gauge theories,''
  {\tt hep-th/0207027}.
  %%CITATION = HEP-TH/0207027;%%

%\cite{Frampton:2002ta}
\bibitem{Frampton:2002ta}
  P.~H.~Frampton and P.~Minkowski,
  %``Perturbative inaccessibility of conformal fixed points in nonsupersymmetric
  %quiver theories,''
  {\tt hep-th/0208024}.
  %%CITATION = HEP-TH/0208024;%%

%\cite{Hollowood:2002zk}
\bibitem{Hollowood:2002zk}
  T.~J.~Hollowood and T.~Kingaby,
  %``The phase structure of mass-deformed SU(2) x SU(2) quiver theory,''
  JHEP {\bf 0301} (2003) 005
  {\tt hep-th/0210096}.
  %%CITATION = JHEPA,0301,005;%%

%\cite{He:2002fp}
\bibitem{He:2002fp}
  Y.~H.~He,
  %``G(2) quivers,''
  JHEP {\bf 0302} (2003) 023
  {\tt hep-th/0210127}.
  %%CITATION = JHEPA,0302,023;%%

%\cite{Seki:2002ti}
\bibitem{Seki:2002ti}
  S.~Seki,
  %``Comments on quiver gauge theories and matrix models,''
  Nucl.\ Phys.\  B {\bf 661} (2003) 257
  {\tt hep-th/0212079}.
  %%CITATION = NUPHA,B661,257;%%

%\cite{AitBenHaddou:2003ew}
\bibitem{AitBenHaddou:2003ew}
  M.~Ait Ben Haddou and E.~H.~Saidi,
  %``Explicit analysis of Kaehler deformations in 4D N = 1 supersymmetric quiver
  %theories,''
  Phys.\ Lett.\  B {\bf 575} (2003) 100.
  %%CITATION = PHLTA,B575,100;%%

%\cite{Frampton:2003zt}
\bibitem{Frampton:2003zt}
  P.~H.~Frampton,
  %``Quiver gauge theory and unification at about 4-TeV,''
  {\tt hep-th/0302057}.
  %%CITATION = HEP-TH/0302057;%%

%\cite{Casero:2003gf}
\bibitem{Casero:2003gf}
  R.~Casero and E.~Trincherini,
  %``Quivers via anomaly chains,''
  JHEP {\bf 0309} (2003) 041
  {\tt hep-th/0304123}.
  %%CITATION = JHEPA,0309,041;%%

%\cite{Dai:2003ak}
\bibitem{Dai:2003ak}
  J.~Dai and Y.~S.~Wu,
  %``Quiver mechanics for deconstructed matrix string,''
  Phys.\ Lett.\  B {\bf 576} (2003) 209
  {\tt hep-th/0306216}.
  %%CITATION = PHLTA,B576,209;%%

%\cite{Casero:2003gr}
\bibitem{Casero:2003gr}
  R.~Casero and E.~Trincherini,
  %``Phases and geometry of the N = 1 A(2) quiver gauge theory and matrix
  %models,''
  JHEP {\bf 0309} (2003) 063
  {\tt hep-th/0307054}.
  %%CITATION = JHEPA,0309,063;%%

%\cite{Benhaddou:2003nd}
\bibitem{Benhaddou:2003nd}
  M.~A.~Benhaddou and E.~H.~Saidi,
  %``Explicit analysis of Kahler deformations in 4D N = 1 supersymmetric  quiver
  %theories,''
  {\tt hep-th/0307103}.
  %%CITATION = HEP-TH/0307103;%%

%\cite{Narayan:2003et}
\bibitem{Narayan:2003et}
  K.~Narayan and M.~R.~Plesser,
  %``Coarse-graining quivers,''
  {\tt hep-th/0309171}.
  %%CITATION = HEP-TH/0309171;%%

%\cite{Mukhopadhyay:2003ky}
\bibitem{Mukhopadhyay:2003ky}
  S.~Mukhopadhyay and K.~Ray,
  %``Seiberg duality as derived equivalence for some quiver gauge theories,''
  JHEP {\bf 0402} (2004) 070
  {\tt hep-th/0309191}.
  %%CITATION = JHEPA,0402,070;%%

%\cite{Narayan:2003rd}
\bibitem{Narayan:2003rd}
  K.~Narayan,
  %``Block spins and D-branes: Coarse-graining matrices and quivers,''
%\href{http://www.slac.stanford.edu/spires/find/hep/www?irn=6477887}{SPIRES entry}
{\it Prepared for 3rd International Symposium on Quantum Theory and Symmetries (QTS3), Cincinnati, Ohio, 10-14 Sep 2003}

%\cite{Walcher:2003rh}
\bibitem{Walcher:2003rh}
  J.~Walcher,
  %``Quivers, dibaryons, and exceptional collections,''
%\href{http://www.slac.stanford.edu/spires/find/hep/www?irn=6477925}{SPIRES entry}
{\it Prepared for 3rd International Symposium on Quantum Theory and Symmetries (QTS3), Cincinnati, Ohio, 10-14 Sep 2003}

%\cite{Belhaj:2003qa}
\bibitem{Belhaj:2003qa}
  A.~Belhaj,
  %``On geometric engineering of N = 1 ADE quiver models,''
  {\tt hep-th/0310230}.
  %%CITATION = HEP-TH/0310230;%%

%\cite{Chiantese:2003qb}
\bibitem{Chiantese:2003qb}
  S.~Chiantese, A.~Klemm and I.~Runkel,
  %``Higher order loop equations for A(r) and D(r) quiver matrix models,''
  JHEP {\bf 0403} (2004) 033
  {\tt hep-th/0311258}.
  %%CITATION = JHEPA,0403,033;%%

%\cite{Dai:2003dy}
\bibitem{Dai:2003dy}
  J.~Dai and Y.~S.~Wu,
  %``Quiver matrix mechanics for IIB string theory. I: Wrapping membranes  and
  %emergent dimension,''
  Nucl.\ Phys.\  B {\bf 684} (2004) 75
  {\tt hep-th/0312028}.
  %%CITATION = NUPHA,B684,75;%%

%\cite{Belhaj:2004ws}
\bibitem{Belhaj:2004ws}
  A.~Belhaj and E.~H.~Saidi,
  %``Non simply laced quiver gauge theories in superstrings compactifications,''
  Afr.\ J.\ Math.\ Phys.\  {\bf 1} (2004) 29.
  %%CITATION = 00451,1,29;%%

%\cite{Frampton:2004xb}
\bibitem{Frampton:2004xb}
  P.~H.~Frampton and T.~Takahashi,
  %``Candidates for inflaton in quiver gauge theory,''
  Phys.\ Rev.\  D {\bf 70} (2004) 083530
  {\tt hep-ph/0402119}.
  %%CITATION = PHRVA,D70,083530;%%

%\cite{Dai:2004ke}
\bibitem{Dai:2004ke}
  J.~Dai and Y.~S.~Wu,
  %``Quiver matrix mechanics for IIB string theory. II: Generic dual tori,
  %fractional matrix membrane and SL(2,Z) duality,''
  Nucl.\ Phys.\  B {\bf 708} (2005) 72
  {\tt hep-th/0402201}.
  %%CITATION = NUPHA,B708,72;%%

%\cite{Ita:2004yn}
\bibitem{Ita:2004yn}
  H.~Ita, H.~Nieder, Y.~Oz and T.~Sakai,
  %``Topological B-model, matrix models, c-hat = 1 strings and quiver gauge
  %theories,''
  JHEP {\bf 0405} (2004) 058
  {\tt hep-th/0403256}.
  %%CITATION = JHEPA,0405,058;%%

%\cite{AhlLaamara:2004yz}
\bibitem{AhlLaamara:2004yz}
  R.~Ahl Laamara, M.~Ait Ben Haddou, A.~Belhaj, L.~B.~Drissi and E.~H.~Saidi,
  %``RG cascades in hyperbolic quiver gauge theories,''
  Nucl.\ Phys.\  B {\bf 702} (2004) 163
  {\tt hep-th/0405222}.
  %%CITATION = NUPHA,B702,163;%%

%\cite{Robles-Llana:2004dd}
\bibitem{Robles-Llana:2004dd}
  D.~Robles-Llana and M.~Rocek,
  %``Quivers, quotients, and duality,''
  {\tt hep-th/0405230}.
  %%CITATION = HEP-TH/0405230;%%

%\cite{DiNapoli:2004rm}
\bibitem{DiNapoli:2004rm}
  E.~Di Napoli, V.~S.~Kaplunovsky and J.~Sonnenschein,
  %``Chiral rings of deconstructive (SU(n(c)))**N quivers,''
  JHEP {\bf 0406} (2004) 060
  {\tt hep-th/0406122}.
  %%CITATION = JHEPA,0406,060;%%

%\cite{Halmagyi:2004ju}
\bibitem{Halmagyi:2004ju}
  N.~Halmagyi, C.~Romelsberger and N.~P.~Warner,
  %``Inherited duality and quiver gauge theory,''
  Adv.\ Theor.\ Math.\ Phys.\  {\bf 10} (2006) 159
  {\tt hep-th/0406143}.
  %%CITATION = 00203,10,159;%%

%\cite{Halmagyi:2004jy}
\bibitem{Halmagyi:2004jy}
  N.~Halmagyi, K.~Pilch, C.~Romelsberger and N.~P.~Warner,
  %``The complex geometry of holographic flows of quiver gauge theories,''
  JHEP {\bf 0609} (2006) 063
  {\tt hep-th/0406147}.
  %%CITATION = JHEPA,0609,063;%%

%\cite{Franco:2004wp}
\bibitem{Franco:2004wp}
  S.~Franco and A.~Hanany,
  %``On the fate of tachyonic quivers,''
  JHEP {\bf 0503} (2005) 031
  {\tt hep-th/0408016}.
  %%CITATION = JHEPA,0503,031;%%

%\cite{Fucito:2004gi}
\bibitem{Fucito:2004gi}
  F.~Fucito, J.~F.~Morales and R.~Poghossian,
  %``Instantons on quivers and orientifolds,''
  JHEP {\bf 0410} (2004) 037
  {\tt hep-th/0408090}.
  %%CITATION = JHEPA,0410,037;%%

%\cite{Robles-Llana:2004nq}
\bibitem{Robles-Llana:2004nq}
  D.~Robles-Llana,
  %``On N = 2 Seiberg duality for quiver theories,''
  {\tt hep-th/0411059}.
  %%CITATION = HEP-TH/0411059;%%

%\cite{Benvenuti:2004dw}
\bibitem{Benvenuti:2004dw}
  S.~Benvenuti and A.~Hanany,
  %``New results on superconformal quivers,''
  JHEP {\bf 0604} (2006) 032
  {\tt hep-th/0411262}.
  %%CITATION = JHEPA,0604,032;%%

%\cite{Benvenuti:2004dy}
\bibitem{Benvenuti:2004dy}
  S.~Benvenuti, S.~Franco, A.~Hanany, D.~Martelli and J.~Sparks,
  %``An infinite family of superconformal quiver gauge theories with
  %Sasaki-Einstein duals,''
  JHEP {\bf 0506} (2005) 064
  {hep-th/0411264}.
  %%CITATION = JHEPA,0506,064;%%

%\cite{Bertolini:2004xf}
\bibitem{Bertolini:2004xf}
  M.~Bertolini, F.~Bigazzi and A.~L.~Cotrone,
  %``New checks and subtleties for AdS/CFT and a-maximization,''
  JHEP {\bf 0412} (2004) 024
  [arXiv:hep-th/0411249].
  %%CITATION = JHEPA,0412,024;%%



%\cite{Benvenuti:2004wx}
\bibitem{Benvenuti:2004wx}
  S.~Benvenuti, A.~Hanany and P.~Kazakopoulos,
  %``The toric phases of the Y(p,q) quivers,''
  JHEP {\bf 0507} (2005) 021
  {\tt hep-th/0412279}.
  %%CITATION = JHEPA,0507,021;%%

%\cite{DiNapoli:2005ma}
\bibitem{DiNapoli:2005ma}
  E.~Di Napoli,
  ``Quiver gauge theories, chiral rings and random matrix models,''
  %%CITATION = UMI-32-03116;%%

%\cite{Billo:2005jw}
\bibitem{Billo:2005jw}
  M.~Billo, M.~Frau, F.~Lonegro and A.~Lerda,
  %``N = 1/2 quiver gauge theories from open strings with R-R fluxes,''
  JHEP {\bf 0505} (2005) 047
 {\tt hep-th/0502084}.
  %%CITATION = JHEPA,0505,047;%%

%\cite{Bergman:2005ba}
\bibitem{Bergman:2005ba}
  A.~Bergman,
  %``Undoing orbifold quivers,''
  JHEP {\bf 0703} (2007) 112
  {\tt hep-th/0502105}.
  %%CITATION = JHEPA,0703,112;%%

%\cite{Hanany:2005hq}
\bibitem{Hanany:2005hq}
  A.~Hanany, P.~Kazakopoulos and B.~Wecht,
  %``A new infinite class of quiver gauge theories,''
  JHEP {\bf 0508} (2005) 054
  {\tt hep-th/0503177}.
  %%CITATION = JHEPA,0508,054;%%

%\cite{Popov:2005ik}
\bibitem{Popov:2005ik}
  A.~D.~Popov and R.~J.~Szabo,
  %``Quiver gauge theory of nonabelian vortices and noncommutative  instantons
  %in higher dimensions,''
  J.\ Math.\ Phys.\  {\bf 47} (2006) 012306
  {\tt hep-th/0504025}.
  %%CITATION = JMAPA,47,012306;%%

%\cite{Franco:2005rj}
\bibitem{Franco:2005rj}
  S.~Franco, A.~Hanany, K.~D.~Kennaway, D.~Vegh and B.~Wecht,
  %``Brane dimers and quiver gauge theories,''
  JHEP {\bf 0601} (2006) 096
  {\tt hep-th/0504110}.
  %%CITATION = JHEPA,0601,096;%%

%\cite{Burrington:2005zd}
\bibitem{Burrington:2005zd}
  B.~A.~Burrington, J.~T.~Liu, M.~Mahato and L.~A.~Pando Zayas,
  %``Towards supergravity duals of chiral symmetry breaking in  Sasaki-Einstein
  %cascading quiver theories,''
  JHEP {\bf 0507} (2005) 019
  {\tt hep-th/0504155}.
  %%CITATION = JHEPA,0507,019;%%

%\cite{Bertolini:2005di}
\bibitem{Bertolini:2005di}
  M.~Bertolini, F.~Bigazzi and A.~L.~Cotrone,
  %``Supersymmetry breaking at the end of a cascade of Seiberg dualities,''
  Phys.\ Rev.\  D {\bf 72} (2005) 061902
  [arXiv:hep-th/0505055].
  %%CITATION = PHRVA,D72,061902;%%


%\cite{Benvenuti:2005ja}
\bibitem{Benvenuti:2005ja}
  S.~Benvenuti and M.~Kruczenski,
  %``From Sasaki-Einstein spaces to quivers via BPS geodesics: L(p,q|r),''
  JHEP {\bf 0604} (2006) 033
  {\tt hep-th/0505206}.
  %%CITATION = JHEPA,0604,033;%%

%\cite{Aspinwall:2005ur}
\bibitem{Aspinwall:2005ur}
  P.~S.~Aspinwall and L.~M.~Fidkowski,
  %``Superpotentials for quiver gauge theories,''
  JHEP {\bf 0610} (2006) 047
  {\tt hep-th/0506041}.
  %%CITATION = JHEPA,0610,047;%%

%\cite{Szendroi:2005ct}
\bibitem{Szendroi:2005ct}
  B.~Szendroi,
  %``Sheaves on fibered threefolds and quiver sheaves,''
  arXiv:math.ag/0506301.
  %%CITATION = MATH.AG/0506301;%%

%\cite{Takahashi:2005qu}
\bibitem{Takahashi:2005qu}
  A.~Takahashi,
  %``Matrix factorizations and representations of quivers. I,''
  arXiv:math.ag/0506347.
  %%CITATION = MATH.AG/0506347;%%

%\cite{Zhu:2005ki}
\bibitem{Zhu:2005ki}
  X.~y.~Zhu,
  %``Finite representations of a quiver arising from string theory,''
  arXiv:math.ag/0507316.
  %%CITATION = MATH.AG/0507316;%%

%\cite{Fujii:2005dk}
\bibitem{Fujii:2005dk}
  S.~Fujii and S.~Minabe,
  %``A Combinatorial Study on Quiver Varieties,''
  arXiv:math.ag/0510455.
  %%CITATION = MATH.AG/0510455;%%

%\cite{Hanany:2005ss}
\bibitem{Hanany:2005ss}
  A.~Hanany and D.~Vegh,
  %``Quivers, tilings, branes and rhombi,''
  {\tt hep-th/0511063}.
  %%CITATION = HEP-TH/0511063;%%

%\cite{Kajiura:2005yu}
\bibitem{Kajiura:2005yu}
  H.~Kajiura, K.~Saito and A.~Takahashi,
  %``Matrix Factorizations and Representations of Quivers II: type ADE case,''
  arXiv:math.ag/0511155.
  %%CITATION = MATH.AG/0511155;%%

%\cite{Feng:2005gw}
\bibitem{Feng:2005gw}
  B.~Feng, Y.~H.~He, K.~D.~Kennaway and C.~Vafa,
  %``Dimer models from mirror symmetry and quivering amoebae,''
  {\tt hep-th/0511287}.
  %%CITATION = HEP-TH/0511287;%%

%\cite{Wijnholt:2005mp}
\bibitem{Wijnholt:2005mp}
  M.~Wijnholt,
  %``Parameter space of quiver gauge theories,''
  {\tt hep-th/0512122}.
  %%CITATION = HEP-TH/0512122;%%

%\cite{Proudfoot:2005mz}
\bibitem{Proudfoot:2005mz}
  N.~J.~Proudfoot and A.~Bergman,
  %``Moduli spaces for Bondal quivers,''
  arXiv:math.ag/0512166.
  %%CITATION = MATH.AG/0512166;%%

%\cite{Israel:2005zp}
\bibitem{Israel:2005zp}
  D.~Israel,
  %``Non-critical string duals of N = 1 quiver theories,''
  JHEP {\bf 0604} (2006) 029
  {\tt hep-th/0512166}.
  %%CITATION = JHEPA,0604,029;%%

%\cite{Antebi:2005hr}
\bibitem{Antebi:2005hr}
  Y.~E.~Antebi, Y.~Nir and T.~Volansky,
  %``Solving flavor puzzles with quiver gauge theories,''
  Phys.\ Rev.\  D {\bf 73} (2006) 075009
  {\tt hep-ph/0512211}.
  %%CITATION = PHRVA,D73,075009;%%

%\cite{Butti:2005sw}
\bibitem{Butti:2005sw}
  A.~Butti, D.~Forcella and A.~Zaffaroni,
  %``The dual superconformal theory for L(p,q,r) manifolds,''
  JHEP {\bf 0509} (2005) 018
  [arXiv:hep-th/0505220].
  %%CITATION = JHEPA,0509,018;%%


%\cite{Butti:2005ps}
\bibitem{Butti:2005ps}
  A.~Butti and A.~Zaffaroni,
  %``From toric geometry to quiver gauge theory: The equivalence of
  %a-maximization and Z-minimization,''
  Fortsch.\ Phys.\  {\bf 54} (2006) 309
  {\tt hep-th/0512240}.
  %%CITATION = FPYKA,54,309;%%

%\cite{Nakayama:2005mf}
\bibitem{Nakayama:2005mf}
  Y.~Nakayama,
  %``Index for orbifold quiver gauge theories,''
  Phys.\ Lett.\  B {\bf 636} (2006) 132
  {\tt hep-th/0512280}.
  %%CITATION = PHLTA,B636,132;%%

%\cite{Intriligator:2005aw}
\bibitem{Intriligator:2005aw}
  K.~Intriligator and N.~Seiberg,
  %``The runaway quiver,''
  JHEP {\bf 0602} (2006) 031
  {\tt hep-th/0512347}.
  %%CITATION = JHEPA,0602,031;%%

%\cite{Belhaj:2006wh}
\bibitem{Belhaj:2006wh}
  A.~Belhaj, J.~Rasmussen, A.~Sebbar and M.~B.~Sedra,
  %``On ADE quiver models and F-theory compactification,''
  J.\ Phys.\ A  {\bf 39} (2006) 9339
  {\tt hep-th/0512321}.
  %%CITATION = JPAGB,A39,9339;%%
  
  \bibitem{moremodels4}
T.~W.~Kephart, C.~A.~Lee and Q.~Shafi,
  %``Family unification, exotic states and light magnetic monopoles,''
  JHEP {\bf 0701}, 088 (2007)
  [arXiv:hep-ph/0602055].

%\cite{Burrington:2006uu}
\bibitem{Burrington:2006uu}
  B.~A.~Burrington, J.~T.~Liu and L.~A.~Pando Zayas,
  %``Finite Heisenberg groups in quiver gauge theories,''
  Nucl.\ Phys.\  B {\bf 747} (2006) 436
  {\tt hep-th/0602094}.
  %%CITATION = NUPHA,B747,436;%%

\bibitem{DiNapoli:2006wz}
  E.~Di Napoli and P.~H.~Frampton, Phys. Lett. {\bf B638}, 374 (2006).
  {\tt hep-th/0603065}.

%\cite{Garcia-Etxebarria:2006aq}
\bibitem{Garcia-Etxebarria:2006aq}
  I.~Garcia-Etxebarria, F.~Saad and A.~M.~Uranga,
  %``Quiver gauge theories at resolved and deformed singularities using
  %dimers,''
  JHEP {\bf 0606} (2006) 055
  {\tt hep-th/0603108}.
  %%CITATION = JHEPA,0606,055;%%

%\cite{Burrington:2006aw}
\bibitem{Burrington:2006aw}
  B.~A.~Burrington, J.~T.~Liu and L.~A.~Pando Zayas,
  %``Central extensions of finite Heisenberg groups in cascading quiver gauge
  %theories,''
  Nucl.\ Phys.\  B {\bf 749} (2006) 245
  {\tt hep-th/0603114}.
  %%CITATION = NUPHA,B749,245;%%

%\cite{Lechtenfeld:2006wu}
\bibitem{Lechtenfeld:2006wu}
  O.~Lechtenfeld, A.~D.~Popov and R.~J.~Szabo,
  %``Rank two quiver gauge theory, graded connections and noncommutative
  %vortices,''
  JHEP {\bf 0609} (2006) 054
  {\tt hep-th/0603232}.
  %%CITATION = JHEPA,0609,054;%%

%\cite{Burrington:2006pu}
\bibitem{Burrington:2006pu}
  B.~A.~Burrington, J.~T.~Liu, M.~Mahato and L.~A.~Pando Zayas,
  %``Finite Heisenberg groups and Seiberg dualities in quiver gauge theories,''
  Nucl.\ Phys.\  B {\bf 757} (2006) 1
  {\tt hep-th/0604092}.
  %%CITATION = NUPHA,B757,1;%%

%\cite{Giedt:2006dd}
\bibitem{Giedt:2006dd}
  J.~Giedt,
  %``Quiver lattice supersymmetric matter, D1/D5 branes and AdS(3)/CFT(2),''
  {\tt hep-lat/0605004}.
  %%CITATION = HEP-LAT/0605004;%%

%\cite{Herzog:2006bu}
\bibitem{Herzog:2006bu}
  C.~P.~Herzog and R.~L.~Karp,
  %``On the geometry of quiver gauge theories: Stacking exceptional
  %collections,''
  {\tt hep-th/0605177}.
  %%CITATION = HEP-TH/0605177;%%

%\cite{Ueda:2006jn}
\bibitem{Ueda:2006jn}
  K.~Ueda and M.~Yamazaki,
  %``A Note on Brane Tilings and McKay Quivers,''
  arXiv:math.ag/0605780.
  %%CITATION = MATH.AG/0605780;%%

%\cite{Park:2006va}
\bibitem{Park:2006va}
  J.~Park and W.~Sim,
  %``Recursive relations for a quiver gauge theory,''
  JHEP {\bf 0610} (2006) 026
  {\tt hep-th/0607075}.
  %%CITATION = JHEPA,0610,026;%%

%\cite{Butti:2006nk}
\bibitem{Butti:2006nk}
  A.~Butti, D.~Forcella and A.~Zaffaroni,
  %``Deformations of conformal theories and non-toric quiver gauge theories,''
  JHEP {\bf 0702} (2007) 081
  {\tt hep-th/0607147}.
  %%CITATION = JHEPA,0702,081;%%

%\cite{Benvenuti:2006qr}
\bibitem{Benvenuti:2006qr}
  S.~Benvenuti, B.~Feng, A.~Hanany and Y.~H.~He,
  %``Counting BPS operators in gauge theories: Quivers, syzygies and
  %plethystics,''
  {\tt hep-th/0608050}.
  %%CITATION = HEP-TH/0608050;%%

%\cite{Florea:2006si}
\bibitem{Florea:2006si}
  B.~Florea, S.~Kachru, J.~McGreevy and N.~Saulina,
  %``Stringy instantons and quiver gauge theories,''
  {\tt hep-th/0610003}.
  %%CITATION = HEP-TH/0610003;%%

%\cite{Oota:2006eg}
\bibitem{Oota:2006eg}
  T.~Oota and Y.~Yasui,
  %``New example of infinite family of quiver gauge theories,''
  Nucl.\ Phys.\  B {\bf 762} (2007) 377
  {\tt hep-th/0610092}.
  %%CITATION = NUPHA,B762,377;%%

%\cite{Berenstein:2006pk}
\bibitem{Berenstein:2006pk}
  D.~Berenstein and S.~Pinansky,
  %``The minimal quiver standard model,''
  Phys.\ Rev.\  D {\bf 75} (2007) 095009
  {\tt hep-th/0610104}.
  %%CITATION = PHRVA,D75,095009;%%

%\cite{Volansky:2006wn}
\bibitem{Volansky:2006wn}
  T.~Volansky,
  %``Extracting flavor from quiver gauge theories,''
  AIP Conf.\ Proc.\  {\bf 903} (2007) 393
  {\tt hep-ph/0611018}.
  %%CITATION = APCPC,903,393;%%

%\cite{Zhu:2006va}
\bibitem{Zhu:2006va}
  X.~Zhu,
  %``Representations of N = ADE quivers via reflection functors,''
  Michigan Math.\ J.\  {\bf 54} (2006) 671.
  %%CITATION = MIMJA,54,671;%%


%\cite{Butti:2006au}
\bibitem{Butti:2006au}
  A.~Butti, D.~Forcella and A.~Zaffaroni,
  %``Counting BPS baryonic operators in CFTs with Sasaki-Einstein duals,''
  JHEP {\bf 0706} (2007) 069
  [arXiv:hep-th/0611229].
  %%CITATION = JHEPA,0706,069;%%


%\cite{AhlLaamara:2006zj}
\bibitem{AhlLaamara:2006zj}
  R.~Ahl Laamara, A.~Belhaj, L.~B.~Drissi and E.~H.~Saidi,
  %``Black holes in type IIA string on Calabi-Yau threefolds with affine ADE
  %geometries and q-deformed 2d quiver gauge theories,''
  {\tt hep-th/0611289}.
  %%CITATION = HEP-TH/0611289;%%

%\cite{Burrington:2007mj}
\bibitem{Burrington:2007mj}
  B.~A.~Burrington, J.~T.~Liu and L.~A.~P.~Zayas,
  %``Finite Heisenberg groups from nonabelian orbifold quiver gauge theories,''
  {\tt hep-th/0701028}.
  %%CITATION = HEP-TH/0701028;%%

%\cite{Diaconescu:2007ah}
\bibitem{Diaconescu:2007ah}
  D.~E.~P.~Diaconescu, R.~Donagi and B.~Florea,
  %``Metastable quivers in string compactifications,''
  {\tt hep-th/0701104}.
  %%CITATION = HEP-TH/0701104;%%
  
  
  %\cite{Forcella:2007wk}
\bibitem{Forcella:2007wk}
  D.~Forcella, A.~Hanany and A.~Zaffaroni,
  %``Baryonic generating functions,''
  arXiv:hep-th/0701236.
  %%CITATION = HEP-TH/0701236;%%

%\cite{Imamura:2007dc}
\bibitem{Imamura:2007dc}
  Y.~Imamura, H.~Isono, K.~Kimura and M.~Yamazaki,
  %``Exactly marginal deformations of quiver gauge theories as seen from   brane
  %tilings,''
  {\tt hep-th/0702049}.
  %%CITATION = HEP-TH/0702049;%%

%\cite{Antebi:2007xw}
\bibitem{Antebi:2007xw}
  Y.~E.~Antebi and T.~Volansky,
  %``Dynamical supersymmetry breaking from simple quivers,''
  {\tt hep-th/0703112}.
  %%CITATION = HEP-TH/0703112;%%

%\cite{Garcia-Etxebarria:2007vh}
\bibitem{Garcia-Etxebarria:2007vh}
  I.~Garcia-Etxebarria, F.~Saad and A.~M.~Uranga,
  %``Supersymmetry breaking metastable vacua in runaway quiver gauge theories,''
  arXiv:0704.0166 [hep-th].
  %%CITATION = ARXIV:0704.0166;%%

%\cite{Evslin:2007au}
\bibitem{Evslin:2007au}
  J.~Evslin, C.~Krishnan and S.~Kuperstein,
  %``Cascading Quivers from Decaying D-branes,''
  arXiv:0704.3484 [hep-th].
  %%CITATION = ARXIV:0704.3484;%%

%\cite{Jafferis:2007sg}
\bibitem{Jafferis:2007sg}
  D.~L.~Jafferis,
  %``Topological Quiver Matrix Models and Quantum Foam,''
  arXiv:0705.2250 [hep-th].
  %%CITATION = ARXIV:0705.2250;%%

%\cite{Butti:2007jv}
\bibitem{Butti:2007jv}
  A.~Butti, D.~Forcella, A.~Hanany, D.~Vegh and A.~Zaffaroni,
  %``Counting Chiral Operators in Quiver Gauge Theories,''
  arXiv:0705.2771 [hep-th].

\bibitem{Forcella:2007ps}
  D.~Forcella,
  %``BPS Partition Functions for Quiver Gauge Theories: Counting Fermionic
  %Operators,''
  arXiv:0705.2989 [hep-th].

%\cite{Kakushadze:2000mc}
\bibitem{Kakushadze:2000mc}  Z.~Kakushadze, 
%``Conformal brane world and cosmological constant,''
Phys.\ Lett.\ B {\bf 491}, 317 (2000) 
{\tt hep-th/0008041}.
  
%\cite{Dixon:1986jc}
\bibitem{Dixon:1986jc}
L.~J.~Dixon, J.~A.~Harvey, C.~Vafa and E.~Witten,  
%``Strings On Orbifolds. 2,''
Nucl.\ Phys.\  B {\bf 274}, 285 (1986).  

%\cite{Dixon:1985jw}
\bibitem{Dixon:1985jw}
L.~J.~Dixon, J.~A.~Harvey, C.~Vafa and E.~Witten,  
%``Strings On Orbifolds,''
Nucl.\ Phys.\  B {\bf 261}, 678 (1985).

%\cite{Kakushadze:1996iw}
\bibitem{Kakushadze:1996iw}
Z.~Kakushadze and S.~H.~H.~Tye,
%``Three-family SU(5) grand unification in string theory,''
Phys.\ Lett.\  B {\bf 392}, 335 (1997) {\tt hep-th/9609027}.

%\cite{Uranga:1998vf}
\bibitem{Uranga:1998vf}
A.~M.~Uranga,
  %``Brane configurations for branes at conifolds,''
JHEP {\bf 9901}, 022 (1999)  {\tt hep-th/9811004}.

%\cite{Oh:1999sk}
\bibitem{Oh:1999sk}
K.~Oh and R.~Tatar,
%``Branes at orbifolded conifold singularities and supersymmetric gauge  field  theories,''
JHEP {\bf 9910}, 031 (1999) {\tt hep-th/9906012}. 

%\cite{Klebanov:2000hb}
\bibitem{Klebanov:2000hb}
I.~R.~Klebanov and M.~J.~Strassler,
%``Supergravity and a confining gauge theory: Duality cascades and %chiSB-resolution 
of naked singularities,''  JHEP {\bf 0008}, 052 (2000)  {\tt hep-th/0007191}.

%\cite{Hebecker:2003jt}
\bibitem{Hebecker:2003jt}
A.~Hebecker and M.~Ratz,
%``Group-theoretical aspects of orbifold and conifold GUTs,''
Nucl.\ Phys.\  B {\bf 670}, 3 (2003).  
{\tt hep-ph/0306049}.   

%\cite{Benvenuti:2005wi}
\bibitem{Benvenuti:2005wi}
S.~Benvenuti and A.~Hanany,
%``Conformal manifolds for the conifold and other toric field theories,''
JHEP {\bf 0508}, 024 (2005). {\tt hep-th/0502043}.

%\cite{Kakushadze:1998yq}
\bibitem{Kakushadze:1998yq}
Z.~Kakushadze,  
%``Large N gauge theories from orientifolds with NS-NS B-flux,''
Nucl.\ Phys.\  B {\bf 544}, 265 (1999). {\tt hep-th/9808048}.

%\cite{Kakushadze:1998hb}
\bibitem{Kakushadze:1998hb}
Z.~Kakushadze,  
%``Anomaly free non-supersymmetric large N gauge theories from orientifolds,''
Phys.\ Rev.\  D {\bf 59}, 045007 (1999). 
{\tt hep-th/9806091}.

%\cite{Fayyazuddin:1998fb}
\bibitem{Fayyazuddin:1998fb}
A.~Fayyazuddin and M.~Spalinski,
%``Large N superconformal gauge theories and supergravity orientifolds,''
Nucl.\ Phys.\  B {\bf 535}, 219 (1998). {\tt hep-th/9805096}.

%\cite{Kakushadze:1998tz}
\bibitem{Kakushadze:1998tz}
Z.~Kakushadze,  
%``On large N gauge theories from orientifolds,''
Phys.\ Rev.\  D {\bf 58}, 106003 (1998). {\tt hep-th/9804184}. 

%\cite{Kakushadze:1998tr}
\bibitem{Kakushadze:1998tr}
Z.~Kakushadze,
%``Gauge theories from orientifolds and large N limit,''
Nucl.\ Phys.\  B {\bf 529}, 157 (1998). {\tt hep-th/9803214}.
%\cite{Forcella:2007ps}

%\cite{Kakushadze:2001bd}
\bibitem{Kakushadze:2001bd}
Z.~Kakushadze,
  %``Orientiworld,''
JHEP {\bf 0110}, 031 (2001). {\tt hep-th/0109054}.

%\cite{Lalak:1997ti}
\bibitem{Lalak:1997ti}
  Z.~Lalak, A.~Lukas and B.~A.~Ovrut,
  %``Soliton solutions of M-theory on an orbifold,''
  Phys.\ Lett.\  B {\bf 425}, 59 (1998)
  [arXiv:hep-th/9709214].
  %%CITATION = PHLTA,B425,59;%%

%\cite{Faux:1999hm}
\bibitem{Faux:1999hm}
  M.~Faux, D.~Lust and B.~A.~Ovrut,
  %``Intersecting orbifold planes and local anomaly cancellation in  M-theory,''
  Nucl.\ Phys.\  B {\bf 554}, 437 (1999)
  [arXiv:hep-th/9903028].
  %%CITATION = NUPHA,B554,437;%%

%\cite{Faux:2000dv}
\bibitem{Faux:2000dv}
  M.~Faux, D.~Lust and B.~A.~Ovrut,
  %``Local anomaly cancellation, M-theory orbifolds and phase-transitions,''
  Nucl.\ Phys.\  B {\bf 589}, 269 (2000)
  [arXiv:hep-th/0005251].
  %%CITATION = NUPHA,B589,269;%%

%\cite{Faux:2000sp}
\bibitem{Faux:2000sp}
  M.~Faux, D.~Lust and B.~A.~Ovrut,
  %``An M-theory perspective on heterotic K3 orbifold compactifications,''
  Int.\ J.\ Mod.\ Phys.\  A {\bf 18}, 3273 (2003)
  [arXiv:hep-th/0010087].
  %%CITATION = IMPAE,A18,3273;%%

%\cite{Faux:2000mr}
\bibitem{Faux:2000mr}
  M.~Faux, D.~Lust and B.~A.~Ovrut,
  %``Twisted sectors and Chern-Simons terms in M-theory orbifolds,''
  Int.\ J.\ Mod.\ Phys.\  A {\bf 18}, 2995 (2003)
  [arXiv:hep-th/0011031].
  %%CITATION = IMPAE,A18,2995;%%

%\cite{Doran:2001ve}
\bibitem{Doran:2001ve}
  C.~F.~Doran, M.~Faux and B.~A.~Ovrut,
  %``Four-dimensional N = 1 super Yang-Mills theory from an M theory
  %orbifold,''
  Adv.\ Theor.\ Math.\ Phys.\  {\bf 6}, 329 (2003)
  [arXiv:hep-th/0108078].
  %%CITATION = 00203,6,329;%%

 
%\cite{Aldazabal:2000sa}
\bibitem{Aldazabal:2000sa}
G.~Aldazabal, L.~E.~Ibanez, F.~Quevedo and A.~M.~Uranga,
%``D-branes at singularities: A bottom-up approach to the string  embedding of the standard model,''
JHEP {\bf 0008}, 002 (2000)
{\tt hep-th/0005067}.
%%CITATION = HEP-TH 0005067;%%

%\cite{He:2004rn}
\bibitem{He:2004rn}
Y.~H.~He,
%``Lectures on D-branes, gauge theories and Calabi-Yau singularities,''
{\tt hep-th/0408142}.

%\cite{Glashow:1984gc}
\bibitem{Glashow:1984gc}
S.L. Glashow, in Proceedings of the Fifth Workshop on Grand Unification,
Editors: K.Kang, H. Fried and P.H. Frampton, World Scientific (1984). pages 88-94.

\bibitem{PS}
J.C. Pati and A. Salam, Phys. Rev. {\bf D10,} 275 (1974).

%\cite{Blazek:2003wz}
\bibitem{Blazek:2003wz}
T.~Blazek, S.~F.~King and J.~K.~Parry,
%``Global analysis of a supersymmetric Pati-Salam model,''
JHEP {\bf 0305}, 016 (2003). {\tt hep-ph/0303192}.  

%\cite{Dent:2007eu}
\bibitem{Dent:2007eu}
J.~B.~Dent and T.~W.~Kephart,
%``Minimal Pati-Salam model from string theory unification,''
{\tt arXiv:0705.1995 [hep-ph]}.

%\cite{Babu:2003nw}
\bibitem{Babu:2003nw}  
K.~S.~Babu, E.~Ma and S.~Willenbrock,
%``Quark lepton quartification,''
Phys.\ Rev.\  D {\bf 69}, 051301 (2004)
{\tt hep-ph/0307380}.

%%%\cite{Chen:2004jz}
\bibitem{Chen:2004jz}
S.~L.~Chen and E.~Ma,
%``Exotic fermions and bosons in the quartification model,'
Mod.\ Phys.\ Lett.\  A {\bf 19}, 1267 (2004)
{\tt hep-ph/0403105}.

%%%\cite{Demaria:2005gk}
\bibitem{Demaria:2005gk}
A.~Demaria, C.~I.~Low and R.~R.~Volkas,
%``Unification via intermediate symmetry breaking scales with the quartification gauge group,''
Phys.\ Rev.\  D {\bf 72}, 075007 (2005)  
[Erratum-ibid.\  D {\bf 73}, 079902 (2006)]  
{\tt hep-ph/0508160}.

%%\cite{Demaria:2006uu}
\bibitem{Demaria:2006uu}  
A.~Demaria, C.~I.~Low and R.~R.~Volkas,
%``Neutrino masses in quartification schemes,''  
Phys.\ Rev.\  D {\bf 74}, 033005 (2006).
{\tt hep-ph/0603152}.

%\cite{Demaria:2006bd}
\bibitem{Demaria:2006bd}
A.~Demaria and K.~L.~McDonald,
%``Quartification on an orbifold,''
 Phys.\ Rev.\  D {\bf 75}, 056006 (2007)  {\tt hep-ph/0610346}.

%\cite{Babu:2007fx}
\bibitem{Babu:2007fx}
  K.~S.~Babu, T.~W.~Kephart and H.~Pas,
  %``Leptonic color models from Z_8 orbifolded AdS/CFT,''
  arXiv:0709.0765 [hep-ph].
  %%CITATION = ARXIV:0709.0765;%%

\bibitem{FK03}
P.H. Frampton and T.W. Kephart, Int. J. Mod. Phys. {\bf A19,} 593 (2004). {\tt hep-th/0306207}.

\bibitem{WS}
P.H. Frampton and W. F. Shively, Phys. Lett. {\bf B454,} 49 (1999).
{\tt hep-th/9902168}.

\bibitem{vafa}
P.H. Frampton and C. Vafa. {\tt hep-th/9903226}.

\bibitem{F2}
P.H. Frampton, Phys. Rev. {\bf D60,} 085004 (1999).
{\tt hep-th/9905042}.

\bibitem{F3}
P.H. Frampton, Phys. Rev. {\bf D60,} 121901 (1999). 
{\tt hep-th/9907051}.

\bibitem{nonabelian}
P.H. Frampton and T.W. Kephart, Phys. Lett. {\bf B485,} 403 (2000). {\tt hep-th/9912028}.

\bibitem{nonabelian1}
P.H. Frampton, J. Math. Phys. {\bf 42,} 2915 (2001).
{\tt hep-th/0011165}.

\bibitem{nonabelian2}
P.H. Frampton and T.W. Kephart, Phys. Rev. {\bf D64,} 086007 (2001). {\tt hep-th/0011186}.

\bibitem{nonabelian3}
P.H. Frampton, R.N. Mohapatra and S. Suh, Phys. Lett. {\bf B520,} 331 (2001).
{\tt hep-ph/0104211}.

\bibitem{KS}
S. Kachru and E. Silverstein, Phys. Rev. Lett. {\bf 80,} 4855 (1998).
{\tt hep-th/9802183}.

\bibitem{Kephart:2001qu}  T.~W.~Kephart and H.~P\"{a}s, 
%``Three family N = 1 SUSY models from Z(n) orbifolded AdS/CFT,''
Phys.\ Lett.\ B {\bf 522}, 315 (2001) 
{\tt hep-ph/0109111}. 

\bibitem{Unification}
P.H. Frampton, Mod. Phys. Lett. {\bf A18,} 1377 (2003). {\tt hep-ph/0208044}.

\bibitem{Unification2}
P.H. Frampton, R.M. Rohm and T. Takahashi, Phys. Lett. {\bf B570,} 67 (2003).
{\tt hep-ph/0302074}.

\bibitem{moremodels2}
P.H. Frampton and T.W. Kephart, Phys. Lett. {\bf B585,} 24 (2004). {\tt hep-th/0306053}.  

\bibitem{Lawrence}
A. Lawrence, N. Nekrasov and C. Vafa, Nucl. Phys. {\bf B533,} 199 (1998).
{\tt hep-th/9803015}.

 



\bibitem{moremodels}
T.W. Kephart and H. Paes, Phys. Lett. {\bf B522,} 315 (2001). {\tt hep-ph/0109111}.

%\cite{Frampton:2002st}
\bibitem{F2002}
P.~H.~Frampton,
%``Strong-electroweak unification at about 4-TeV,''
{\tt hep-ph/0208044}.

\bibitem{moremodels3}
T.W. Kephart and H. Paes, Phys. Rev. {\bf D70,} 086009 (2004). {\tt hep-ph/0402228}.

%\cite{Pickering:2001aq}
\bibitem{Pickering:2001aq}
A.~G.~Pickering, J.~A.~Gracey and D.~R.~Jones,
%``Three loop gauge beta-function for the most general single  gauge-coupling theory,''
Phys.\ Lett.\ B {\bf 510}, 347 (2001)
[Phys.\ Lett.\ B {\bf 512}, 230 (2001)]
{\tt hep-ph/0104247}.
%%CITATION = HEP-PH 0104247;%%

%\cite{Kachru:1998ys}
\bibitem{Kachru:1998ys}
S.~Kachru and E.~Silverstein,
%``4d conformal theories and strings on orbifolds,''
Phys.\ Rev.\ Lett.\  {\bf 80}, 4855 (1998)
{\tt hep-th/9802183}.

 
\bibitem{Kephart:2001iu}
T.~W.~Kephart and H.~P\"{a}s,
%``Muon anomalous magnetic moment in string inspired extended family  models,''
{\tt hep-ph/0102243}.
%%CITATION = HEP-PH 0102243;%%



\bibitem{FK}
P.H. Frampton and T.W. Kephart, Int. J. Mod. Phys. {\bf A10,} 4689
(1995).

\bibitem{books}
D.E. Littlewood, {it The Theory of Group Characters and Matrix
Representations
of Groups}
(Oxford 1940).

\bibitem{books2}
M. Hamermesh, {\it Group Theory and Its Applications to Physical
Problems}
(Addison-Wesley, 1962).

\bibitem{books3}
J.S. Lomont, {\it Applications of Finite Groups} (Academic, 1959),
reprinted by
Dover (1993).

\bibitem{books4}
A.D. Thomas and G.V. Wood, {\it Group Tables} (Shiva, 1980).

\bibitem{antoniadis}
I. Antoniadis, Phys. Lett. {\bf B246,} 377 (1990).

\bibitem{antoniadis2}
I. Antoniadis and K. Benakli, Phys. Lett. {\bf B326,} 69 (1994).

\bibitem{antoniadis3}
I. Antoniadis, K. Benakli and M. Quiros, Phys. Lett. {\bf B331,} 313 (1994).

\bibitem{antoniadis4}
J. D. Lykken, Phys. Rev. {\bf D54,} 3693 (1996).

\bibitem{antoniadis5}
I. Antoniadis, N. Arkani-Hamed, S. Dimopoulos and G. Dvali, Phys. Lett.
{\bf B436,} 257 (1998).

\bibitem{antoniadis6}
I. Antoniadis, S. Dimopoulos, A. Pomarol and M. Quiros, Nucl. Phys. {\bf
B544,} 503 (1999).

\bibitem{antoniadis7}
K. Dienes, E. Dudas and T. Gherghetta, Phys. Lett. {\bf B436,} 55 (1998);
Nucl. Phys. {\bf 537,} 47 (1999).

\bibitem{antoniadis8}
K. Dienes, E. Dudas and T. Gherghetta, Phys. Lett. {\bf B436,} 55 (1998).
Nucl. Phys. {\bf B543,} 387 (1999).

\bibitem{antoniadis9}
P.H. Frampton and A. Rasin, Phys. Lett. {\bf 460B,} 313 (1999).

\bibitem{antoniadis10}
D. Ghilencea and G.G. Ross, Phys. Lett. {\bf B442,} 165 (1998).

\bibitem{antoniadis11}
C.D. Carone, Phys. Lett. {\bf B454,} 70 (1999).

\bibitem{antoniadis12}
C. Bachas, JHEP {\bf 981,} 023 (1998).

\bibitem{antoniadis13}
G. Shiu and S.-H. H. Tye, Phys. Rev. {\bf D58,} 106007 	(1998).

\bibitem{antoniadis14}
Z. Kakushadze and S.H. Tye, Nucl. Phys. {\bf B548,} 180 (1999).

\bibitem{antoniadis15}
Z. Kakushadze, Nucl. Phys. {\bf B551,} 549 (1999).

\bibitem{antoniadis16}
N. Arkani-Hamed, S. Dimopoulos and G. Dvali, Phys. Lett {\bf 429,} 506 (1998).

\bibitem{Pak}
K.S. Babu, X.G. He and S. Pakvasa, Phys. Rev. {\bf D33,} 763 (1986).

\bibitem{quadratic}
X. Calmet, P.H. Frampton and R.M. Rohm, Phys. Rev. {\bf D72,} 055005 (2005).  {\tt hep-th/0412176}

\bibitem{global}
P.H. Frampton, Mod. Phys. Lett. {\bf A21,} 893 (2006).
{\tt hep-th/0511265}.

\bibitem{BSS}
L. Brink, J. Scherk and J.H. Schwarz, Nucl. Phys. {\bf B121,} 77 (1977).

\bibitem{Dienes}
K.R. Dienes, Nucl. Phys. {\bf B429,} 533 (1994).
{\tt hep-th/9402006}.

\bibitem{Nambu}
Y. Nambu, Phys. Lett. {\bf B26,} 626 (1968).

\bibitem{LCP}
P.H. Frampton, Mod. Phys. Lett. {\bf 22A,} 931 (2007). 
{\tt astro-ph/0607391}.

\bibitem{GG}
H. Georgi and S.L. Glashow, Phys. Rev. Lett. {\bf 32,} 438 (1974);

\bibitem{GG2}
H. Georgi, H.R. Quinn and S. Weinberg,
Phys. Rev. Lett. {\bf 33,} 451 (1974).

\bibitem{proton}
P.H. Frampton, Mod. Phys. Lett. {\bf 22A,} 347 (2007).
{\tt hep-ph/0610116}.

%\cite{Dent:2007ed}
\bibitem{Dent:2007ed}  
J.~B.~Dent and T.~W.~Kephart,
%``Proton Decay Constraints on Low Scale AdS/CFT Unification,''
{\tt  arXiv:0704.1451 [hep-ph]}, Phys. Rev {\bf D}, to appear.  

\bibitem{AdamsSilverstein}
A. Adams and E. Silverstein,
Phys. Rev. {\bf D64,} 086001 (2001).
{\tt hep-th/0103220}.

\bibitem{AdamsPolchinskiSilverstein}
A. Adams, J. Polchinski and E. Silverstein,
JHEP 0110:029 (2001). {\tt hep-th/0108075}.

\bibitem{Vafa1}
M. Bershadsky and A. Johansen,
Nucl Phys. {\bf B536,} 141 (1998).
{\tt hep-th/9803249}.

\bibitem{Vafa2}
M. Bershadsky, Z. Kakushadze and C. Vafa,
Nucl. Phys. {\bf B523,} 59 (1998).
{\tt hep-th/9803076}.

\bibitem{DKR1}
A. Dymarsky, I.R. Klebanov and R. Roiban,
JHEP 0508:011 (2005).
{\tt hep-th/0505099}.

\bibitem{DKR2}
A. Dymarsky, I.R. Klebanov and R. Roiban,
JHEP 0511:038 (2005).
{\tt hep-th/0509132}.

\end{thebibliography}
\end{document}